\documentclass[]{aa}
\usepackage{graphicx}
\usepackage{txfonts}
\usepackage{lscape}
\def\be{\begin{equation}}
\def\ee{\end{equation}}
\def\bea{\begin{eqnarray}}
\def\eea{\end{eqnarray}}
\def\btable{\begin{table}}
\def\etable{\end{table}}
\def\btabular{\begin{tabular}}
\def\etabular{\end{tabular}}
\def\bdoc{\begin{document}}
\def\edoc{\end{document}}
\def\bfig{\begin{figure}}
\def\efig{\end{figure}}
\def\ben{\begin{enumerate}}
\def\een{\end{enumerate}}
\def\bit{\begin{itemize}}
\def\eit{\end{itemize}}
\def\bitem{\begin{itemize}}
\def\eitem{\end{itemize}}
\def\bquote{\begin{quote}}
\def\equote{\end{quote}}

\def\bcenter{\begin{center}}
\def\ecenter{\end{center}}

\def\bsmall{\begin{small}}
\def\esmall{\end{small}}
\def\blarge{\begin{large}}
\def\elarge{\end{large}}
\def\bLarge{\begin{Large}}
\def\eLarge{\end{Large}}
\def\bhuge{\begin{huge}}
\def\ehuge{\end{huge}}
\def\bHuge{\begin{Huge}}
\def\eHuge{\end{Huge}}
\def\bfoot{\begin{footnotesize}}
\def\efoot{\end{footnotesize}}

\def\n{\noindent}
\def\bb{\left(}
\def\eb{\right)}
\def\bbs{\left[}
\def\ebs{\right]}
\def\bba{\left<}
\def\eba{\right>}
\def\bbsq{\left[}
\def\ebsq{\right]}
\def\bbcu{\left\{}
\def\ebcu{\right\}}
\def\bml{\left|}
\def\emr{\right|}

\def\etal{et al.\ }
\def\ie{i.\,e.\,, }
\def\eg{e.\,g.\,} 
\def\etc{etc.\ }
\def\viz{viz.\,, }

\def\gne #1#2{\ \vphantom{S}^{\raise-0.5pt\hbox{$\scriptstyle#1$}}_
{\raise0.5pt \hbox{$\scriptstyle#2$}}}
\def\gneq{\gne > \sim}
\def\lneq{\gne < \sim}

\def\mod #1{|#1|}
\def\vec#1{{\bf #1}}
\def\dotvec#1{\dot\vec{#1}}
\def\mvec#1{\mod{\vec{#1}}}
\def\mean #1{\bba #1 \eba}
\def\meanvsq#1{\mean{\vec{#1}^2}}
\def\dotp#1#2{\vec{#1}\cdot\vec{#2}}
\def\meandotp#1#2{\mean{\dotp{#1}{#2}}}
\def\crossp#1#2{\vec{#1}\times\vec{#2}}
\def\vecp{\vec{p}}
\def\vecpone{\vec{p}_{1}}
\def\vecn{\vec{n}}
\def\vecnone{\vec{n}_{1}}

\def\reci #1{\frac {1} {#1}}
\def\ten#1#2{#1{\times10^{#2}}}
\def\onlyten#1{10^{#1}}
\def\deg{\unit{deg}}
\def\degsq{\unit{deg^2}}
\def\ster{\unit{steradian}}
\def\perc{\unit{percent}}
\def\percent{\unit{percent}}

\def\deri#1#2{\frac {d#1} {d#2}}
\def\deriemp#1{\frac {d} {d#1}}
\def\pderi#1#2{\frac {\partial#1} {\partial#2}}
\def\pderiemp#1{\frac {\partial} {\partial#1}}
\def\secderi#1#2{\frac {d^2#1} {d#2^2}}
\def\secpderi#1#2{\frac {\partial^2#1} {\partial#2^2}}

\def\grad{\nabla}

\def\unit #1{\,{\rm #1}} 

\def\angst{\unit{\AA}}

\def\gm{\unit{gm}}
\def\gmi{\gm^{-1}}
\def\mg{\unit{mg}}
\def\kg{\unit{kg}}
\def\nm{\unit{nm}}
\def\mum{\unit{\mu m}}
\def\mumi{\unit{\mu\,m^{-1}}}
\def\mm{\unit{mm}}
\def\cm{\unit{cm}}
\def\mtr{\unit{m}}
\def\mtri{\unit{m^{-1}}}
\def\km{\unit{km}}

\def\pc{\unit{pc}}
\def\kpc{\unit{kpc}}
\def\mpc{\unit{Mpc}}
\def\gpc{\unit{Gpc}}

\def\cmi{\unit{cm^{-1}}}
\def\cmsq{\unit{cm^2}}
\def\cmsqi{\unit{cm^{-2}}}
\def\cmcu{\unit{cm^3}}
\def\cmcui{\unit{cm^{-3}}}
\def\kmi{\unit{km^{-1}}}
\def\pci{\unit{pc^{-1}}}
\def\kpci{\unit{kpc^{-1}}}
\def\mtrsq{\unit{m^2}}
\def\mtrsqi{\unit{m^{-2}}}
\def\mtrcu{\unit{m^3}}
\def\mtrcui{\unit{m^{-3}}}
\def\mpccu{\unit{cm^3}}
\def\mpccui{\unit{cm^{-3}}}
\def\gpccu{\unit{cm^3}}
\def\gpccui{\unit{Gpc^{-3}}}
\def\kpsi{\unit{km \, s^{-1}}}

\def\ksec{\unit{ksec}}
\def\msec{\unit{msec}}
\def\musec{\unit{\mu\,sec}}
\def\min{\unit{min}}
\def\hr{\unit{hr}}
\def\day{\unit{day}}
\def\days{\unit{days}}
\def\week{\unit{week}}
\def\yr{\unit{yr}}
\def\myr{\unit{Myr}}
\def\gyr{\unit{Gyr}}
\def\micron{\unit{\mu m}}
\def\delto{\Delta t_{\rm o}}

\def\seci{\unit{s^{-1}}}
\def\dayi{\unit{day^{-1}}}
\def\yri{\unit{yr^{-1}}}

\def\cms{\unit{cm\,s^{-1}}}
\def\mts{\unit{m\,s^{-1}}}
\def\mtss{\unit{m\,s^{-2}}}
\def\kms{\unit{km\,s^{-1}}}
\def\kmh{\unit{km\,h^{-1}}}

\def\mas{\unit{mas}}
\def\muas{\unit{{\mu}}as}
\def\masyr{\unit{mas\,yr^{-1}}}
\def\arcs{\unit{arcsec}}
\def\arcsec{\unit{arcsec}}
\def\arcsecsq{\unit{arcsec^2}}
\def\arcsecsqi{\unit{arcsec^{-2}}}
\def\arcmin{\unit{arcmin}}
\def\arcm{\unit{arcmin}}
\def\arcmsq{\unit{arcmin^2}}
\def\degsq{\unit{deg^2}}
\def\degsqi{\unit{deg^{-2}}}

\def\vlos{\vec{v}_{\rm los}}

\def\hz{\unit{Hz}}
\def\khz{\unit{kHz}}
\def\mhz{\unit{MHz}}
\def\ghz{\unit{GHz}}
\def\thz{\unit{Thz}}
\def\ev{\unit{eV}}
\def\kev{\unit{keV}}
\def\mev{\unit{MeV }}
\def\gev{\unit{GeV }}
\def\tev{\unit{TeV }}

\def\kel{\unit{K}}

\def\newt{\unit{N}}

\def\funit{\rm\,erg\,cm^{-2}\,sec^{-1}}
\def\pfunit{\unit{cm^{-2}\,sec^{-1}}}
\def\funith{\rm\,erg\,cm^{-2}\,sec^{-1}\,Hz^{-1}}
\def\funitb{\rm\,erg\,cm^{-2}\,sec^{-1}\ster^{-1}}
\def\funitbh{\rm\,erg\,cm^{-2}\,sec^{-1}\Hz^{-1}\ster^{-1}}
\def\funitba{\rm\,erg\,cm^{-2}\,sec^{-1}\AA^{-1}\ster^{-1}}
\def\punitb{\rm\,photons\,cm^{-2}\,sec^{-1}\ster^{-1}}
\def\lunit{\rm\,erg\,sec^{-1}}
\def\lunith{\rm\,erg\,sec^{-1}\,Hz^{-1}}
\def\whz{\unit{W\,Hz^{-1}}}
\def\whzs{\unit{W\,Hz^{-1}\,str^{-1}}}
\def\wmtrsqi{\unit{W\,m^{-2}}}
\def\jy{\unit{Jy}}
\def\mjy{\unit{mJy}}
\def\mujy{\unit{\mu Jy}}
\def\jyu{\unit{W\,m^{-2}\,Hz}}
\def\erg{\unit{erg}}
\def\ergs{\unit{erg\,s^{-1}}}
\def\ergsh{\unit{erg\,s^{-1}\,Hz^{-1}}}
\def\mcrab{\unit{milliCrab}}
\def\ergsc{\unit{erg\,s^{-1}\,cm^{-3}}}
\def\ergcmcu{\unit{erg\,cm^{-3}}}
\def\ergscmcu{\unit{erg\,s^{-1}\,cm^{-3}}} 
\def\ergcmt{\unit{erg\,cm^{-3}}}
\def\ergcmcui{\unit{erg\,cm^{-3}}}

\def\gmcmcui{\gm\cmcui}
\def\kgmtrcui{\kg\mtrcui}
 
\def\melec{m_{\rm e}}
\def\mectwo{\melec c^2}
\def\mprot{m_{\rm p}}
\def\nelec{n_{\rm e}}
\def\nion{n_{\rm i}}
\def\nprot{n_{\rm p}}
\def\pelec{p_{\rm e}}
\def\pfermi{p_{\rm f}}
\def\efermi{E_{\rm f}}
\def\mue{\mu_{\rm e}}

\def\gntff{\overline{g}_{\rm ff}}

\def\eneutrino{\nu_{\rm e}}
\def\eaneutrino{\overline{\nu}_{\rm e}}
\def\mneutrino{\nu_{\mu}}
\def\maneutrino{\\overline{nu}_{\mu}}
\def\tneutrino{\nu_{\tau}}
\def\maneutrino{\\overline{nu}_{\tau}}

\def\fine{\alpha_{\rm f}}
\def\comptw{\lambda_{\rm e}}

\def\threehe{^3{\rm He}}
\def\fourhe{^4{\rm He}}
\def\sixli{^6{\rm Li}}
\def\sevenli{^7{\rm Li}}
\def\ninebe{^9{\rm Be}}
\def\tenb{^{10}{\rm B}}
\def\elevenb{^{11}{\rm B}}
\def\twelvec{^{12}{\rm C}}
\def\sixteeno{^{16}{\rm O}}

\def\fourhebyh{\fourhe/{\rm H}}
\def\threehebyh{\threehe/{\rm H}}
\def\sevenlibyh{^7{\rm Li}/{\rm H}}
\def\bsqfh{{\rm B^2FH}}

\def\mag{\unit{magnitude}}
\def\mags{\unit{magnitudes}}
\def\absm#1{M_{\rm#1}}
\def\appm#1{m_{\rm#1}}
\def\Delm#1{\Delta m_{#1}}
\def\magarcsecsqi{\unit{mag\arcsecsqi}}

\def\lsol{L_{\odot}}
\def\msol{M_{\odot}}
\def\rsol{R_{\odot}}
\def\zsol{Z_{\odot}}

\def\aunit{\unit{AU}}

\def\mear{M_{\oplus}}
\def\rear{R_{\oplus}}
\def\aear{a_{\oplus}}
\def\pear{P_{\oplus}}

\def\mjup{M_{\rm Jupiter}}
\def\ajup{a_{\rm Jupiter}}
\def\pjup{P_{\rm Jupiter}}
\def\msat{M_{\rm Saturn}}

\def\amer{a_{\rm Mercury}}
\def\pmer{P_{\rm Mercury}}

\def\aplu{a_{\rm Pluto}}
\def\pplu{P_{\rm Pluto}}

\def\meight{\bb\frac{M}{\onlyten{8}\msol}\eb}

\def\mbymsol{\bb\frac{M}{\msol}\eb}

\def\aop{\alpha_{\rm op}}
\def\ar{\alpha_{\rm R}}
\def\ax{\alpha_{\rm x}}
\def\ag{\alpha_{\gamma}}
\def\aire{\alpha_{2.2-25\micron}}
\def\asm{\alpha_{\rm sm}}
\def\onemal{1-\alpha}
\def\nuone{\nu_1}
\def\nutwo{\nu_2}
\def\nuz{\nu_0}
\def\nug{\nu_g}
\def\nuc{\nu_c}
\def\nua{\nu_a}
\def\nup{\nu_p}
\def\nux{\nu_{\rm x}}
\def\nuop{\nu_{\rm op}}
\def\nupz{\nu_{p0}}
\def\nuaop{\nu^{-\alpha_{op}}}
\def\nualpha{\nu^{-\alpha}}
\def\alnu{\alpha_{\nu}}
\def\alnua{\alpha_{\nu_a}}
\def\alnusc{\alnu^{\rm sc}}
\def\jnu{j_{\nu}}
\def\jnua{j_{\nu_a}}
\def\tstar{\tau_{*}}
\def\tnu{\tau_{\nu}}
\def\etnu{e^{\tnu}}
\def\emtnu{e^{-\tnu}}
\def\fnu{F(\nu)}
\def\fnua{F(\nua)}
\def\fr{F_{\rm R}}
\def\fop{F_{\rm op}}
\def\fopl{F_{\rm op}^l}
\def\fx{F_{\rm x}}
\def\fxl{F_{\rm x}^{l}}
\def\frlim{F_{R,lim}}
\def\mv{m_V}
\def\mb{m_B}
\def\mr{m_R}
\def\Mv{M_V}
\def\Mb{M_B}
\def\Mr{M_R}

\def\bnu{B(\nu)}
\def\blnu{B_{\nu}}
\def\inu{I(\nu)}
\def\ilnu{I_{\nu}}
\def\snu{S(\nu)}
\def\slnu{S_{\nu}}
\def\Snu{S(\nu)}
\def\unu{u(\nu)}
\def\ulnu{u_{\nu}}
\def\Snua{S(\nu_a)}
\def\nnu{n(\nu)}
\def\nlnu{n_{\nu}}
\def\nomega{n(\omega)}
\def\nlomega{n_{\omega}}
\def\taunu{\tau_{\nu}}
\def\etaunu{e^{\taunu}}
\def\emtaunu{e^{-\taunu}}
\def\exptaunu{\exp(\taunu)}
\def\expmtaunu{\exp(-\taunu)}
\def\planckinu{\frac{2h\nu^3/c^2}{\exp(h\nu/kT)-1}}
\def\planckunu{\frac{8\pi h\nu^3/c^3}{\exp(h\nu/kT)-1}}
\def\planckilambda{\frac{2hc^2/\lambda^5}{\exp(hc/\lambda kT)-1}}
\def\planckulambda{\frac{8\pi hc/\lambda^5}{\exp(hc/\lambda kT)-1}} 

\def\plisnu{\frac{2h\nu^3/c^2}{\exp(h\nu/kT)-1}}
\def\plusnu{\frac{8\pi h\nu^3/c^3}{\exp(h\nu/kT)-1}}
\def\plislambda{\frac{2hc^2/\lambda^5}{\exp(hc/\lambda kT)-1}}
\def\pluslambda{\frac{8\pi hc/\lambda^5}{\exp(hc/\lambda kT)-1}} 
\def\plnut{\frac{2h\nu^3/c^2}{\exp(h\nu/kT)-1}}
\def\rjnut{\frac{2kT\nu^2}{c^2}}

\def\Pnu{P(\nu)}
\def\lnu{L(\nu)}
\def\llnu{\log \lnu}
\def\lr{L_{\rm R}}
\def\llr{\log L_{\rm R}}
\def\lrlim{L_{R,lim}}
\def\lfive{L(5\ghz)}
\def\lop{L_{{\rm op}}}
\def\llop{\log L_{{\rm op}}}
\def\lx{L_{\rm x}}
\def\llx{\log L_{\rm x}}
\def\lxmin{L_{\rm x,min}}
\def\lxmax{L_{\rm x,max}}
\def\lmax{L_{|rm max}}
\def\lir{L_{\rm IR}}
\def\luv{L_{\rm UV}} 
\def\lrc{L_{\rm rc}}
\def\llrc{\log L_{\rm rc}}
\def\lrext{L_{\rm \rm r,ext}}
\def\llrext{\log L_{r,ext}}
\def\lre{L_{\rm re}}
\def\llre{\log L_{\rm re}}
\def\lxb{L_{\rm xb}}
\def\llxb{\log L_{\rm xb}}
\def\lxu{L_{\rm xu}}
\def\llxu{\log L_{\rm xu}}
\def\lre{L_{\rm re}}
\def\llre{\log L_{\rm re}}
\def\rt{R_{\rm t}}
\def\rx{R_{\rm x}}
\def\rxbu{R_{\rm xbu}}
\def\rtx{R_{\rm tx}}
\def\gxbp{g_{\rm x}(\beta, \psi)}
\def\grbp{g_{\rm r}(\beta, \psi)}
\def\aox{\alpha_{\rm ox}}
\def\aoxl{\alpha_{\rm ox}^l}
\def\aro{\alpha_{\rm ro}}
\def\arx{\alpha_{\rm rx}}
\def\aoxx{\alpha_{\rm oxx}}
\def\auv{\alpha_{\rm uv}}
\def\sigt{\sigma_{\rm T}}
\def\sigkn{\sigma_{\rm KN}}
\def\signu{\sigma_{\nu}}
\def\siggg{\sigma_{\gamma\gamma}}
\def\tausc{\tau_{\rm sc}}
\def\taut{\tau_{\rm T}}
\def\nsc{N_{\rm sc}}
\def\freenu{l_{\nu}}
\def\U #1{U_{\rm #1}}
\def\eps{\epsilon}
\def\epsone{\epsilon_1}
\def\epstwo{\epsilon_2}
\def\epsx{\epsilon_{\rm x}}
\def\epsxs{\eps_{\rm xs}}
\def\epss{\epsilon_{\rm s}}
\def\epsg{\epsilon_{\gamma}}
\def\epsnu{\epsilon_{\nu}}
\def\epsir{\eps_{\rm IR}}
\def\epsi{\eps_{\rm i}}
\def\epsf{\eps_{\rm f}}
\def\DOM{\Delta\Omega}
\def\omegaone{\omega_{1}}

\def\hnu{h\nu}
\def\hnubyc{\frac{h\nu}{c}}
\def\hbarom{\hbar\omega}

\def\lcapc{L_{\rm C}}
\def\lcaps{L_{\rm S}}
\def\lcapi{L{\rm i}}
\def\lcaph{L_{\rm h}}
\def\dtvar{\Delta t_{\rm var}}
\def\tbri{T_{\rm b}}
\def\tsri{T_{\rm i}}
\def\Tvar{T_{\rm var}}
\def\dmv{dm_{\rm V}}
\def\betaa{\beta_{\rm a}}
\def\vela{v_{\rm a}}
\def\te{t_{\rm e}}
\def\dte{dt_{\rm e}}
\def\gammap{\gamma_{\rm p}}
\def\gammab{\gamma_{\rm b}}
\def\gammai{\gamma{\rm i}}
\def\gammaj{\gamma{\rm j}}
\def\gammac{\gamma{\rm c}}

\def\vbyvm{\frac {V} {V_m}}
\def\vbyvmp{\frac {V'} {V_m'}}
\def\mvbyvm{\bba\vbyvm\eba}
\def\mvbyvmp{\bba\vbyvmp\eba}

\def\vbyvml{V/V_m}
\def\vbyvmpl{V'/V_m'}
\def\mvbyvml{\bba\vbyvml\eba}
\def\mvbyvmpl{\bba\vbyvmpl\eba}

\def\bperp{B_{\perp}}
\def\bpar{B_{\parallel}}
\def\bperppar{B_{\perp-\parallel}}
\def\dtvar{\Delta t_{var}}
\def\Tvar{T_{var}}
\def\dmv{dm_v}
\def\betaa{\beta_a}
\def\gammap{\gamma_{\rm p}}
\def\gammab{\gamma_{\rm b}}
\def\gammamax{\gamma_{\rm max}}
\def\scrid{{\cal D}}
\def\scril{{\cal L}}
\def\scrio{{\cal O}}
\def\scrir{{\cal R}}
\def\fbeam{F_b}
\def\fext{F_{ext}}
\def\fin{F_{\rm in}}
\def\fout{F_{\rm out}}
\def\bcospsi{\beta\cos\psi}
\def\bsinpsi{\beta\sin\psi}
\def\rnu{R_{\nu}}
\def\rnuone{R_{\nu_1}}
\def\rnutwo{R_{\nu_2}}

\def\nh{N_{\rm H}}
\def\mh{M_{\rm H}}
\def\eax{E^{-\ax}}
\def\ex{E_{\rm x}}
\def\ex{E_{\rm xs}}

\def\zmin{z_{\rm min}}
\def\zmax{z_{\rm max}}
\def\rmin{r_{\rm min}}
\def\rmax{r_{\rm max}}
\def\rsc{r_{\rm sc}}
\def\rscm{r_{\rm sc,m}}

\def\corrxz{r_{\rm x,z}}
\def\corrxop{r_{\rm x,op}}
\def\corropz{r_{\rm op,z}}
\def\corrxopz{r_{\rm x,op;z}}
\def\chisq{\chi^2}
\def\chisqr{\chi^{2}_{\nu}}

\def\philx{\Phi{(\lx)}}
\def\phixstar{\Phi_{X}^{*}}
\def\phixstaro{\Phi_{X1}^{*}}
\def\lxstar{\lx^{*}}
\def\lxfour{L_{{\rm x},44}}
\def\lxstarfour{L_{{\rm x},44}^*(0)}

\def\asca{{\em ASCA }}
\def\axaf{{\em AXAF }}
\def\ein{{\em EINSTEIN }}
\def\exosat{{\em EXOSAT }}
\def\ginga{{\em GINGA }}
\def\rosat{{\em ROSAT }}
\def\compton{{\em COMPTON }}
\def\granat{{\em GRANAT }}
\def\iras{{\em IRAS }}
\def\iue{{\em IUE }}

\def\seyone{Seyfert\,1 }
\def\seytwo{Seyfert\,2 }
\def\bllac{BL Lac }
\def\bllacs{BL Lacs }
\def\rbl{{\rm RBL }}
\def\xbl{{\rm XBL }}
\def\ipc{{\rm IPC }}
\def\pspc{{\rm PSPC }}

\def\cfour{{\rm C\,IV }}
\def\cthreesf{{\rm C\,III] }}
\def\fetwo{{\rm Fe\,II }}
\def\halpha{{\rm H\,\alpha }}
\def\hbeta{{\rm H\,\beta }}
\def\hgamma{{\rm H\,\gamma}}
\def\hi{{\rm H\,I }}
\def\lyal{{\rm Ly\,\alpha }}
\def\mgtwo{{\rm Mg\,II }}
\def\ntwo{{\rm [N\,II]}}
\def\nfive{{\rm N\,V }}
\def\otwo{{\rm [O\,II] }}
\def\othree{{\rm [O\,III] }}
\def\kalpha{K_{\alpha} }
\def\ykz{Y^{K}_{Z} }

\def\const{{\rm constant}}
\def\gray{\gamma-{\rm ray}}
\def\grays{\gamma-{\rm rays}}
\def\maxtau{\tausc(1+\tausc)}
\def\maxtnu{\tnu(1+\tnu)}
\def\dm{{\rm DM}}
\def\dmsinmb{\dm\sin\mod{b}}
\def\dmsinb{\dm\sin b}
\def\sbyn{\frac{S}{N}}

\def\noi{\noindent}
\def\vf{\vspace{0.5cm}}
\def\vff{\vspace{1cm}}
\def\pbr{\pagebreak}
\def\pnew{\newpage}
\def\pemp{\pagestyle{empty}}
\def\tpemp{\thispagestyle{empty}}

\newcommand{\onlytitle}[1]
{\title{#1}
\author{ }
\date{ }
\maketitle\tpemp }

\newcommand{\onlyfig}[2]
{\begin{center}\epsfig{figure=#1.ps, width=#2in}\end{center}}

\newcommand{\onlyfige}[2]
{\begin{center}\epsfig{figure=#1.eps, width=#2in}\end{center}}

\def\vbyc{\frac{v}{c}}
\def\vbycsq{\frac{v^2}{c^2}}
\def\omvbycsq{\sqrt{1-\vbycsq}}

\def\gmr{\frac{GM}{r}}
\def\gmcsr{\frac{GM}{c^2r}}
\def\gmcsrl{GM/c^2r}
\def\tgmcsr{\frac{2GM}{c^2r}}
\def\schwarz{r_{\rm S}}
\def\tschwarz{t_{\rm S}}
\def\rergo{r_{\rm E}}
\def\rplus{r_{+}}
\def\mbh{M_{\rm BH}}
\def\mbyl{M/L}

\def\lh{l_{\rm h}}
\def\li{l_{\rm i}}
\def\ls{l_{\rm s}}

\def\omk{\Omega_{\rm K}(r)}
\def\velr{v_{r}}
\def\velp{v_{\phi}}
\def\velz{v_{z}}
\def\sigrt{\Sigma(r,t)}
\def\mach{{\cal M}}
\def\scrir{{\cal R}}
\def\tin{t_{\rm in}}
\def\tl{t_{\rm L}}
\def\ldisk{L_{\rm disk}}
\def\temps{T_{\rm s}}
\def\tempc{T_{\rm c}}
\def\tempb{T_{\rm B}}
\def\bnu{B_{\nu}}
\def\bnuts{B_{\nu}(\temps)}
\def\bnutsr{B_{\nu}(\temps(r))}

\def\rout{r_{\rm out}}
\def\tvis{t_{\rm vis}}
\def\pgas{p_{\rm g}}
\def\prad{p_{\rm r}}
\def\csou{c_{\rm s}}
\def\rstar{r_{*}}
\def\ledd{L_{\rm Edd}}
\def\mdedd{\dot{M}_{\rm Edd}}
\def\tedd{t_{\rm Edd}}
\def\tempedd{T_{\rm Edd}}
\def\bedd{B_{\rm Edd}}
\def\mdot{{\dot M}}
\def\geff{\vec{g}_{\rm eff}}
\def\phir{\Phi_{\rm rot}}
\def\phie{\Phi_{\rm eff}}
\def\zsr{z_{\rm s}(r)}
\def\udzero{u_0}
\def\udphi{u_{\phi}}
\def\uuzero{u^0}
\def\uuphi{u^{\phi}}

\def\isnu{I_{\nu}}
\def\islambda{I_{\lambda}}
\def\fsnu{F_{\nu}}
\def\fslambda{F_{\lambda}}

\def\plisnu{\frac{2h\nu^3/c^2}{\exp(h\nu/kT)-1}}
\def\plusnu{\frac{8\pi h\nu^3/c^3}{\exp(h\nu/kT)-1}}
\def\plislambda{\frac{2hc^2/\lambda^5}{\exp(hc/\lambda kT)-1}}
\def\pluslambda{\frac{8\pi hc/\lambda^5}{\exp(hc/\lambda kT)-1}} 

\def\isnurj{\isnu^{\rm RJ}}
\def\isnuw{\isnu^{\rm W}}

\def\cv{C_V}
\def\cp{C_p}
\def\dqdtv{\bb\pderi{Q}{T}\eb_V}
\def\dqdtp{\bb\pderi{Q}{T}\eb_p}

\def\CH{{\it CH}}

\def\hub{H_0}
\def\hubi{H^{-1}_0}
\def\hubble#1{H_0={#1}\unit{km\,sec^{-1}\,Mpc^{-1}}}
\def\hubbleunit{\unit{km\,sec^{-1}\,Mpc^{-1}}}
\def\hubhund{h_{100}}
\def\hubhundcu{\hubhund^3}
\def\hubhundcui{\hubhund^{-3}}
\def\hubhundi{\hubhund^{-1}}
\def\hubhundis{\hubhund^{-2}}
\def\cuponh{\bb \frac {c} {H_0} \eb}

\def\qz{q_0}
\def\qzsq{\qz^2}
\def\fqz{\bbcu\reci{\qzsq}\bbsq \qz z + (q_0-1)(\sqrt{1+2q_0z}-1)
\ebsq\ebcu}
\def\fqzno{\qz z + (q_0-1)(\sqrt{1+2q_0z}-1)}
\def\omb{\Omega_{{\rm b}}}
\def\omhi{\Omega_{{\rm H\,I}}}
\def\omlls{\Omega_{{\rm LLS}}}
\def\omlb{\Omega_{{\rm lb}}}
\def\omigm{\Omega_{{\rm IGM}}}
\def\omdlyal{\Omega_{{\rm DLy\,\alpha}}}
\def\zpr{z^{\prime}}
\def\zabs{z_{{\rm abs}}}
\def\zem{z_{{\rm em}}}
\def\zlim{z_{\rm lim}}
\def\zlimg{z_{\rm lim,G}}
\def\zlimgq{z_{\rm lim,GQ}}
\def\dlum{D_{\rm L}}
\def\dlumsq{D^2_{\rm L}}
\def\dmet{d_{\rm m}}
\def\dang{d_{\rm A}}

\def\lamo#1{\lambda_{\rm o_{#1}}}
\def\lame#1{\lambda_{\rm e_{#1}}}

\def\re{r_{\rm e}}
\def\rbyre{\frac{r}{\re}}
\def\rbyredev{\bb\rbyre\eb^{1/4}}
\def\rbyren{\bb\rbyre\eb^{1/n}}
\def\rbyrel{r/\re}
\def\rbyredevl{(\rbyrel)^{1/4}}
\def\rbyrenl{(\rbyrel)^{1/n}}
\def\rebyrd{\frac{\re}{\rd}}
\def\rebyrdl{\re/\rd}

\def\izero{I(0)}
\def\ire{I(\re)}
\def\ir{I(r)}
\def\icr{I_{\rm c}(r)}

\def\muzero{\mu(0)}
\def\mure{\mu(\re)}
\def\mur{\mu(r)}

\def\ibzero{I_{\rm b}(0)}
\def\ibre{I_{\rm b}(\re)}
\def\mibre{\mean{\ibre}}
\def\iblre{I_{\rm b}(<\re)}
\def\ibr{I_{\rm b}(r)}
\def\iblr{I_{\rm b}(<r)}
\def\ibtot{I_{\rm b,tot}}

\def\mubzero{\mu_{\rm b}(0)}
\def\mubre{\mu_{\rm b}(<\re)}
\def\mmubre{\mean{\mubre}}
\def\mubr{\mu_{\rm b}(r)}
\def\mulim{\mu_{\rm lim}}

\def\mudzero{\mu_{\rm d}(0)}
\def\mugzero{\mu_{\rm G}(0)}

\def\rd{r_{\rm d}}
\def\rbyrd{\frac{r}{\rd}}
\def\rbyrdl{r/\rd}
\def\idzero{I_{\rm d}(0)}
\def\idr{I_{\rm d}(r)}
\def\idlr{I_{\rm d}(<r)}
\def\idtot{I_{\rm d,tot}}

\def\imax{I_{\rm max}}
\def\imin{I_{\rm min}}

\def\dev{\onlyten{-3.33\rbyredevl}}
\def\devn{\onlyten{-\cn\rbyren}}
\def\exprbyrd{e^{-(\rbyrdl)}}
\def\cn{c_n}

\def\dbyb{\frac{D}{B}}
\def\bbyd{\frac{B}{D}}
\def\dbybl{D/B}
\def\bbydl{B/D}

\def\absgal{M_{\rm G}}
\def\absbul{M_{\rm B}}
\def\absdisk{M_{\rm d}}
\def\absq{M_{\rm Q}}

\def\lstar{L_{*}}

\def\mbyl{\bb\frac{M}{L}\eb}

\def\labchap #1{\label{chap:#1}}
\def\labequn #1{\label{eq:#1}}
\def\labfig #1{\label{fig:#1}}
\def\labsecn #1{\label{sec:#1}}
\def\labsubsecn #1{\label{subsecn:#1}}
\def\labsubsubsecn #1{\label{subsubsecn:#1}}
\def\labtablem #1{\label{tab:#1}}
\def\labapp #1{\label{app:#1}}
\def\labboxl #1{\label{box:#1}}

\def\boxl #1{box~\ref{box:#1}}
\def\chap #1{Chapter~\ref{chap:#1}}
\def\equn #1{Equation~\ref{eq:#1}}
\def\fig #1{Figure~\ref{fig:#1}}
\def\relation #1{relation~\ref{eq:#1}}
\def\secn #1{Section~\ref{sec:#1}}
\def\subsecn #1{Section~\ref{subsecn:#1}}
\def\subsubsecn #1{Section~\ref{subsubsecn:#1}}
\def\app#1{Appendix~\ref{app:#1}}
\def\tablem #1{Table~\ref{tab:#1}}

\def\dequn#1#2{Equations~{\ref{eq:#1}}~and~{\ref{eq:#2}}}
\def\tequn#1#2#3{Equations~{\ref{eq:#1}},~{\ref{eq:#2}} and ~{\ref{eq:#3}}}
\def\mequn#1#2{Equations~{\ref{eq:#1}}~to~{\ref{eq:#2}}}
\def\dfig#1#2{Figures~{\ref{fig:#1}}~and~{\ref{fig:#2}}}
\def\mfig#1#2{Figures~{\ref{fig:#1}}~to~{\ref{fig:#2}}}
\def\dtablem#1#2{Tables~{\ref{tab:#1}}~and~{\ref{tab:#2}}}
\def\dsubsecn#1#2{Subsections~{\ref{subsecn:#1}}~and~{\ref{subsecn:#2}}}

\def\tobs#1{{\it The Observatory}{\bf #1}}

\def\cfa#1{{\it Center for Astrophysics Preprint }{\bf #1}}
\def\asp{Astron. Soc. of the Pacific, San Francisco.}
\def\clar{Clarendon Press, Oxford.} 
\def\cup{Cambridge University Press, Cambridge.}
\def\free{W. H. Freeman, New York.}
\def\klu{Kluwer, Dordrecht.}
\def\nrao{NRAO, Greenbank.}
\def\plenumn{PLenum Press, New York.}
\def\prince{Princeton University Press, Princeston}
\def\ridel{Ridel, Dordrecht.}
\def\springerb{Springer-Verlag, Berlin.}
\def\springern{Springer-Verlag, New York.}
\def\wiley{Wiley, New York.}
\def\yale{Yale University Press, New Haven.}

\def\noin{\noindent}

\def\sqr#1#2{{\vcenter{\vbox{\hrule height.#2pt
  \hbox {\vrule width.#2pt height#1pt \kern#1pt
  \vrule width.#2pt}
  \hrule height.#2pt}}}}

\def\square{\mathchoice\sqr68\sqr68\sqr{2.1}3\sqr{1.5}3}

\def\CH{{\it CH }}

\newcommand{\myruleb}[1]{\vspace{0.5cm}\noindent\rule{#1}{0.05in}}
\newcommand{\myrulee}[1]{\newline\noindent\rule{#1}{0.05in\vspace{0.5cm}}}

\newenvironment{mybox}[3]
{\begin{minipage}{4.5in}
\setlength{\parindent}{0.5cm}
\myruleb{4.5in}
\label{box:#1}
\begin{center}\fbox{\textbf{#2}}\end{center}
{\begin{small}#3\end{small}}
\myrulee{4.5in}
\end{minipage}}

\newenvironment{myboxp}[4]
{\begin{minipage}{4.5in}
\setlength{\parindent}{0.5cm}
\myruleb{4.5in}
\label{box:#1}
\begin{center}\fbox{\textbf{#2}}\end{center}
\begin{center}\epsfig{figure=#3.ps,width=2in}\end{center}
{\begin{small}#4\end{small}}
\myrulee{4.5in}
\end{minipage}}

\newcommand{\myfig}[2]
{\begin{figure}[ht]
\begin{center}
\label{fig:#1}
\epsfig{figure=#1.ps,width=5in}
\caption{#2}
\end{center}
\end{figure}}

\newcommand{\myplotone}[2]
{\begin{figure}[ht]
\begin{center}
\label{fig:#1}
\plotone{#1.ps}
\caption{#2}
\end{center}
\end{figure}}

\newcommand{\myequn}[2]
{\begin{equation}
{#2}
\labequn{#1}
\end{equation}}

\newcommand{\myequna}[2]
{\begin{eqnarray}
{#2}
\labequn{#1}
\end{eqnarray}}

\newenvironment{headnoff}[1]
{\pnew\bHuge\bcenter{\bf\underline{#1}}\ecenter\eHuge}

\newcommand{\state}[1]{\noindent{#1}\vf}

\newcommand{\mnote}[1]{{{\small\it\marginpar{#1}}}}

\begin{document}
   \title{Unravelling the morphologies of Luminous Compact Galaxies using the HST/ACS GOODS survey}
   \author{A. Rawat \inst{1,2}
          \and Ajit K. Kembhavi \inst{2}
          \and F. Hammer \inst{1}
          \and H. Flores \inst{1}
          \and S. Barway \inst{2}
          }

   \offprints{rawat@iucaa.ernet.in\\ 
}

   \institute{
 GEPI, Observatoire de Paris-Meudon, 92195 Meudon, France
 \and
 Inter-University Centre for Astronomy and Astrophysics, Post Bag 4, Ganeshkhind, Pune 411007, India             
              }
   \date{Received  / Accepted }

\abstract{Luminous Compact Galaxies (LCGs) ($M_B \leq -20$, $R_{1/2} \leq$ 4.5\kpc~ and $EW_{0}$(OII) $\geq$15\AA) constitute one of the most rapidly evolving galaxy populations over the last $\sim$8\gyr~history of the universe. Due to their inherently compact sizes, any detailed quantitative analysis of their morphologies has proved to be difficult in the past. Hence, the morphologies and thereby the local counterparts of these enigmatic sources have been hotly debated.}
{Our aim is to use the high angular resolution, deep, multiband HST/ACS imaging data, from the HST/ACS GOODS survey, to study the quantitative morphology of a complete sample of LCGs in the redshift range $0.5\leq z \leq1.2$.}
{We have derived structural parameters for a representative sample of 39 LCGs selected from the GOODS-S HST/ACS field, using full 2-dimensional surface brightness profile fitting of the galaxy images in each of the four filters available. $B_{435W}-z_{850LP}$ color maps are constructed for the sample to aid in the morphological classification. We then use the rest frame B band bulge flux fraction (B/T) to determine the morphological class of galaxies which are well fit by a bulge+disk two dimensional structure. Mergers were essentially identified visually by the presence of multiple maxima of comparable intensity in the rest frame B band images, aided by the color maps to distinguish them from HII regions. We also make use of the Spitzer 24$\mu m$ source catalog of sources in the CDFS to derive the dust enshrouded star formation rates (SFR) for some of the sample LCGs}
{We derive the following morphological mix for our sample of intermediate redshift LCGs:\\ Mergers: $\sim$36\%, Disk dominated: $\sim$22\%, S0: $\sim$20\%, Early types: $\sim$7\%, Irr/tadpole: $\sim$15\%.
We establish that our sample LCGs are intermediate mass objects with stellar mass ranging from $9.44 \leq Log_{10}(M/M_{\odot}) \leq 10.96$, with a median mass of $Log_{10}(M/M_{\odot})=10.32$. We also derive SFR values ranging from a few to $\sim$ 65 $M_{\odot}$/year as expected for this class of objects. We find that LCGs account for $\sim$26\% of the $M_{B}\leq -20$ galaxy population in the redshift range $0.5 \leq z \leq 1.2$. We estimate a factor $\sim$11 fall in the comoving number density of blue LCGs from redshifts $0.5\leq z \leq1.2$ to the current epoch, even though this number is subject to large uncertainities given the small sample size at zero redshift available from the literature.}
{The strong redshift evolution exhibited by LCGs, and the fact that a significant fraction of LCGs are in
merging systems, seem to indicate that LCGs might be an important phase in the hierarchical evolution of galaxies. We envisage that some of the LCGs that are classified as merging systems, might go on to rebuild their disks and evolve into disk galaxies in the local universe.         }

   \titlerunning{Morphologies of Luminous Compact Galaxies}


   \maketitle

%
\section{Introduction}
 
Understanding the true nature of Luminous Compact Galaxies (LCGs) is one of the outstanding 
problems in observational extragalactic astronomy. LCGs constitute one of the most rapidly 
evolving galaxy populations 
at intermediate redshifts, with upto a factor $\sim$10 fall in the comoving number density 
from a redshift of $\sim$1 to the present epoch (Phillips et al. \cite{phillips1997}; 
Werk et al. \cite{jessica2004}). They were first reported by Koo and Kron (\cite{koo1988}) as possible 
blue compact galaxies due to the presence of strong narrow emission lines in the spectra of point-like sources, which were otherwise believed to be QSO candidates. Hubble Space Telescope (HST) follow up observations (Koo et al. \cite{koo1994}) of these 
compact narrow emission line galaxies (CNELGs) gave typical half light radii of $\sim$0.65\arcsec. The observations also provided evidence for exponential light profiles for these objects (as opposed to $r^{1/4}$ profiles). 
Using high resolution spectroscopy from HIRES on the Keck, Koo et al. (\cite{koo1995}) reported 
that these CNELGs have roughly Gaussian profile, narrow emission lines with velocity widths ranging between $\sigma=28-157\kms$. They suggested that these CNELGs might be the progenitors of local spheroidal galaxies by fading of upto 4-5 mags.

Guzman et al. (\cite{guzman1997}) and Phillips et al. (\cite{phillips1997}) identified 
a large population of such compact galaxies, defined as those with half-light radii $r_{1/2} \leq$ 0.5\arcsec~and high surface brightness within half light radius ($\mu_{I814} \leq$ 22.2 mag$\arcsecsqi$) in the flanking fields 
of the Hubble Deep Field (HDF). Using high resolution Keck spectroscopy, they determined the emission line profiles 
to be roughly Gaussian with velocity widths ranging from $\sigma\sim35\kms$ to 
$150\kms$. However, they stated that most compact galaxies at moderate redshifts yielded 
little morphological information even in HST/WFPC2 images. They reported about 2/3 of the galaxies in 
their sample to be consistent with being young star forming HII galaxies.

Guzman et al. (\cite{guzman1998}) found in HST/WFPC images of five CNELGs the presence of blue high surface brightness knots surrounded by a diffuse ``exponential like'' component. The knots were identified as the location of the current star formation due to their bluer colors, whereas the diffuse exponential like component was interpreted to be an older underlying population.

More recently Hammer et al. (\cite{hammer2001}) have used VLT spectra of LCGs to argue in favour 
of their being the progenitors of present day spiral bulges, with tidally pulled in gas from 
interacting systems fueling high rates of star formation, resulting in the completion of the 
bulge formation and beginning of the disk formation. 

Hammer et al. (\cite{hammer2005}) have used multiwavelength observations of 195, z $>$ 0.4 
intermediate mass galaxies to propose a scenario in which a majority of intermediate mass 
spirals have experienced their last major merger event (leading to the disruption of 
the disk) within the last 8\gyr. The merger scenario seems obvious since the merger rate was 
about 10 times more at z$\sim$1 than it is today (Le Fevre et al. \cite{lefevre2000}). This merger phase is followed by enhanced star formation due to the merging, which is termed as the LCG phase, 
and a subsequent inside-out rebuilding of the disk leading finally to the formation of grand 
design spirals that we see today. Hence the LCGs are being proposed as a phase in the 
evolutionary history of galaxies in the hierarchical scenario of galaxy formation.

Noeske et al. (\cite{noeske2006}) have analyzed the morphologies of 26 Luminous 
Compact Blue Galaxies (LCBGs) in the HST/ACS Ultra Deep Field at z $\sim$ 0.2 - 1.3. They 
concluded that the majority of high z LCBGs are small galaxies that will evolve into small disk 
galaxies and low mass spheroidals in the local universe. However, the sample selection of Noeske 
et al. includes galaxies as faint as $M_{B} \leq$ -18.5 and, as per their own opinion, is not fully 
comparable with the sample selection criterion of Hammer et al. (\cite{hammer2001}) (which might 
favor progenitors of larger local galaxies), making any comparison of their results rather 
uncertain.

In this paper we have made use of the high resolution images provided by the Advanced Camera for Surveys (ACS) on board the HST to determine the quantitative morphology of intermediate redshift LCGs using the public data from the HST/ACS GOODS survey (Giavalisco et al. \cite{giavalisco2004}). The unique 
advantage provided by the GOODS survey is that ACS imaging (as opposed to WFPC2 imaging) with it's 
improved drizzled pixel scale of 0.03\arcsec/pixel is a factor of $\sim$2 better than the typical drizzled pixel scale of WFPC2. This better sampling becomes specially important for compact sources such as LCGs with typical $r_{1/2} \leq$ 0.5\arcsec, for which the whole extended emission (eg. disk) may be sampled by just a few pixels of the WFPC2. This in turn allows us to classify galaxy morphology on the basis of their luminosity profile, even for compact sources.
In addition to this, the large area coverage of the 
GOODS survey provides a large enough sample of LCGs to derive robust statistics 
of their properties.

This paper is organized as follows: In Section 2, we 
briefly describe our sample and in Section 3 present the methodology of our morphological 
classification and basic results. In Section 4 we describe the use of data at other wavelengths available in the {\it{Chandra Deep Field South}} (CDFS). In Section 5 we provide individual description for each of the 39 LCGs. In Section 6, we calculate the evolution in number density of LCGs from intermediate redshifts to the current epoch. Our results are discussed 
in Section 7 and our conclusions summarized in Section 8. Throughout this paper we adopt a 
cosmology with $H_0$\,=$70\hubbleunit$,  $\Omega_{\rm  M}$\,=\,0.3 and 
$\Omega_\Lambda$\,=\,0.7 unless specially stated.

\section{The Data}

We have used in our work: 
\begin{enumerate}
\item The publicly available version v1.0 of the reduced, calibrated, stacked and mosaiced images of the CDFS acquired with HST and ACS as part of the {\it{Great Observatories Origins Deep Survey}}, GOODS (Giavalisco et al.~\cite{giavalisco2004}).

\item  The publicly available SExtractor (Bertin and Arnouts \cite{bertin1996}) based version r1.1 of the ACS multi-band source catalogs as released by the GOODS team.

\item Spectroscopic redshifts taken from the publicly available redshift catalog of the Vimos VLT Deep Survey VVDS$\footnote{We have not applied any explicit quality cut on the VVDS redshifts. However due to the requirement of the presence of OII emission line for an LCG, we have a sample of objects with very robust redshift measurements, with $\sim$2/3 of the objects having redshift with better than 95\% confidence, and all but 2 objects having redshifts with better than 75\% confidence.}$ (Le Fevre et al.~\cite{lefevre2004}), and   

\item The near-IR J \& Ks band imaging data from the ESO GOODS/EIS Release Version 1.0 (Vandame et al., \cite{vandame}) which was obtained as part of the GOODS using the ISAAC instrument mounted at the Antu Unit Telescope of the VLT at ESO's Paranal Observatory, Chile. This data release includes 21 fully reduced VLT/ISAAC fields in J and Ks bands, covering $131\arcmsq$ of the GOODS/CDFS region. 

The near-IR data, along with the four band HST imaging data, combined with the spectroscopic redshifts mentioned above, were used to derive the rest frame B band absolute magnitudes $M_B$ of the galaxies. A detailed discussion of how the K-corrections were estimated and and how we finally arrived at the absolute magnitudes is provided in Appendix A of this paper.     

\end{enumerate}

\subsection{Sample selection}

In order to select a sample of Luminous Compact Galaxies, we have used the three criteria defined by Hammer et al.~(\cite{hammer2001}) 

First, we apply a luminosity criterion, selecting only galaxies with $M_{B}(AB)\leq -20$. In addition to justifying the class as {\it{Luminous}}, this also ensures that there is sufficiently large signal-to-noise ratio (SNR) in the observed images to allow a detailed bulge-disk decomposition of galaxies even in the highest redshift bin. 

Second, a compactness criterion was applied using a compactness parameter

\begin{equation} 
~~~~~~~~~~~~~~~~~\delta m_{z} = m_{z}(4.5\kpc) - m_{z}(13.5\kpc) \leq 0.75 \\
\label{aperture}
\end{equation}
This ensures that {\it{at least}} half the flux is contained within an aperture of radius 4.5\kpc. The size of the apertures used by us are taken to be equivalent to those of Hammer et al., after correcting for the different cosmology. This correction is applied at the median redshift of our galaxy sample. One important difference is that we have used elliptical instead of circular apertures for the galaxies, using the THETA\_IMAGE and ELLIPTICITY parameters of SExtractor to generate unique elliptical apertures for each object. Hence, we are using the most flexible apertures possible which allow for all possible isophotal shapes, and do not make any assumptions regarding the shapes of the isophotes to be expected. The {\it{radius}} of the aperture mentioned above is therefore understood to mean the length of the semi-major axis of the ellipse. 

Third, we include only those galaxies with known [OII] $\lambda$3727 emission, with rest frame equivalent width $[EW_{0}[OII] \geq 15 \AA]$. This requirement of the presence of [OII] $\lambda$3727 line in the observed spectrum puts a lower limit on the redshift of z $\sim$0.5 for a galaxy in our sample, as the minimum observed frame wavelength in the VVDS is $\sim$5600 $\AA$. We have also put an upper redshift cutoff of z=1.2 as there are few sources beyond this redshift in our catalog and incompleteness becomes a serious issue at higher redshifts.

\begin{figure}[h] \centering
   \includegraphics[angle=0, width=0.5\textwidth]{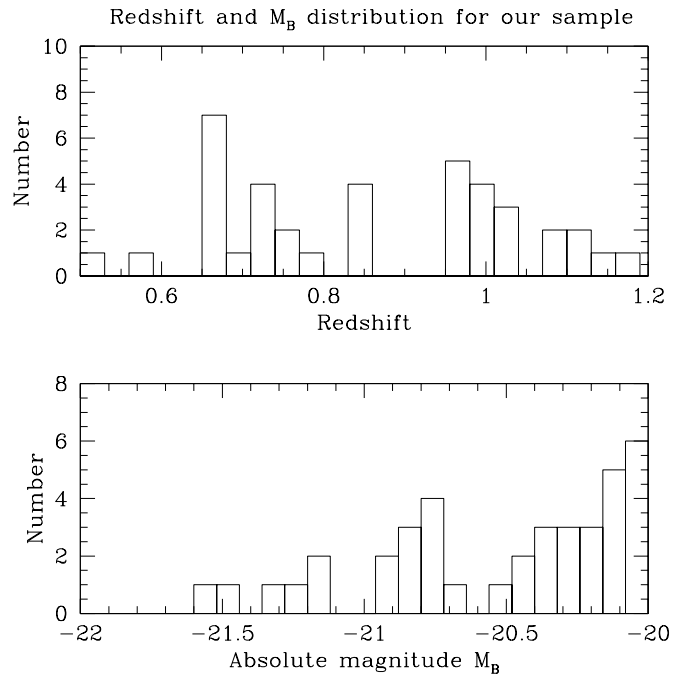}
   \caption{The histogram of the redshift and the absolute $B$ band magnitude distribution for our LCG sample.}
   \label{hisMB}
\end{figure}    

Applying these three selection criteria to the GOODS-S HST/ACS field yields a sample of 39 Luminous Compact Galaxies. This accounts for $\sim$26\% of the $M_{B} \leq -20$ galaxy population in the redshift range $0.5 \leq z \leq 1.2$.

Figure~\ref{hisMB} shows the distribution of the redshift and the absolute $B$ band magnitude of the LCG sample. Table~\ref{tab} lists the three parameters that have been used in the selection criterion for each galaxy, along with their redshift. The ID given in column (1) is used by us throughout this paper, and the GOODS ID given in column (2) is as listed in the r1.1 of the ACS multi-band GOODS source catalogs.

\begin{table}[ht!]
{\scriptsize 
 \caption[]{The LCG sample}
  \label{tab}
  \begin{tabular}{cccccc}
  \hline
  \noalign{\smallskip}
 Our    & GOODS   &  &$M_{B}$ & $\delta m_{z}^{\mathrm{a}}$ & $EW_{0}[OII]$ \\
  ID &ID & Redshift & (Mag.) &  (Mag.)  & ($\AA$)  \\
(1) & (2) & (3) & (4) & (5) & (6)  \\
  \noalign{\smallskip}
  \hline
  \noalign{\smallskip}

904260 &J033211.97-275033.9 & 0.983 & -20.73 & 0.72 & 35.39      \\

904604 &J033212.62-275105.6 & 0.990 & -20.38 & 0.51 & 29.38 \\

904680 &J033212.78-275012.8 & 0.964 & -20.86 & 0.64 & 70.19 \\

905632 &J033214.68-274337.1 & 0.976 & -20.04 & 0.48 & 107.80 \\

905983 &J033215.36-274506.9 & 0.860 & -20.74 & 0.49 & 49.41 \\

906961 &J033217.14-274303.3 & 0.566 & -20.93 & 0.29 & 16.01 \\

907047 &J033217.28-275416.6 & 1.112 & -21.23 & 0.64 & 23.73 \\

907361 &J033217.77-274714.9 & 0.731 & -20.17 & 0.60 & 19.30 \\

907794 &J033218.52-275508.3 & 1.144 & -21.44 & 0.58 & 46.77 \\

908243 &J033219.32-274514.0 & 0.726 & -20.35 & 0.73 & 43.56 \\

909015 &J033220.70-275501.7 & 1.039 & -20.25 & 0.46 & 49.15 \\

909093 &J033220.86-275405.1 & 0.968 & -20.32 & 0.62 & 95.48 \\

909429 &J033221.43-274901.8 & 0.737 & -20.81 & 0.69 & 33.67 \\

910413 &J033223.03-275452.3 & 0.655 & -21.30 & 0.54 & 18.38 \\

911747 &J033225.20-275100.1 & 0.840 & -20.73 & 0.47 & 51.07 \\

911780 &J033225.26-274524.0 & 0.664 & -20.88 & 0.60 & 20.71 \\

911843 &J033225.35-274502.8 & 0.973 & -20.49 & 0.34 & 67.75 \\

912744 &J033226.61-275001.9 & 0.690 & -20.42 & 0.60 & 59.31 \\

913482 &J033227.62-274144.9 & 0.664 & -20.69 & 0.28 & 45.17 \\

914038 &J033228.48-274826.6 & 0.667 & -20.17 & 0.43 & 30.06 \\

915400 &J033230.43-275140.4 & 0.764 & -20.08 & 0.67 & 99.46 \\

916137 &J033231.50-275241.4 & 0.980 & -20.14 & 0.58 & 50.62 \\

916446 &J033231.96-275553.4 & 0.839 & -20.03 & 0.60 & 51.10 \\

916866 &J033232.61-275316.7 & 0.987 & -20.06 & 0.64 & 29.44 \\

918147 &J033234.61-275324.5 & 1.099 & -21.52 & 0.65 & 71.64 \\

919573 &J033236.72-274406.4 & 0.665 & -20.05 & 0.69 & 37.62 \\

919595 &J033236.74-275206.9 & 0.785 & -20.40 & 0.49 & 49.14 \\

920435 &J033238.08-275248.7 & 1.034 & -20.15 & 0.43 & 86.23 \\

921406 &J033239.65-275226.2 & 1.095 & -21.12 & 0.72 & 23.93 \\

922675 &J033241.88-274853.9 & 0.666 & -20.31 & 0.54 & 27.08 \\

922733 &J033242.02-275226.0 & 0.650 & -20.04 & 0.40 & 55.70 \\

922761 &J033242.09-275109.4 & 0.961 & -20.10 & 0.62 & 35.14 \\

923085 &J033242.73-275159.2 & 1.122 & -21.12 & 0.50 & 47.24 \\

923926 &J033244.45-274940.2 & 1.012 & -20.86 & 0.38 & 93.41 \\

924881 &J033246.59-274516.3 & 0.839 & -20.05 & 0.52 & 68.14 \\

926109 &J033249.53-274630.0 & 0.522 & -20.13 & 0.47 & 29.64 \\

926217 &J033249.87-275129.0 & 0.767 & -20.25 & 0.64 & 28.37 \\

907305 &J033217.68-274208.9 & 1.185 & -20.76 & 0.45 & 68.82 \\

914895 &J033229.71-274507.2 & 0.736 & -20.16 & 0.56 & 43.32 \\

  \noalign{\smallskip}
  \hline
  \end{tabular}

 \begin{list}{}{}
  
  \item [$^{\mathrm{a}}$]$\delta m_{z}$ is the aperture magnitude difference given in Equ.\ref{aperture} in section 2.1

  \end{list}

}
\end{table}

\subsection{Completeness of the LCG sample}

\begin{figure}[h] \centering
   \includegraphics[width=0.5\textwidth]{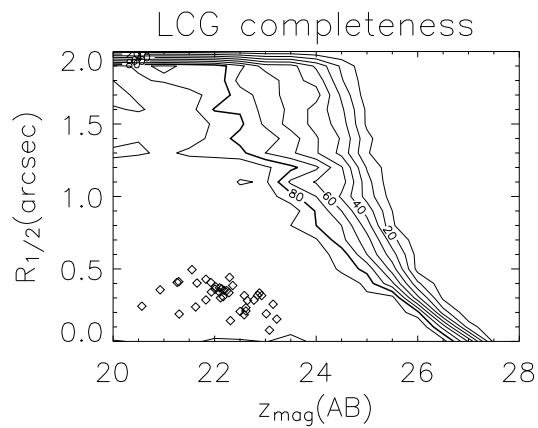}
   \caption{F850LP magnitudes and half light radii ($R_{1/2}$) of our 39 LCGs shown in comparison with the completeness curves of the GOODS survey as depicted on the magnitude-size plane. The percentage completeness is marked on the individual curves. The completeness curves are plotted in intervals of 10\% and were kindly provided by Harry Ferguson.}
   \label{completeness}
\end{figure}

In a flux limited survey like GOODS, one has to be careful to avoid bias caused by incompleteness of the sample, especially when one is working with faint objects. Since we are working with intrinsically luminous and compact sources, they are expected to be high surface brightness objects free of such incompleteness biases. In order to demonstrate this, we used the completeness curves published by the GOODS team as depicted on the magnitude-size plane and plotted our 39 LCGs on the same graph. This is shown in Fig.~\ref{completeness}. Our galaxies occupy a region on the bottom left of the graph as expected by high surface brightness objects. In this region of the magnitude-size plane, the GOODS survey is {\it{at least}} 90\% complete$\footnote{In order to see how the completeness curves were estimated, refer to Giavalisco et al.~(\cite{giavalisco2004})}$. This proves that our sample is not biased by incompleteness issues. 

\section{Morphologies of LCGs}

Using the HST/ACS GOODS images, quantitative two dimensional bulge-disk decomposition has been carried out to derive the structural parameters of the galaxies which are then used to quantify their morphologies. Information derived from the color maps is also used in the morphological classification.

\subsection{Structural parameters}

We used the software {\it galfit} (Peng et al. \cite{peng2002}) to carry out two-dimensional modeling for our sample of 39 LCGs in each of the four HST/ACS filters to obtain their structural parameters. Galfit models the galaxy image as a linear combination of a bulge and a disk, using a well established analytic model for each of the two components. An important caveat in this approach is the underlying assumption that a galaxy can be represented as a linear combination of simple, smooth analytic functions. Real galaxies are known to be much more complex, with the presence of spiral arms, bars and central point sources etc. Despite these reservations, galaxy fitting algorithms have been demonstrated to be successful in the past for the purpose of quantitative morphological classification of galaxies (Peng et al.~\cite{peng2002}, Simard et al.~\cite{simard2002}).   

The intensity profile of the bulge is modeled with the Sersic law (Sersic~\cite{sersic1968}), 
\begin {equation}
\label{sersic}
~~~~~~~~~~~~~~~~~~~~~~\Sigma(r)=\Sigma_e \mbox{exp}\Big[{-2.303b_{n} \Big[\Big({\frac{r}{r_e}}\Big)^{1/n} - 1\Big]}\Big]
\end {equation}
with $r_{e}$ being the half light radius, $\Sigma_e$ the surface brightness at $r_{e}$ and n being the sersic index. 

The disk is modeled as an exponential function (Freeman~\cite{freeman1970}), 
\begin {equation}
\label{exponential}
~~~~~~~~~~~~~~~~~~~~~~~~~~~~~~~~~~\Sigma(r)=\Sigma_0 \mbox{exp}\big[{-\frac{r}{r_d}}\big]
\end {equation}
with $\Sigma_0$ being the central surface brightness and $r_{d}$ the e-folding scale length.

The model galaxy is then constructed as a linear combination of these two components plus other free parameters such as the sky background value $(bkg)_{x,y}$ in pixel position (x,y), which is assumed to be constant across the galaxy image, the ellipticities $\epsilon_{bulge}$ and $\epsilon_{disk}$ of the bulge and the disk isophotes respectively and their position angles $\theta_{bulge}$ and $\theta_{disk}$. The SExtractor based photometric catalog of our sources provided us with valuable starting values for parameters such as sky background, half light radii of the objects and their ellipticity and position angles etc. This analytic model of the galaxy is then convolved with the PSF of the observation and compared with the observed galaxy image. The values of the free parameters are determined iteratively by minimizing the reduced $\chi_{red}^{2}$ defined as usual,

\begin{equation}
\label{redchi2}
~~~~~~~~~~~~~~~~~\chi^2_{red} = \frac{1} {N_{\rm dof}}\sum_{x=1}^{nx}\sum_{y=1}^{ny} \frac {\left(\mbox{galaxy}_{x,y} - \mbox{model}_{x,y}\right)^2} {{\sigma_{x,y}}^2},
\end{equation}
  where $N_{\rm dof}$ is the number of degrees of freedom, $galaxy_{x,y}$ and $model_{x,y}$ are the counts in the pixel (x,y) of the galaxy and the model image respectively, $\sigma_{x,y}$ is the noise in the pixel (x,y) and nx,ny are the number of pixels in the galaxy image in the x and y direction respectively.

Special attention is paid in arriving at a reliable noise model $\sigma_{x,y}$ for use in Eq.~\ref{redchi2}. In case of the HST/ACS GOODS fields, the noise characteristics of the science images are stored in the form of weight maps, $w_{x,y}$ which are images produced as part of the data reduction process called {\it drizzling} (Fruchter \& Hook~\cite{fruchter1997};\cite{fruchter2002}), which give a measure of the background plus instrumental noise per pixel in the science data. Since the GOODS images are mosaics consisting of a different number of overlapping pointings at different areas, the effective exposure time varies across the field leading to variable {\it depth} across the field. The weight maps take into account this variation and should provide a faithful estimation of the inverse variance (1/$\sigma_{x,y}^2$) per pixel. However, the interpolations introduced by {\it drizzling} result in correlations between pixels in the drizzled science images. The weight maps are normalized to show the expected noise per pixel that the images would have in the absence of these correlations. Therefore, the apparent $\sigma_{x,y}$ that one measures in the science image is smaller than that given by the $(1/w_{x,y})^{1/2}$,because the apparent $\sigma_{x,y}$ is suppressed by the effects of these correlations. This necessitates the use of a corrective factor when deriving the $\sigma_{x,y}$ from the weight maps $w_{x,y}$ such that:
\begin{equation}
\label{fudge}
~~~~~~~~~~~~~~~~~~~~~~~~~~\sigma_{x,y} = F \times \frac{1}{\sqrt{w_{x,y}}}
\end{equation}
where F depends upon the drizzling parameters SCALE (s) and PIXFRAC (p) (see Fruchter \& Hook~\cite{fruchter1997};\cite{fruchter2002}). For the GOODS dataset, s=0.6 for all the four filters, whereas p=0.8 for F435W, F606W and F775 filters, and p=0.7 for F850LP filter (Anton Koekemoer, private communication). Using the prescription given in Casertano et al.~(\cite{casertano2000}), we derive a value of F=0.61 for the F850LP filter and F=0.56 for the other three filters. These derived values of F, in conjunction with Eq. \ref{fudge} and the supplied weight maps provided the noise model for our sample of galaxies. A more detailed discussion on weight map conventions and noise correlation in drizzling can be found in Casertano et al.~(\cite{casertano2000}).      

We have determined the sky background using SExtractor and held it fixed at this value during the minimization. All other parameters like $r_e$, $r_d$, n, position angle etc. were allowed to be free in the fitting process, so that galfit can explore the full range of parameter space to find the global minimum and is not pulled into a local minimum due to constraints put in by hand.

Once the best fit galaxy model has been obtained, we calculate the {\it bulge fraction} B/T, defined as:

\begin{equation}
~~~~~~~~~~~~~~~~~~~~~~~~~\frac{B}{T}=\frac{flux_{bulge}} {flux_{bulge} + flux_{disk}}    
\end{equation}
where the $flux_{bulge}$ \& $flux_{disk}$ are calculated by integrating the bulge and the disk profiles respectively over all values of the semi-major axis using the best fit parameters obtained by galfit. This bulge fraction B/T is found to be broadly correlated with the traditional Hubble type of a galaxy in the sense that an early type galaxy has a high B/T ratio and a late type spiral has a low value. It is this ratio that we use for characterizing the morphology of our galaxy sample.    

In the above scheme, an accurate estimation of the PSF is crucial to the robust determination of the derived structural parameters of a galaxy. In our work, the star closest to a given galaxy is used to determine the PSF for the purpose of bulge-disk decomposition. Stars are the most pristine characterization of the PSF in any given observation, as they have gone through the same optics as well as the reduction pipeline as the program galaxies. Only bright (non saturated) and well behaved stars, free of neighbours and other contamination were used as candidate PSF stars in our work. A master list of such stars was prepared by visually examining all candidate stars in the GOODS field and rejecting any unsuitable star based on the basic criterion outlined above. This yielded $\sim$60 stars scattered all across the CDFS which could be used to determine the PSF, from among $\sim$400 candidate stars. 

\begin{figure}[h] \centering
   \includegraphics[angle=270, width=0.5\textwidth]{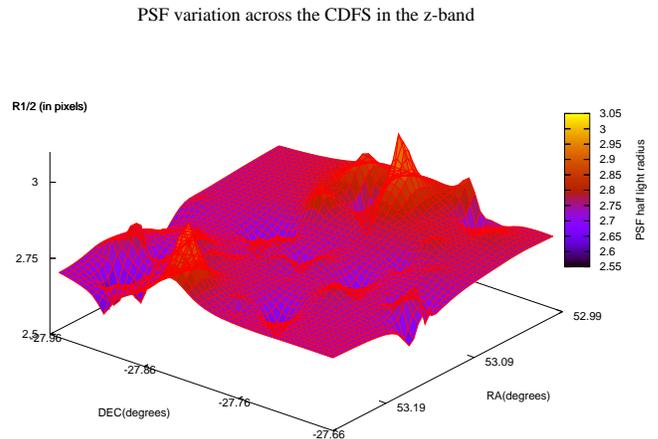}
   \caption{The variation of half-light radius(in units of ACS pixels) of stars with RA and DEC across the GOODS-S field in the F850LP filter. This plot uses $\sim$60 stars spread across the field.}
   \label{psf}
\end{figure}

We looked for the variation of the PSF half-light radius with RA/DEC for the $\sim$60 PSF stars that we had shortlisted. A "surface plot" of the PSF half light radius variation is shown in Fig.~\ref{psf}. The PSF is seen to be extremely constant across the GOODS-S field of view. Barring a couple of "peaks" in the plot (which are explained on the basis of the {\it{red halo}} effect below), the $R_{1/2}$ of the PSF stars is stable to within 5\%. 
This is because of the extremely stable detector/optical system offered by HST/ACS which yields almost identical PSF even for two images taken at different epochs or for two different exposures (as in GOODS). In addition, the lack of an atmosphere takes away a major source of variability in the PSF from one epoch to another seen from the ground.    

 One important point to consider while interpreting Fig.~\ref{psf} is that whatever variation of PSF is seen in this figure is due to the combined effects of both spatial variation of the PSF, plus any other factors such as the {\it{red halo}} (Sirianni et al.~\cite{sirianni2005}, Gilliland and Riess~\cite{gilliland2002}) effect in the F850LP band ACS PSF which depends on the color of the star. In fact, we find that all of the ``peaks'' that are seen in Fig.~\ref{psf} belong to stars that are extremely red, and therefore the red halo might be responsible for their large $R_{1/2}$ rather than any spatial variation$\footnote{For a more detailed discussion of the red halo effect on the F850LP PSF stars used by us and how we dealt with it, see Appendix B}$. 

\begin{figure}[h] \centering
   \includegraphics[angle=270, width=0.5\textwidth]{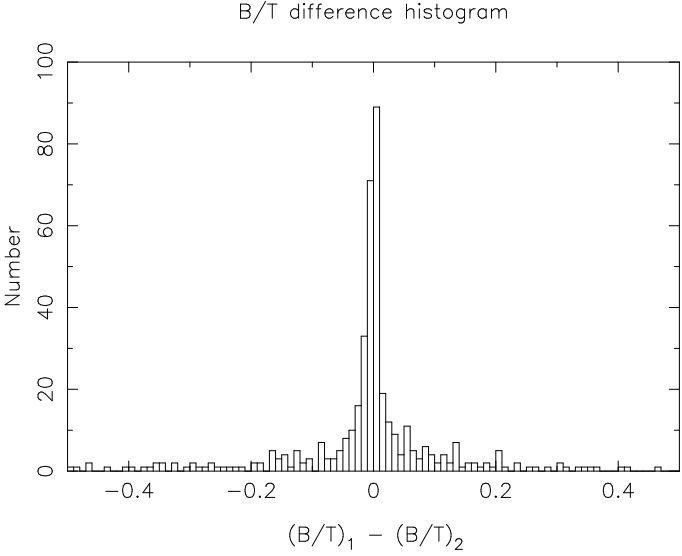}
   \caption{The histogram of the difference between the B/T ratios obtained by using the nearest and the 2nd nearest star to the galaxy as the PSF.}
   \label{BbyT}
\end{figure}
Extensive tests were carried out to rule out the possibility of variations in the determination of the structural parameters of galaxies due to the use of different stars as the PSF.
For a sample of $\sim$450 galaxies in the CDFS, we used galfit in batch mode to derive the structural parameters and B/T ratio {\it {twice}} for each galaxy. In the first run, we used the star nearest to the galaxy being modeled as the PSF. In the second run, we used the second nearest star to the galaxy as the PSF, everything else remaining the same as in the first run$\footnote{The median distance from the sample galaxy to the nearest PSF star is 48\arcsec~while the median distance to the second nearest PSF star is 80\arcsec.}$. In Fig.~\ref{BbyT} we show the histogram of the difference between the B/T ratio obtained in the two cases. As is evident from the figure, the estimation of B/T ratio is quite robust irrespective of which star is used as the PSF. About 70\% of the sources have their B/T value stable to within 10\%. The sharp peak around zero means that the B/T ratios obtained in the two run agree quite well, and the symmetrical shape of the peak at zero points towards random errors as the cause for any dispersion in the B/T values obtained in the two runs rather than any systematic effect.      

We conclude that the final value of B/T is stable to within 10\% irrespective of which star is used as the PSF. Keeping this in mind, the star closest to a given galaxy is used as the PSF for the purpose of bulge-disk decomposition of our sample of LCGs. The important thing to notice is that in Fig.~\ref{BbyT}, the differences in B/T ratio that we obtain on using two different PSFs is due to the {\it{cumulative}} effects of PSF variation due to all possible reasons such as spatial variation of the PSF, red halo effect etc. Since the cumulative error in the derived B/T ratio is shown to be within $\sim$10\%, our results are stable to within this error irrespective of the fact as to which particular factor contributes to the variation in the PSF. This should convince the reader that our final results are insensitive to the choice of PSF star.

\begin{figure}[t]
\includegraphics[height=4.5cm,clip]{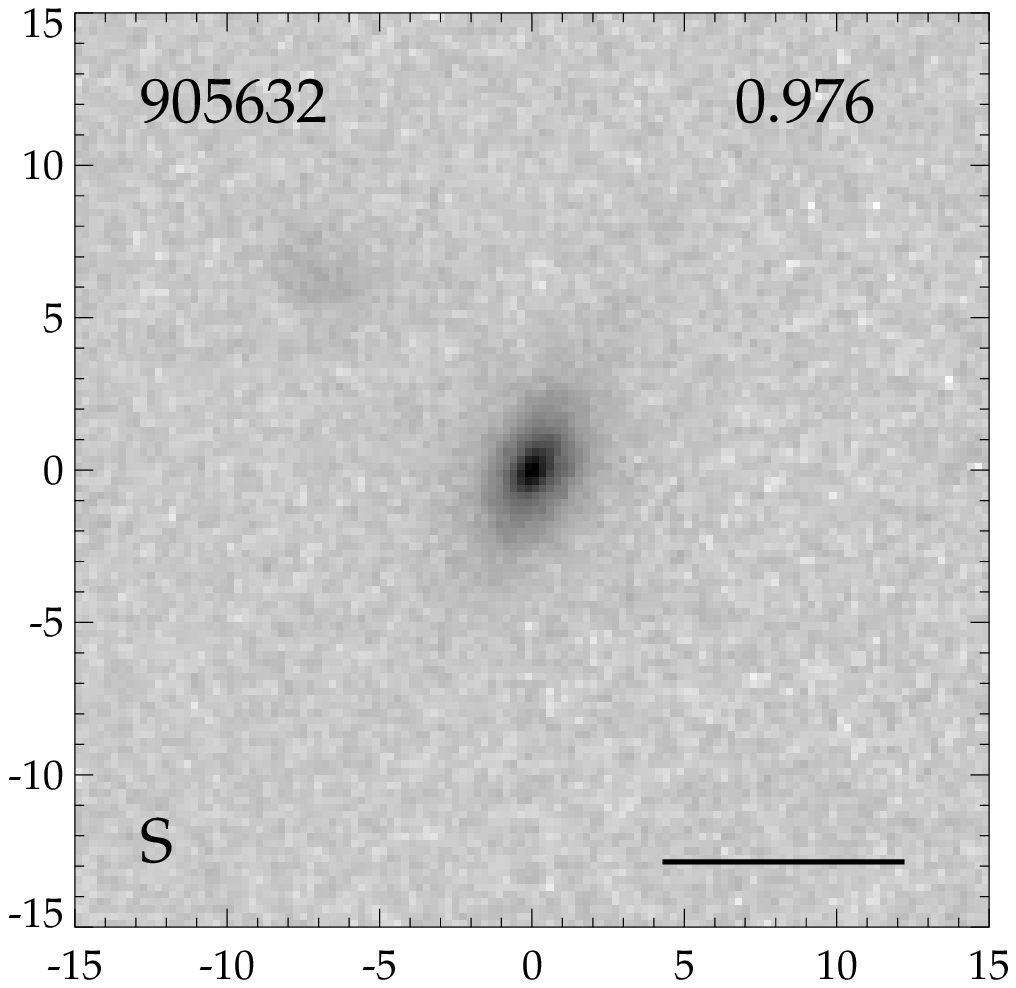}%
\includegraphics[height=4.5cm,clip]{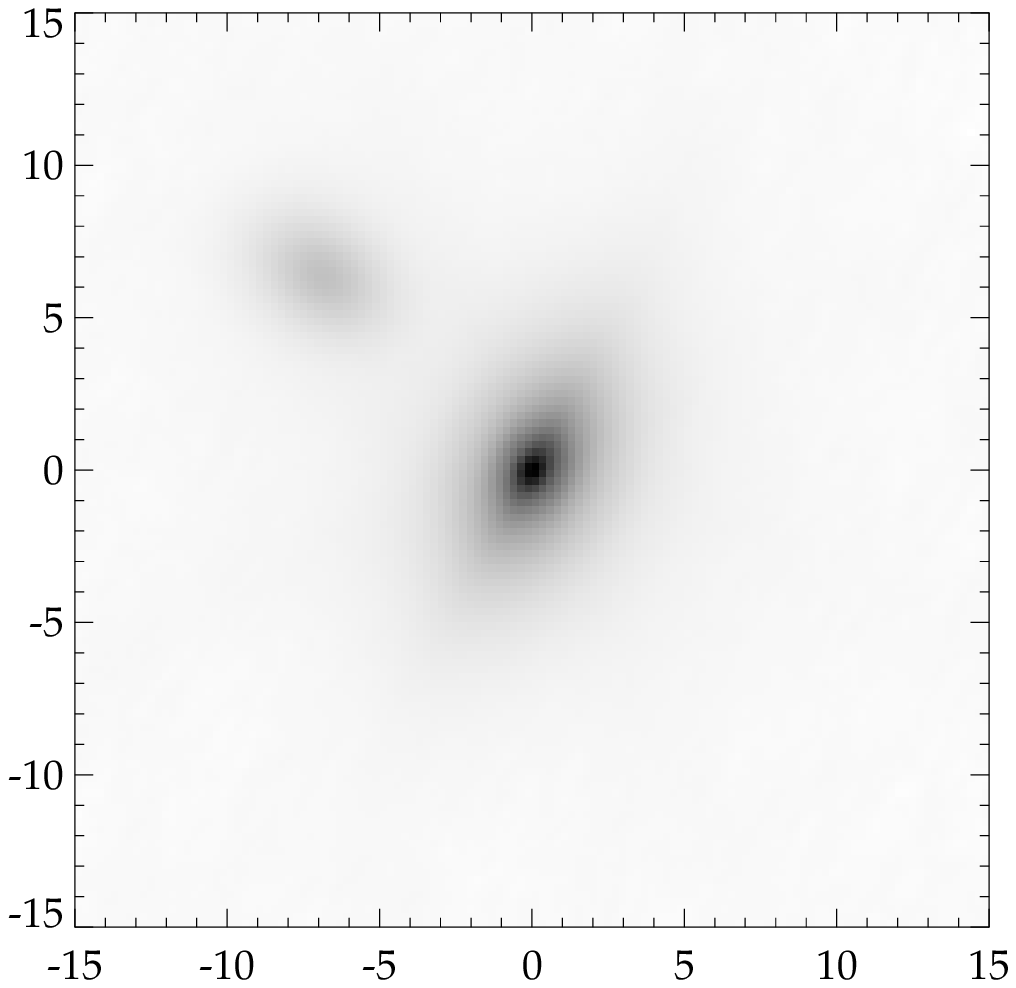}
\includegraphics[height=4.5cm,clip]{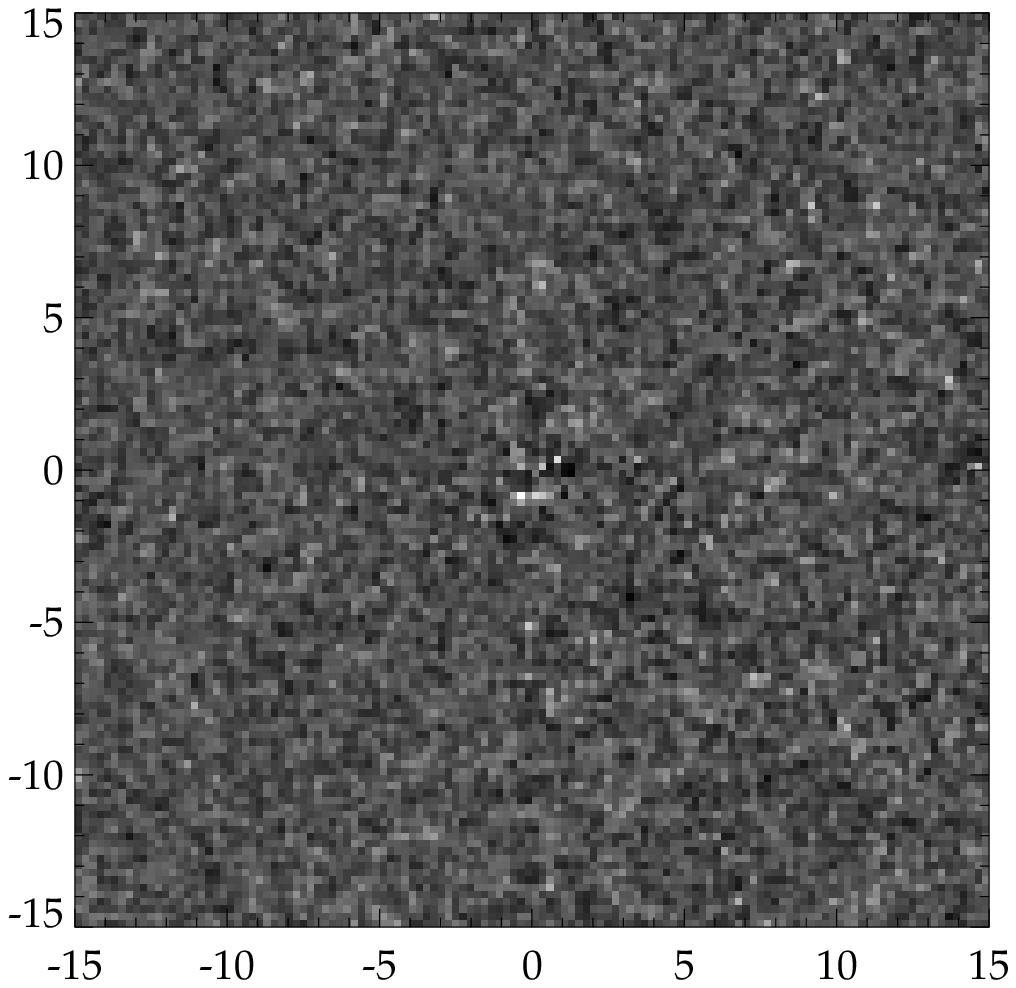}%
\includegraphics[height=4.5cm,clip]{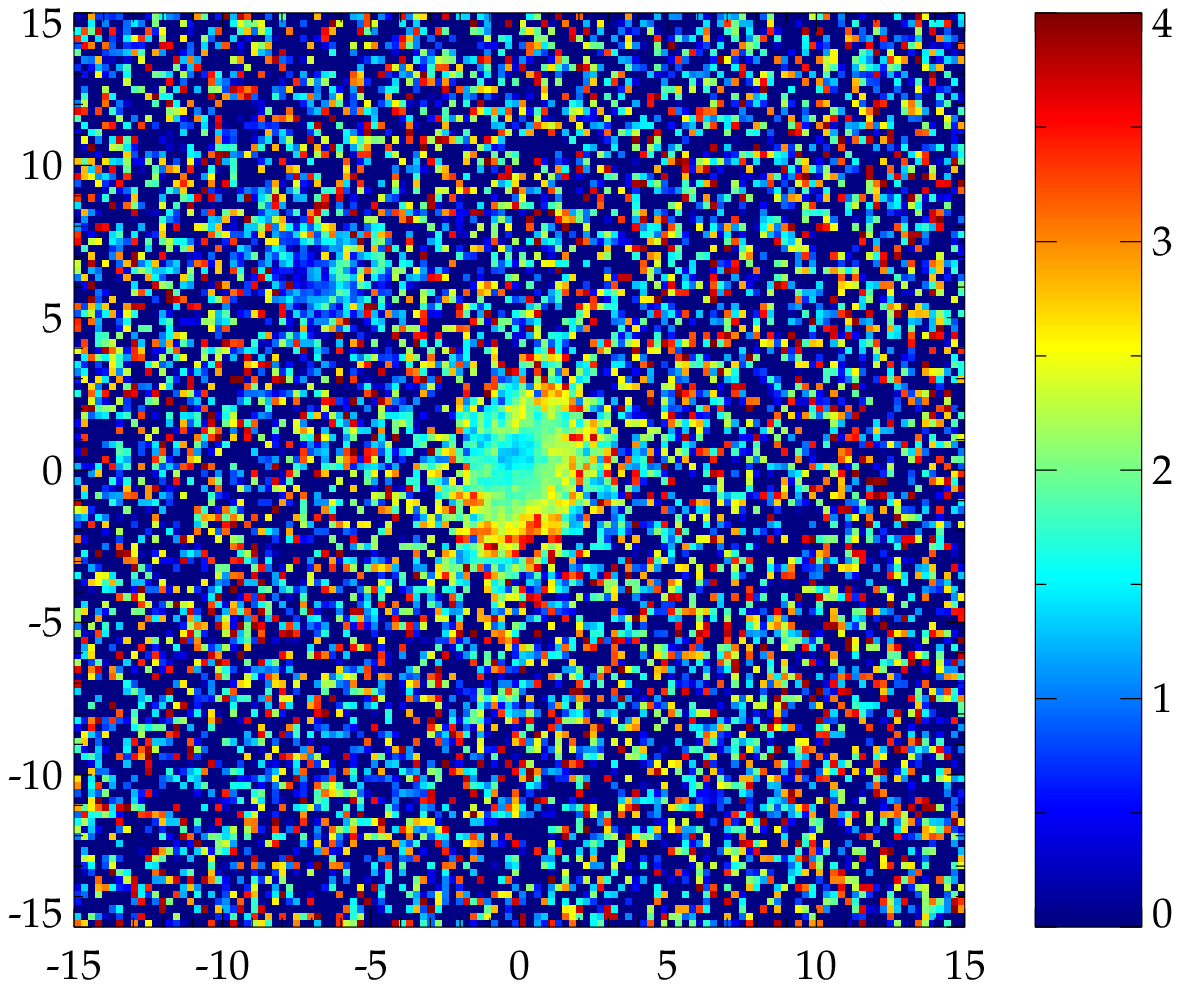}
\caption{From top-left to bottom-right $30\times30\kpc$ F850LP image, Galfit Model image, Residual image and the B-z color map of the LCG 905632. North is up, East is to the left. The galaxy name, redshift and morphological classification are indicated in the F850LP galaxy image. The horizontal bar at the bottom right of the galaxy image indicates $1\arcsec$. The intensity scaling is the same (squareroot) in each of the three grayscale images.}
\label{sample}
\end{figure}

Figure~\ref{sample} shows the result of our quantitative bulge-disk decomposition for one of the LCG sample galaxies at a redshift of 0.976. The upper left frame shows a $30\kpc \times 30\kpc$ F850LP galaxy image with the galaxy name, redshift and morphological classification indicated. The upper right frame shows the best fit model image having $\chi^{2}_{Red}$=1.093. Notice that the galaxy in the neighbourhood of the primary target (north-east) is also being fitted simultaneously. In this work, we have taken the approach of fitting a single component sersic function to all neighbouring objects within $\sim$5.0\arcsec~of the program galaxy. This approach, though more computation intensive, gives vastly improved results compared to the traditional approach of masking the neighbouring objects in the frame (which does nothing to account for the flux contributed to the program object by the extended wings of the neighbours). This effect can be particularly severe in case of bright neighbours and would lead to an inaccurate estimation of the structural parameters of the program galaxy. At the bottom left, the residual image is shown, which is the difference between the galaxy image and the best fit model. The residual image is devoid of any structures and consists solely of Poisson noise, which is an additional indication of the good quality of the fitting. Finally, the bottom right frame shows the B-z color map of the galaxy in AB magnitude system; a blue core is seen. This is a disk dominated galaxy with a B/T ratio of 0.33 and a disk scale length $r_{d}=1.16\pm0.02\kpc$.

\begin{figure}[h] \centering
   \includegraphics[width=0.5\textwidth]{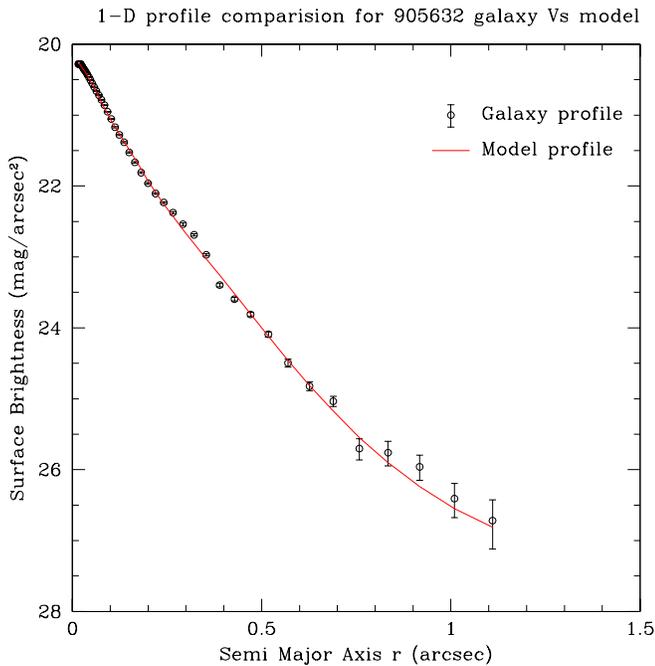}
   \caption{A comparison of the azimuthally averaged F850LP band radial surface brightness profile of the galaxy 905632(points with error bars), as compared to the surface brightness profile of the model image produced by galfit(red smooth line) produced using the IRAF task ELLIPSE.}
   \label{compare_extend}
\end{figure}

Figure~\ref{compare_extend} shows a comparison between the azimuthally averaged radial surface brightness profile of the same LCG 905632 (points with error bars), and the surface brightness profile of the model image produced by galfit (red smooth line), produced using the IRAF$\footnote{IRAF  is distributed by the National Optical Astronomy Observatories, which are operated by the Association of Universities for Research in Astronomy, Inc.   under   cooperative  agreement   with   the  National   Science Foundation.}$ task ELLIPSE. The two profiles agree within errorbars, which serves as an additional test of the accuracy of the fit. 

\subsection{Morphological classification}

We have carried out bulge-disk decomposition on all the 39 LCGs in our sample. The galaxies that are well fit by a bulge+disk two-dimensional structure, are classified into three `Hubble types' mainly based on the {\it rest frame} B band $B/T$ ratio: early types E (0.8\,$<B/T\leq$\,1), intermediate types S0 (0.4\,$<B/T\leq$\,0.8) and disk dominated S (0.0\,$<B/T\leq$\,0.4). This segregation of galaxies into different Hubble types based on their B/T ratio is made using the criterion described by Zheng et al. (\cite{zheng2004}, \cite{zheng2005}), even though we have grouped our galaxies more broadly in B/T as compared to Zheng et al. Since our sample of LCGs range in redshift from 0.5 to 1.2, the rest frame B band falls in different observed filter bandpasses, depending upon the redshift of the object. We made use of the multiband nature of the HST/ACS GOODS dataset to apply a {\it morphological k-correction}. What this means is that we use the structural parameters from the {\it observed frame} band that corresponds most closely to the {\it rest frame} B band, given the redshift of the object. For this, the redshift range of the galaxies and the observed frame filter used for obtaining the structural parameters are paired as:\\
$0.5 \leq z \leq 0.6$ (F606W); $0.6 \leq z \leq 0.85$ (F775W); $0.85 \leq z \leq 1.2$ (F850LP). Since the minimum redshift of a galaxy in our sample is $\sim$0.5, the {\it observed frame} F435W fits were not used in our classification scheme. The structural parameters obtained by galfit for galaxies which are well fit by bulge+disk two-dimensional structure in F850LP, F775W and F606W bands are given in Table~\ref{rawparamz}, \ref{rawparami} and \ref{rawparamv} respectively. We find that galaxies that have redshifts at the border between any two of the three redshift bins described above have very comparable structural parameters in both the filters. We are unable to obtain acceptable fits for some galaxies in the shorter wavelength filters even though they have reasonable fits in the F850LP band. This is primarily due to the fact that at intermediate redshifts, the bluer filters are probing the UV region of galaxies which have non-uniform and clumpy light distribution, making it difficult to obtain a meaningful fit.

In addition to the three 'Hubble types' mentioned above, an additional type, tadpole/irregular (Irr) is included in our classification scheme to describe the objects without a clear bulge+disk structure. A single nucleus offset from the ``center'' of the image (i.e. having a one sided elongated light distribution) prompted us to tag some objects as ``tadpole''. We designated a galaxy as irregular if it had a patchy, blue light distribution with no clear nucleus at all.

A quality factor running from 1 (secure fit) to 4 (fit failed) is used to give the confidence of the galfit fit. The quality flags were decided using the following guidelines:\\
Q=1: Perfect fit with $\chi^2_{red} \leq 1.3$ and bulge mag error $\leq$ 0.1 and disk mag error $\leq$ 0.1 and fractional error in bulge effective radius $r_e$, disk scale length $r_d$ and Sersic index n $\leq$ 10\%\\
Q=2: One or two of the above conditions are violated or the presence of an AGN is indicated due to CXO X-ray detection which might complicate the bulge-disk fitting.\\
Q=3: More than two of the above conditions are violated or a fit was overruled by visual inspection to classify the galaxy as Irr/Tadpole.\\
Q=4: The fit fails completely due to the presence of multiple nuclei or patchy light distribution and fits the bright nuclei rather than the underlying component, or if the galaxy is classified as a merger by visual inspection. 

In some cases, the above quality flag had to be overruled where one of the components is very weak (eg. 905983), resulting in large uncertainties in the determination of parameters of the weaker component (bulge in case of 905983). This is indicative of the fact that there are not enough counts in that particular component leading to large errorbars on the determined parameters. In such a case, the B/T ratio has large fractional errors, but it still provides a robust discrimination between a bulge dominated and a disk dominated system, and it is not proper to tag it as a poor fit.

We have used the method of elimination to classify a galaxy as a merger. If a galaxy has a sufficiently {\it{smooth}} brightness profile allowing for a good fit to be obtained by galfit, it is classified as E/S0/S depending on it's B/T ratio. If however galfit fails to provide a good fit (as per the criteria given above), it is supposed that the brightness distribution of the galaxy is too distorted to be put in either of the above three morphological classes. It is at this point that we invoke the use of colormaps and/or visual examination of ACS images to discriminate between irregulars and mergers. Mergers were essentially identified visually by the presence of multiple maxima of comparable intensity in the rest frame B band images. This was aided by the use of color maps to distinguish multiple nuclei (generally redder) from HII regions (generally bluer). 
A label is employed to identify possible merging systems (M1/M2); see the footnotes of Table~\ref{table} for details.

Even though in the first run, galfit was used in automated batch mode, the results were examined visually for obvious deficiencies such as badly fitted neighbours, or a badly fitted program galaxy due to patchy light distribution or presence of bright HII regions in the galaxy. In many cases where bad fits were obtained, a decision had to be made, for example as to whether the patchy light distribution is due to the presence of HII regions, or due to the presence of double nuclei. Galfit, by design, cannot provide the necessary information to identify a merger. Hence it is not sensitive to subtle differences between multiple nuclei and HII regions (for example) and visual examination is needed in such cases. In order to avoid individual biases, in the classification of such objects, visual inspection of the galaxy's appearance in HST/ACS imaging and color distribution were performed by two of the authors (A.R. and S.B.) independently. 
A final classification was made after discussing any inconsistent cases. Table~\ref{table} lists the results of our morphological classification, including galaxy type (Col. 9), quality factor (Col. 10) and merging type (Col. 12), and the individual description for each object is given in Section 5.

\subsection{The morphological mix of LCGs}
We have derived the following morphological mix for our sample of intermediate redshift LCGs:\\ Mergers: $\sim$36\%, Disk dominated: $\sim$22\%, S0: $\sim$20\%, Early types: $\sim$7\%, Irr/tadpole: $\sim$15\%.\\
The detailed results of our morphological classification are listed in Table~\ref{table}, and individual comments for each galaxy are given in Section 5. The significance of the relative fractions of morphological types that we obtained is discussed in Section 7.

\subsection{Color maps}

B-z color maps were produced for our entire sample of 39 LCGs. We directly used the F435W and F850LP v1.0 images, which are astrometrically aligned with each other to an accuracy of better than a fraction of a pixel. $30\kpc \times 30\kpc$ cutouts of these color maps and the associated F850LP grayscale images for each of the 39 LCGs are shown in Fig.~\ref{colormap}.

\begin{figure}[h] \centering
   \includegraphics[width=0.5\textwidth]{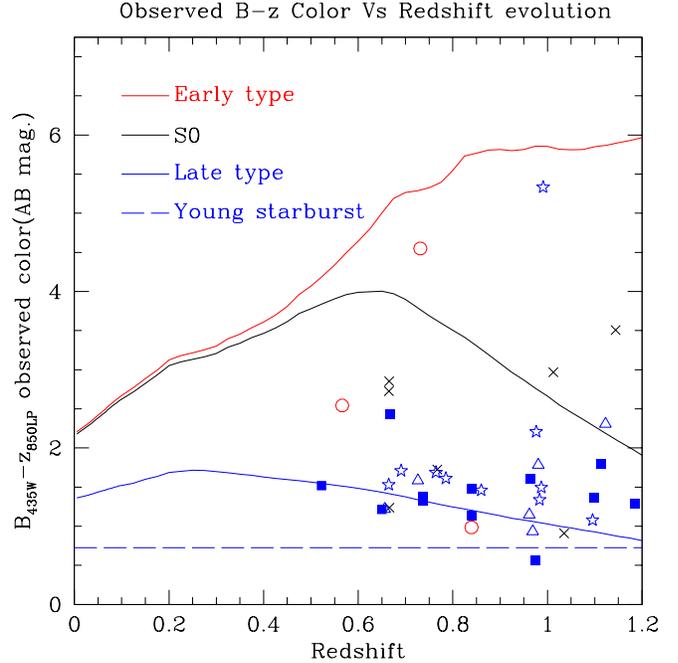}
   \caption{B-z observed color vs redshift evolution for three, Solar metallicity, formed at redshift=5.0 spectrophotometric galaxy models assuming instantaneous burst (early type), exponentially decaying SFR with an e-folding time scale of 1\gyr~(S0) and 7\gyr~(late type). A young starburst has also been plotted for comparison, assuming a power law spectra $L_{\nu} \propto \nu^{-1}$. Our sample LCGs are plotted using different symbols for different morphological class, with Early Types (E) as red open circles, S0 as black diagonal crosses, Disk dominated (S) as blue stars, Irr/Tadpole as blue open triangles and Mergers(M1/M2) as blue filled squares.}
   \label{lcg.color}
\end{figure}

We also made use of the integrated {\it observed} colors of these galaxies by comparing them with the color-redshift evolution curves as predicted by Bruzual-Charlot stellar population synthesis code GALAXEV 2003 (Bruzual and Charlot \cite{bruzual1993};\cite{bruzual2003}) for different galaxy types. We have shown in Fig.~\ref{lcg.color} three different evolution models, all for galaxies with Solar metallicity formed at redshift=5.0: an instantaneous burst corresponding to early types, an exponentially decaying SFR with an e-folding time scale of 1\gyr~corresponding to S0, and another with a time scale of 7\gyr~corresponding to late types. The e-folding timescales for different Hubble types are the same as those used by Zheng et al. (\cite{zheng2004}). The redshift evolution of their B-z observed color is plotted along with the known B-z color and redshift of the 39 LCGs segregated by their morphological class.  A young starburst has also been shown for comparison, assuming a power law spectrum $L_{\nu} \propto \nu^{-1}$.  

In Fig.~\ref{lcg.color} many LCGs lie close to the curve for late type galaxies. It is also seen that most galaxies classified as mergers are close to the late type curve (i.e. they are bluer), as opposed to E/S0 classified galaxies which are much redder (even though there is a lot of scatter). Hence, our classification acts as a crude {\it{spectrophotometric}} galaxy classification scheme in which the galaxies designated by us as E/S0 are generally redder and lie closer to the Early type or S0 curve, than those designated as disk dominated/mergers/Irr. Every one of the galaxies classified as disk dominated (S) by us is redder than the late type model curve. This might be due to the effect of dust reddening. Also, some of them can be early type spirals, which are expected to be redder than the late type curve. Most LCGs are seen to be redder than what is expected of a young starburst which is bluer than a typical late type galaxy (as shown). This might be due to the presence of an older underlying stellar population as established by Hammer et al. (\cite{hammer2001}). Also, one must remember that our B-z observed color for individual LCGs are not corrected for dust extinction, where as the Bruzual and Charlot generated curves are dust free models. The high rate of detection in mid-IR Spitzer (8/39 LCGs are detected in Spitzer, a $\sim$20\% detection rate) indicates that these are dust enshrouded starbursts where the UV light from young O/B stars is being reprocessed by the dust and is being emitted in mid-IR. This might help to explain the redder than expected colors for some of the LCGs in our sample.

\begin{table*}[h]
{\scriptsize 
\caption[]{{\it Galfit} results in F850LP band for those LCGs which are well fit by bulge+disk two-dimensional structure.}
\label{rawparamz}
\vskip.08in
\begin{center}
\begin{tabular}{cccccccc}
\hline
\hline
ID & Bulge Mag$^{\mathrm{a}}$ & Bulge $r_e$ & Bulge n & Disk mag$^{\mathrm{a}}$  & Disk $r_d$ & $\chi^{2}_{Red}$ & B/T \\
    &   & (pixels)$^{\mathrm{b}}$ &      &    &(pixels)$^{\mathrm{b}}$    &     & \\
\hline
904604   &   23.66$\pm$0.02  &     1.92$\pm$0.03  & 1.21$\pm$0.08 & 22.87$\pm$0.01   &   10.56$\pm$0.15 & 1.063 & 0.33$\pm$0.01\\
905632   &   24.04$\pm$0.08 &  2.56$\pm$0.31   &  4.29$\pm$0.43 & 23.29$\pm$0.03   &  4.87$\pm$0.07  & 1.093 & 0.33$\pm$0.03 \\
905983   &   24.55$\pm$0.23 & 19.52$\pm$12.92 &  12.01$\pm$3.70 & 21.92$\pm$0.00   & 8.00$\pm$0.06 & 1.129 & 0.08$\pm$0.02 \\
907361  &    21.63$\pm$0.01 & 5.86$\pm$0.11 & 2.71$\pm$0.04 & 23.42$\pm$0.07 & 17.33$\pm$1.13 & 1.259 & 0.84$\pm$0.07  \\
907794  &    22.76$\pm$0.11   &    2.67$\pm$0.52  &  4.64$\pm$0.70 & 23.53$\pm$0.19   &  8.05$\pm$0.46 & 1.099 & 0.67$\pm$0.09 \\
908243  &    24.07$\pm$ 0.04  &    3.65$\pm$0.11    &   0.82$\pm$0.05 & 22.21$\pm$0.01   &   10.46$\pm$0.07 & 1.168 & 0.15$\pm$0.01 \\
909093  &    23.10$\pm$ 0.02  &    8.16$\pm$0.07    &   0.53$\pm$0.02 & 23.76$\pm$0.04   &   11.33$\pm$0.41 & 1.093 & 0.65$\pm$0.02 \\
911780   &   22.72$\pm$0.02  &   7.44$\pm$0.05   &  0.73$\pm$0.01 & 21.49$\pm$0.00  &   11.27$\pm$0.06 & 1.185 & 0.24$\pm$0.00 \\
913482   &   22.04$\pm$0.00  &   1.61$\pm$0.01   &  0.98$\pm$0.01 & 21.83$\pm$0.00  &   6.64$\pm$0.05 & 1.111 & 0.45$\pm$0.00 \\
915400   &  23.55$\pm$0.42 & 51.35$\pm$73.45 & 17.69$\pm$8.14 & 22.25$\pm$0.01 & 10.27$\pm$0.08 &1.260 & 0.23$\pm$0.01  \\
916137   &  23.29$\pm$0.03 & 14.44$\pm$0.12 & 0.20$\pm$0.01 & 23.73$\pm$0.05 & 6.83$\pm$0.17 & 1.094 & 0.60$\pm$0.02  \\
916866  &    23.75$\pm$0.05  & 11.88$\pm$0.30   &    0.71$\pm$0.03 & 23.39$\pm$0.03  &     5.76$\pm$0.09 & 1.137 & 0.42$\pm$0.03 \\
919573  &    22.13$\pm$0.05  & 8.68$\pm$0.90   &    7.43$\pm$0.35 & 22.25$\pm$0.03  &     10.89$\pm$0.13 & 1.131 & 0.53$\pm$0.03 \\
919595  &    -- &  --   &  -- & --   &    -- & -- & -- \\
920435  &    23.81$\pm$0.02   &    3.36$\pm$0.03   &    0.07$\pm$0.06 & 23.77$\pm$0.01  &     4.38$\pm$0.14 & 1.116 & 0.49$\pm$0.01 \\
921406   &   23.47$\pm$0.08    &   7.03$\pm$0.55 &      2.03$\pm$0.12 & 22.61$\pm$0.04   &    9.90$\pm$0.12 & 1.125 & 0.31$\pm$0.03 \\ 
922675  &    22.75$\pm$0.02  &    17.14$\pm$0.27   &    1.14$\pm$0.02 & 22.47$\pm$0.01   &    4.85$\pm$0.03 & 1.089 & 0.44$\pm$0.01 \\
922761  &    24.33$\pm$0.05  &    8.86$\pm$0.17   &    0.23$\pm$0.03 & 23.05$\pm$0.02   &    7.64$\pm$0.10 & 1.095 & 0.24$\pm$0.01 \\
923085  &    22.70$\pm$0.01  &    8.25$\pm$0.08   &    0.93$\pm$0.02 & 24.75$\pm$0.04   &    3.51$\pm$0.14 & 1.212 & 0.87$\pm$0.01 \\
923926   &   22.58$\pm$0.02   &     3.31$\pm$0.19  &     7.76$\pm$0.77 & 23.22$\pm$0.05   &    1.97$\pm$0.02 & 1.113 & 0.64$\pm$0.02 \\
924881   &   22.59$\pm$0.01  &     6.03$\pm$0.15  &     2.99$\pm$0.07 & 23.93$\pm$0.03   &    2.07$\pm$0.05 & 1.167 & 0.77$\pm$0.01 \\
926217   &   22.70$\pm$0.05  &    11.09$\pm$1.03  &     5.48$\pm$0.27 & 22.80$\pm$0.03  &     8.21$\pm$0.12 & 1.097 & 0.52$\pm$0.03 \\

\hline

\end{tabular}
 \begin{list}{}{}
  
  \item[$^{\mathrm{a}}$]
Integrated Bulge/Disk magnitude given by {\it galfit}.
  \item[$^{\mathrm{b}}$]
$1 pixel = 0.03\arcsec$

  \end{list}

\end{center}
}
\end{table*}

\begin{table*}
{\scriptsize 
\caption[]{{\it Galfit} results in F775W band for those LCGs which are well fit by bulge+disk two-dimensional structure. The columns are the same as in Table~\ref{rawparamz}.}
\label{rawparami}
\vskip.08in
\begin{center}
\begin{tabular}{cccccccc}
\hline
\hline
ID & Bulge Mag & Bulge $r_e$ & Bulge n & Disk mag  & Disk $r_d$ & $\chi^{2}_{Red}$ & B/T \\
    &   & (pixels) &      &    &(pixels)    &     & \\
\hline
904604   &   24.67$\pm$0.04 &  1.83$\pm$0.07    & 1.31$\pm$0.14 & 23.91$\pm$0.01   &   10.43$\pm$0.33 & 1.088 & 0.33$\pm$0.01 \\
905632   &   24.24$\pm$0.06 & 2.71$\pm$0.23 & 3.29$\pm$0.27 & 23.97$\pm$0.04  &   4.36$\pm$0.09 & 1.132 & 0.44$\pm$0.03 \\
905983   &   24.73$\pm$0.30 &   19.08$\pm$17.34 & 15.77$\pm$5.67 & 22.19$\pm$0.00   &  8.33$\pm$0.05 & 1.200 & 0.09$\pm$0.02 \\
907361  &    22.26$\pm$0.03  &  5.27$\pm$0.17 &  2.50$\pm$0.05 & 23.76$\pm$0.10   &    14.45$\pm$1.43 & 1.322 & 0.80$\pm$0.04 \\
907794  &    -- &  --   &  -- & --   &    -- & -- & -- \\
908243  &    24.45$\pm$0.03    & 3.52$\pm$0.09  & 1.01$\pm$0.05 & 22.43$\pm$0.01  &    10.85$\pm$0.07 & 1.195 & 0.13$\pm$0.00 \\
909093  &    23.42$\pm$0.02    & 11.87$\pm$0.23  & 1.43$\pm$0.04 & 23.89$\pm$0.02  &    3.92$\pm$0.06 & 1.195 & 0.61$\pm$0.01 \\
911780   &   22.81$\pm$0.01 &   7.19$\pm$0.04  & 0.76$\pm$0.01 & 21.70$\pm$0.00  &    11.16$\pm$0.05 & 1.295 & 0.26$\pm$0.00 \\
913482   &   22.29$\pm$0.00 &   1.74$\pm$0.01  & 1.05$\pm$0.01 & 22.22$\pm$0.00  &    7.08$\pm$0.06 & 1.221 & 0.48$\pm$0.00 \\
915400   &   25.00$\pm$0.34   &  7.50$\pm$6.83    &  11.23$\pm$4.06 & 22.41$\pm$0.01   &   11.03$\pm$0.07 & 1.333 & 0.08$\pm$0.01 \\
916137   &   23.60$\pm$0.04   &  13.92$\pm$0.10    &  0.22$\pm$0.01 & 24.27$\pm$0.07   &   7.52$\pm$0.18 & 1.124 & 0.65$\pm$0.03 \\
916866  &    -- &  -- &  -- & --  &     -- & -- & -- \\
919573   &   22.19$\pm$0.08   &  41.48$\pm$10.85    &  17.97$\pm$1.39 & 22.68$\pm$0.02   &   9.38$\pm$0.10 & 1.172 & 0.61$\pm$0.05 \\
919595   &   24.51$\pm$0.01   &  8.59$\pm$2.31    &  0.05$\pm$0.04 & 22.32$\pm$0.00   &   7.66$\pm$0.04 & 1.182 & 0.12$\pm$0.00 \\
920435  &    24.20$\pm$0.03  &   4.99$\pm$1.02   &  0.02$\pm$0.03 & 24.17$\pm$0.02  &  2.53$\pm$0.04 & 1.143 & 0.49$\pm$0.02 \\
921406  &    24.53$\pm$0.12 &  5.51$\pm$0.74  &  2.63$\pm$0.29 & 22.74$\pm$0.02  &   9.68$\pm$0.09 & 1.186 & 0.16$\pm$0.02 \\
922675  &    22.78$\pm$0.01  &    17.64$\pm$0.22  & 1.11$\pm$0.02 & 22.66$\pm$0.01  &     4.86$\pm$0.03 & 1.142 & 0.47$\pm$0.01 \\
922761  &    24.81$\pm$0.03  &    8.77$\pm$0.13  & 0.13$\pm$0.03 & 23.28$\pm$0.01  &     8.38$\pm$0.09 & 1.124 & 0.20$\pm$0.01 \\
923085  &    23.44$\pm$0.01  &    10.05$\pm$0.08  & 0.40$\pm$0.01 & 25.66$\pm$0.04  &     2.87$\pm$0.15 & 1.248 & 0.89$\pm$0.01 \\
923926   &   23.26$\pm$0.04  &     6.42$\pm$1.44   &   10.47$\pm$1.89 & 23.84$\pm$0.05  &     1.79$\pm$0.03 & 1.163 & 0.63$\pm$0.03 \\
924881   &   22.81$\pm$0.01  &     5.94$\pm$0.17    &   3.49$\pm$0.09 & 24.18$\pm$0.03   &    2.13$\pm$0.04 & 1.277 & 0.78$\pm$0.01 \\
926217  &    23.00$\pm$0.09  &     8.41$\pm$1.03  &     3.64$\pm$0.23 & 23.00$\pm$0.07  &    10.65$\pm$0.25 & 1.119 & 0.50$\pm$0.05 \\
 
\hline

\end{tabular}

\end{center}
}
\end{table*}

\begin{table*}
{\scriptsize 
\caption[]{{\it Galfit} results in F606W band for those LCGs which are well fit by bulge+disk two-dimensional structure. The columns are the same as in Table~\ref{rawparamz} }
\label{rawparamv}
\vskip.08in
\begin{center}
\begin{tabular}{cccccccc}
\hline
\hline
ID & Bulge Mag & Bulge $r_e$ & Bulge n & Disk mag  & Disk $r_d$ & $\chi^{2}_{Red}$ & B/T \\
    &   & (pixels) &      &    &(pixels)    &     & \\
\hline
904604  &    26.32$\pm$0.05  &  1.99$\pm$0.12   &  0.91$\pm$0.25 & 25.58$\pm$0.03  &    12.27$\pm$0.81 & 1.316 & 0.34$\pm$0.02 \\
905632  &    24.91$\pm$0.02  &  1.73$\pm$0.03  &  0.71$\pm$0.08 & 24.69$\pm$0.01  &     4.55$\pm$0.14 & 1.355 & 0.45$\pm$0.01 \\
905983   &   -- &  --  &    -- & --   &    -- & -- & -- \\ 
907361   &  23.67$\pm$0.02 & 6.33$\pm$0.11 & 2.69$\pm$0.05 & 27.04$\pm$0.29 & 13.1$\pm$3.07 & 1.408 & 0.96$\pm$0.21 \\
907794   &   --  &  -- & -- & --  &    -- & -- & -- \\
908243  &    23.65$\pm$0.06 &   28.62$\pm$3.88    &  6.63$\pm$0.42 & 23.41$\pm$0.01   &    9.57$\pm$0.06 & 1.571 & 0.44$\pm$0.03 \\
909093   &   --  &  -- & -- & --  &    -- & -- & -- \\
911780  &    23.29$\pm$0.01  &  7.24$\pm$0.04    &  0.88$\pm$0.01 & 22.56$\pm$0.00 &     11.66$\pm$0.05 & 1.616 & 0.34$\pm$0.00 \\
913482  &    23.40$\pm$0.00  &  1.77$\pm$0.01    &  0.78$\pm$0.01 & 23.43$\pm$0.01 &     7.16$\pm$0.08 & 1.395 & 0.51$\pm$0.00 \\
915400  &    -- &  --  &  -- & --  &     -- & -- & -- \\
916137  &    -- &  --  &  -- & --  &     -- & -- & -- \\
916866  &    --   &   --     &  -- & --   &   -- & -- & -- \\
919573  &    23.72$\pm$0.10  &  19.81$\pm$5.93    &  15.22$\pm$1.53 & 23.93$\pm$0.02 &     9.20$\pm$0.13 & 1.365 & 0.55$\pm$0.06 \\
919595  &    --   &   --     &  -- & --   &   -- & -- & -- \\
920435   &   24.28$\pm$0.01  &  2.35$\pm$0.03   &  1.74$\pm$0.07 & 25.17$\pm$0.02   &    1.13$\pm$0.03 & 1.453 & 0.69$\pm$0.01 \\
921406   &   24.76$\pm$0.06 &  5.59$\pm$0.38  &  2.32$\pm$0.13 & 23.23$\pm$0.01  &     9.58$\pm$0.08 & 1.470 & 0.20$\pm$0.01 \\
922675  &    23.37$\pm$0.01    &  17.13$\pm$0.15  &   0.98$\pm$0.01 & 23.31$\pm$0.01  &  5.01$\pm$0.03 & 1.398 & 0.49$\pm$0.01 \\
922761  &    24.37$\pm$0.05    &  13.31$\pm$0.13  &   0.27$\pm$0.01 & 24.65$\pm$0.07  &  8.20$\pm$0.15 & 1.316 & 0.56$\pm$0.03 \\
923085  &    24.39$\pm$0.01    &  9.28$\pm$0.08  &   0.23$\pm$0.01 & 26.22$\pm$0.03  &  1.34$\pm$0.06 & 1.592 & 0.84$\pm$0.01 \\
923926  &    --  &    --  &    -- & --   &   -- & -- & -- \\
924881  &    23.53$\pm$0.01   &    3.02$\pm$0.02   &    1.98$\pm$0.03 & 25.04$\pm$0.01   &    1.21$\pm$0.02 & 1.565 & 0.80$\pm$0.01 \\
926217   &   --  &    --   &    -- & --  &   -- & -- & -- \\
 
\hline

\end{tabular}

\end{center}
}
\end{table*}

\section{Input from other wavelengths}
We have looked for the detection of our sample LCGs in other wavelength bands, to calculate their stellar masses, star formation rates (SFR) \& to look for possible AGN activity etc. Since CDFS is one of the most data rich regions on the sky, we have access to public data on objects in this region at various wavelengths such as radio, X-ray and mid-infrared. The results of our analysis in these wavelength bands are given below.

\subsection{Radio emission from LCGs}

Koekemoer et al. (\cite{koekemoer2006}) have prepared a comprehensive catalog of all radio sources in an area of $1.2\degsq$ covering the CDFS to a limiting (1$\sigma$) sensitivity of $\sim$14$\mu$Jy at $1.4\ghz$ using the Australia Telescope Compact Array (ATCA). Afonso et al. (\cite{afonso2006}) have published optical and X-ray identifications of faint radio sources in the GOODS-CDFS ACS field. From this catalog we find that only 1 out of 39 LCGs of our sample had any radio emission. The radio flux of this sole detected galaxy 906961 is 79$\mu$Jy (see individual description of this galaxy in Section 5). This puts an upper limit of $\sim$14$\mu$Jy at $1.4\ghz$ for the radio flux from the other 38 LCGs in our sample. Assuming a power law radio spectrum $L_{\nu} \propto \nu^{-\alpha}$, with a spectral index $\alpha=0.5$ (which is typically used as the dividing line between steep and flat spectrum radio sources), and using the highest redshift for our LCG sample (z=1.2), puts a strict upper limit of $\sim$$10^{39}\lunit$ on the $1.4\ghz$ luminosity for the 38 non-detected LCGs. Since the VVDS spectrum does not have enough spectral coverage to discriminate between AGN and starbursts using line ratios, we decided to use X-ray and/or radio detection as a sign of AGN activity, so as to make sure that our LCG sample does not suffer from any serious AGN contamination. The limit that we derived for the $1.4\ghz$ luminosity rules out any radio loud AGN contamination of our sample. 

It must be mentioned here that the beam size of the ATCA radio survey was rather broad at $16.8\arcsec \times 6.95\arcsec$. Since the typical size of our galaxies is of the order of $\sim$1$\arcsec$, this restricts the usefulness of this radio catalog for our purpose.               

\subsection{CXO X-ray detection}

Giacconi et al. (\cite{giacconi2002}) have published a X-ray source catalog obtained from the $\sim$1Msec exposure of the {\it{Chandra Deep Field South}} (CDFS), using the Advanced CCD Imaging Spectrometer (ACIS) on the Chandra X-ray Observatory, as well as their optical counterparts.

We used a very simple prescription of looking for potential X-ray counterparts for our sample of 39 galaxies by searching for X-ray sources within a $6\arcsec$ search radius of the program galaxy using the NASA Extragalactic Database (NED). Each potential X-ray counterpart was then checked visually using optical counterpart image cutouts published by Giacconi et al. (\cite{giacconi2002}). 

This simple method resulted in X-ray emission being confirmed from four of the 39 LCGs. All four X-ray sources are confirmed to be AGN by Zheng W. et al. (\cite{zhengw2004}). The hardness ratio (HR) for these galaxies as given by Giacconi et al. (\cite{giacconi2002}) is listed in Table~\ref{table}. These are mainly provided for the reader who might want to distinguish between Type 1 and Type 2 AGNs, although such a distinction was not necessary for our work (see Zheng W. et al. \cite{zhengw2004}). Interestingly, we find that all the 4 LCGs with X-ray emission are E/S0 type galaxies. Two of these LCGs show clear blue point like nucleus in the colormaps (Fig.~\ref{colormap}), while in the other two cases, the AGN might be obscured. We also noticed that 3 of these X-ray detected LCGs are detected by Spitzer MIPS (see next section). This is not surprising given the known AGN contribution to mid-infrared surveys (Fadda et al.~\cite{fadda2002}).

\subsection{Spitzer MIPS detection and SFR for LCGs}
The GOODS Spitzer Legacy program observations cover two fields on the sky. One of these fields (GOODS-N) coincides with the Hubble Deep Field North, while the other (GOODS-S) coincides with the Chandra Deep Field South.

We made use of the public data made available as part of {\it{The Great Observatories Origins Deep Survey}} (GOODS), Spitzer Legacy Data Products, Third Data Release (DR3). In addition to other data products, this release consists of the "best-effort" reductions of 24 micron data for the southern GOODS field taken with the Multiband Imaging Photometry for Spitzer (MIPS), and  a 24 micron v0.91 source list of all sources brighter than 80$\mu$Jy. Dickinson et al. (\cite{dickinson2006}) describe the {\it{Spitzer}} observations, data reductions, and the data products. Chary et al. (\cite{chary2006}) explain the preparation of the MIPS catalog in detail.

We performed a simple positional cross correlation of our 39 galaxy sample with the v0.91 MIPS source list with a tolerance radius of $2.5\arcsec$, which was chosen keeping in mind the $\sim$ 5$\arcsec$ PSF of the Spitzer MIPS observations. This yielded 24$\mu$m fluxes for 8 of the 39 LCGs. It was found that the maximum separation between the HST-Spitzer counterparts in case of a true match was of the order of $\sim$0.5$\arcsec$. Visual examination of the images was employed to ensure a one-to-one correspondence between the HST and the Spitzer sources. The other 31 LCGs for which no counterpart was found in the Spitzer MIPS catalog, have an upper limit of $80\mu$Jy for their 24$\mu$m fluxes. We calculated the infra-red luminosities $L_{IR}$ for the 8 galaxies using the approximations given in Chary and Elbaz (\cite{chary2001}), using the {\it{rest frame}} 12 and 15 $\mu$m luminosities, which were in turn determined using the 24$\mu$m MIPS fluxes. We used the approximation of 15$\mu$m for galaxies with $0.5 \leq z \leq 0.8$ and that of 12$\mu$m for galaxies with $0.8 \leq z \leq 1.2$ to calculate the IR luminosity. The IR luminosities obtained using the two approximations agree to within 0.1 dex. We also calculated the upper limits for the infra-red luminosities $L_{IR}$ for the other 31 LCGs with no counterparts in Spitzer/MIPS catalog. These values for the $L_{IR}$ are listed in Table~\ref{table}.

Kennicutt (\cite{kennicutt1998}) has transformed the IR luminosity of young (age $<$ $10^{8}$yr) starburst galaxies to a star formation rate (SFR). He gives an approximate estimate of the dust-enshrouded SFR $\rho$          
\begin{equation}
 ~~~~~~~~~~~~~~~~~ \rho (M_{\odot}yr^{-1}) = 1.71 \times 10^{-10}L_{IR}(L_{\odot})
\end{equation}
Using this relation, we obtain SFRs varying from $20 - 65 M_{\odot}yr^{-1}$ for our sample LCGs with 24$\mu$m MIPS fluxes, which is the range expected for galaxies of this class (eg. Hammer et al.~\cite{hammer2001}). For the objects which have no detection in Spitzer/MIPS catalog, this method yielded an upper limit on their SFR. These values are listed in Table~\ref{table}.  

In order to be able to better constrain the SFR of those objects that are not detected in the Spitzer/MIPS catalog, we used their rest frame UV continuum flux to derive an estimate of the SFR. This method assumes that the rest frame UV flux that is observed for the objects is directly emitted by the O/B type stars in a starburst and there is no dust extinction (see Kennicutt~\cite{kennicutt1998} for details). Since galaxies are dusty, the UV flux undergoes varying amount of extinction, and hence the SFR yielded by this method is to be treated as a lower limit. Hence the SFR derived by this method are listed as lower limits in Table~\ref{table}.   

\subsection{Stellar masses of our sample LCGs}
Stellar masses for our sample LCGs were calculated on the basis of their absolute K band magnitudes, corrected for the presence of red supergiant stars, using the empirical relationship of Bell et al. (\cite{bell2003}). A variation of the Salpeter Initial Mass Function (Salpeter~\cite{salpeter1955}), called {\it{diet}} Salpeter IMF was used as explained in Bell et al. and the values were calculated in a consistent manner as in Hammer et al. (\cite{hammer2005}). The derived values of stellar mass for our sample are listed in Table~\ref{table}. To derive the values of stellar mass using a classical Salpeter IMF, it is necessary to add 0.15 dex to the stellar mass values listed in Table~\ref{table}. 
Accuracy is better than 0.1-0.2 dex as it has been shown to correlate particularly well (for rotating disks) with log($V_{max}$) at z=0.6 by Flores et al. (\cite{flores2006}).


\begin{table*}[h]
 {\footnotesize
  \caption[]{Catalog of derived parameters for LCGs}
  \label{table}
  \begin{tabular}{ccccccccllcl}
  \hline
  \noalign{\smallskip}
& &Stellar Mass$^{\mathrm{a}}$ & &SFR($\rho$)& &\multicolumn{2}{c}{\hrulefill Rest Frame B \hrulefill}&& && \\
  Our ID & $z$ & $Log_{10}(M/M_{\odot})$ & $L_{IR}(10^{10} L_{\odot})$ &  ($M_{\odot}yr^{-1}$)& HR$^{\mathrm{b}}$  &  $B/T$  & $\chi^{2}_{red}$    &  Type$^{\mathrm{c}}$ & $Q^{\mathrm{d}}$ &$R_{d}(kpc)^{\mathrm{e}}$ & M1/M2$^{\mathrm{f}}$  \\
(1) & (2) & (3) & (4) & (5) & (6) & (7) & (8) & (9) & (10) & (11) & (12)\\
  \noalign{\smallskip}
  \hline
  \noalign{\smallskip}

904260 & 0.983 & 10.32  &  $<$ 10.37    & 3.62 $<$ $\rho$ $<$ 17  & &  --     &  --   &  S & 3 & & M2      \\
904604 & 0.990 & 10.94 & $<$ 10.59     & 0.13 $<$ $\rho$ $<$ 18  & &  0.33$\pm$0.01     &  1.063   &  S & 1 &2.53$\pm$0.04 &    \\
904680 & 0.964 & 10.28  & $<$ 9.83      & 3.28 $<$ $\rho$ $<$ 16  & &  --     &  --   &   & 4 & & M2   \\
905632 & 0.976 & 10.23  & $<$ 10.18     & 0.87 $<$ $\rho$ $<$ 17	& &  0.33$\pm$0.03   &  1.093  &  S & 1 &1.16$\pm$0.02 &     \\	
905983 & 0.860 & 10.36  & $<$ 7.22      & 3.87 $<$ $\rho$ $<$ 12	& &  0.08$\pm$0.02   &  1.129  &  S & 2 &1.85$\pm$0.01	&     \\
~906961$^{\mathrm{g}}$ & 0.566 & 10.81  &     70.36     & $<$ 120	 &-0.55$\pm$0.03 & -- &  --   &  E & -- &	&   \\	
907047 & 1.112 & 10.65  & $<$ 14.49     & 3.32 $<$ $\rho$ $<$ 24	& &  --	   &  --   &   & 4 &  &	M1     \\
907361 & 0.731 & 10.96 & $<$ 6.69      & 0.18 $<$ $\rho$ $<$ 11	& &  0.80$\pm$0.04   &  1.322  &  E & 2 &	&    \\	
907794 & 1.144 & 10.92 &     20.82     & 35& & 0.67$\pm$0.09   &  1.099  &  S0 & 1 &	&     \\
908243 & 0.726 & 10.23  & $<$ 6.57      & 2.41 $<$ $\rho$ $<$ 11	& &  0.13$\pm$0.00   &  1.195  &  Tad & 3 &   &  \\	
909015 & 1.039 & 10.25  & $<$ 12.07     & 0.29 $<$ $\rho$ $<$ 20	& &  --	   &  --   &   & 4 &	   &  \\
909093 & 0.968 & 9.71  & $<$ 9.97      & 3.37 $<$ $\rho$ $<$ 17	& &  0.65$\pm$0.02   &  1.093  &  Tad & 3 &	&     \\
909429 & 0.737 & 10.32  & $<$ 6.81      & 4.48 $<$ $\rho$ $<$ 11	& &  --	   &  --   &   & 4 &  &	M2     \\	
910413 & 0.655 & 10.67  &     15.84     & 27& &  --   &  --  &  Tad & 4 &  &	M2     \\
911747 & 0.840 & 10.18  & $<$ 6.79      & 4.81 $<$ $\rho$ $<$ 11	 & & --	   &  --   &   & 4 &  &	M2     \\	
911780 & 0.664 & 10.38  & $<$ 5.29      & 3.69 $<$ $\rho$ $<$ 9	 & & 0.26$\pm$0.00   &  1.295  &  S & 2 &2.34$\pm$0.01    &	     \\
911843 & 0.973 & 9.75  & $<$ 10.09     & 5.50 $<$ $\rho$ $<$ 17	 &  &--	   &  --   &   & 4 &	& M2     \\
912744 & 0.690 & 10.40  & $<$ 5.81      & 2.31 $<$ $\rho$ $<$ 9	 &  &--   &  --  &  S & 3 &	  &   \\
913482 & 0.664 & 10.71  & 27.17         & $<$ 46 &0.52$\pm$0.06 & 0.48$\pm$0.00	   &  1.221 &  S0 & 3 &	 &   \\	
914038 & 0.667 & 10.50  & 12.00         & 20& & --	   &  --   &   & 4 &	& M2     \\
915400 & 0.764 & 10.42  & $<$ 7.44      & 1.73 $<$ $\rho$ $<$ 12	 & & 0.08$\pm$0.01 &  1.333  &  S & 2 &2.45$\pm$0.02	&     \\	
916137 & 0.980 & 10.33  & $<$ 10.29     & 1.40 $<$ $\rho$ $<$ 17	 & & 0.60$\pm$0.02   &  1.094  &  Irr & 3 &   &	     \\
916446 & 0.839 & 10.10  & $<$ 6.77      & 1.75 $<$ $\rho$ $<$ 11	 & & --	   &  --   &   & 4 &	& M2     \\
916866 & 0.987 & 10.22  & $<$ 10.48     & 1.68 $<$ $\rho$ $<$ 17	 & & 0.42$\pm$0.03   &  1.137  &  S0 & 2 &	  &   \\
918147 & 1.099 & 10.77  & 39.48         & 67&  &--	   &  --   &   & 4 &	& M1     \\
919573 & 0.665 & 10.61  & 5.89          & $<$ 10&-0.40$\pm$0.03  & 0.61$\pm$0.05	   &  1.172 &  S0 & 3 &	&   \\	
919595 & 0.785 & 10.24  & $<$ 7.95      & 2.43 $<$ $\rho$ $<$ 13	 &  &0.12$\pm$0.00   &  1.182  &  S & 3 &	   &  \\
920435 & 1.034 & 9.44  & $<$ 11.91     & 2.66 $<$ $\rho$ $<$ 20	 &  &0.49$\pm$0.01   &  1.116  &  S0 & 2 &    &	     \\	
921406 & 1.095 & 10.37  & $<$ 13.89     & 5.22 $<$ $\rho$ $<$ 23	 &  &0.31$\pm$0.03   &  1.125  &  S & 1 &2.43$\pm$0.03	  &   \\
922675 & 0.666 & 10.13  & $<$ 5.32      & 2.79 $<$ $\rho$ $<$ 9	 &  &0.47$\pm$0.01   &  1.142   &  S0 & 1 &	&     \\
922733 & 0.650 & 10.05  & $<$ 5.01      & 2.24 $<$ $\rho$ $<$ 8	 &  &--	   &  --   &   & 4 &	& M2     \\
922761 & 0.961 & 9.71  & $<$ 9.77      & 2.56 $<$ $\rho$ $<$ 16	 &  &0.24$\pm$0.01   &  1.095&  Irr & 3 & &      \\	
923085 & 1.122 & 10.68  & 15.74         & 26&  &0.87$\pm$0.01   &  1.212  &  Irr & 3 &    &      \\
923926 & 1.012 & 10.74  & $<$ 11.23     & 1.17 $<$ $\rho$ $<$ 19	 &0.12$\pm$0.07  &0.64$\pm$0.02   &  1.113   &  S0 & 3 &	&    \\	
924881 & 0.839 & 9.80  & $<$ 6.77      & 2.95 $<$ $\rho$ $<$ 11	 &  &0.78$\pm$0.01   &  1.277   &  E & 2 &	&     \\
926109 & 0.522 & 10.19  & $<$ 2.95      & 1.55 $<$ $\rho$ $<$ 5	 &  &--	   &  --   &   & 4 &	& M1     \\	
926217 & 0.767 & 10.27  & $<$ 7.50      & 1.84 $<$ $\rho$ $<$ 12	 &  &0.50$\pm$0.05   &  1.119   &  S0 & 1 &	&     \\
907305 & 1.185 & 10.11  & $<$ 17.21     & 3.17 $<$ $\rho$ $<$ 29	 &  &--	   &  --   &   & 4 &	& M1     \\	
914895 & 0.736 & 10.09  & $<$ 6.78      & 2.36 $<$ $\rho$ $<$ 11	 &  &--	   &  --   &   & 4 &	& M1     \\

  \noalign{\smallskip}
  \hline
  \end{tabular}
  \begin{list}{}{}
  
  \item[$^{\mathrm{a}}$]
Derived using a diet Salpeter IMF. In order to derive values using a classic Salpeter IMF, add 0.15 dex to the values listed here.		
  \item[$^{\mathrm{b}}$]
X-ray hardness ratio, defined as (H-S)/(H+S) where H and S are the counts in the hard and soft bands respectively.
  \item[$^{\mathrm{c}}$]
Galaxy type --- {\bf E}: 0.8\,$<B/T\leq$\,1; {\bf S0}: 0.4\,$<B/T\leq$\,0.8; {\bf S}: 0.0\,$<B/T\leq$\,0.4; {\bf Irr}: irregular; {\bf Tad}: Tadpole. 
  \item[$^{\mathrm{d}}$]
$Q$ quality factor --- 1: secure; 2: possibly secure; 3: insecure; 4: fit failed.
  \item[$^{\mathrm{e}}$]
Exponential disk scale length(kpc) for disk dominated galaxies.
  \item[$^{\mathrm{f}}$]
Merging? --- M1: obvious merging; M2: possible merging.
 \item[$^{\mathrm{g}}$]
This object could not be fitted properly due to the presence of a strong point source at the center. 
  \end{list}
}
\end{table*}






\section{Individual description}

In Fig.~\ref{colormap}, color map stamps (right) of the 39 LCGs are shown, along with $F850LP$ imaging graylevel stamps (left). The color bar in each color map stamp shows the B-z color range from 0 to 4, except for objects ID 904604, 907361 and 909015 where it ranges from 0 to 6. A description of each target is presented below. The FITS files of all our 39 LCGs in each of the four filters can be downloaded from the following url http://www.iucaa.ernet.in/$\sim$rawat/lcg\_1.html. They are far more informative than the postscript versions depicted here.    

\begin{description}

\item[{\bf  904260}] This is a possible ongoing merger. Spiral arms are discernible in one of the constituent galaxies, specially in the B band image. The nucleus of the other member is extremely blue indicating star formation. The colormap also shows the spiral arms to be bluer than the rest of the galaxy.

\item[{\bf 904604}] This is a featureless edge on disk galaxy. It practically drops out of the blue filter and is hence quite a red object with a rest frame $(U-B)_{0}=1.47$ mag. 

\item[{\bf 904680}] This object has two nuclei which becomes apparent in the shorter wavelength images as well as in the color map. Both the nuclei are bluer compared to the surrounding area. This is possibly an ongoing merger.

\item[{\bf 905632}] This is a relatively featureless disk galaxy. The colormap shows that the core is bluer compared to the outer edges of the galaxy. Also the OII emission line is very strong at 107.8$\AA$. However, there is no detectable X-ray/radio emission from this galaxy as described in the text. Also there are no signs of nuclear activity, hence there are no indications for an AGN.
 
\item[{\bf 905983}] This looks like a face on spiral galaxy. The spiral arms are easily visible in the summed image as well as in the galfit residuals, especially in the shorter wavelength fits. Bright HII regions can be seen embedded inside the spiral arms in B and V band images. The colormap shows that the spiral arms are bluer than the rest of the galaxy.

\item[{\bf 906961}] This has the appearance of an early type galaxy with a distinct blue point like core as shown by the color map. A satisfactory bulge-disk decomposition could not be obtained due to the presence of the strong point source at the center. This galaxy has X-ray detection by the CXO with a HR of -0.55, and radio detection with $f_{\nu_{0}}$=79$\mujy$ at 1.4Ghz. A Spitzer MIPS flux density of 1580$\mujy$ at 24$\mu$m gives it an $L_{IR}$=$7 \times 10^{11}L_{\odot}$ putting it in the category of Luminous Infra Red Galaxies (LIRGs). All the evidence points to the presence of an AGN.

\item[{\bf 907047}] This galaxy is an ongoing merger and morphological classification was not possible. At least two nuclei are clearly seen in the summed image, and the colormap shows the galaxy to be rather blue.

\item[{\bf 907305}] These are actually two merging galaxies with a separation of about 0.35\arcsec. The fit failed. There are two distinct nuclei plus a halo that connects the two galaxies. The nuclei are bluer than the halo. 

\item[{\bf 907361}] This is an early type bulge dominated galaxy. The galaxy is extremely red with a rest frame $(U-B)_{0} = 1.65$ mag. The galaxy disappears in the B band.

\item[{\bf 907794}] This galaxy is compact, isolated and featureless. The fit quality is very good and the appearance is that of an E/S0 galaxy with red color as shown by the color map. The inferred rest frame $(U-B)_{0} = 1.34$ magnitude. This galaxy has a Spitzer MIPS $f_{v_{0}}$=104$\mujy$ at 24$\mu$m yielding a $L_{IR}$=$2 \times 10^{11}L_{\odot}$ making it a LIRG. The SFR for this galaxy is 35 $M_{\odot}$/year.

\item[{\bf 908243}] A linear/edge on galaxy with a bright nucleus quite offset from the center of the galaxy. The nucleus and the northern part of the galaxy are bluer than the rest of the galaxy. This is designated a tadpole.

\item[{\bf 909015}] This is a point source with the morphology undetermined. The object is red as seen from the color map. The integrated B-z color is $\sim$5 magnitudes. Also, with a $M_{B}$ of only -20.25, it is unlikely to be a quasar. Since this is a non-detection in Spitzer MIPS, it has an upper limit of 80$\mujy$ at 24$\mu$m. Perhaps this is a star misidentified by VVDS.

\item[{\bf 909093}] This galaxy is shaped like a tadpole. There is an intensity maximum located towards the head of the tadpole (due south-west), ie {\it {not}} at the center of the galaxy. The fit is unreliable due to the peculiar morphology. Again, the head of the tadpole is bluer than the rest of the galaxy.

\item[{\bf 909429}] This is a possible ongoing merger. At least two nuclei are distinctly visible in the i-band image. The nucleus at the northern part of the system is blue, whereas the other nucleus is red. The fit failed.

\item[{\bf 910413}] This galaxy is shaped like an elongated tadpole. Two distinct intensity maxima are clearly seen. Both the maxima are bluer than the rest of the galaxy. The fit is unreliable. This might be an ongoing merger M2. This galaxy has a Spitzer MIPS $f_{v_{0}}$=248$\mujy$ at 24$\mu$m, $L_{IR}=1.6 \times 10^{11}L_{\odot}$, making it a LIRG with a SFR of 27$M_{\odot}$/year.

\item[{\bf 911747}] This is a possible ongoing merger system M2. Two intensity maxima are distinctly seen in the z band image, embedded in an elongated structure(visible in the summed image), which might be the tidal tail. Interestingly, one of the maxima is blue while the other is red. The fit failed.

\item[{\bf 911780}] This looks like an edge on disk dominated galaxy. The B band image looks like a tadpole and is markedly offset with respect to the z band image. This is obvious from the color map, which shows that the blue region is offset from the center of the galaxy. This might be a result of dust attenuation in the B band or alternatively off center star formation.

\item[{\bf 911843}] A double nucleus is suspected in this system. The galaxy is shaped like a kidney. The north-west part of the galaxy is brighter and bluer than the rest of the galaxy. This brighter part is in fact composed of two nuclei which are barely resolved and are extremely blue with the rest frame $(U-B)_{0} = 0.49$ magnitude. In some pixels the B-z color is negative, but this could partly be a consequence of the ACS z band red leak which transfers z band flux away from the center of bright nuclei and makes the center look bluer than it really is (see Appendix B for details). This is a possible merger system. The fit failed.

\item[{\bf 912744}] A face on disk galaxy with spiral arms clearly discernible in the summed image. There is a very bright and blue HII region on the south-west side. The patchy light distribution due to the presence of HII regions as well as the spiral arms make it difficult to obtain a meaningful fit. The colormap clearly shows the HII region to be bluer than the rest of the galaxy.

\item[{\bf 913482}] This is a compact, isolated and featureless E/S0 galaxy with an AGN at the center. The core is point like and bluer than the outer regions of the galaxy. The fit is unreliable up due to the presence of the central point source. A visual examination of the VVDS spectra of this galaxy confirms that this is a type 2 AGN. This galaxy is a X-ray source with HR=0.52, as well as a Spitzer MIPS source with $f_{v_{0}}=412.0\mujy$ at 24$\mu$m and $L_{IR}=2.7 \times 10^{11}L_{\odot}$ making it a LIRG. Since this is an AGN, our prescription for calculating the SFR only gives an {\it{upper limit}} of 46$M_{\odot}$/year.

\item[{\bf 914038}] The morphology of this galaxy is knotty and it is probably a merger system M2. Double nuclei is suspected at the south-east corner of the galaxy evident from the i band image. The color map shows a blue region on the south-west side of the galaxy. This galaxy is a Spitzer MIPS source with $f_{v_{0}}=180.0\mujy$ at 24$\mu$m and $L_{IR}=1.2 \times 10^{11}L_{\odot}$ making it a LIRG with a SFR of 20$M_{\odot}$/year.

\item[{\bf 914895}] This is most probably an ongoing merger. There is possibly an underlying disk galaxy with another galaxy merging with it (due north). The nucleus of the secondary member is very blue as seen in the colormap as well as the B band image.

\item[{\bf 915400}] This is a flocculent spiral galaxy seen edge on as confirmed by the fit. The color map shows blue colors on the outer edges of the disk, which might be indicative of inside-out disk formation.

\item[{\bf 916137}] This is a linear and flocculent galaxy. The light distribution is patchy and the center is not obvious. The fit to this irregular galaxy is unreliable. The north-east side of the galaxy is bluer compared to the other parts of the galaxy as seen in the color map. The summed image shows a more robust structure of the galaxy.

\item[{\bf 916446}] This system is a possible ongoing merger. There is a strong peak at the south-east side and a very faint peak at north-west side as seen in the i band image. There is some saturation in the pixels in the B band image, hence we have to disregard the color map for this galaxy. The fit failed.

\item[{\bf 916866}] This is an intermediate type galaxy (S0). The nucleus is slightly offset from the center of the galaxy. The colormap shows that the south-west side is bluer than the rest of the galaxy. This galaxy is the host galaxy for the type Ia SN 2002ga (Strolger et al. \cite{strolger2004}).

\item[{\bf 918147}] This system is an ongoing merger M1. The two nuclei are distinctly visible in both B band and z band. The fit failed. The colormap shows that the south-east side nucleus is blue, and the north-west side nucleus is redder. This system has a Spitzer MIPS detection with $f_{v_{0}}=206.0\mujy$ at 24$\mu$m giving it an $L_{IR}=3.9 \times 10^{11}L_{\odot}$, making it a LIRG with a SFR of 67$M_{\odot}$/year.

\item[{\bf 919573}] This is an isolated galaxy with a suspected AGN due to the presence of a bright point like nucleus. The fit suggests this to be an S0 galaxy, but is unreliable due to suspected AGN activity. The colormap shows that the south side of the galaxy is redder. This might be indicative of off-center star formation in the north side of this galaxy. The intensity maximum in the z band is offset from the center of the galaxy. This galaxy is an CXO X-ray source with HR=-0.40, and has a Spitzer MIPS detection with $f_{v_{0}}=88.8\mujy$ at 24$\mu$m, giving it an $L_{IR}=5.9 \times 10^{10}L_{\odot}$. Due to the suspected AGN activity, our prescription for calculating the SFR gives an {\it{upper limit}} of 10$M_{\odot}$/year.

\item[{\bf 919595}] This seems to be a disk galaxy with evidence of patchy star formation as seen in the color map as well as in the B band image. At least four prominent star formation HII regions can be seen in the B band image. The fit is unreliable due to the patchy nature of this galaxy. The z band image shows a single nucleus, hence the data does not suggest a recent merger.

\item[{\bf 920435}] {\it Galfit} suggests this to be an S0. The color map shows double nuclei (both are blue). The double nucleus is also seen in the B band image, but in the z band, the second nucleus disappears. This might be due to the presence of a dust patch which causes attenuation in the B band but not in the z band. Otherwise the galaxy is isolated with no signs of interaction/merger. Designated an S0.

\item[{\bf 921406}] The fit is excellent and this is designated to be a disk galaxy. It is an edge on disk galaxy with a low surface brightness extension whose reason is not clear. The whole galaxy is blue in color. The disky nature of the galaxy is most clear in the summed image. 

\item[{\bf 922675}] The galaxy is elongated. There is only one nucleus with a light distribution that is somewhat elongated to one side of the nucleus. The "tail" is bluer than the nucleus of the galaxy. The fit is excellent and the galaxy is designated to be a S0. 

\item[{\bf 922733}] There is probable evidence for two nuclei specially in the summed image. The fit failed. This object is shaped like an elongated chain with multiple (two) peaks in intensity. This is possibly a merging system. The whole chain is blue in color.

\item[{\bf 922761}] The z band image of this galaxy is irregular with no obvious center. The fit is unreliable. The color map is blue throughout but is patchy with no obvious central intensity maximum. This is a blue irregular galaxy.

\item[{\bf 923085}] An irregular galaxy with multiple intensity maxima seen specially in the summed image. The fit is unreliable. There is a red patch at the core of the galaxy(dust?) surrounded by a relatively blue region. This galaxy has a Spitzer MIPS detection with $f_{v_{0}}=84.4\mujy$ at 24$\mu$m giving it an $L_{IR}=1.6 \times 10^{11}L_{\odot}$ making it a LIRG with a SFR of 26$M_{\odot}$/year.

\item[{\bf 923926}] Very compact, isolated and featureless galaxy. This galaxy has CXO X-ray detection with a HR of 0.12 which might be indicative of AGN activity, even though there are no discernible signs of point like activity from visual examination of the galaxy image. It is designated S0 but the fit is unreliable due to the possible presence of the AGN activity. The galaxy as a whole is red with a rest frame $(U-B)_{0} = 1.22$ mag.

\item[{\bf 924881}] This galaxy has extensions at low surface brightness levels. The nucleus is offset from the center of the galaxy. The fit is reasonable and this is designated to be an E type galaxy. The colormap shows the center of the galaxy to be bluer compared to the rest of the galaxy.

\item[{\bf 926109}] A simple visual examination of the B band image confirms that this is an ongoing merger. There are at least four intensity maxima (star forming regions) with no tendency to form any regular pattern. These star forming regions also show up in the color map. The fit failed.

\item[{\bf 926217}] This is an isolated lenticular galaxy. The fit is excellent. The core of the galaxy is redder than the rest of the galaxy. There is some indication of a faint outer disk in the summed image, but that is not confirmed. This is designated as a lenticular.

\end{description}

\begin{figure*}[h!] \centering
\includegraphics[height=0.22\textwidth,clip]{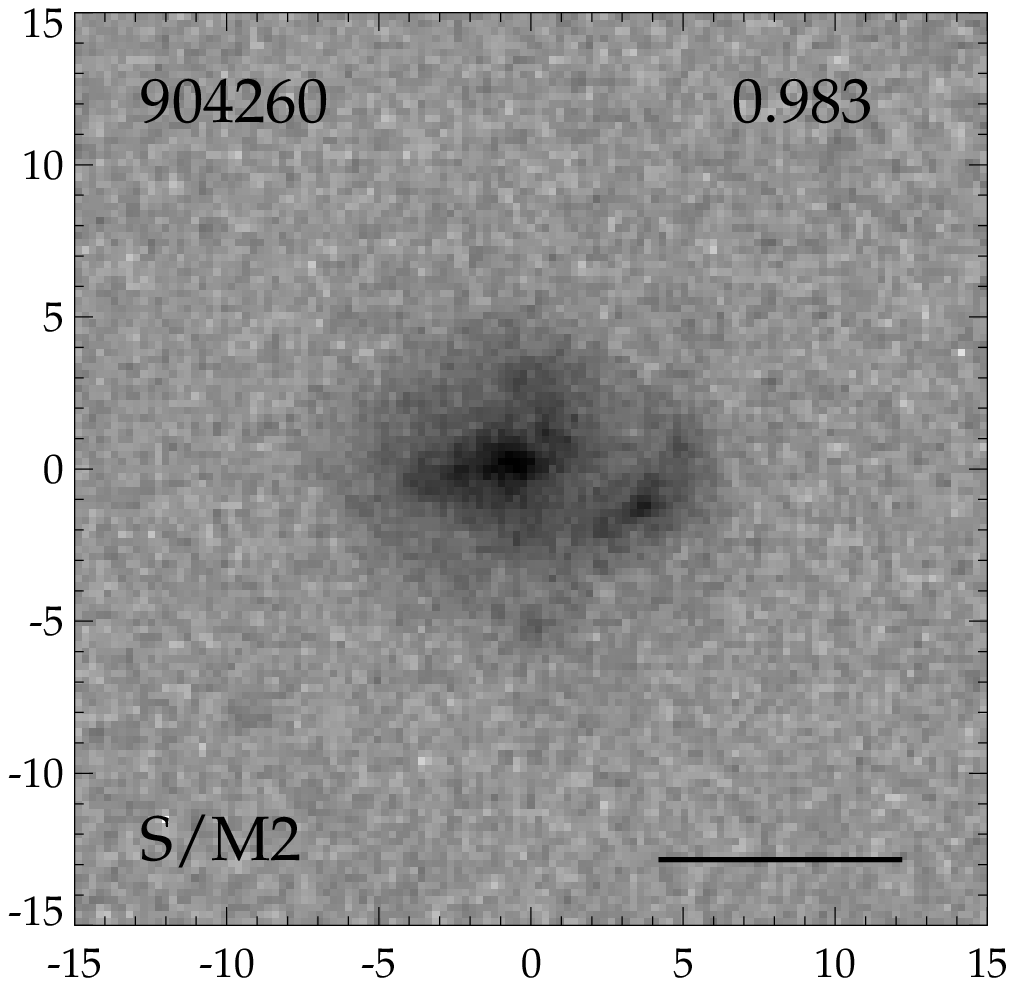} \includegraphics[height=0.22\textwidth,clip]{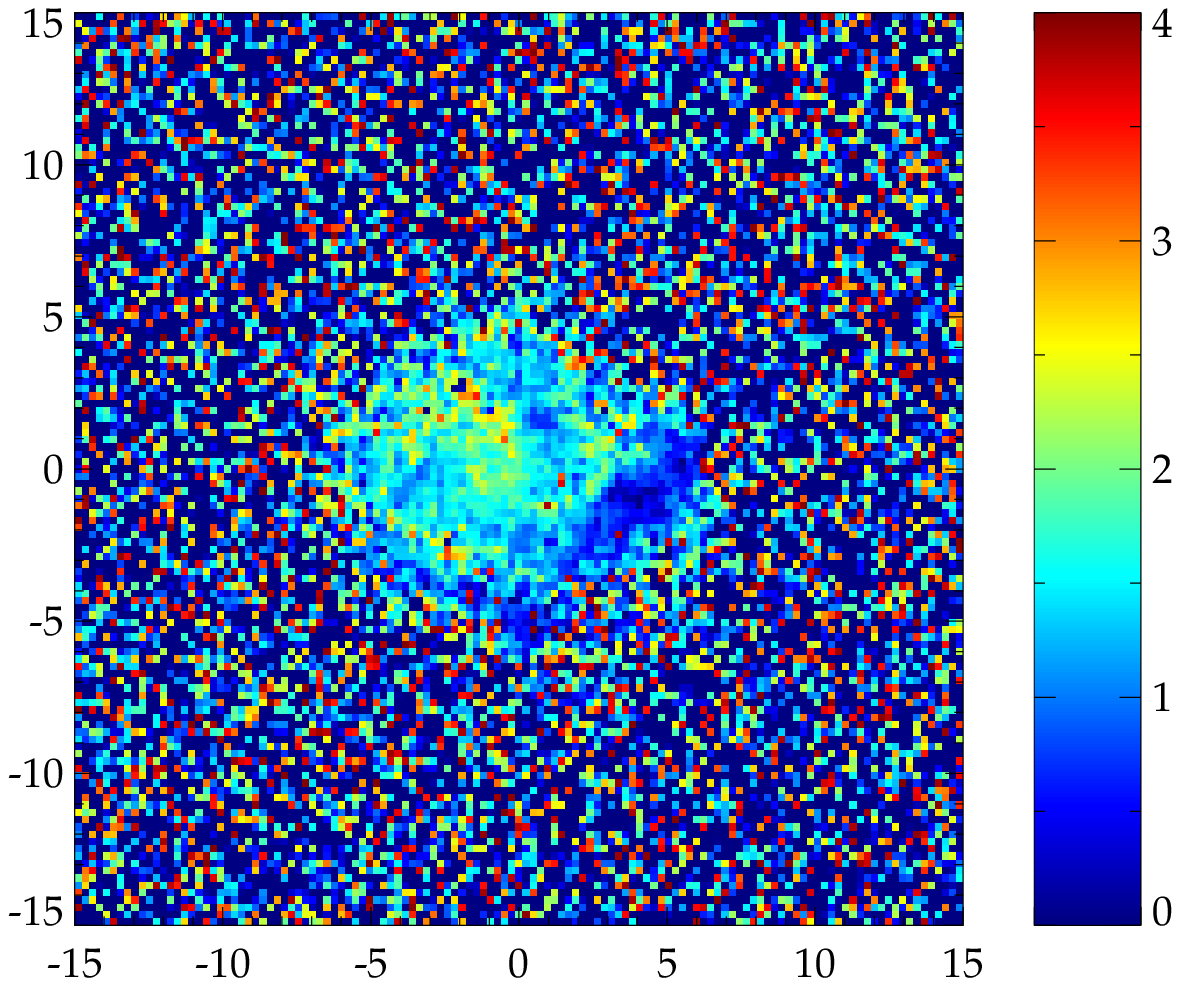}
\includegraphics[height=0.22\textwidth,clip]{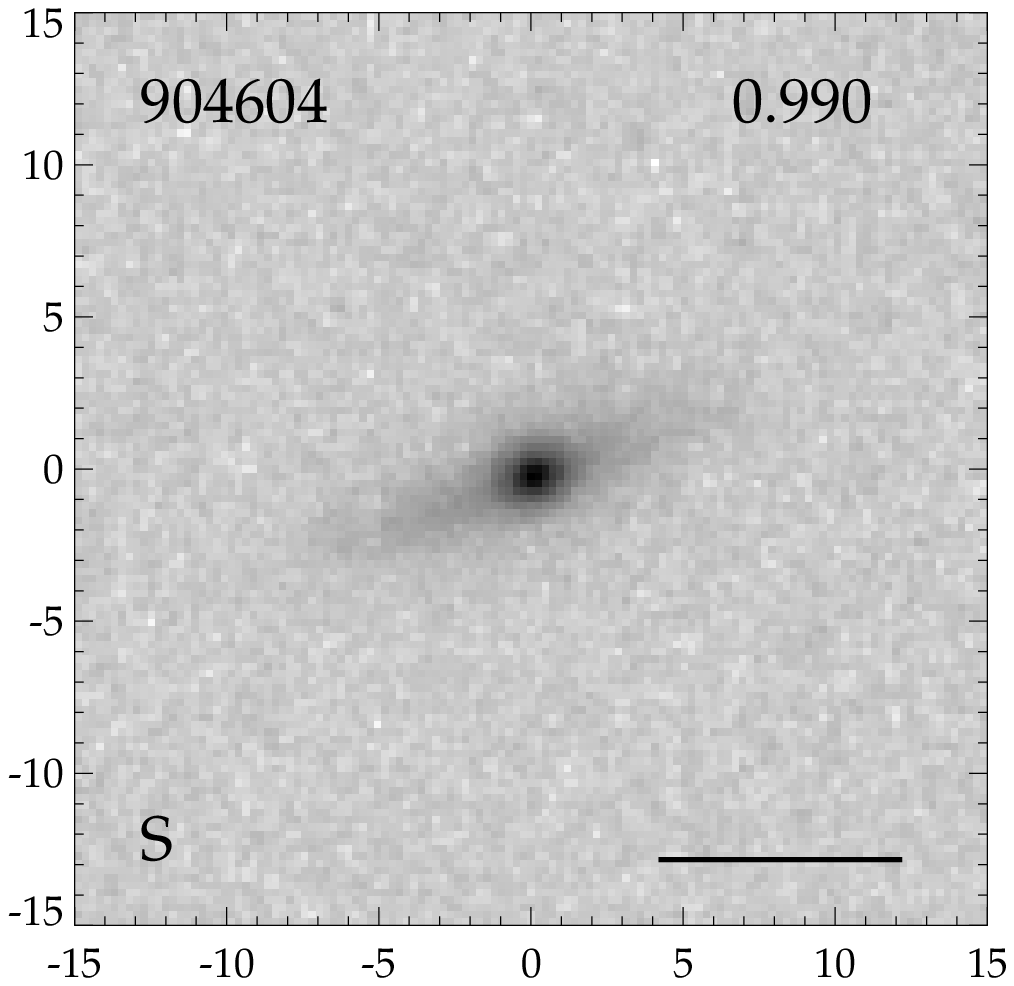} \includegraphics[height=0.22\textwidth,clip]{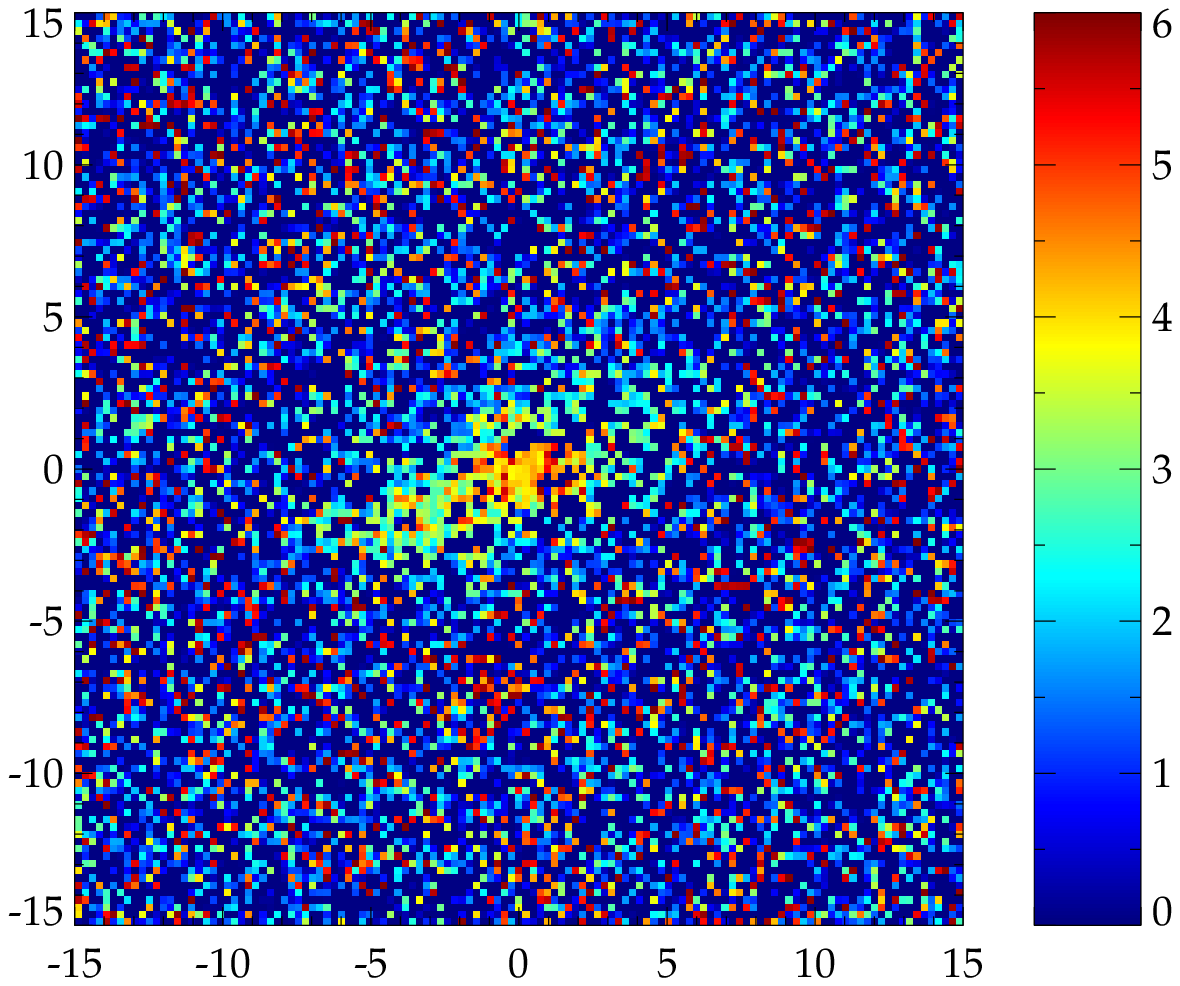}
\includegraphics[height=0.22\textwidth,clip]{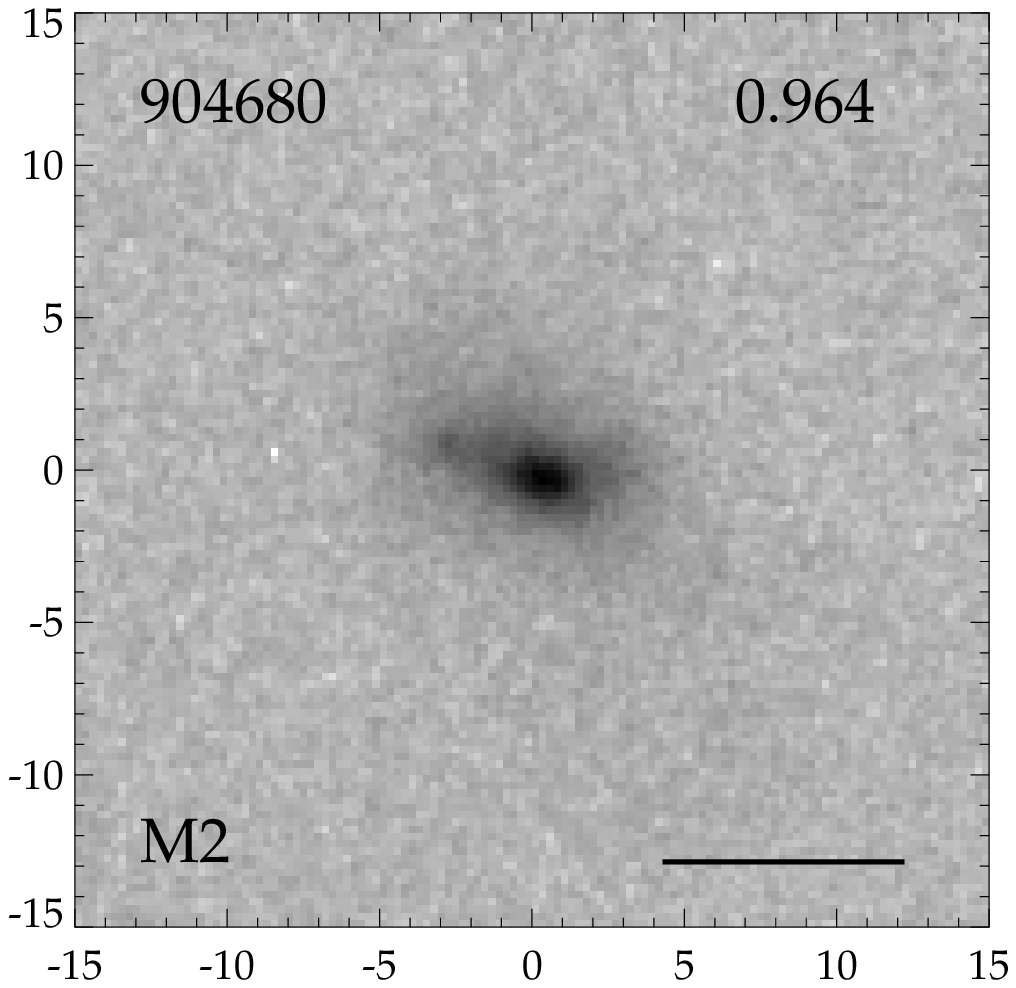} \includegraphics[height=0.22\textwidth,clip]{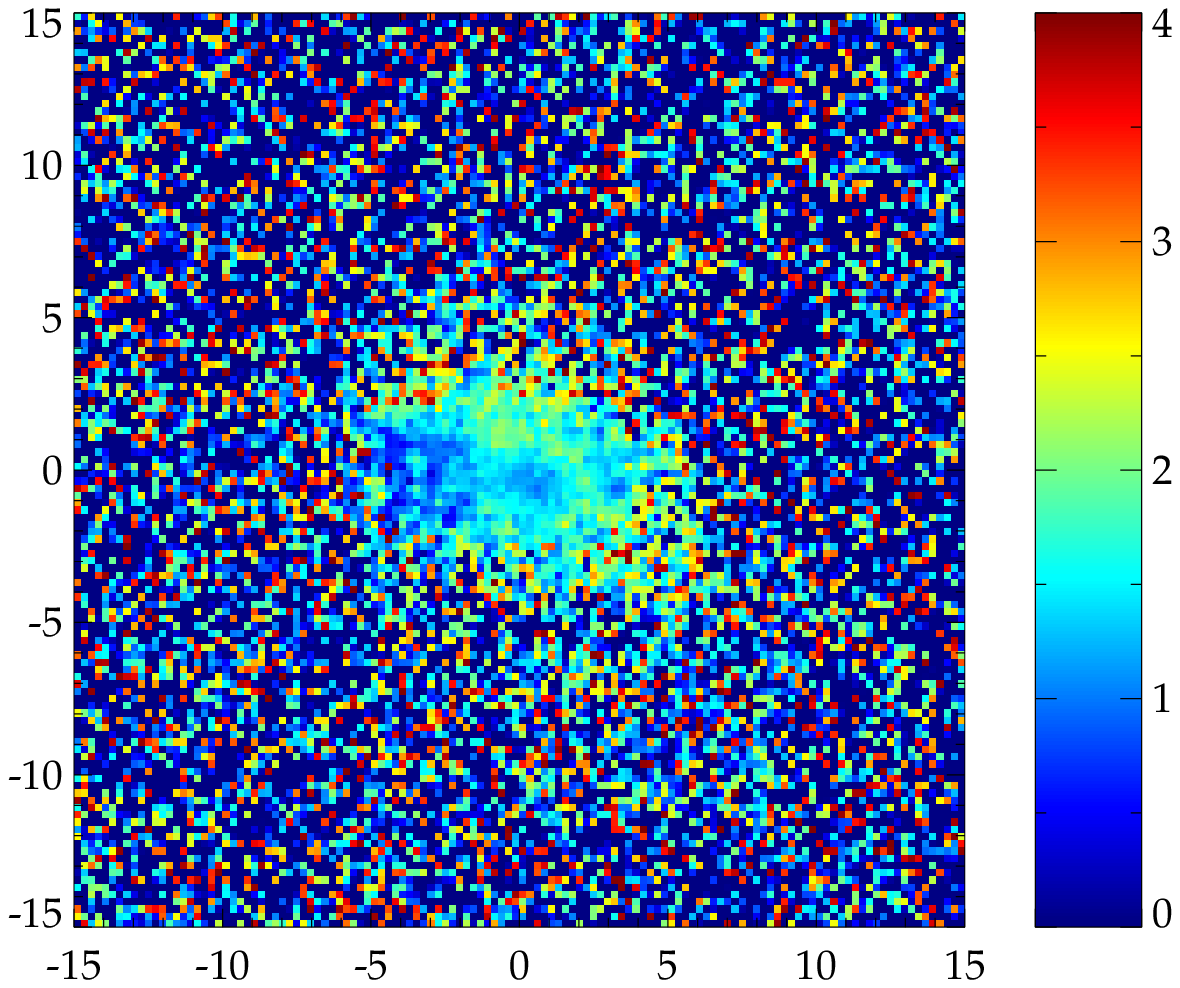}
\includegraphics[height=0.22\textwidth,clip]{h_sz_905632_drz.eps} \includegraphics[height=0.22\textwidth,clip]{h_sb-z_905632_color.eps}
\includegraphics[height=0.22\textwidth,clip]{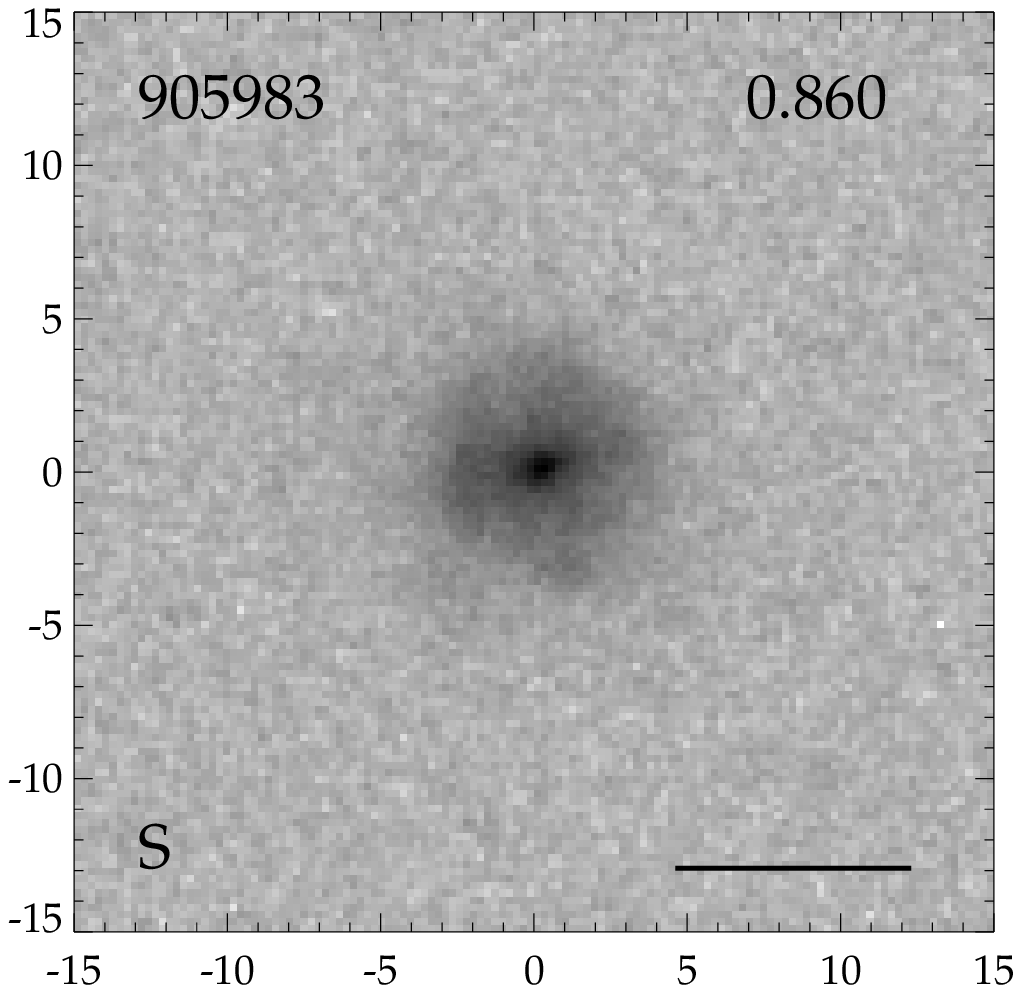} \includegraphics[height=0.22\textwidth,clip]{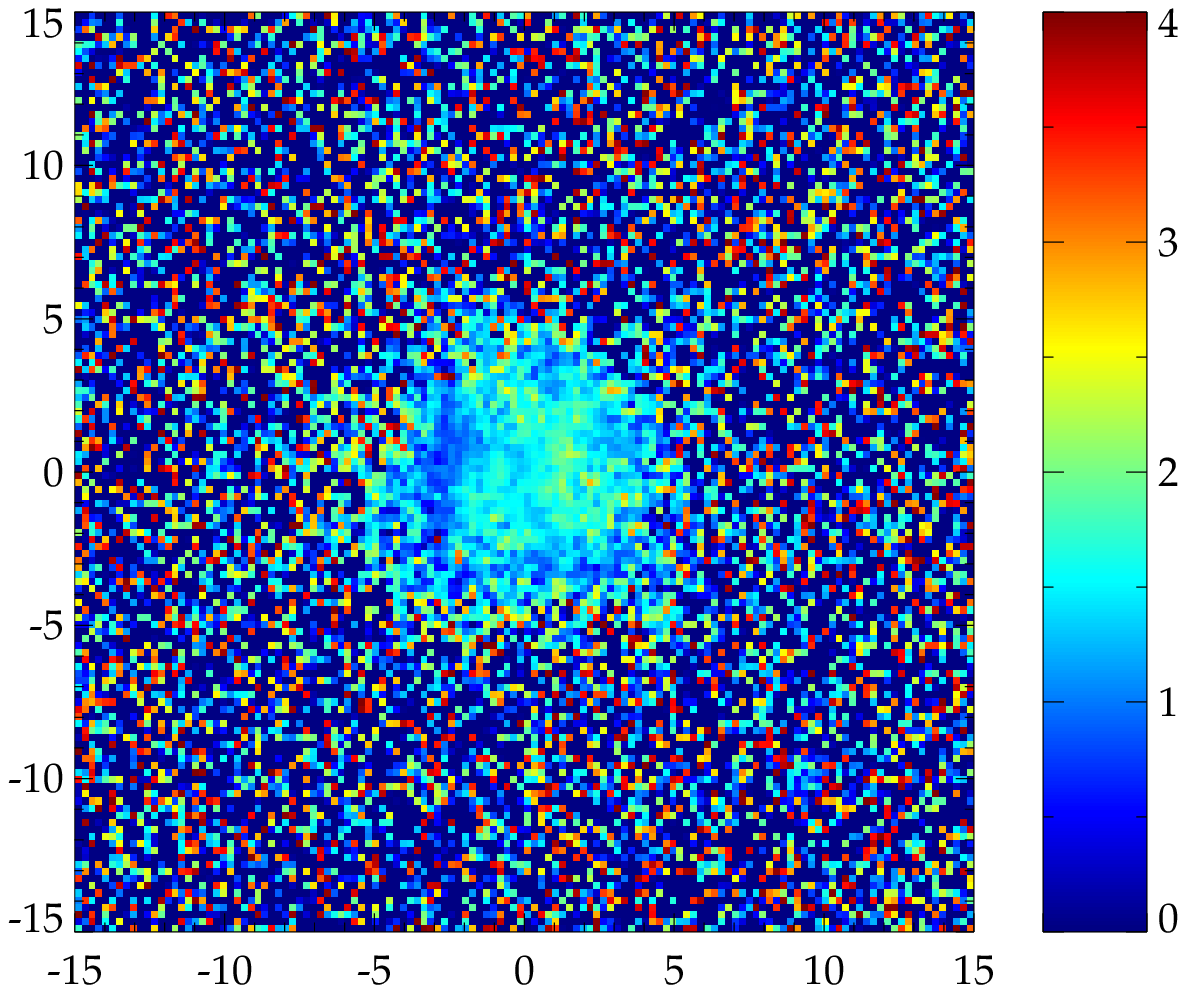}
\includegraphics[height=0.22\textwidth,clip]{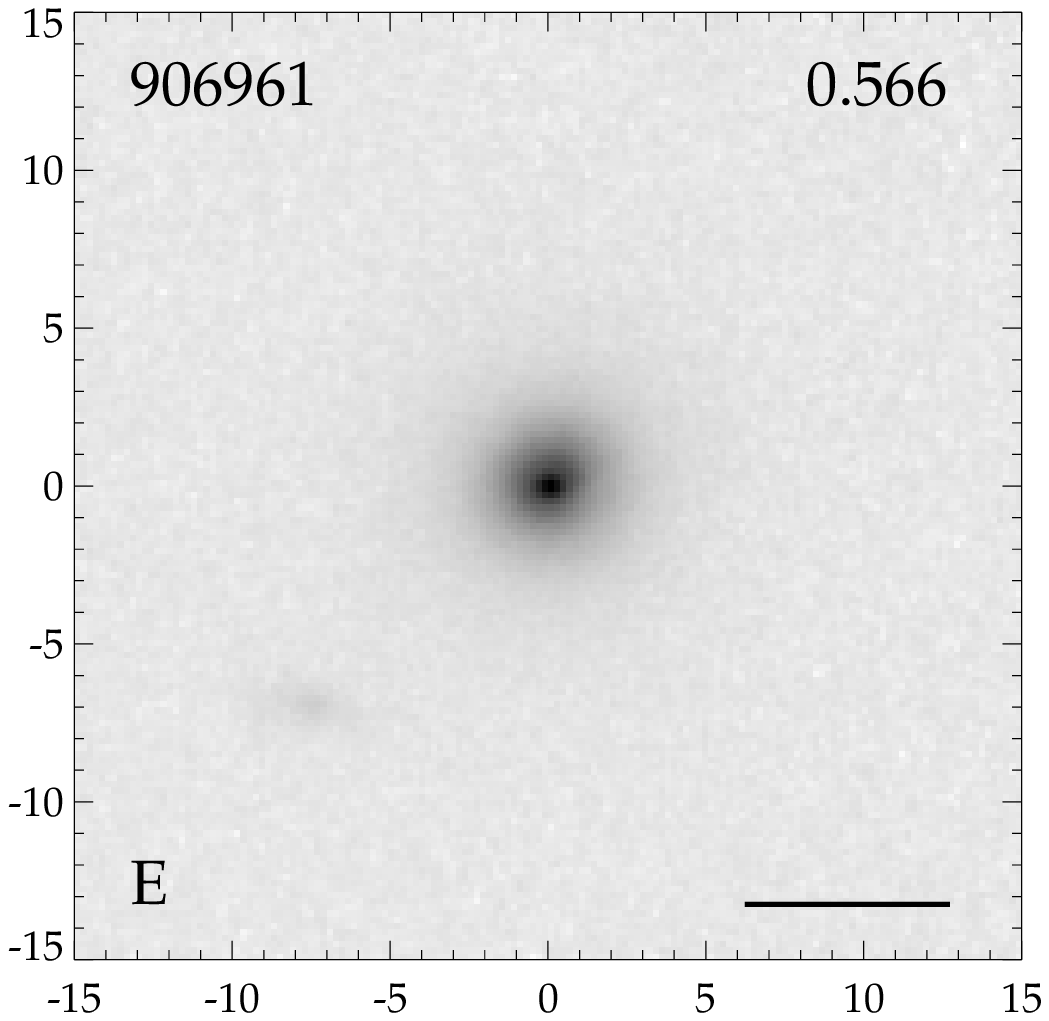} \includegraphics[height=0.22\textwidth,clip]{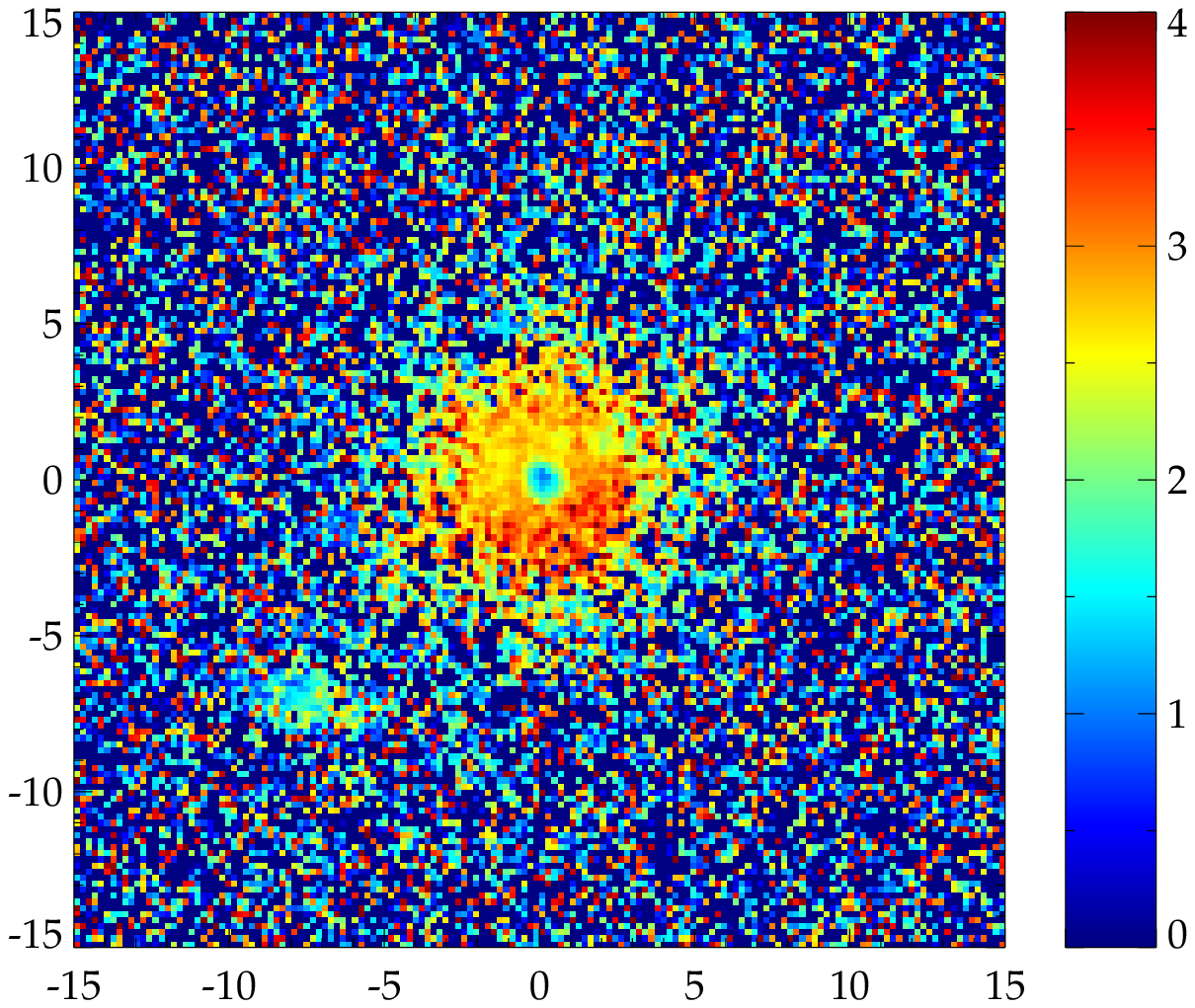}
\includegraphics[height=0.22\textwidth,clip]{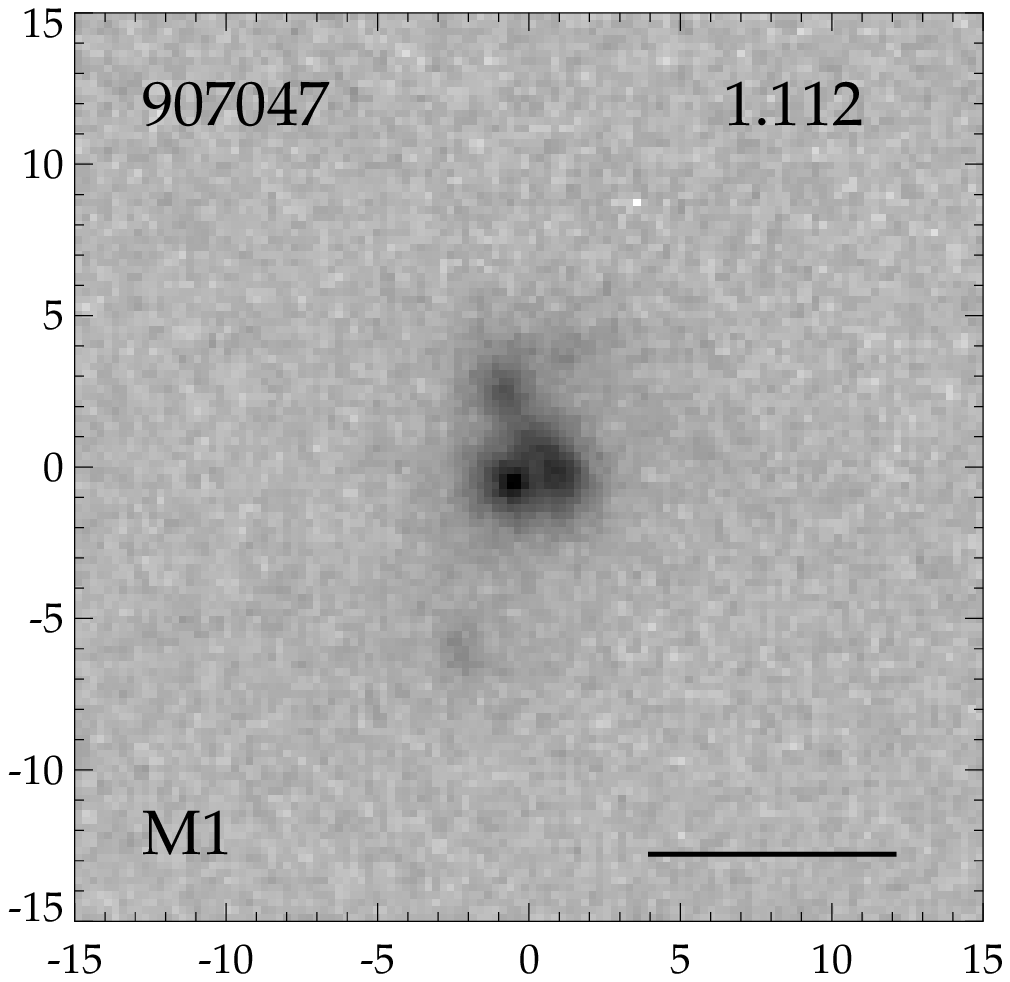} \includegraphics[height=0.22\textwidth,clip]{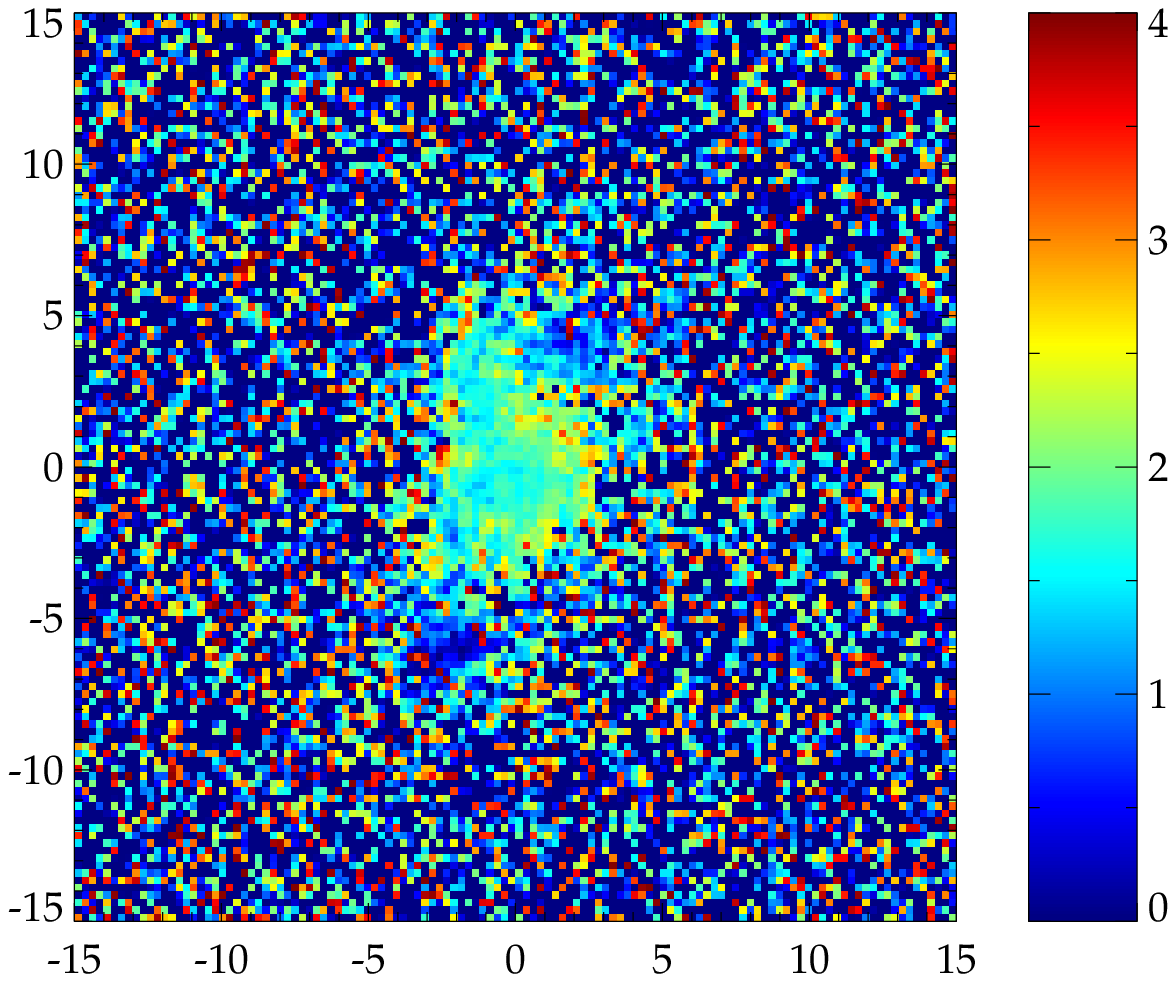}
\includegraphics[height=0.22\textwidth,clip]{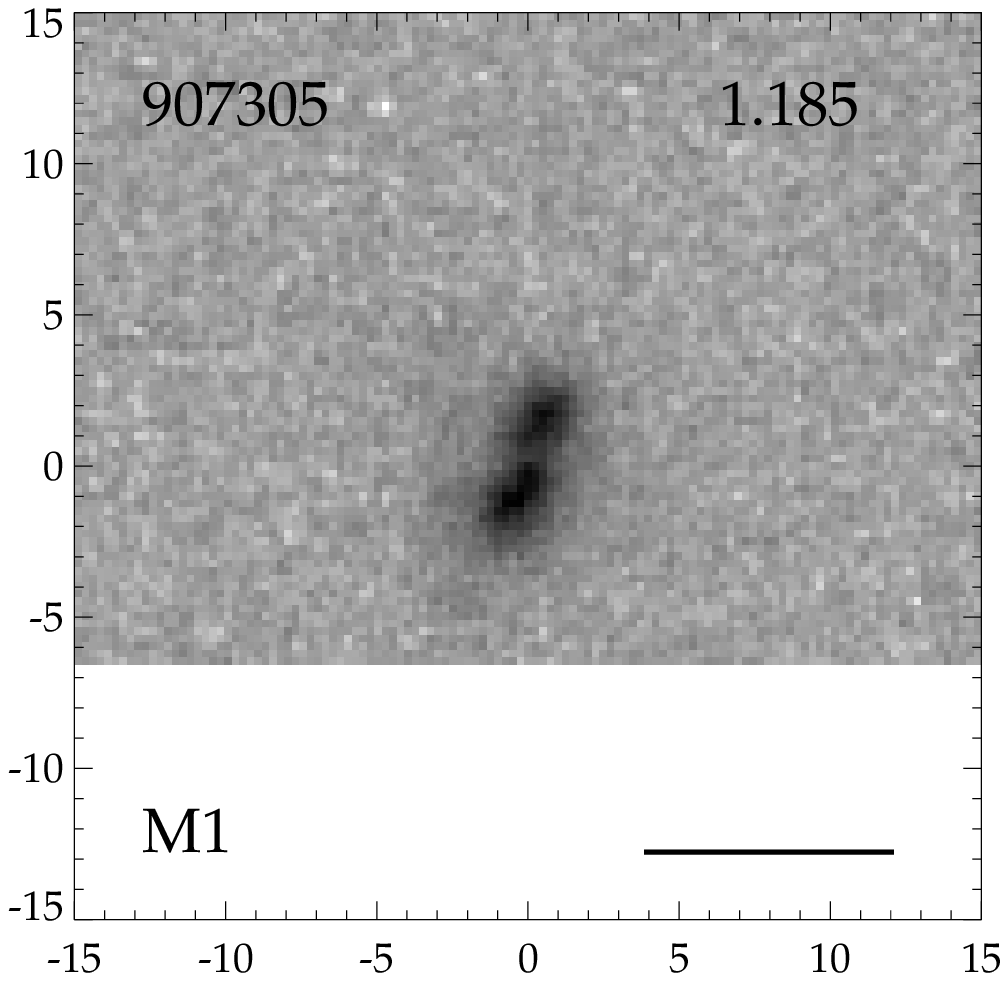} \includegraphics[height=0.22\textwidth,clip]{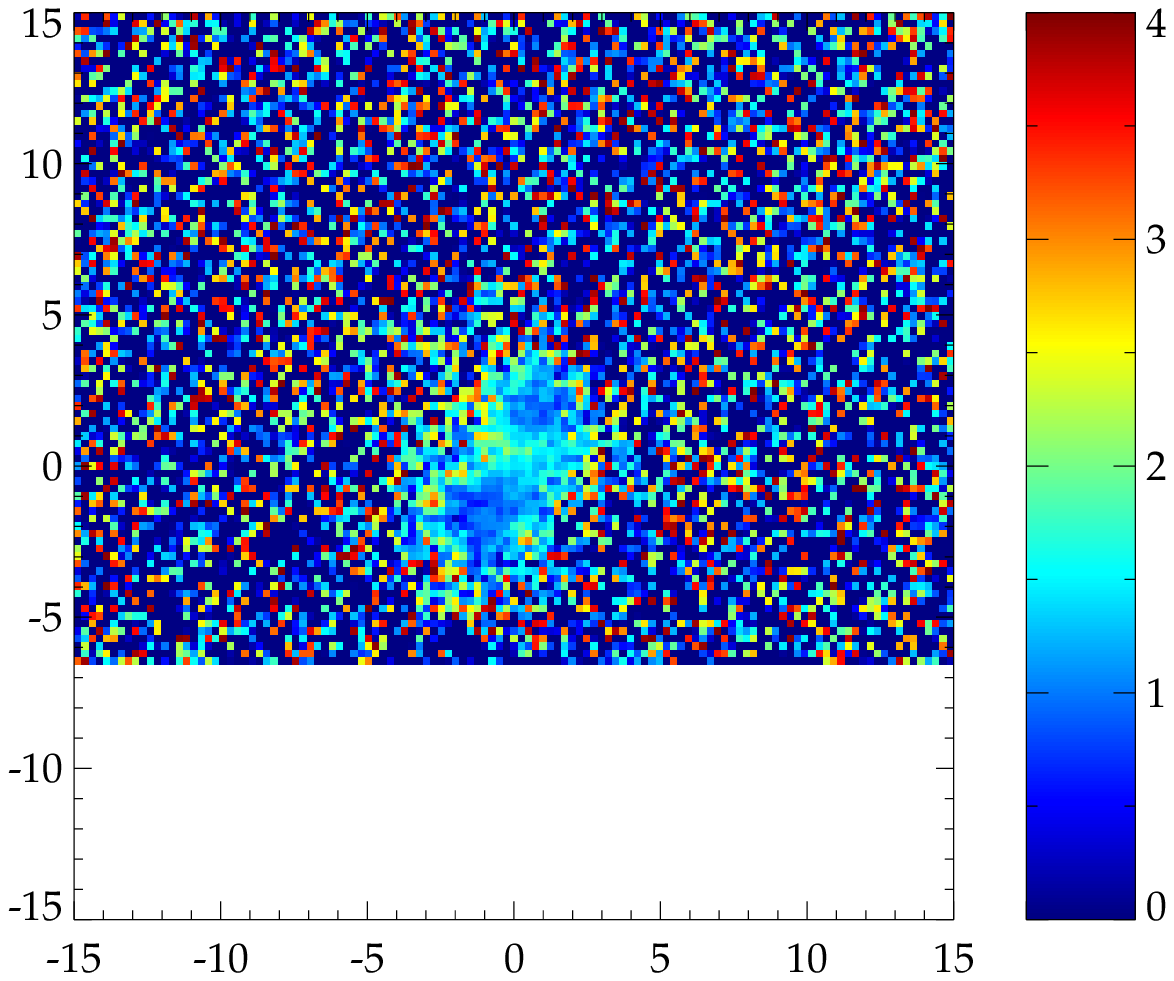}
\includegraphics[height=0.22\textwidth,clip]{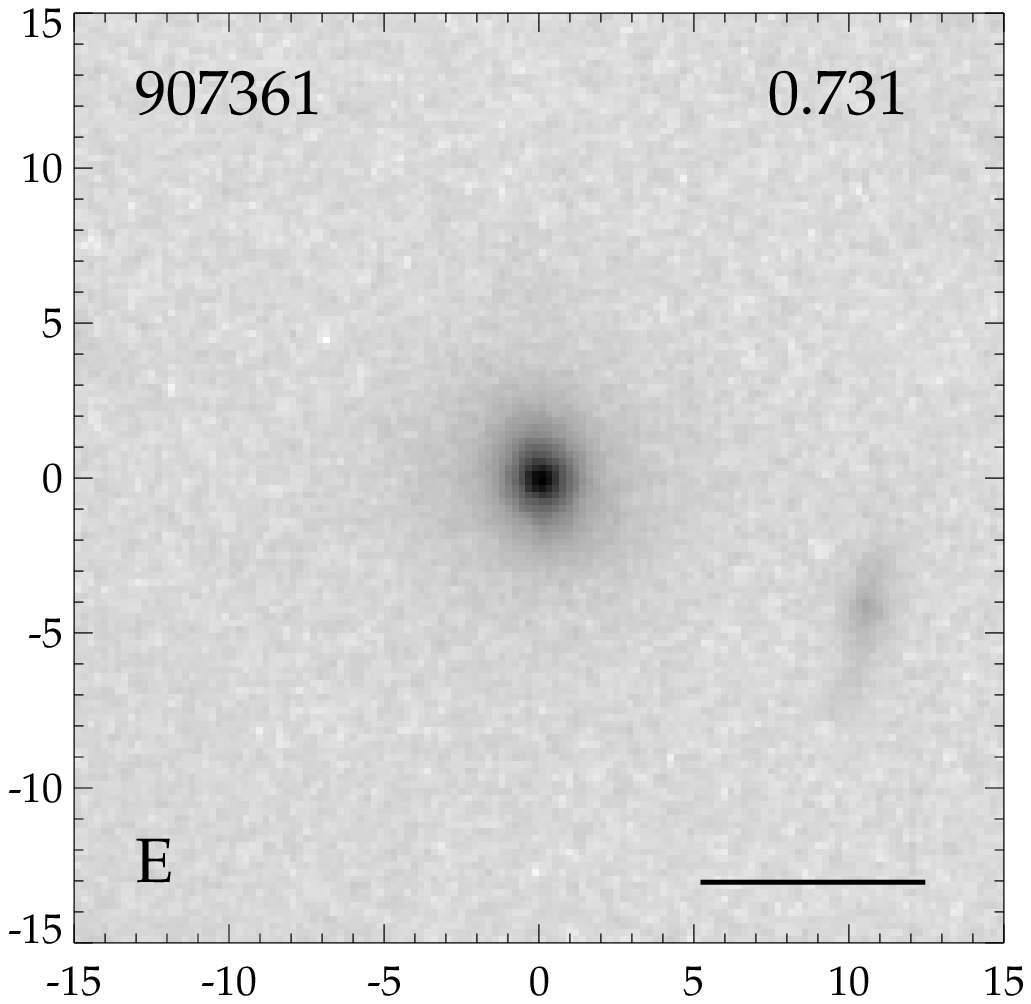} \includegraphics[height=0.22\textwidth,clip]{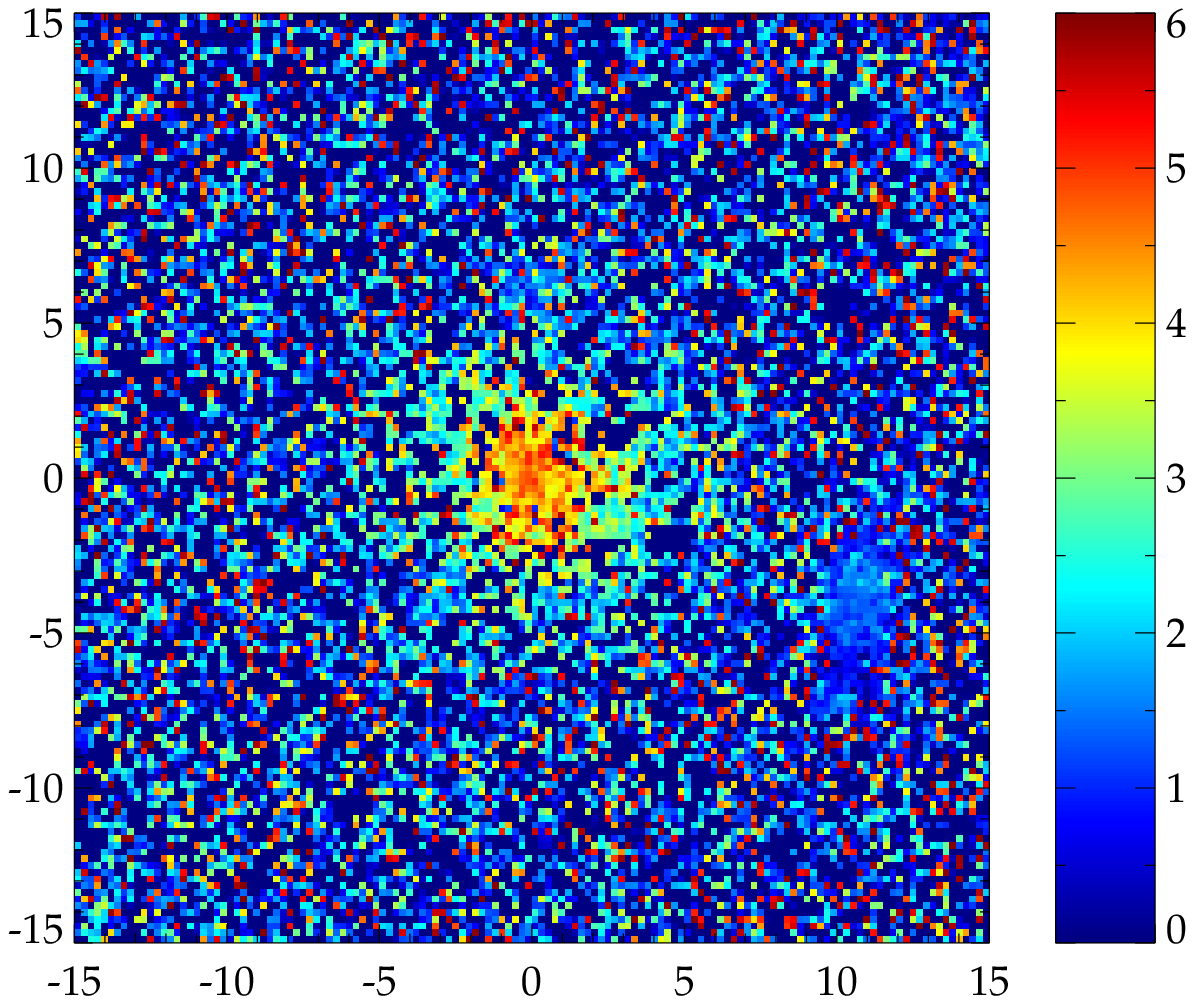}
\includegraphics[height=0.22\textwidth,clip]{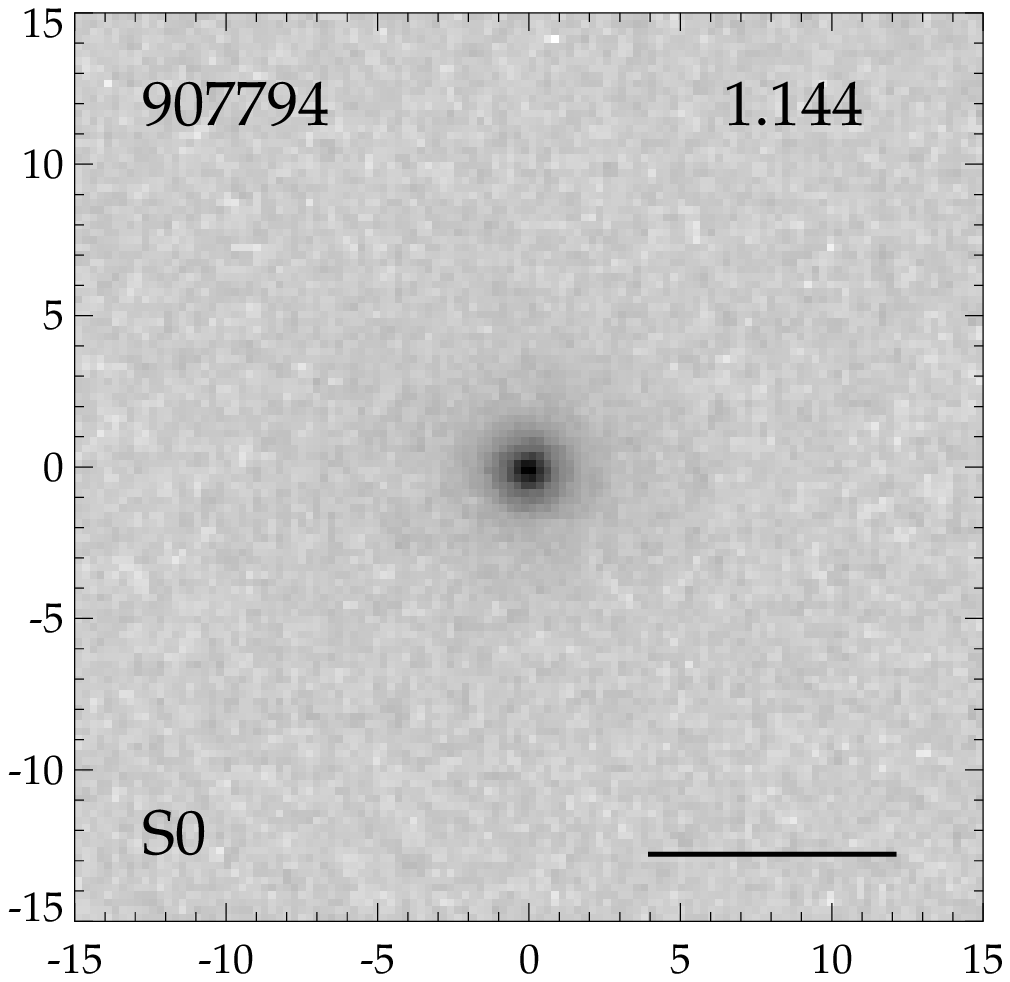} \includegraphics[height=0.22\textwidth,clip]{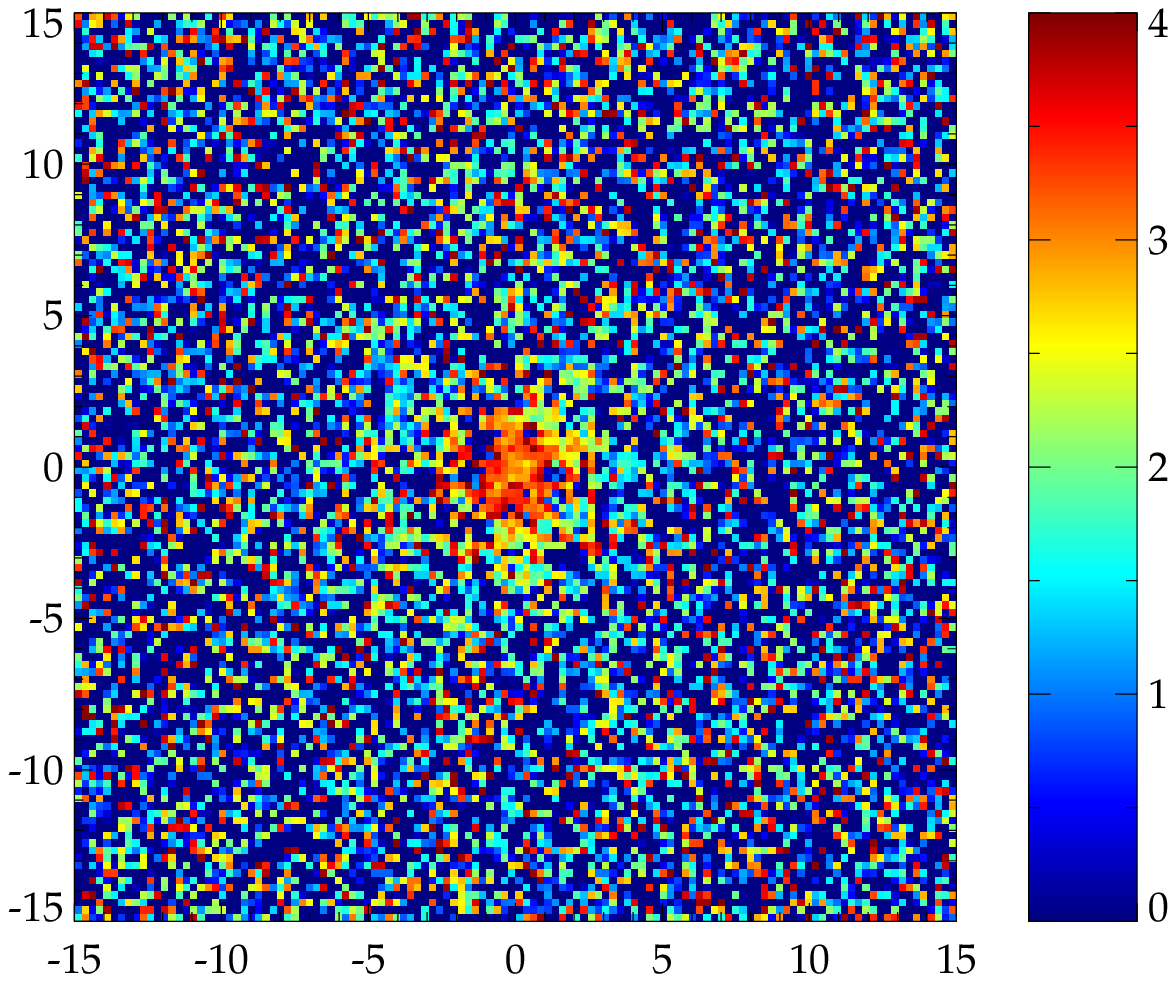}
\includegraphics[height=0.22\textwidth,clip]{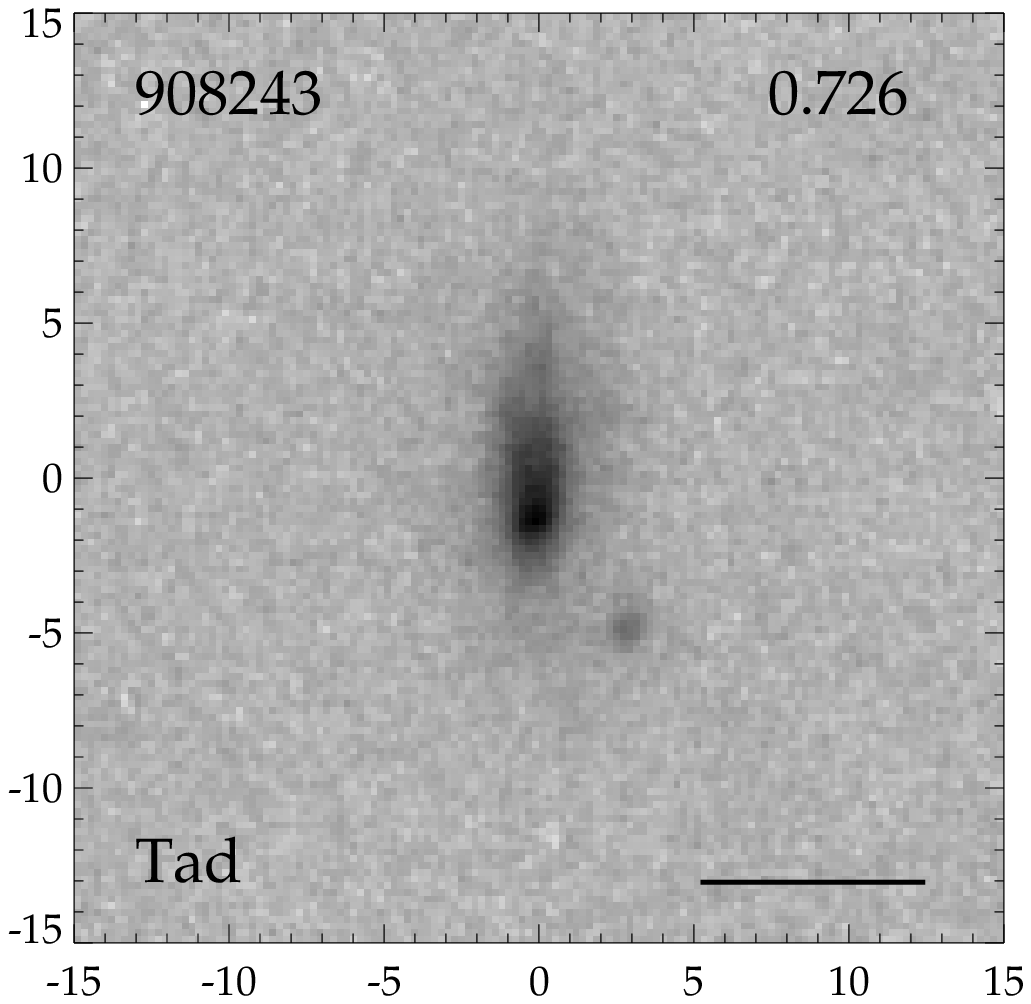} \includegraphics[height=0.22\textwidth,clip]{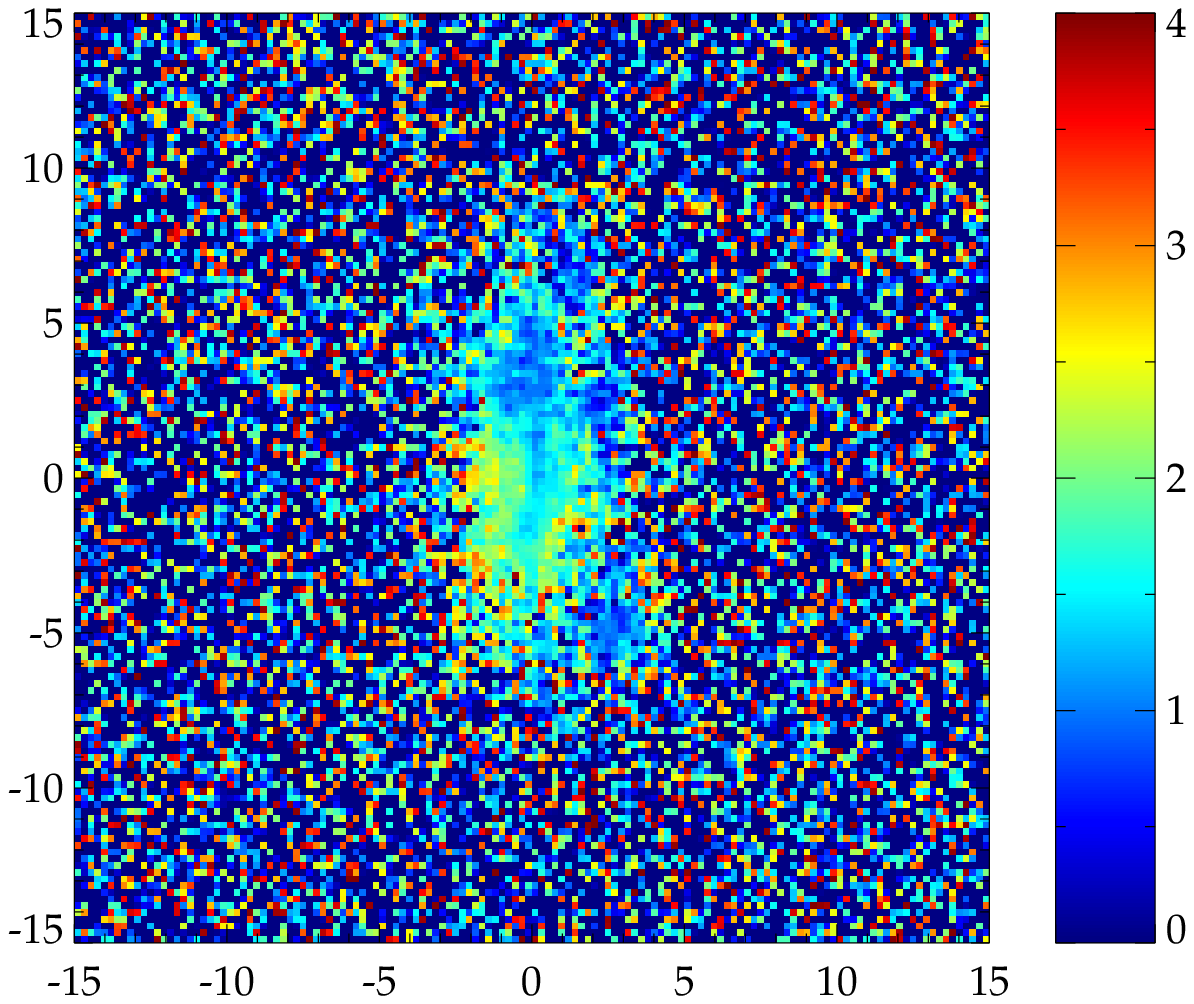}
\includegraphics[height=0.22\textwidth,clip]{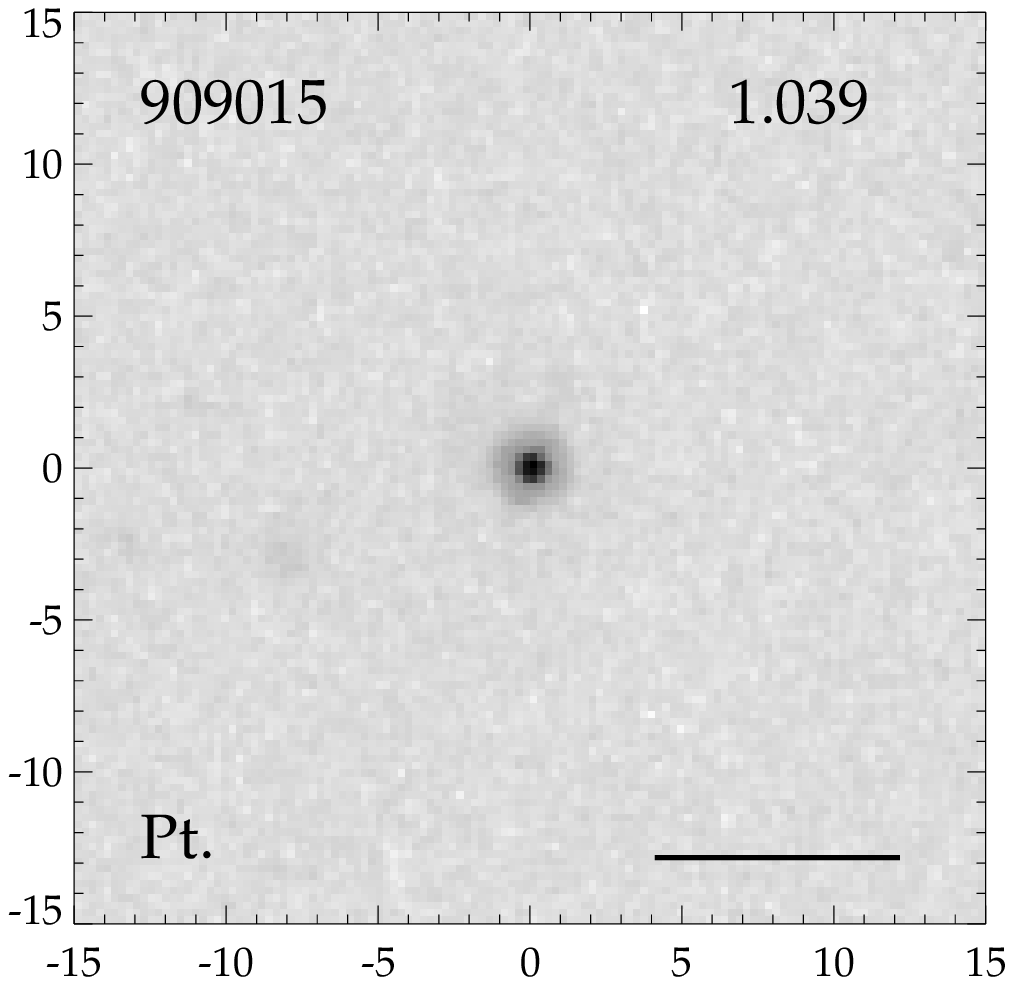} \includegraphics[height=0.22\textwidth,clip]{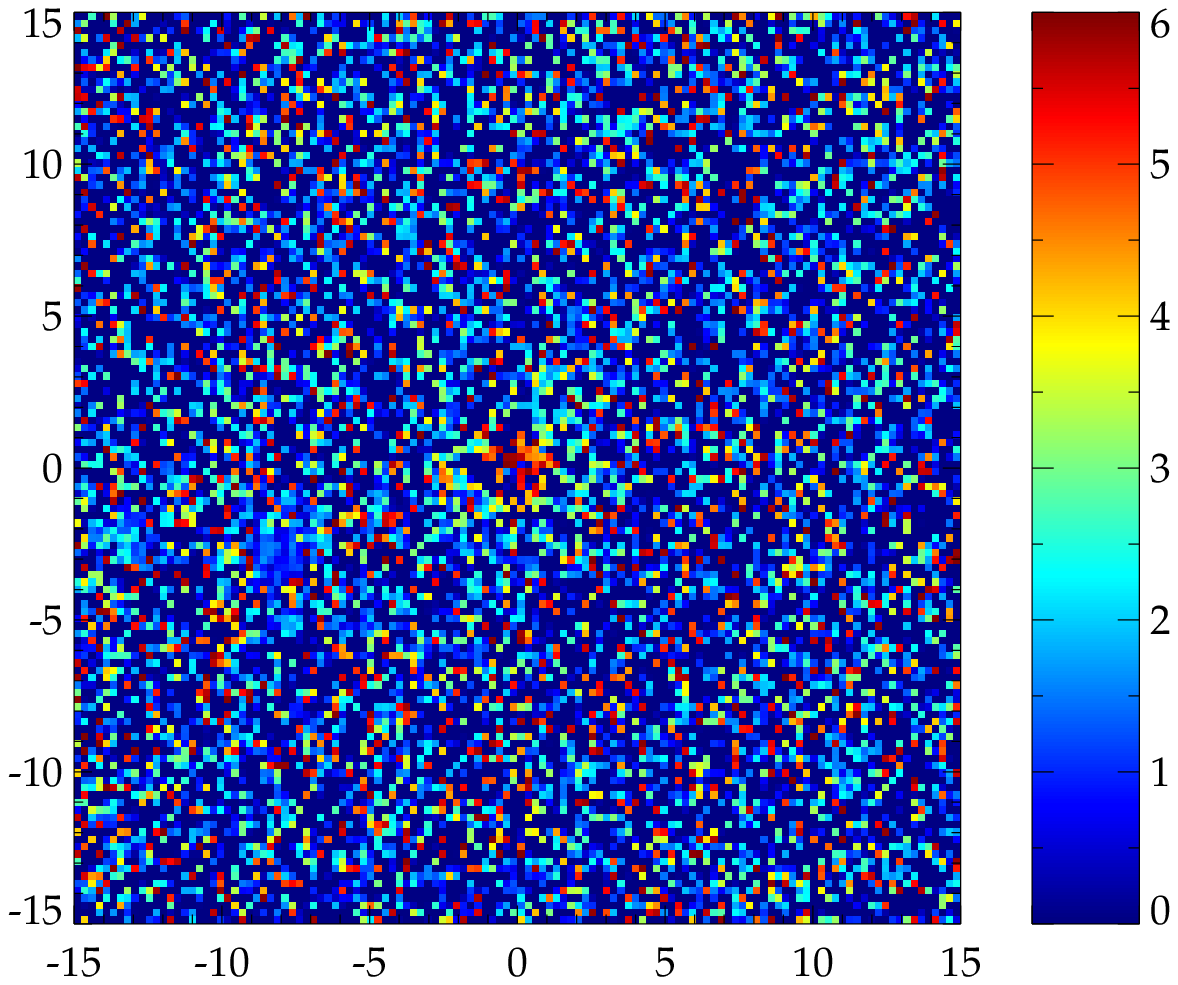}
\caption{$F850LP$ and the B-z color map images. Explanation is given at the end of this figure.} \label{colormap} \end{figure*}

\addtocounter{figure}{-1}
\begin{figure*} \centering
\includegraphics[height=0.22\textwidth,clip]{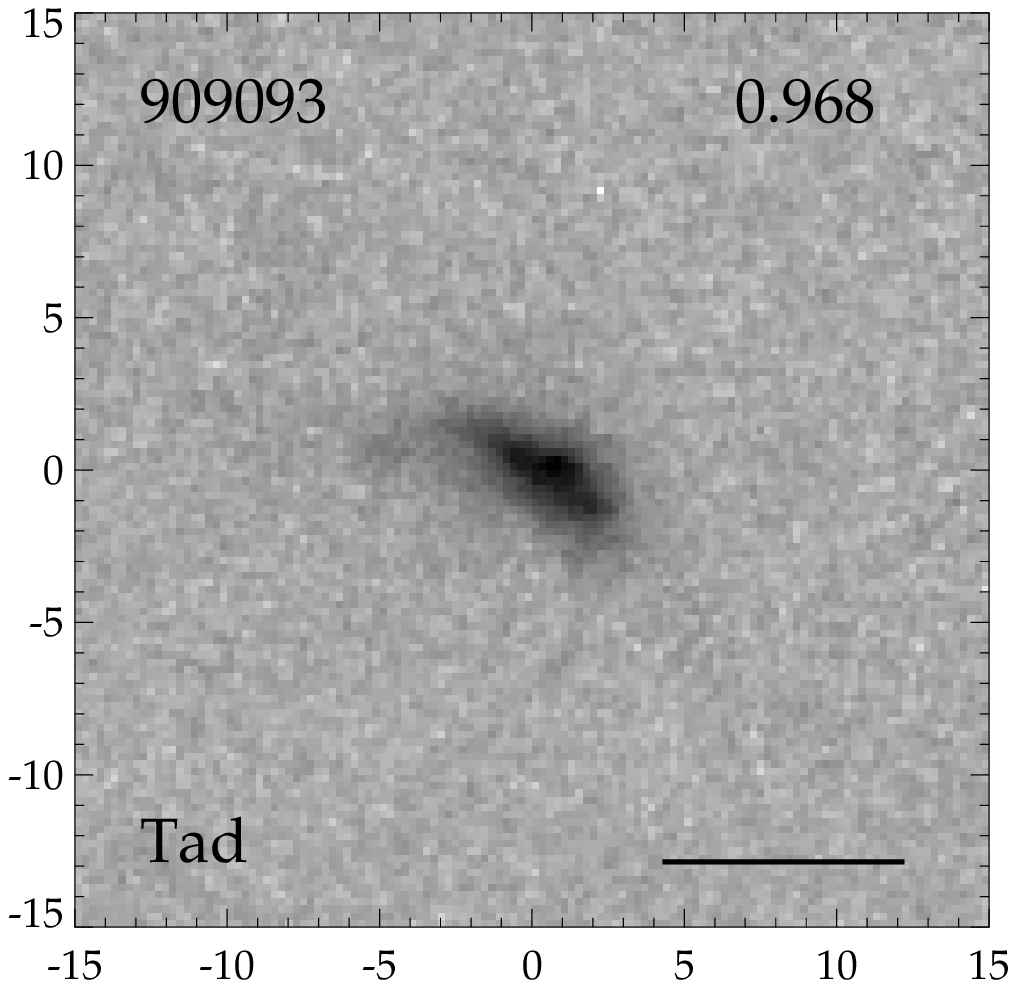} \includegraphics[height=0.22\textwidth,clip]{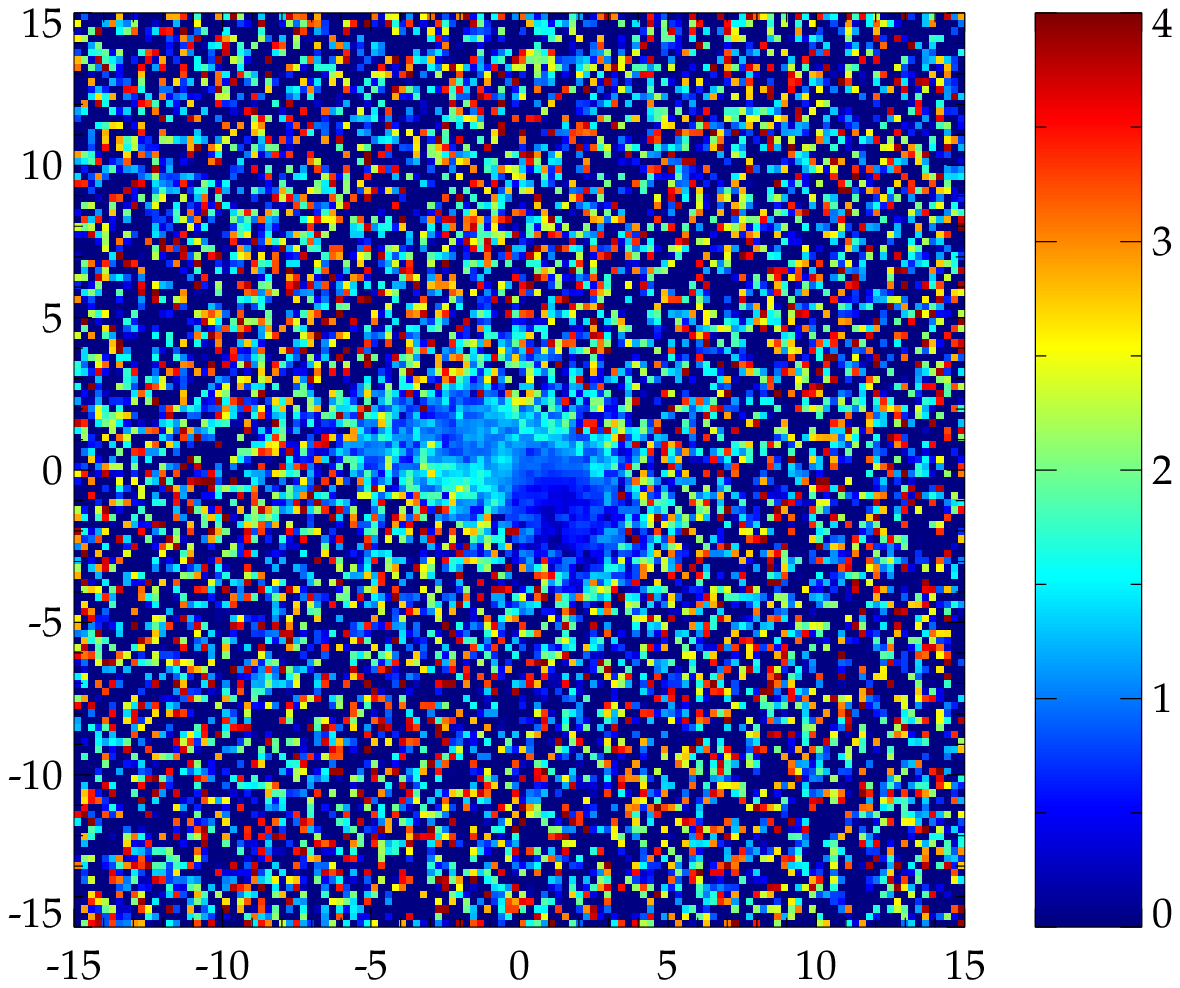}
\includegraphics[height=0.22\textwidth,clip]{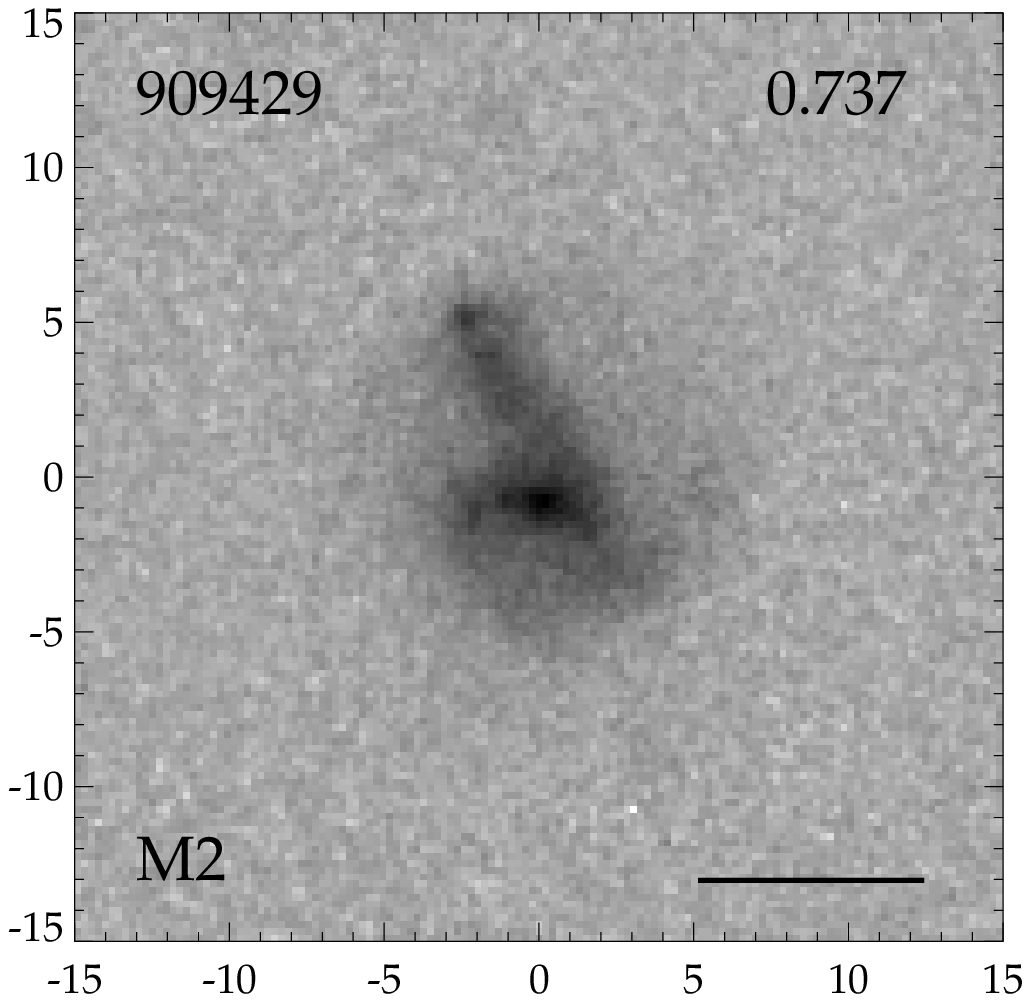} \includegraphics[height=0.22\textwidth,clip]{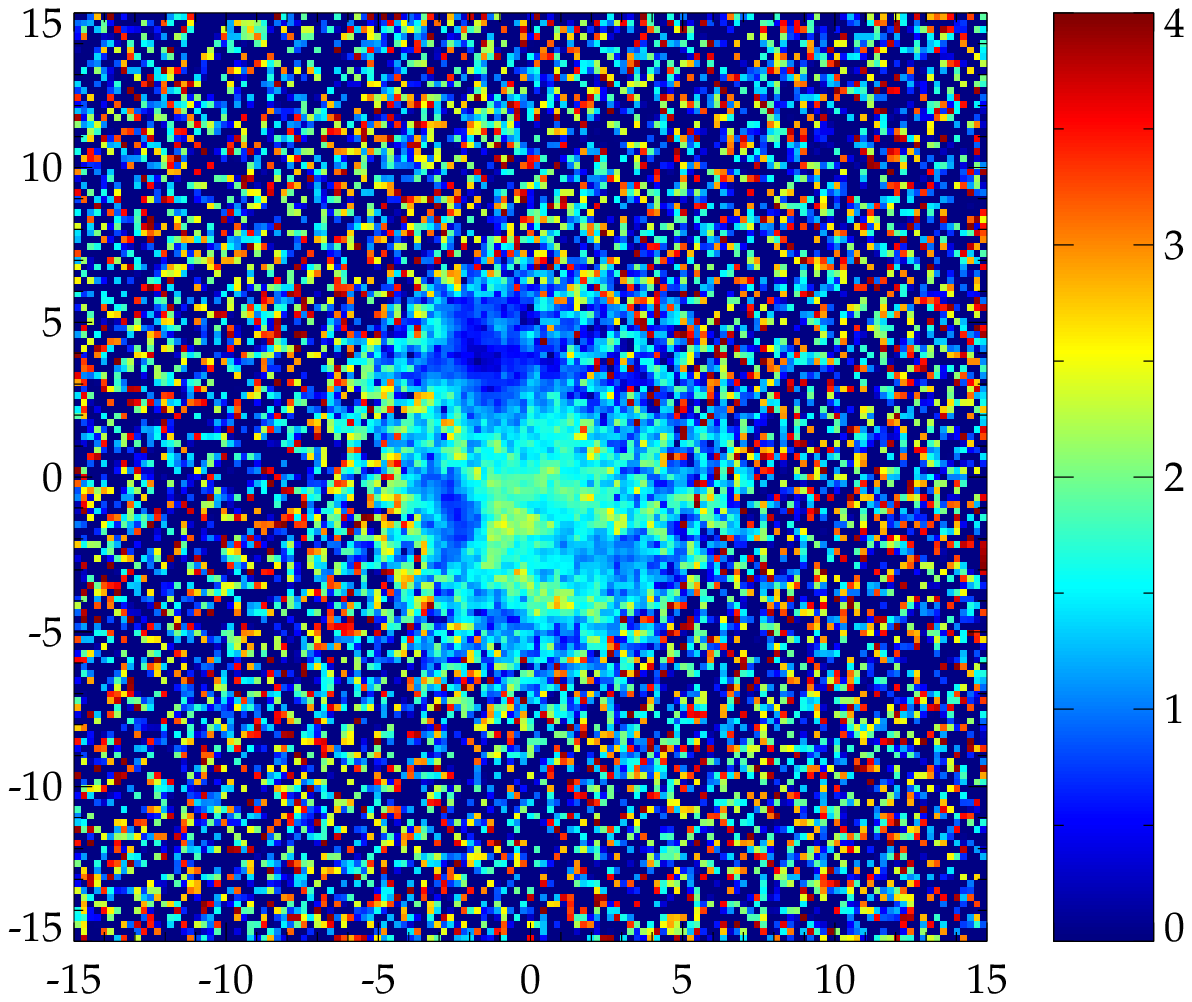}
\includegraphics[height=0.22\textwidth,clip]{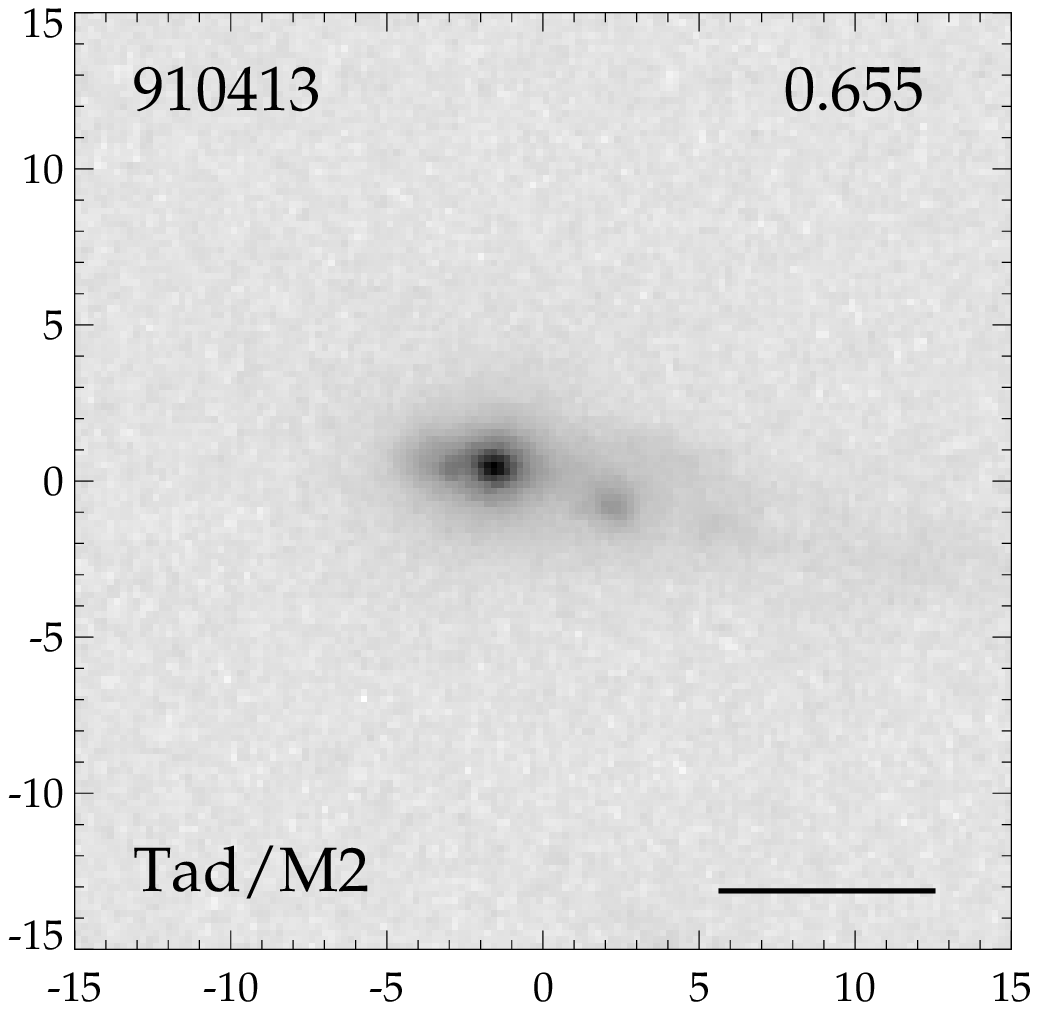} \includegraphics[height=0.22\textwidth,clip]{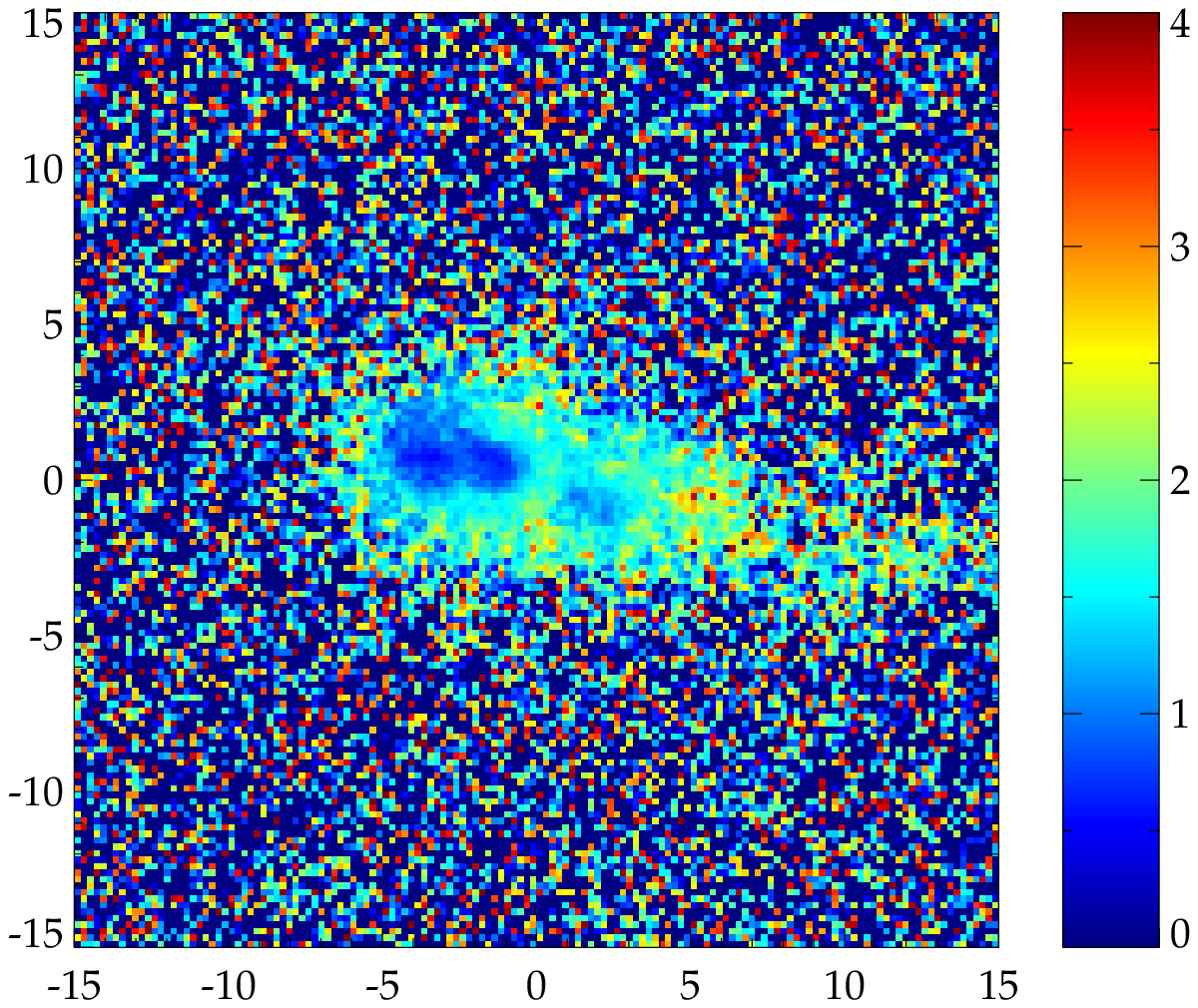}
\includegraphics[height=0.22\textwidth,clip]{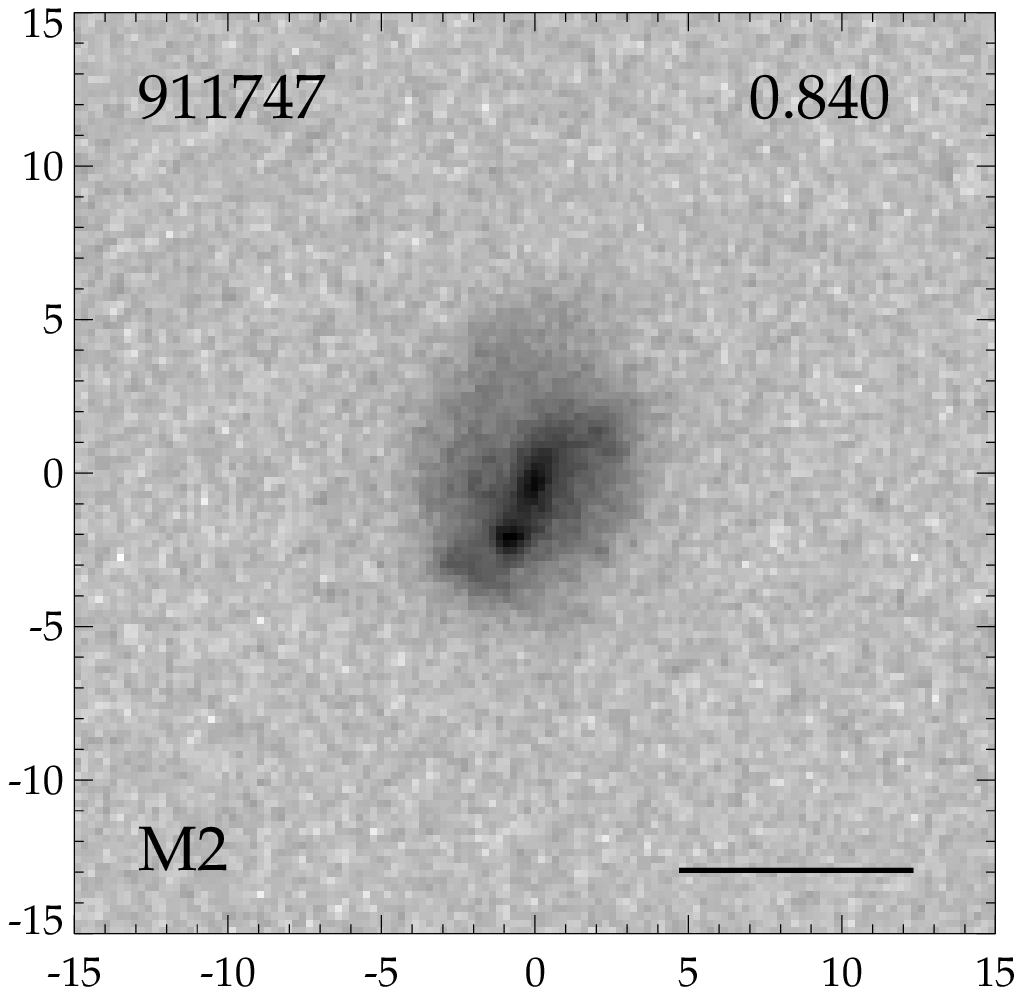} \includegraphics[height=0.22\textwidth,clip]{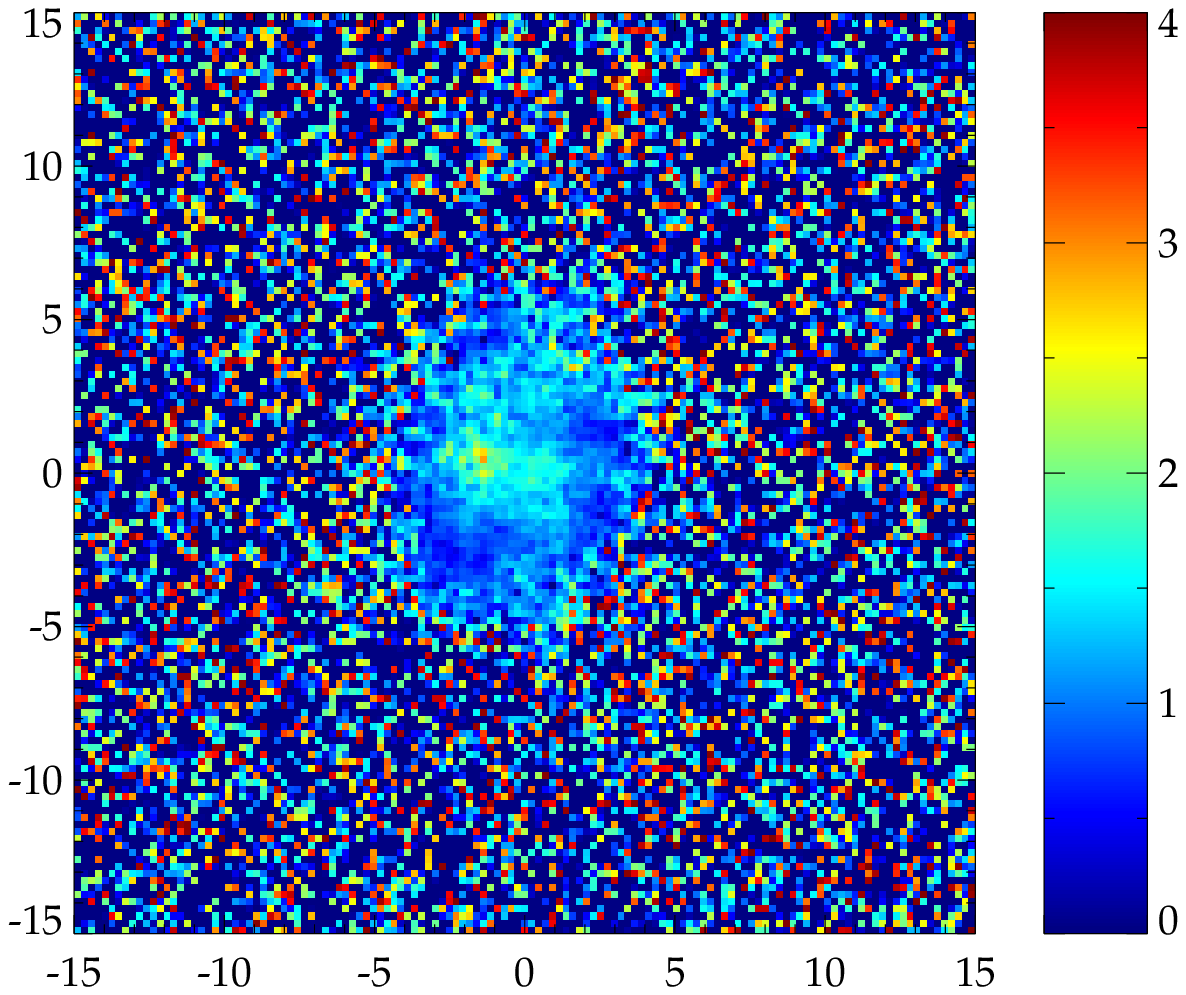}
\includegraphics[height=0.22\textwidth,clip]{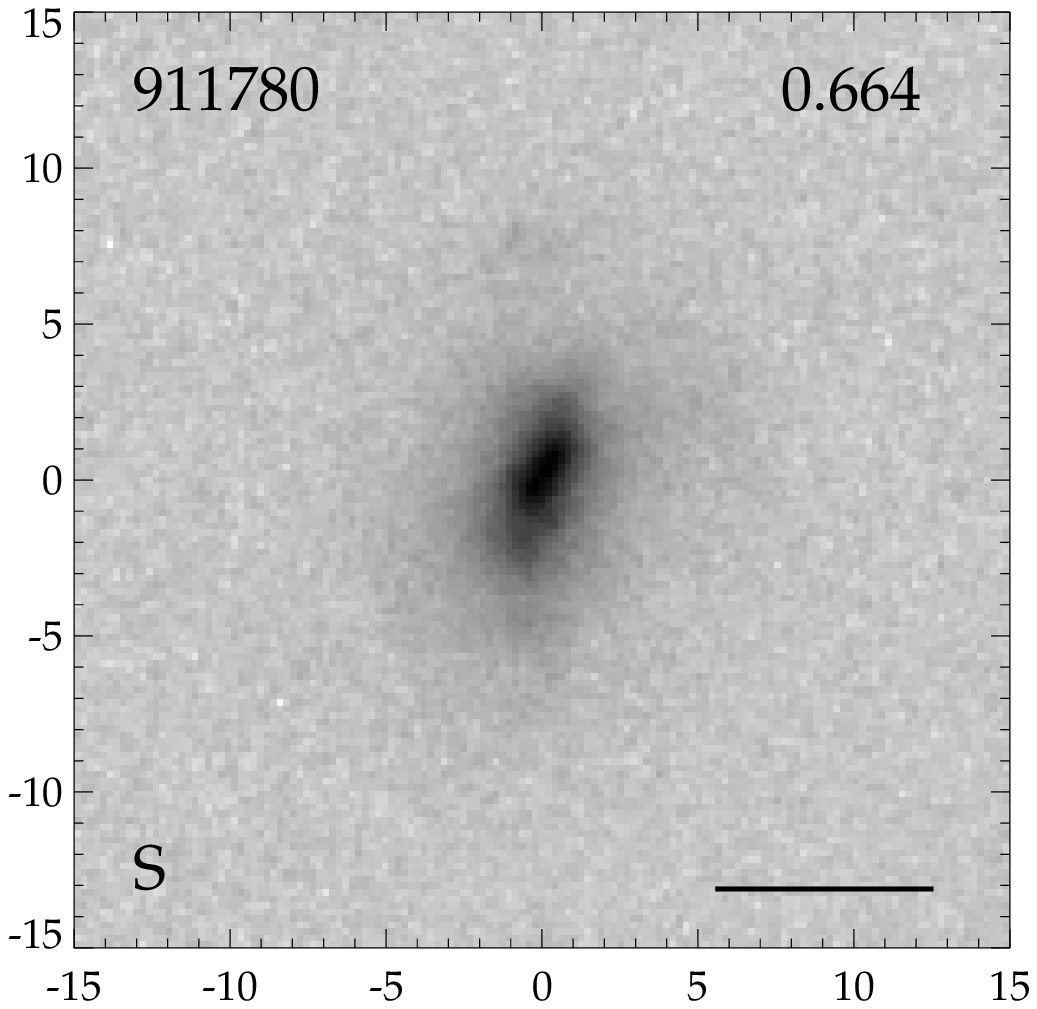} \includegraphics[height=0.22\textwidth,clip]{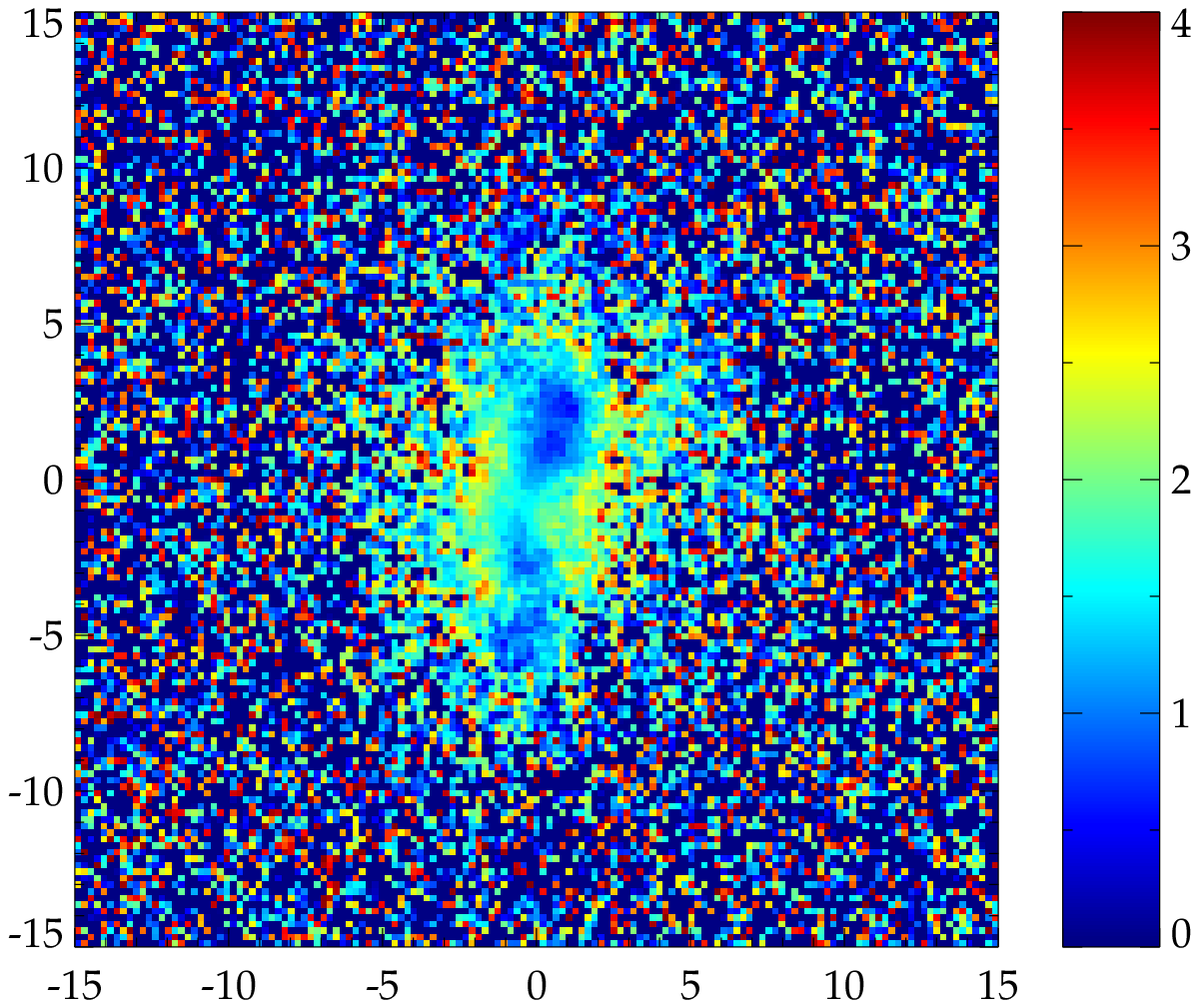}
\includegraphics[height=0.22\textwidth,clip]{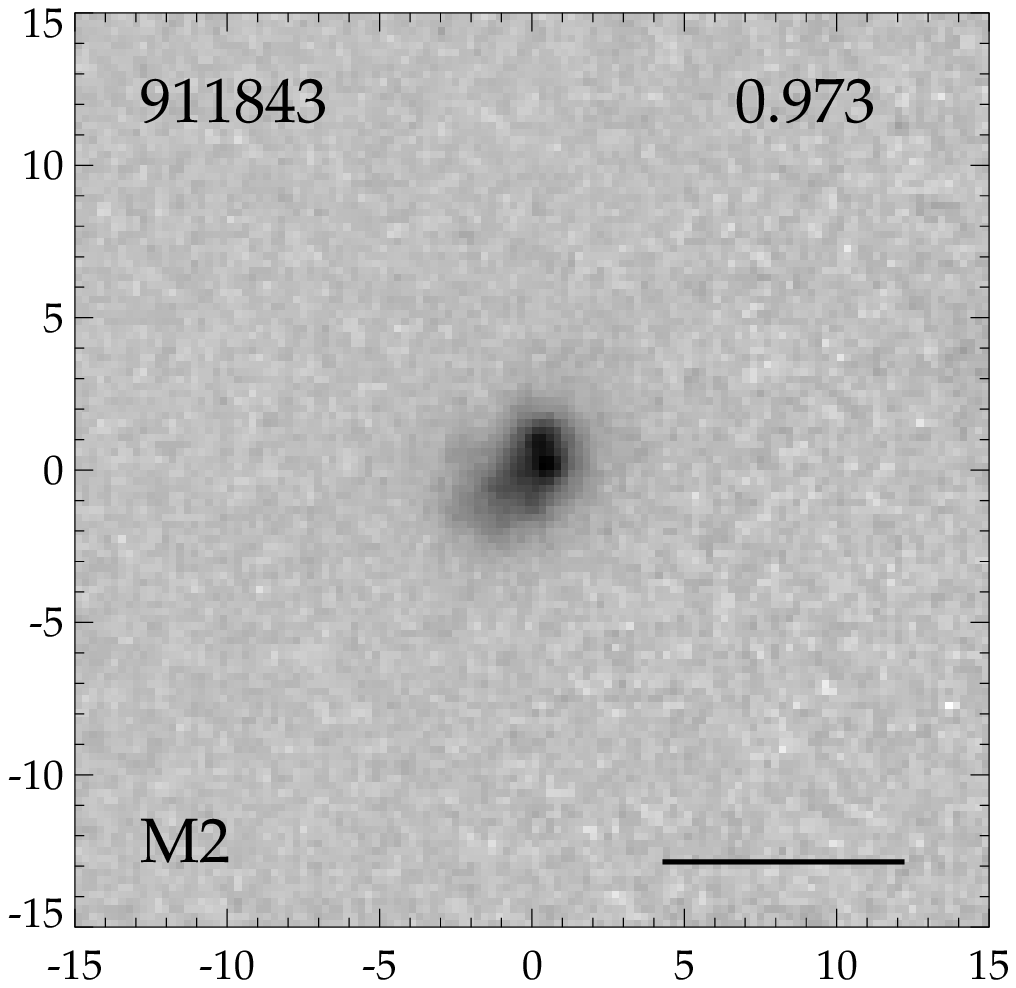} \includegraphics[height=0.22\textwidth,clip]{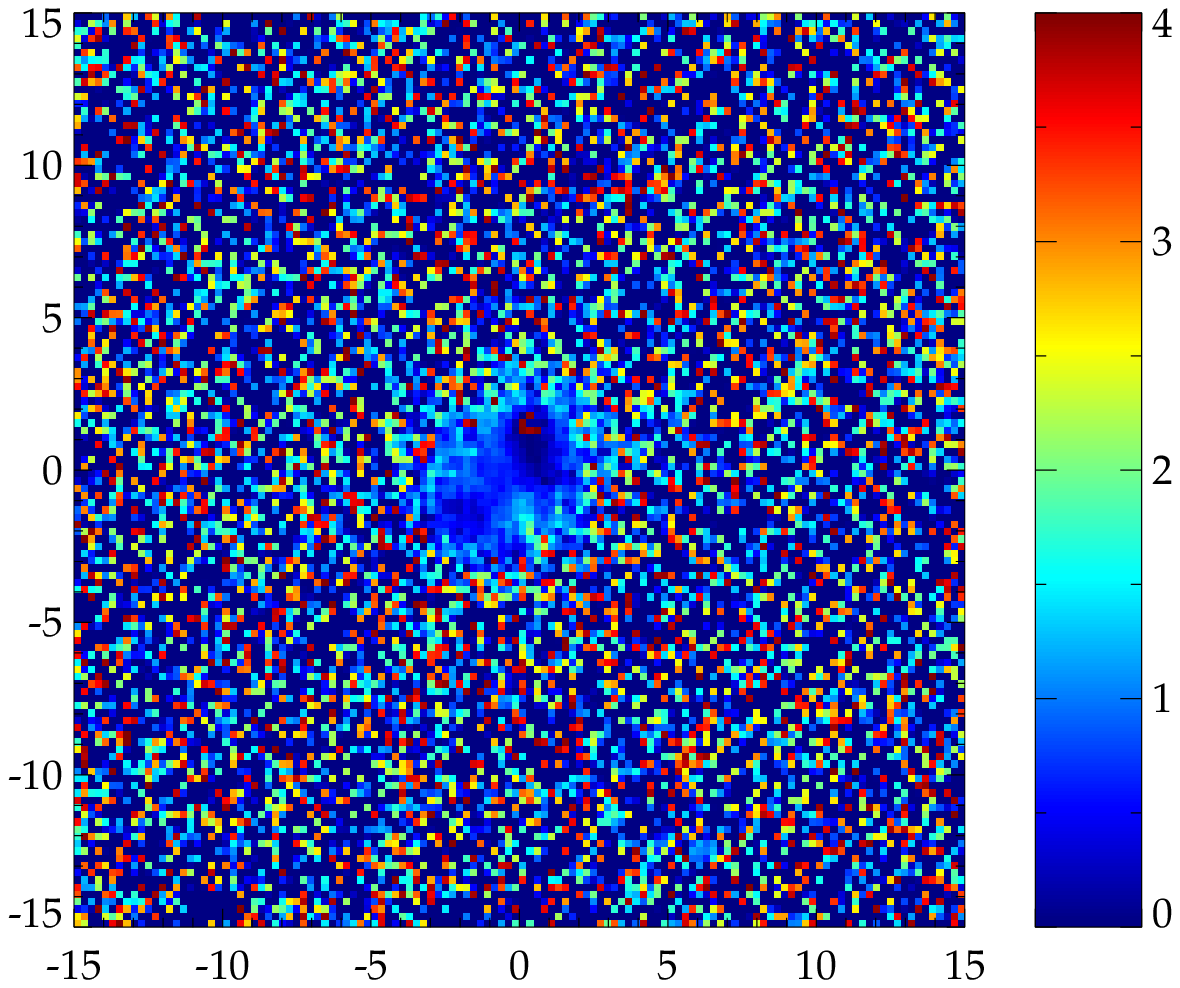}
\includegraphics[height=0.22\textwidth,clip]{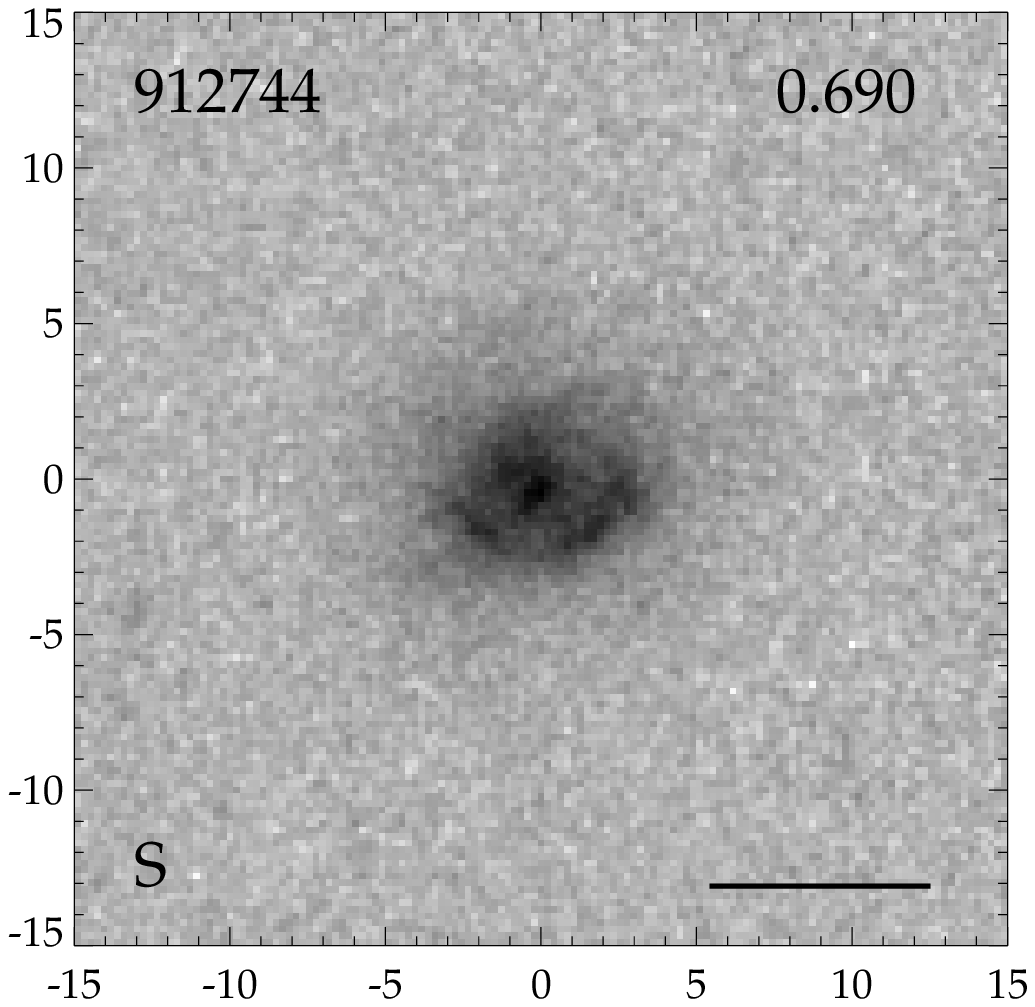} \includegraphics[height=0.22\textwidth,clip]{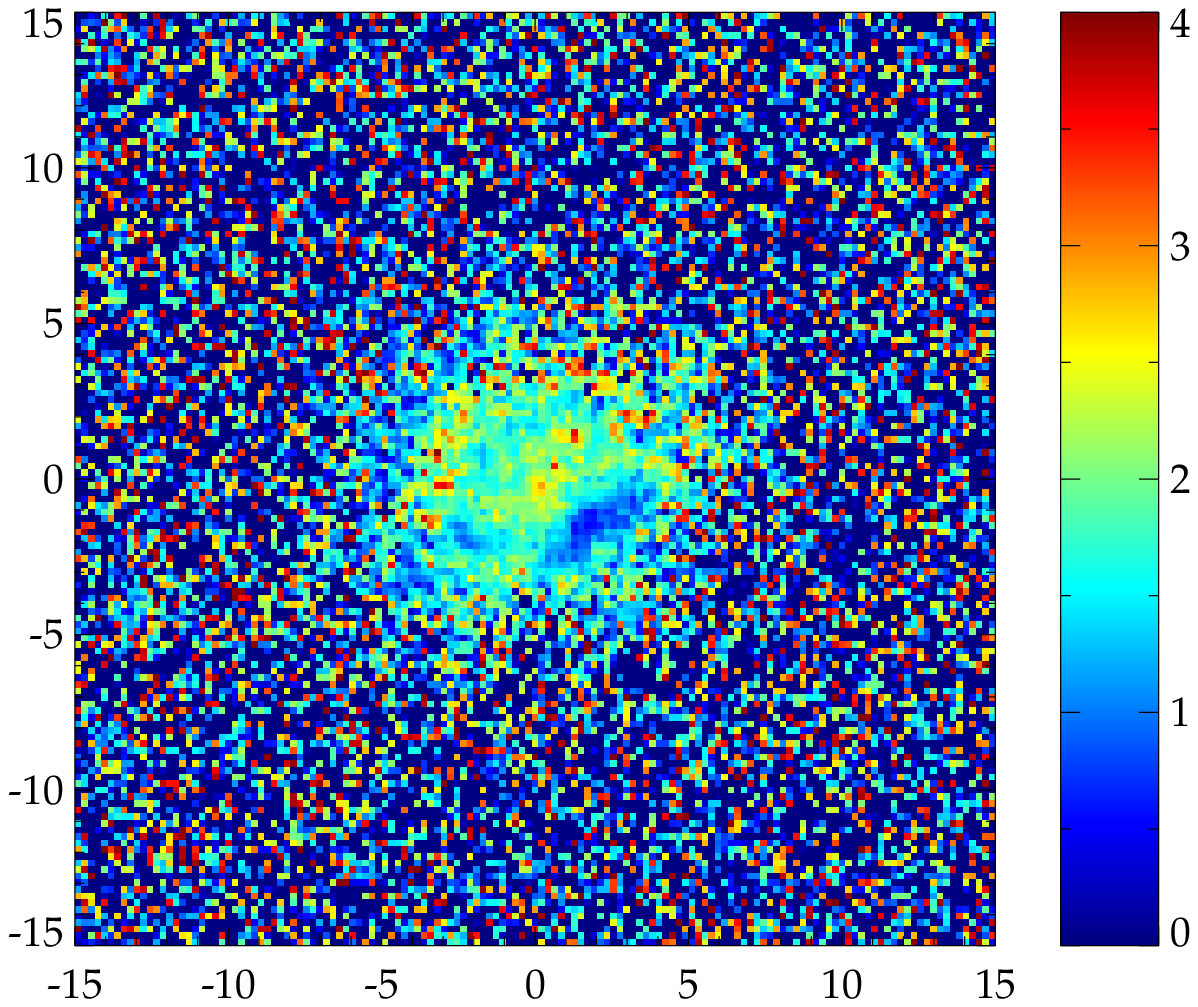}
\includegraphics[height=0.22\textwidth,clip]{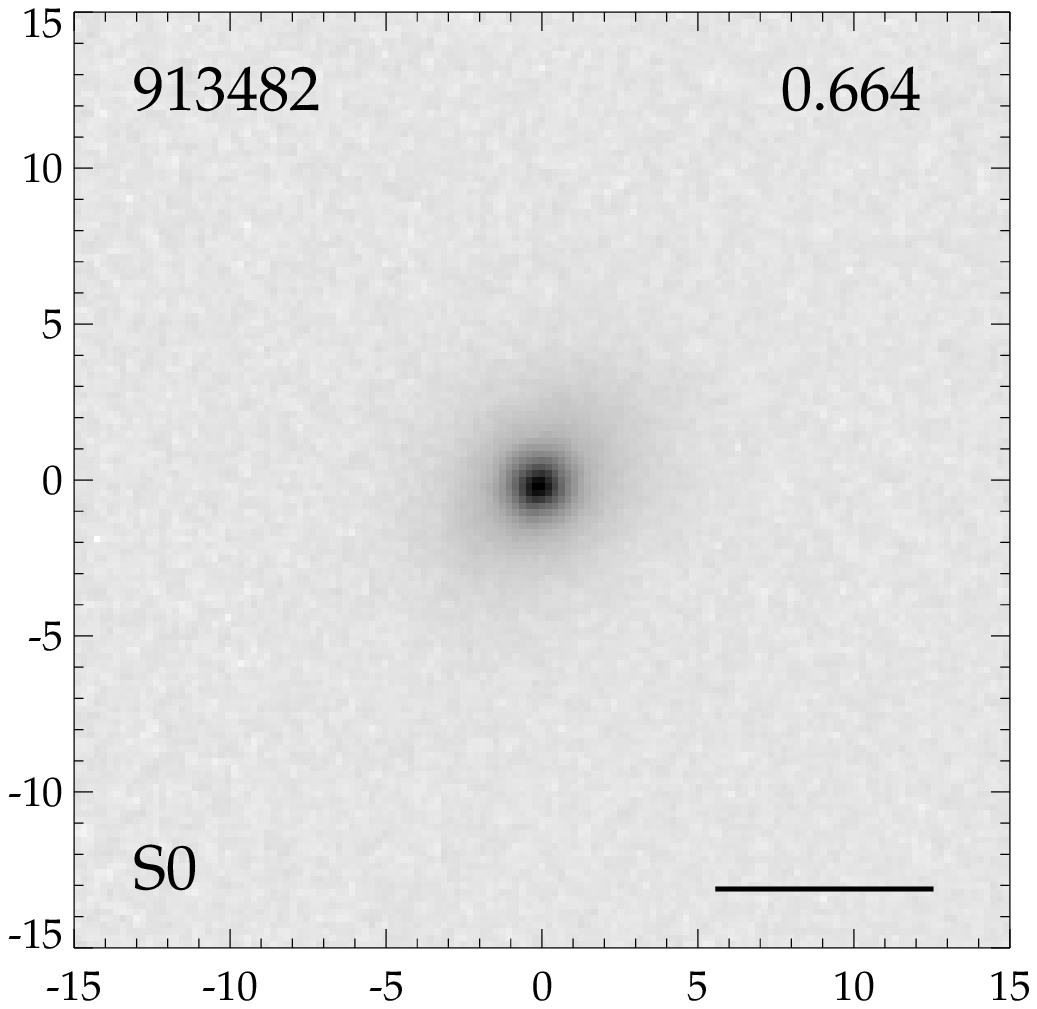} \includegraphics[height=0.22\textwidth,clip]{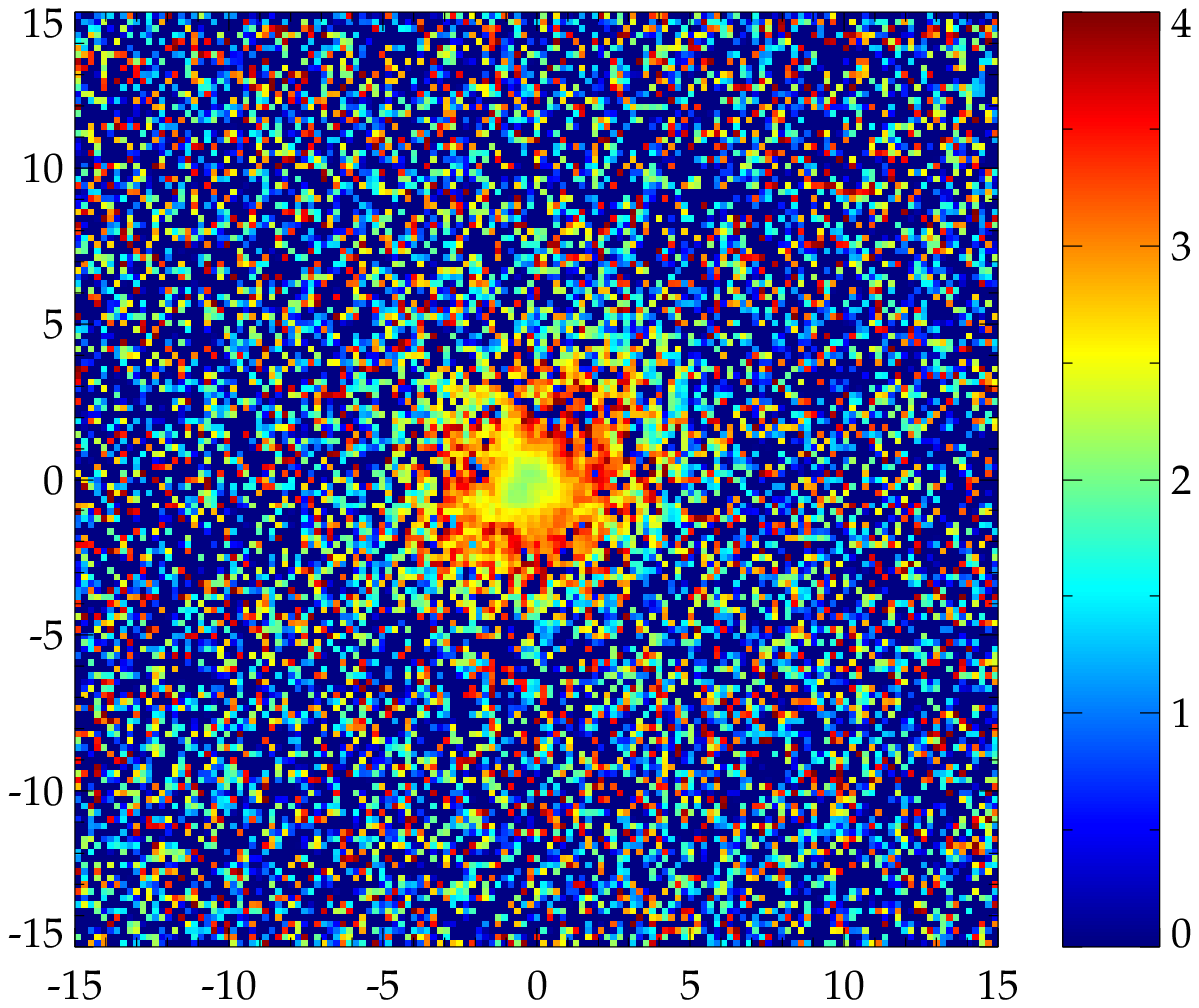}
\includegraphics[height=0.22\textwidth,clip]{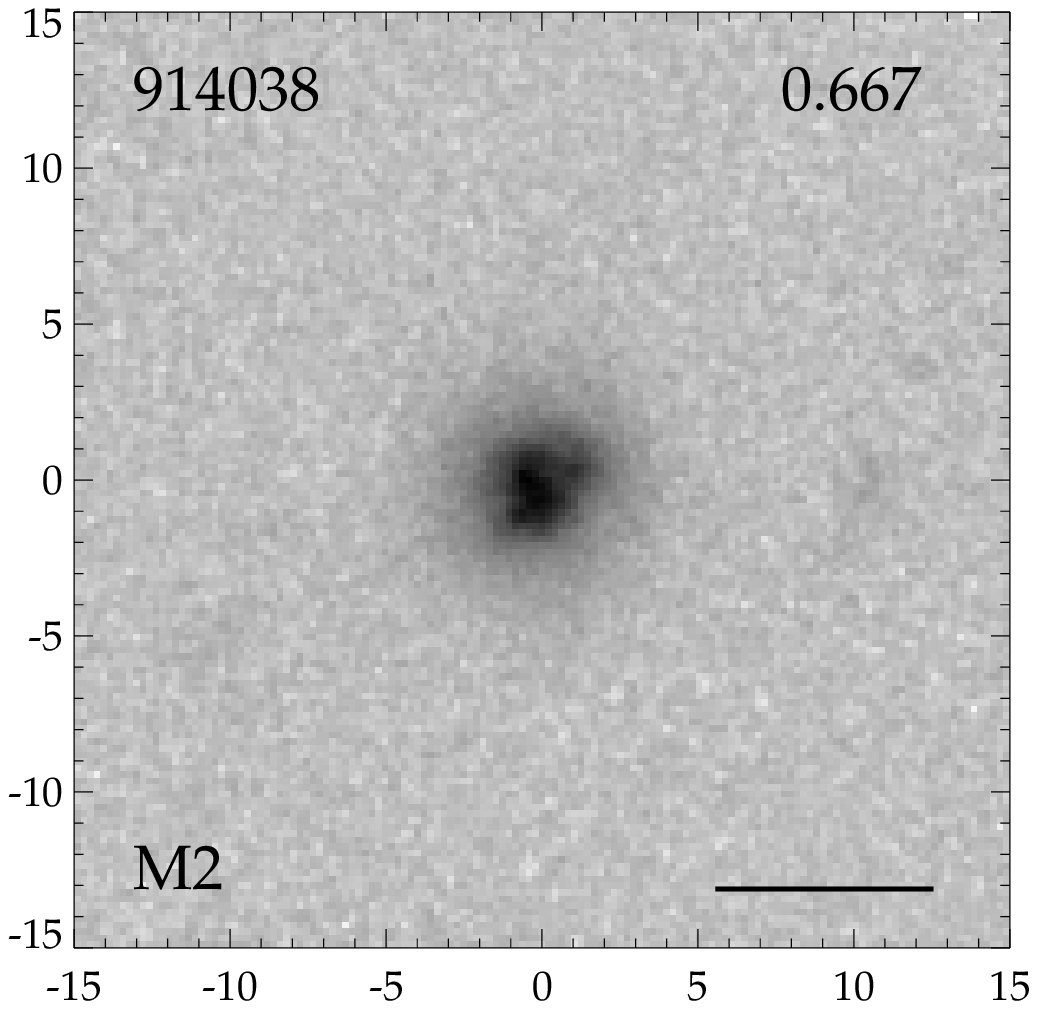} \includegraphics[height=0.22\textwidth,clip]{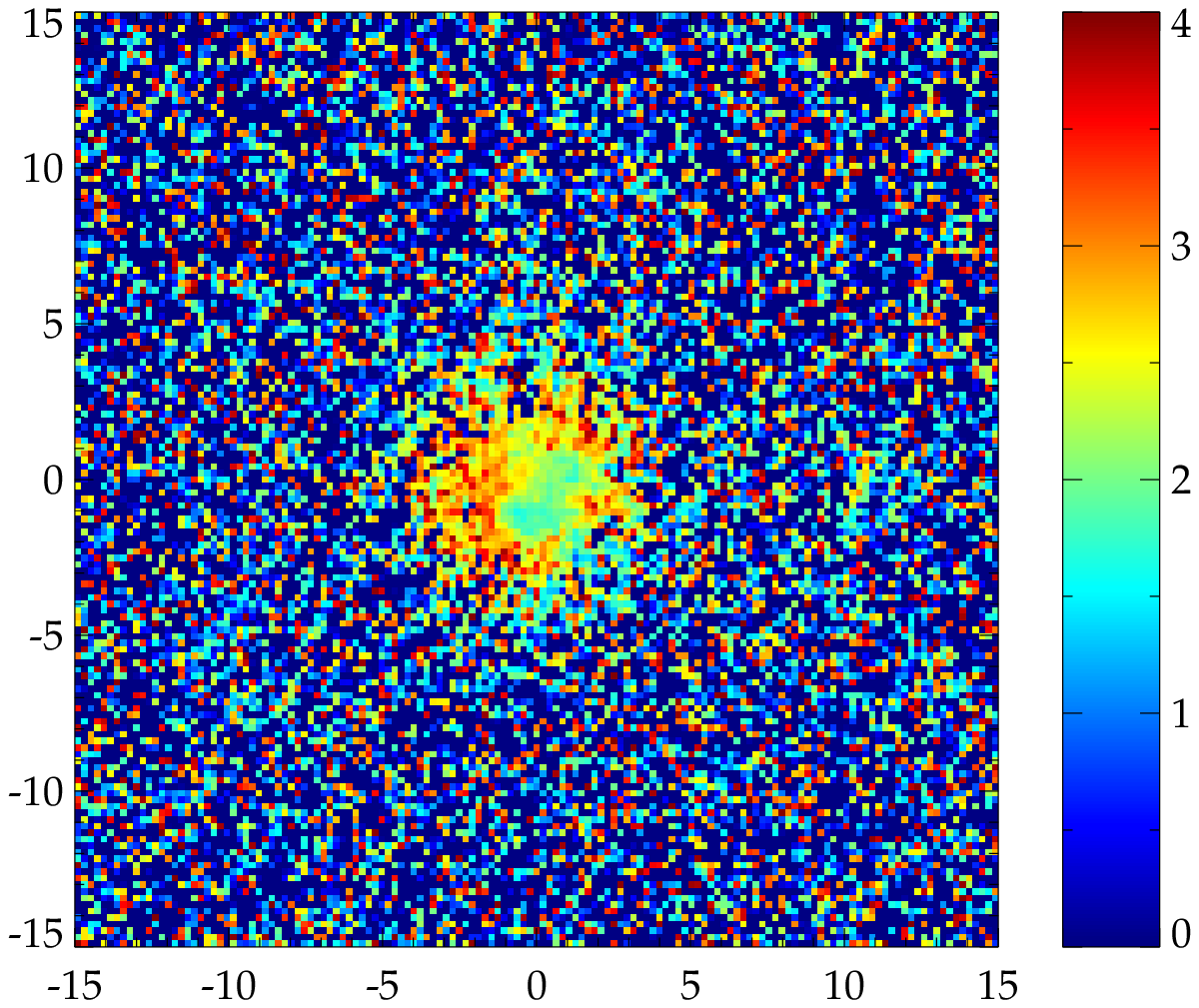}
\includegraphics[height=0.22\textwidth,clip]{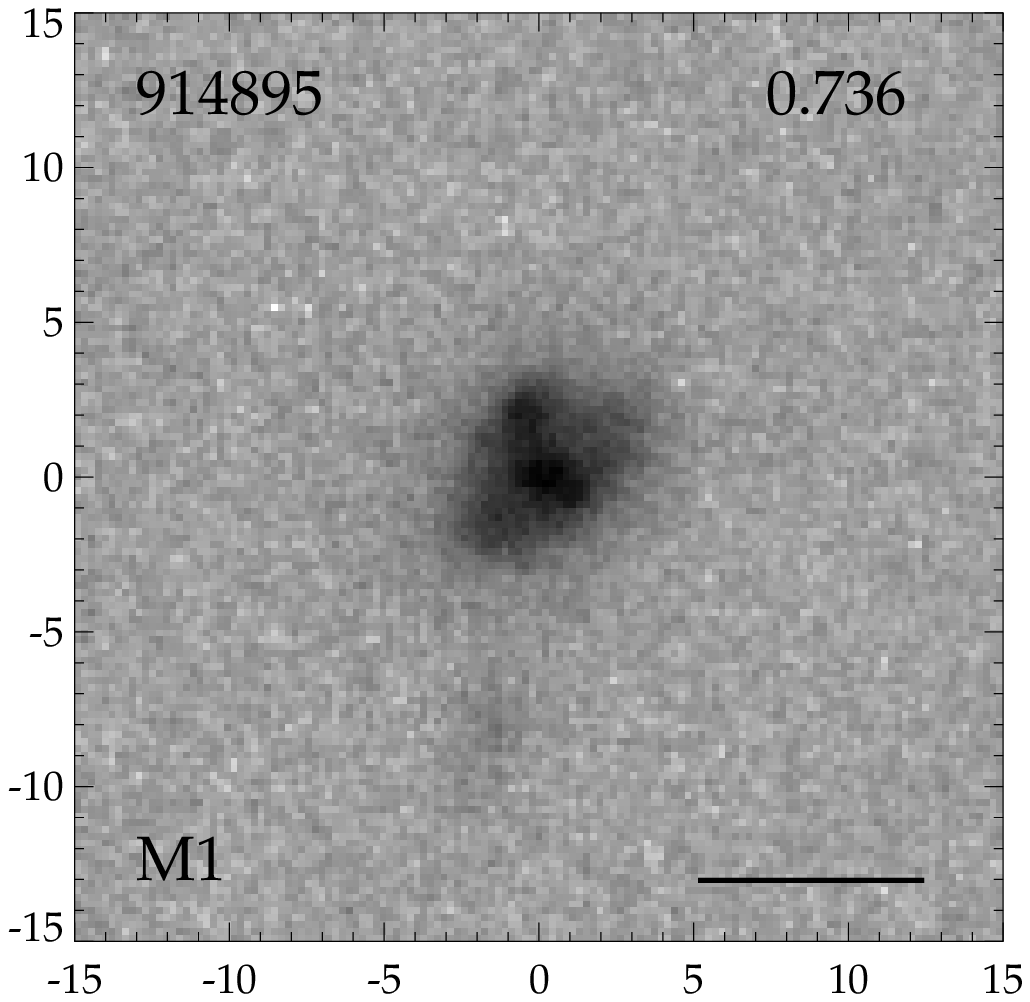} \includegraphics[height=0.22\textwidth,clip]{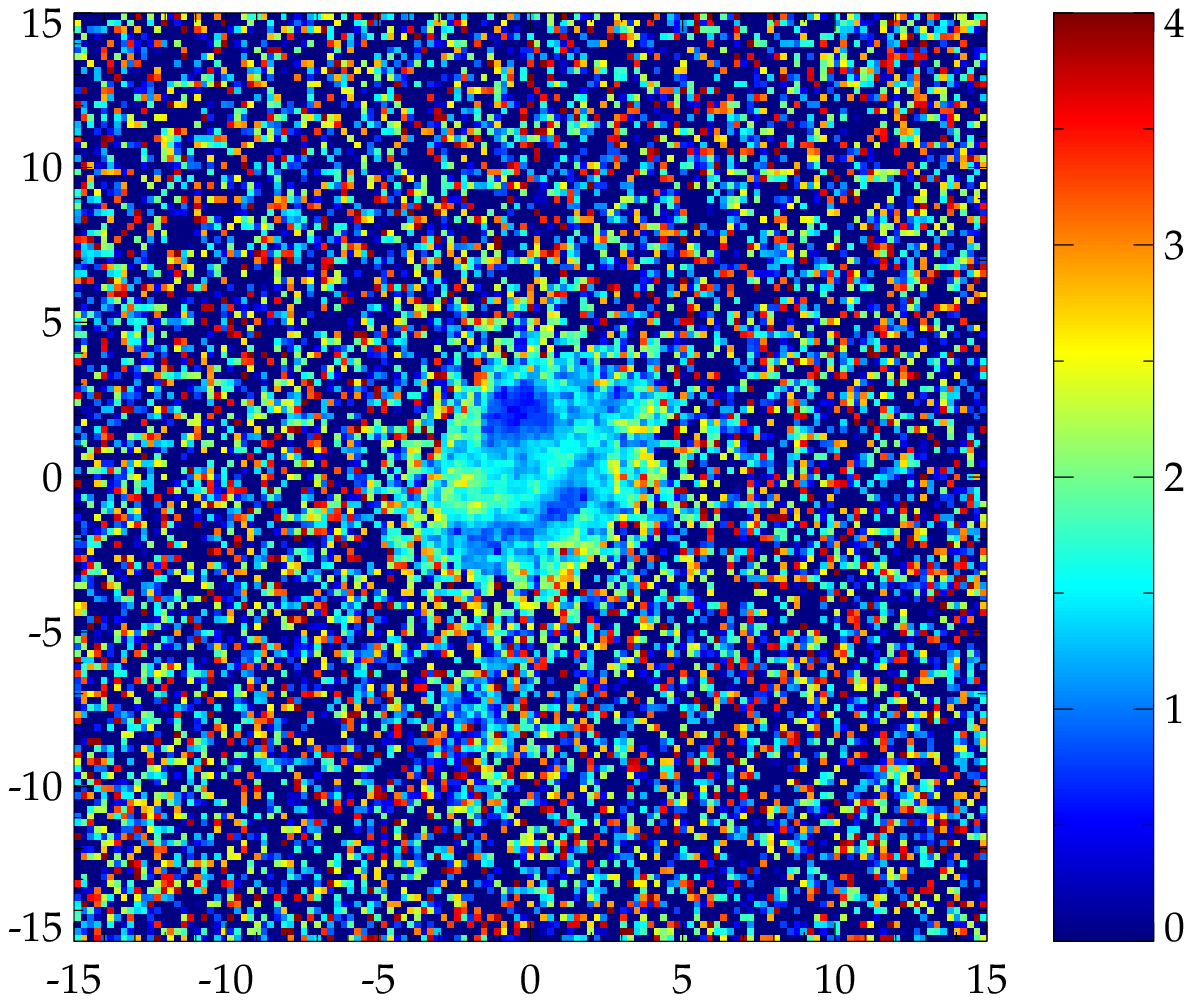}
\includegraphics[height=0.22\textwidth,clip]{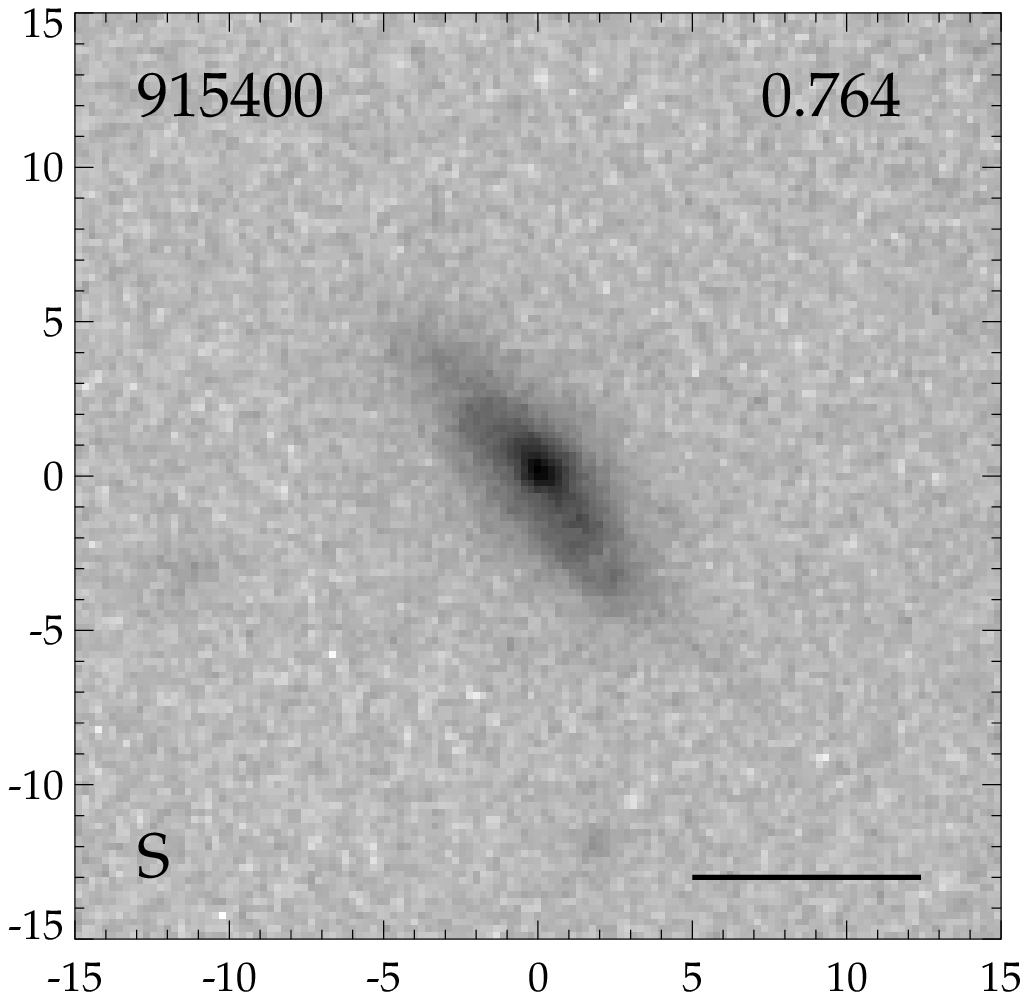} \includegraphics[height=0.22\textwidth,clip]{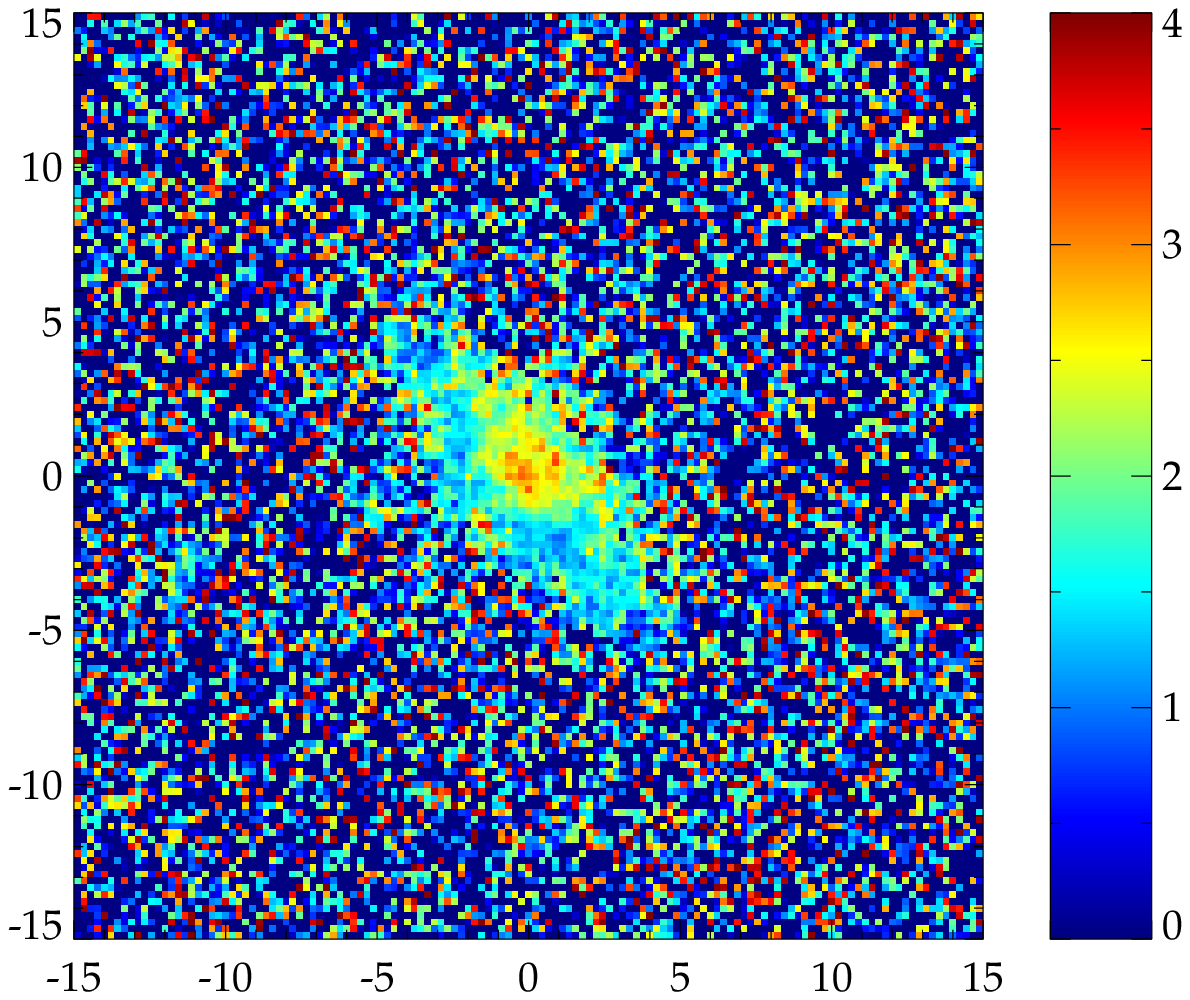}
\includegraphics[height=0.22\textwidth,clip]{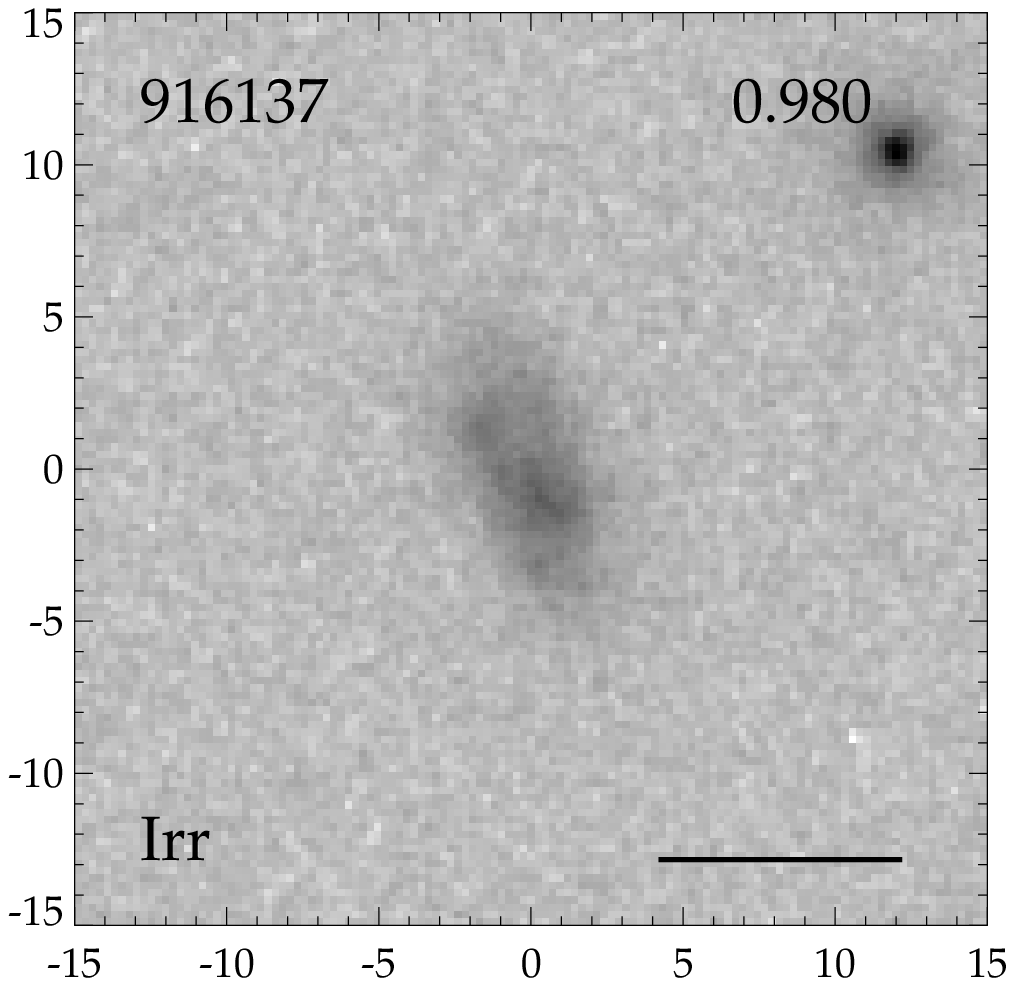} \includegraphics[height=0.22\textwidth,clip]{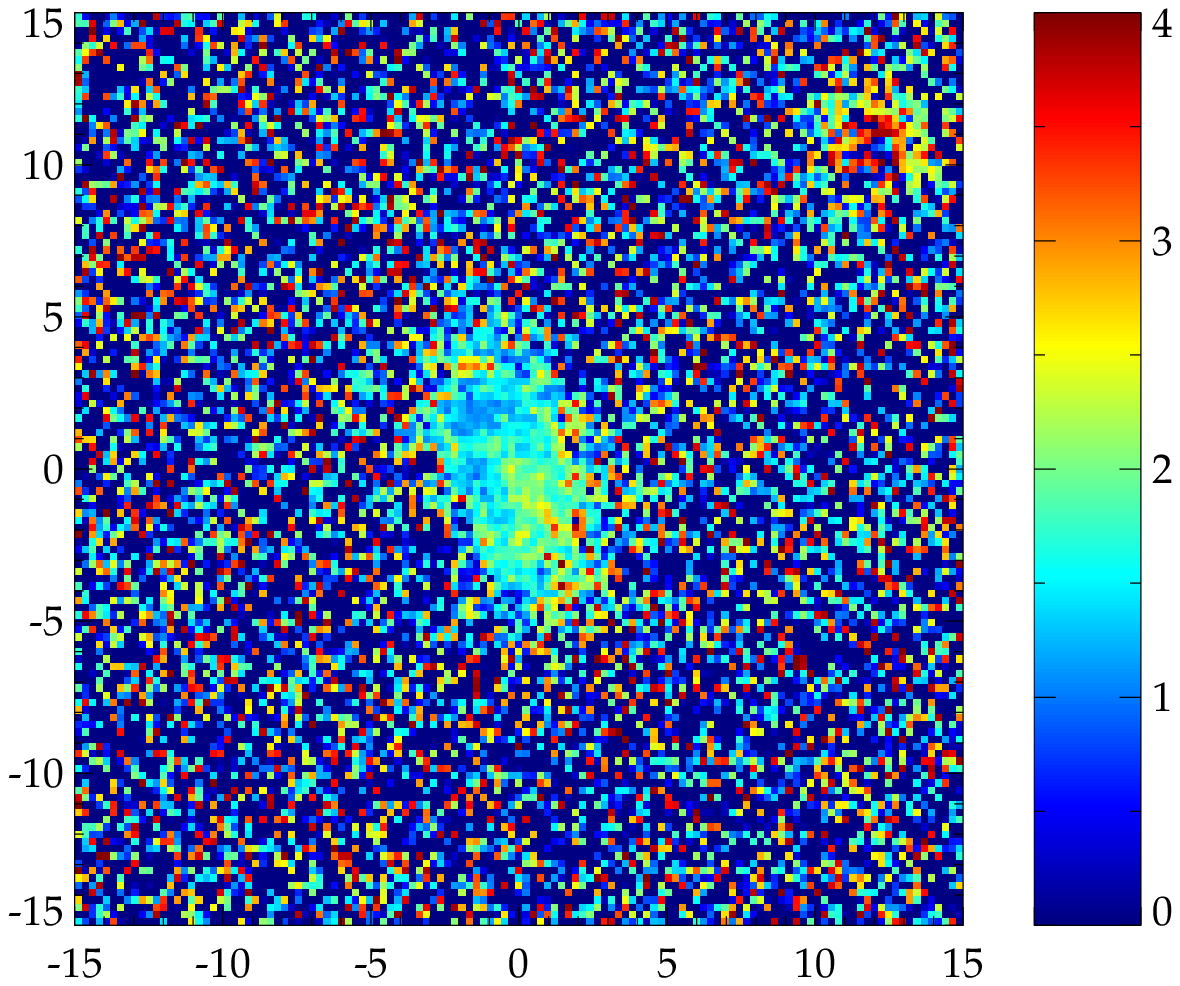}
\caption{Continued.} \end{figure*}

\addtocounter{figure}{-1}
\begin{figure*} \centering
\includegraphics[height=0.22\textwidth,clip]{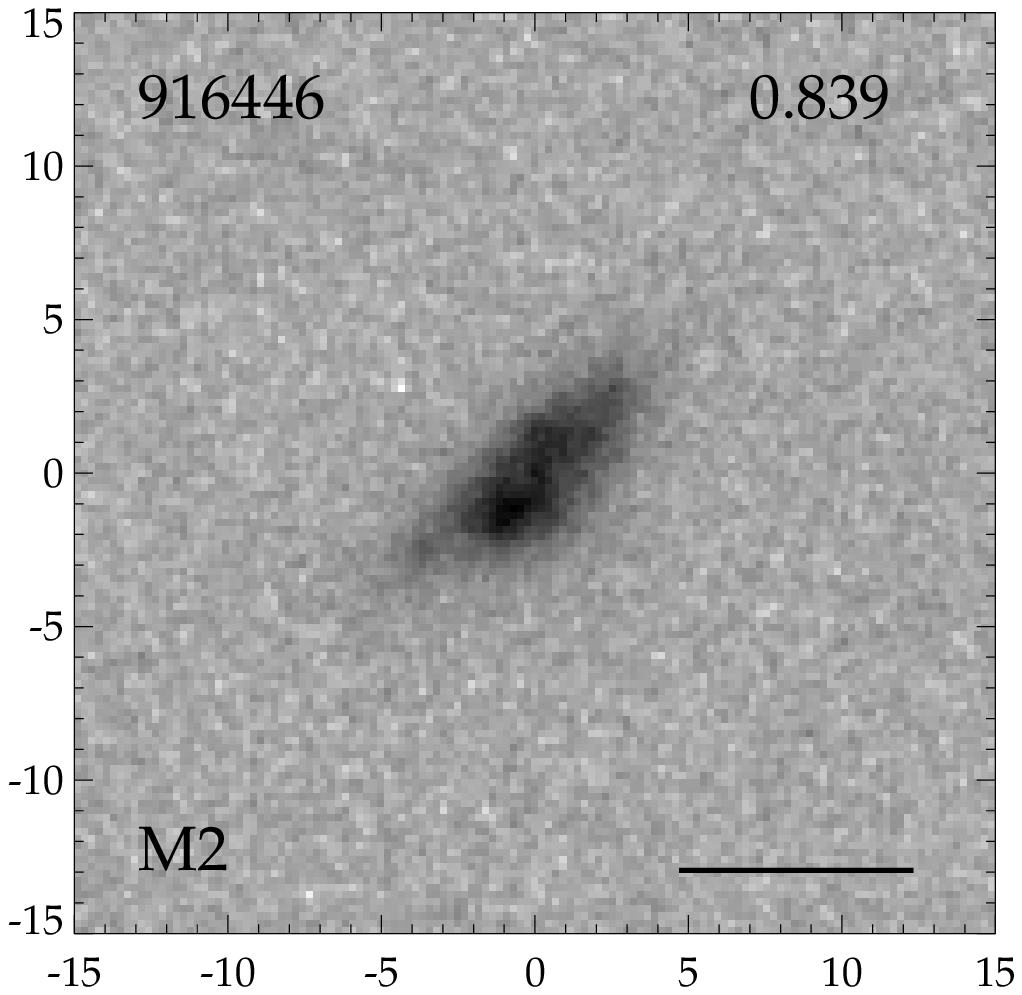} \includegraphics[height=0.22\textwidth,clip]{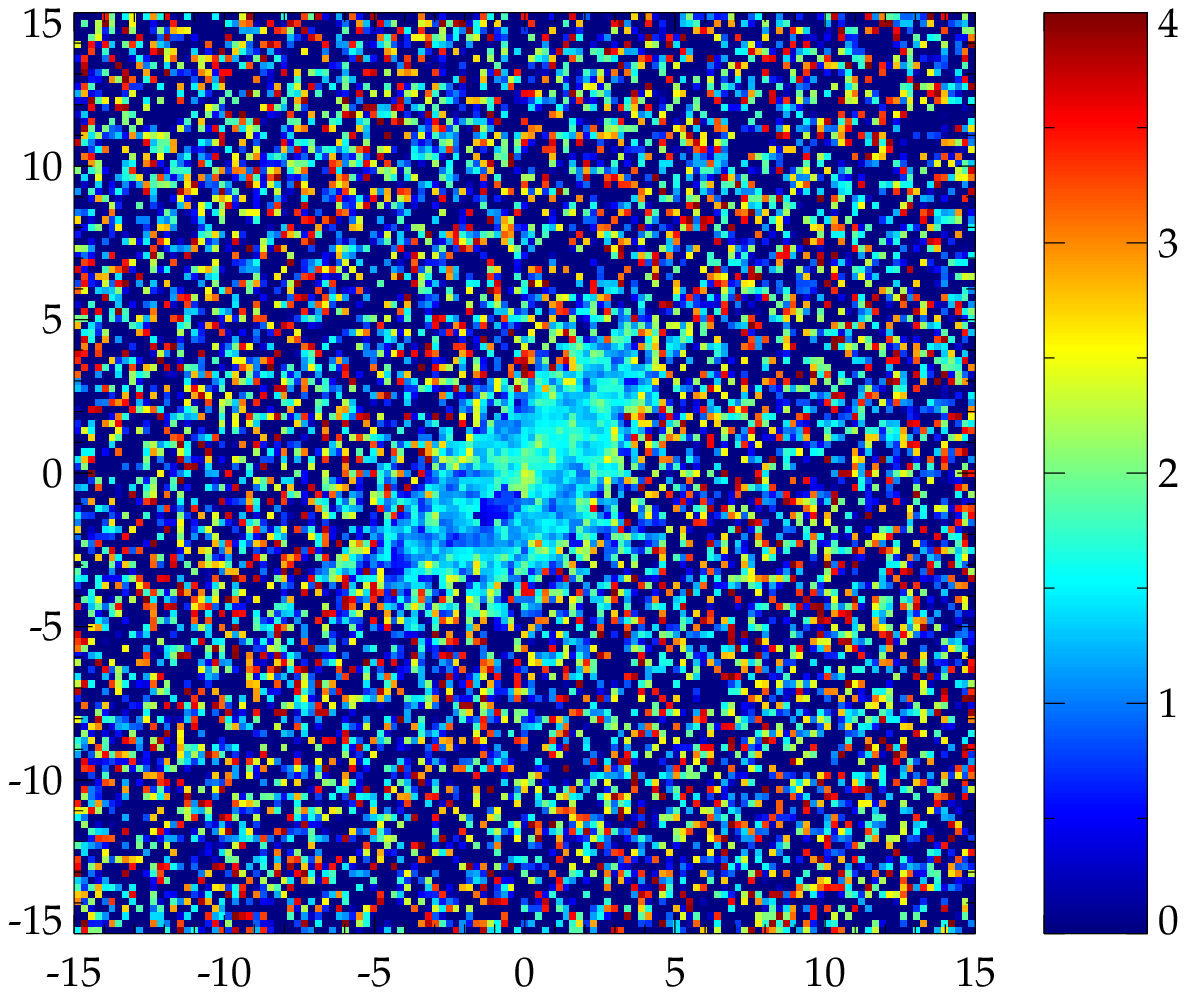}
\includegraphics[height=0.22\textwidth,clip]{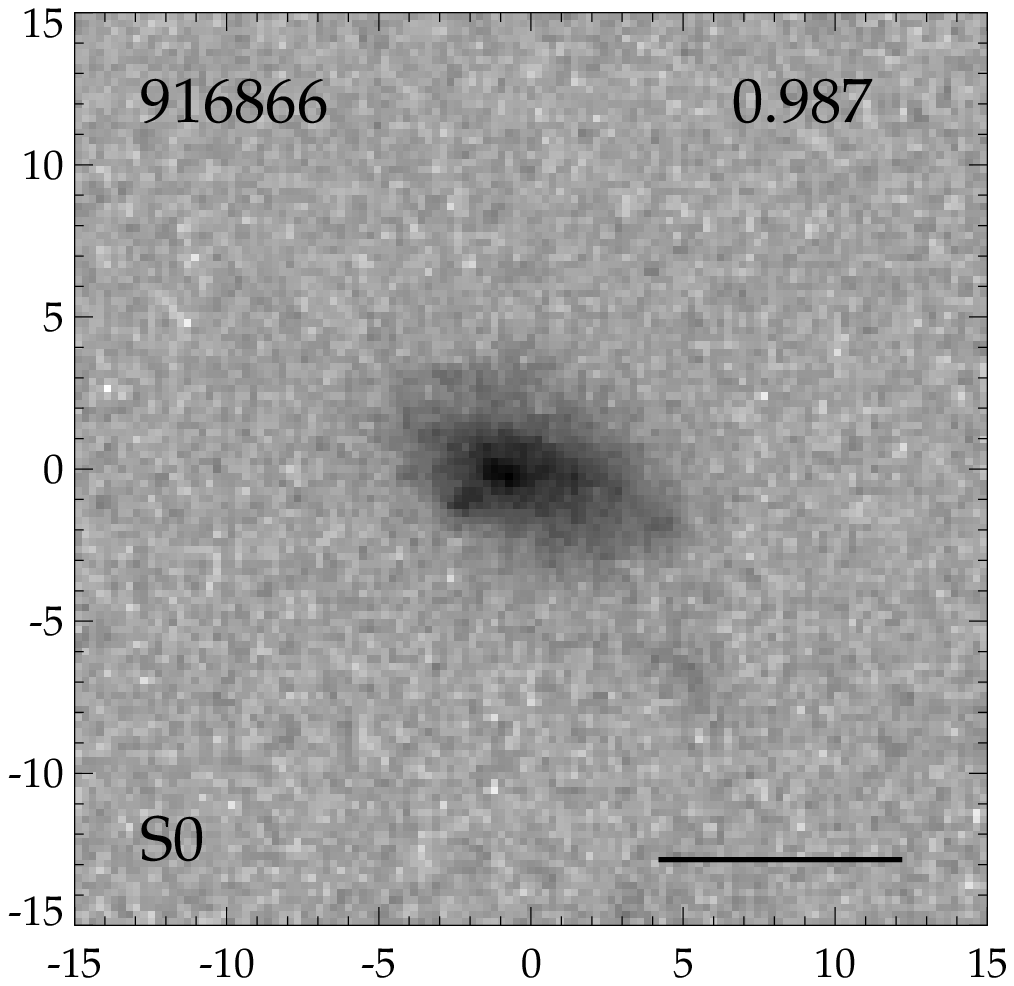} \includegraphics[height=0.22\textwidth,clip]{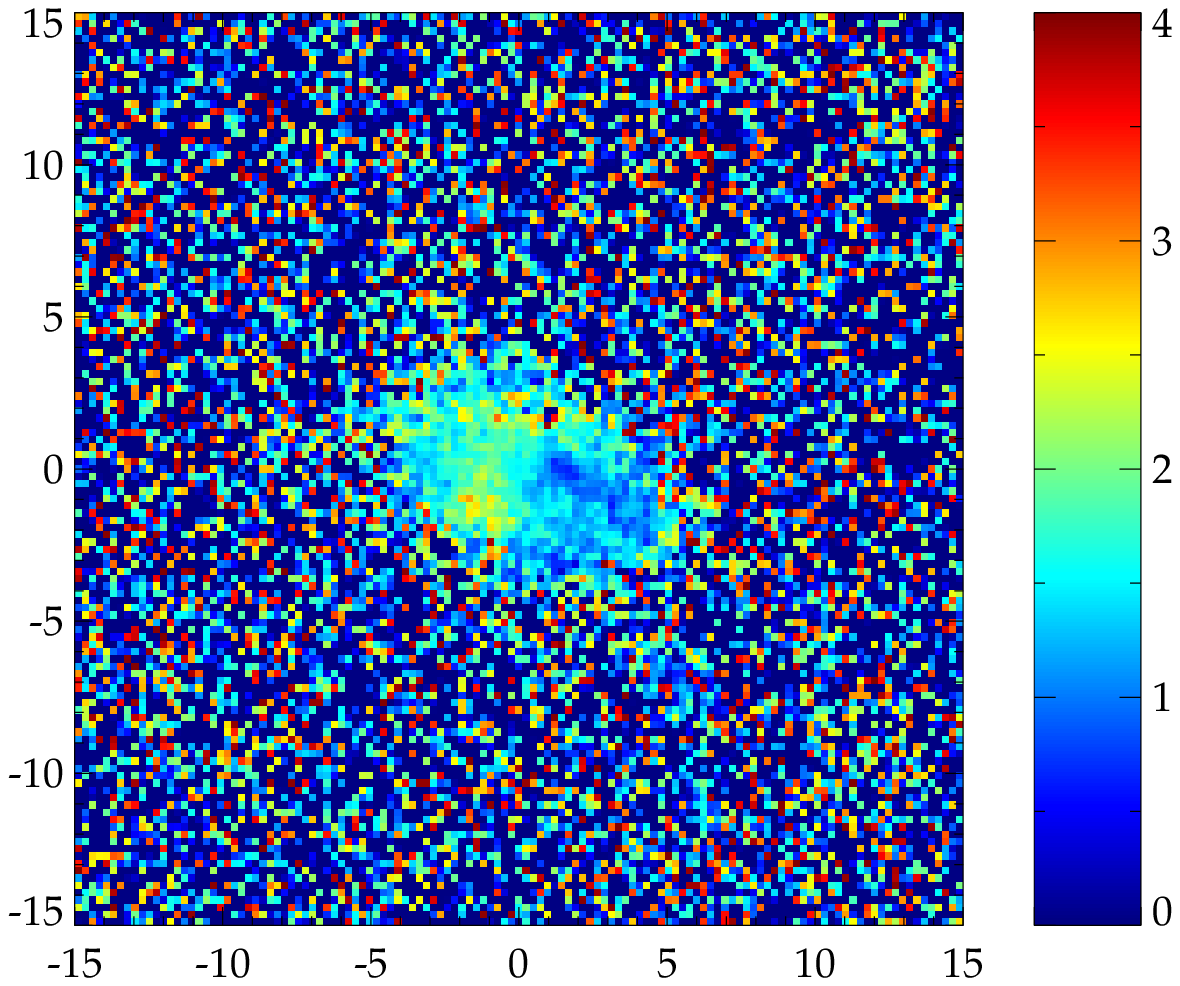}
\includegraphics[height=0.22\textwidth,clip]{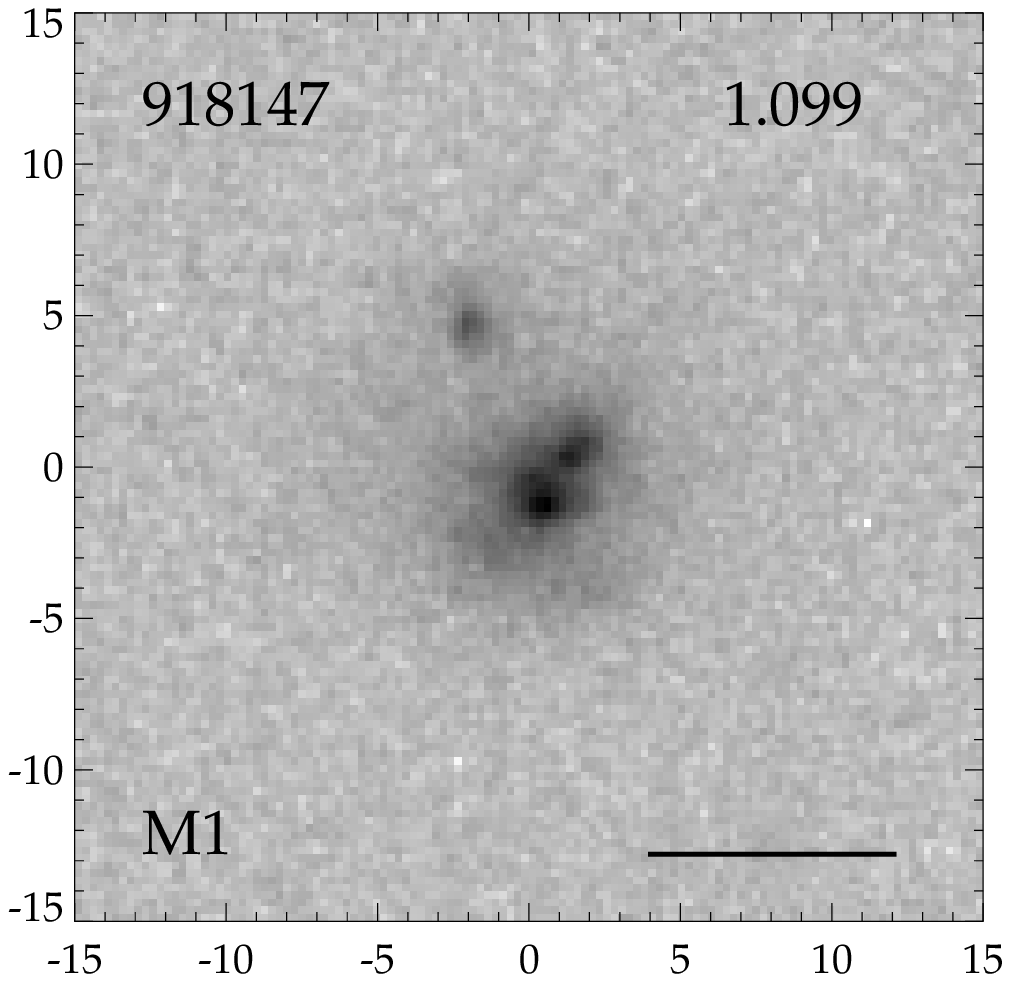} \includegraphics[height=0.22\textwidth,clip]{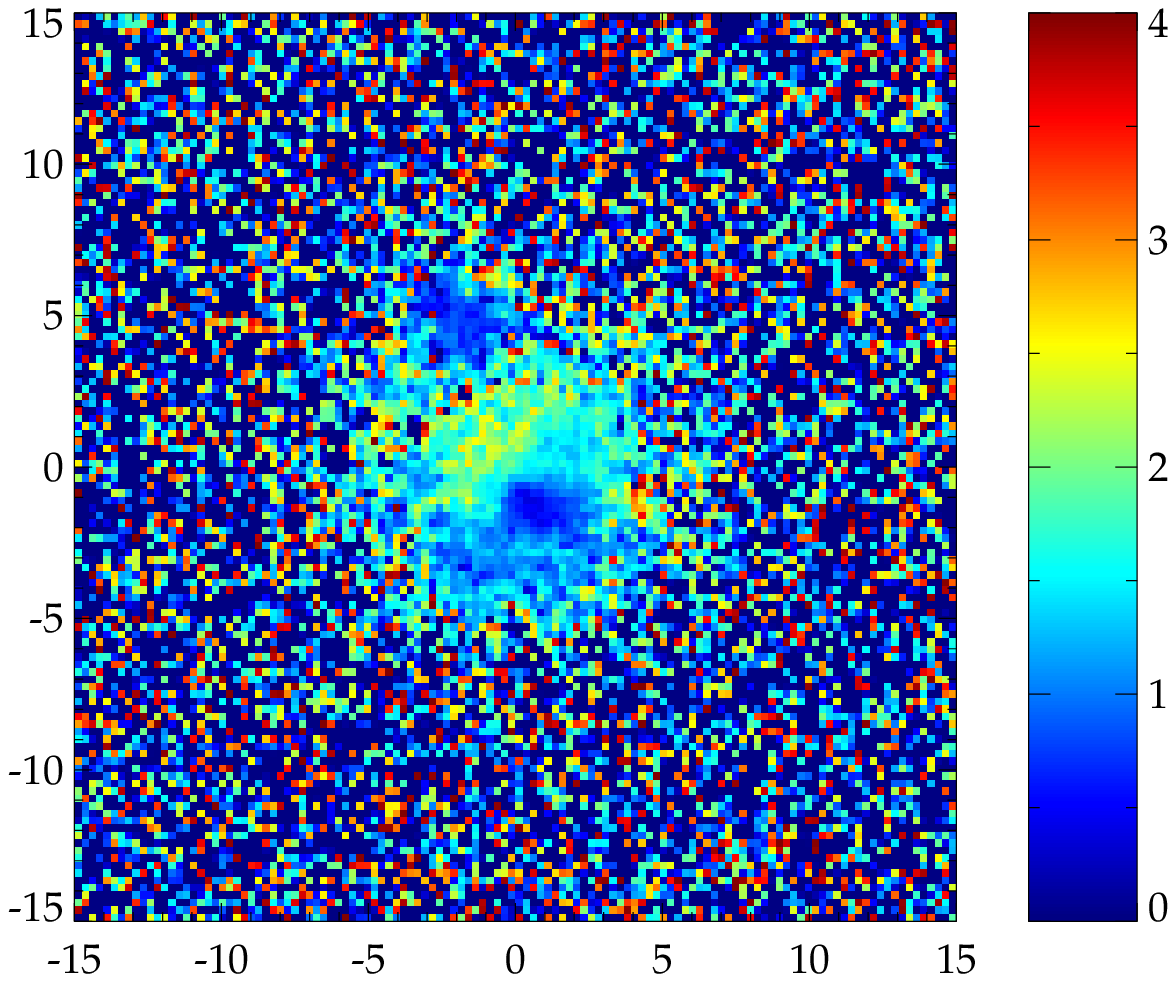}
\includegraphics[height=0.22\textwidth,clip]{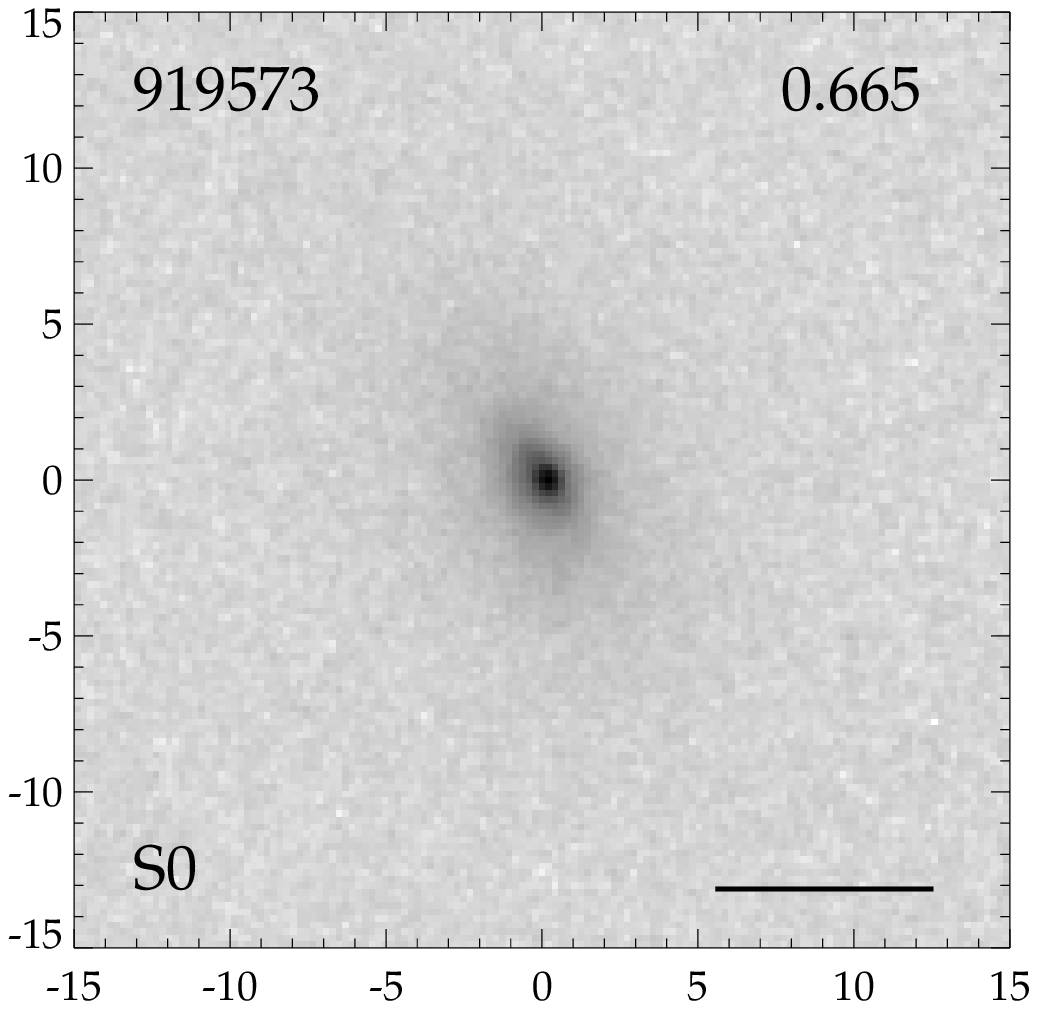} \includegraphics[height=0.22\textwidth,clip]{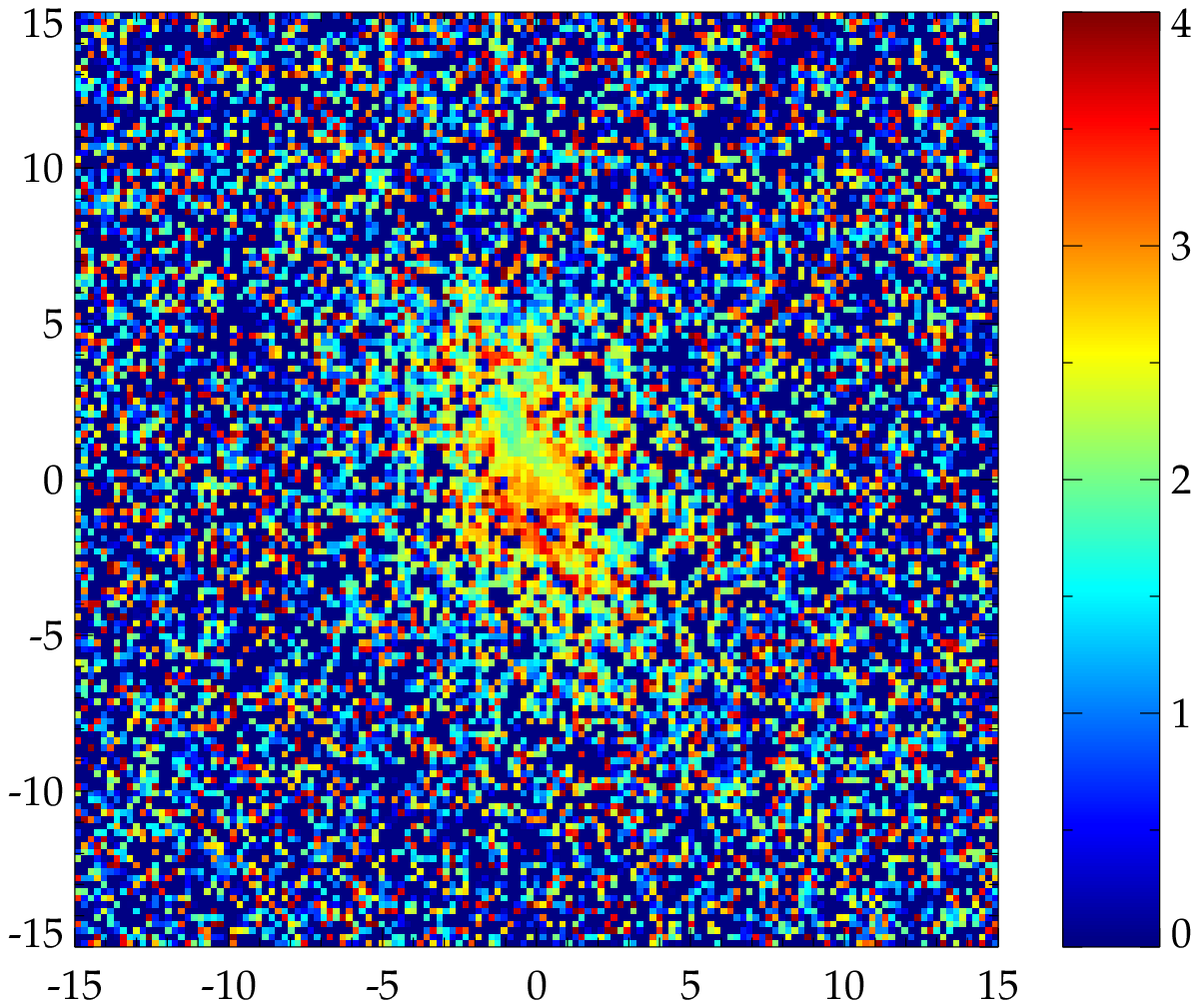}
\includegraphics[height=0.22\textwidth,clip]{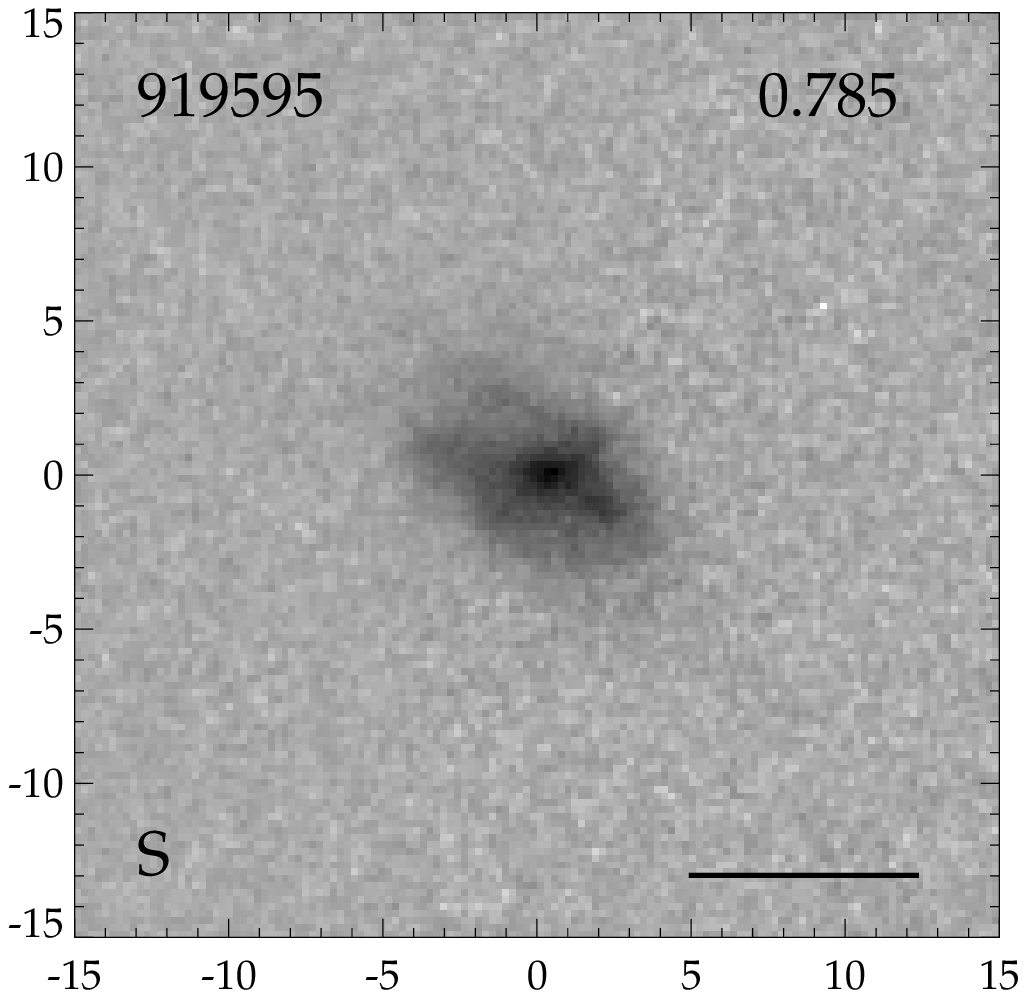} \includegraphics[height=0.22\textwidth,clip]{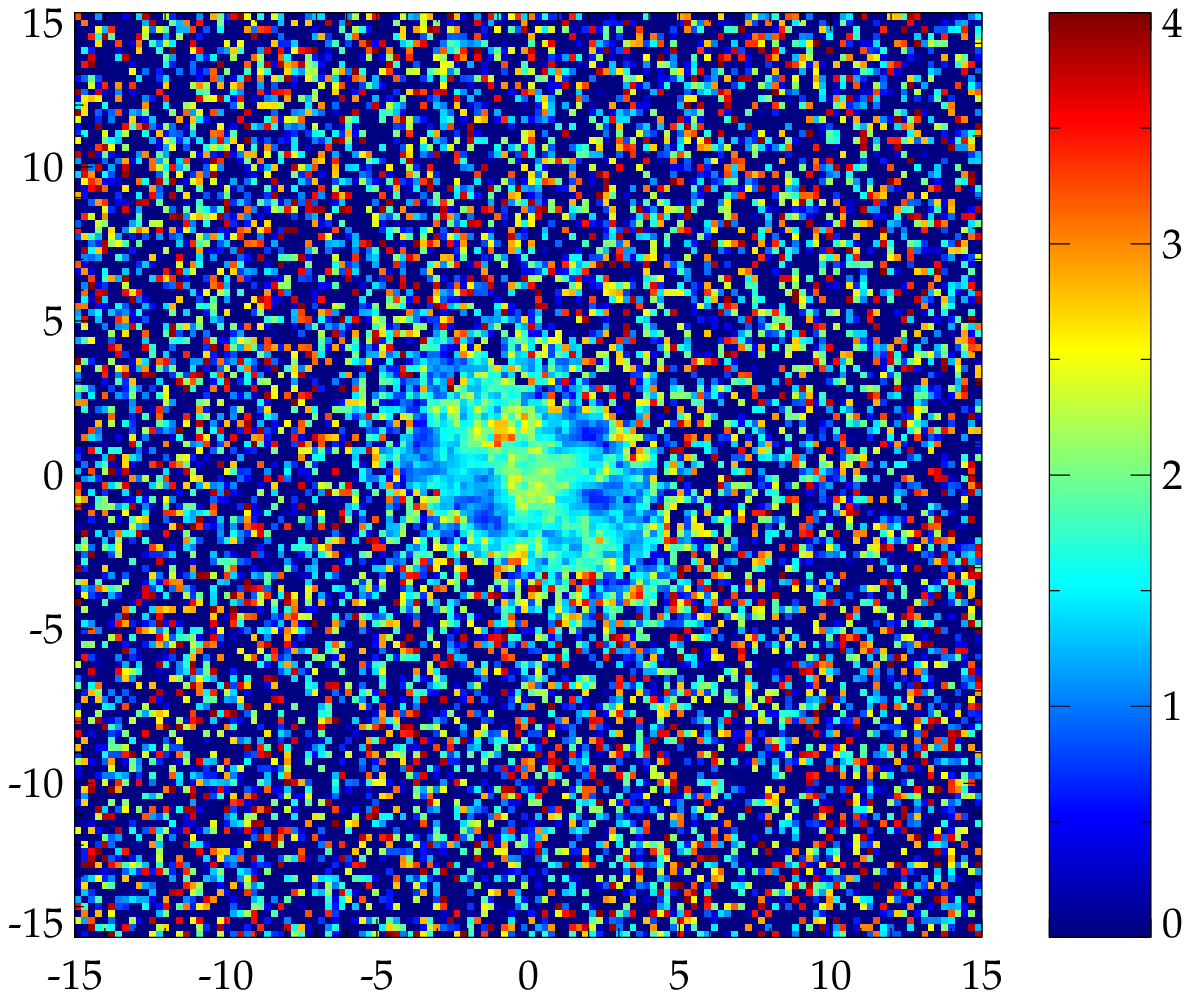}
\includegraphics[height=0.22\textwidth,clip]{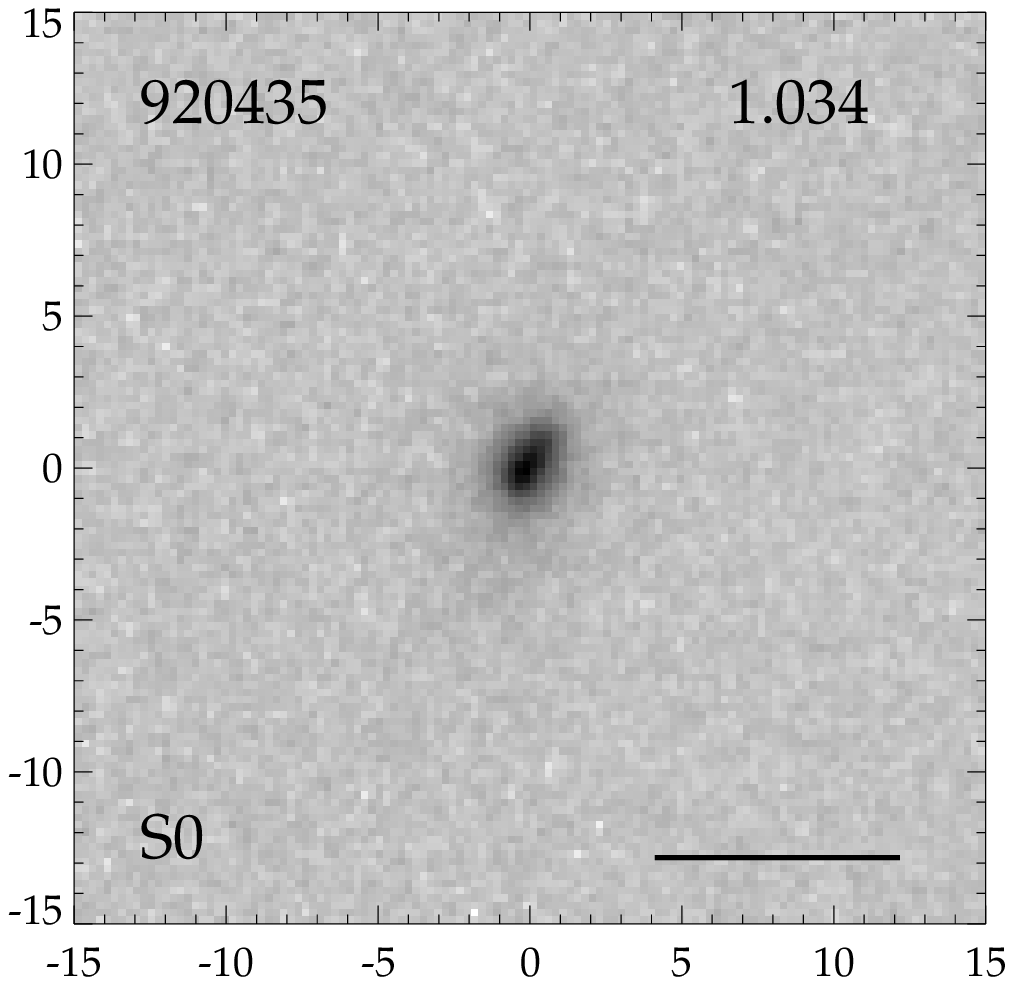} \includegraphics[height=0.22\textwidth,clip]{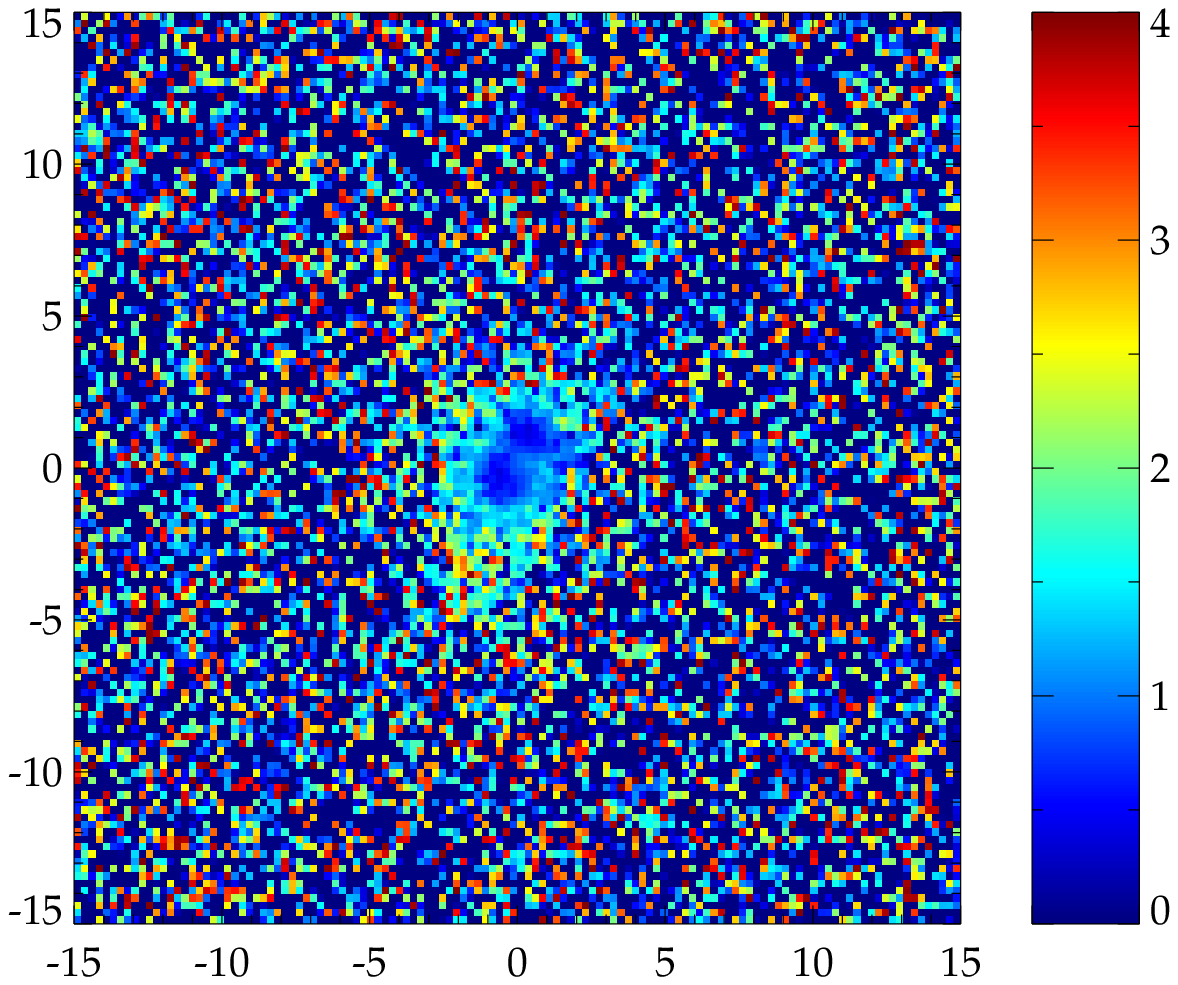}
\includegraphics[height=0.22\textwidth,clip]{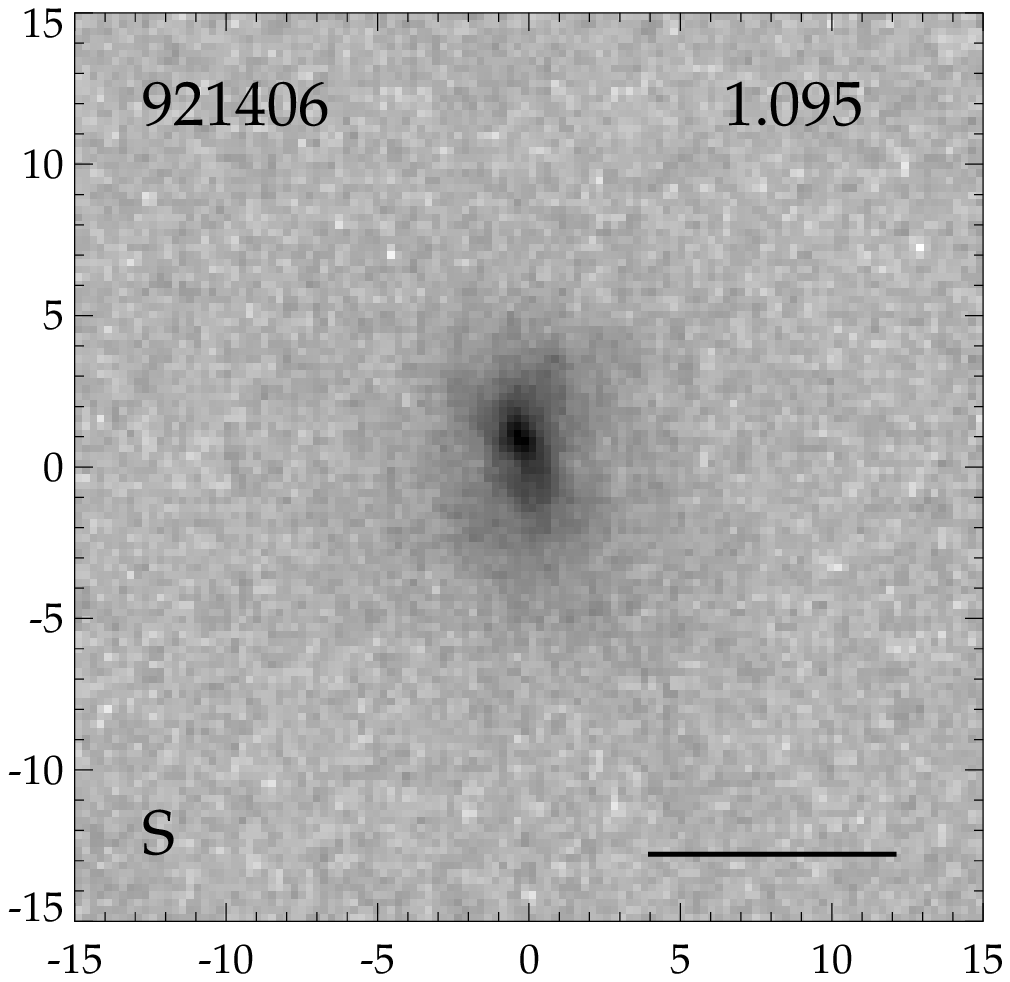} \includegraphics[height=0.22\textwidth,clip]{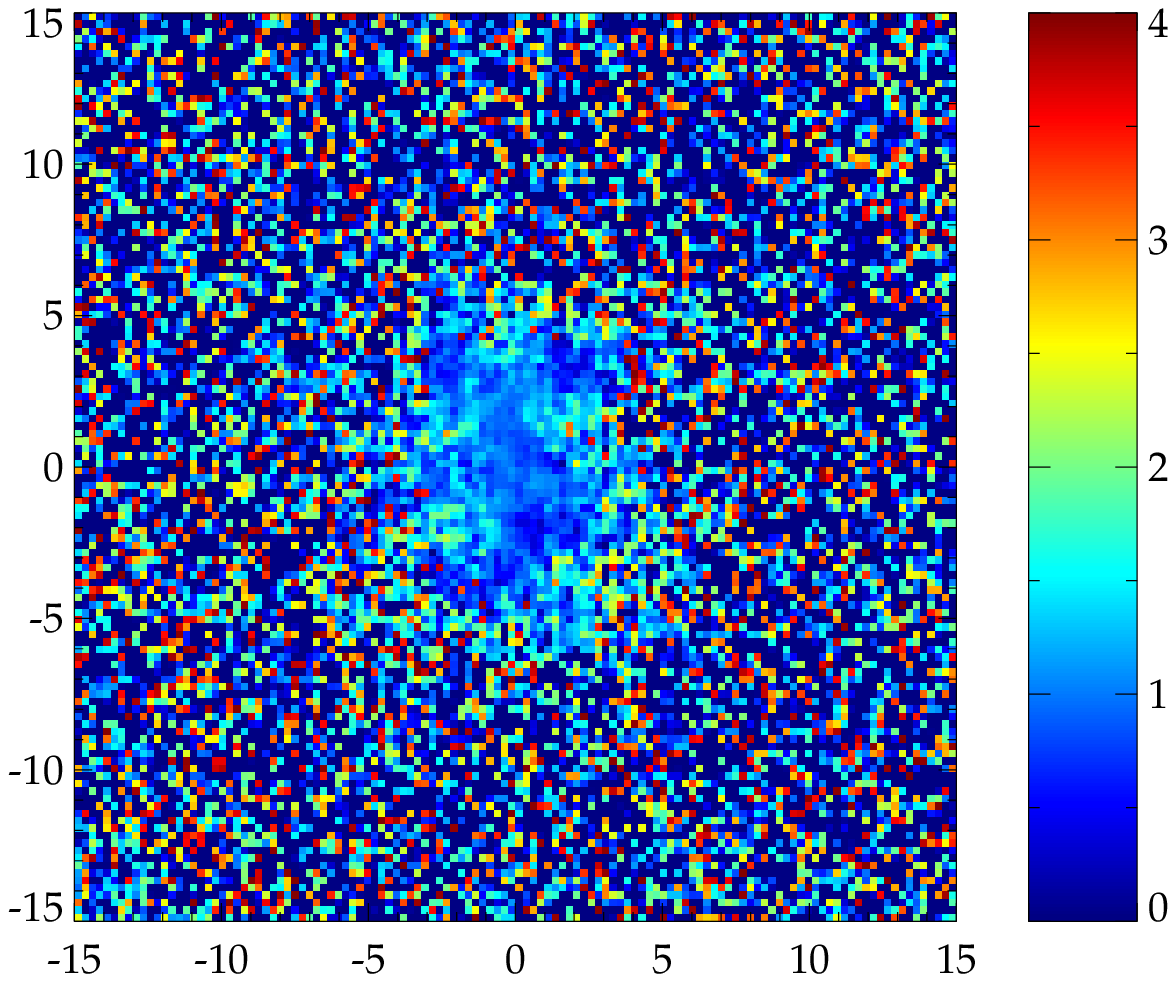}
\includegraphics[height=0.22\textwidth,clip]{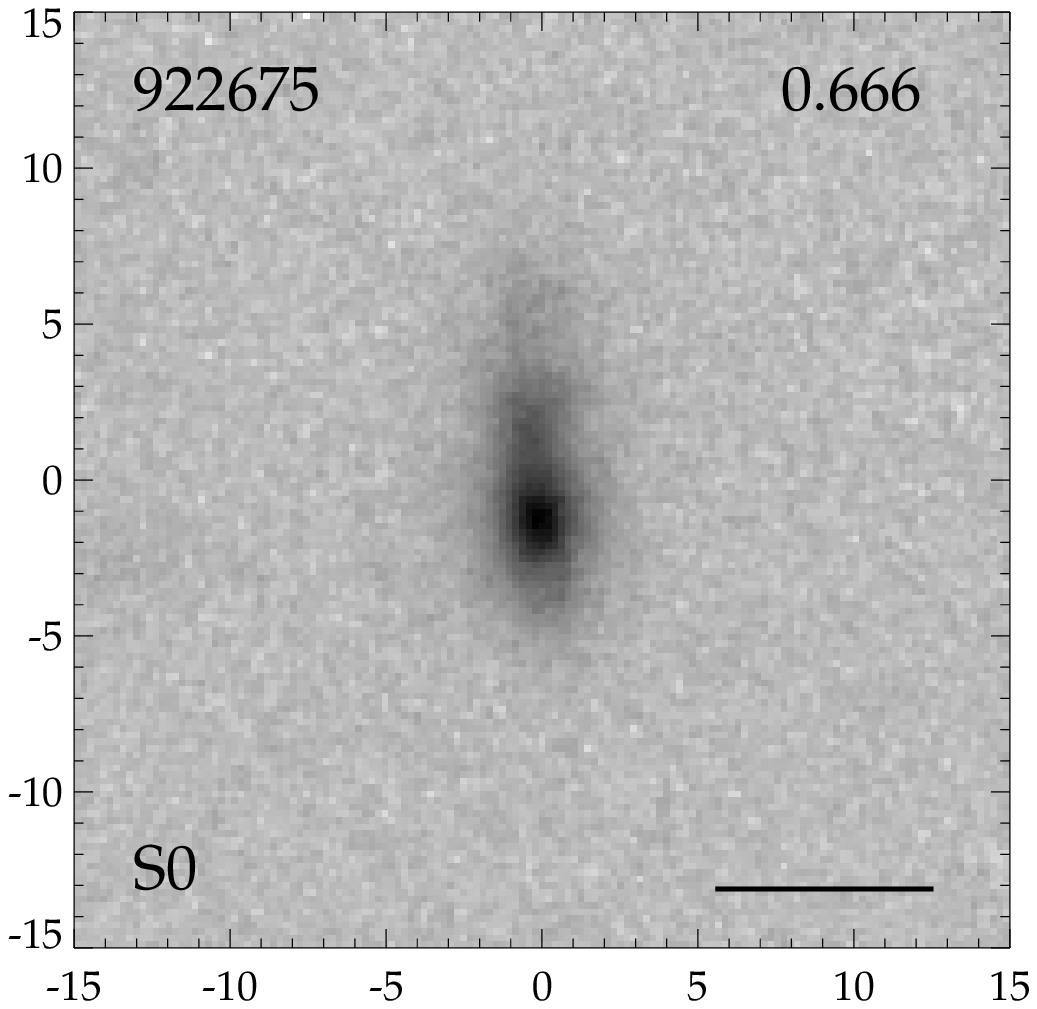} \includegraphics[height=0.22\textwidth,clip]{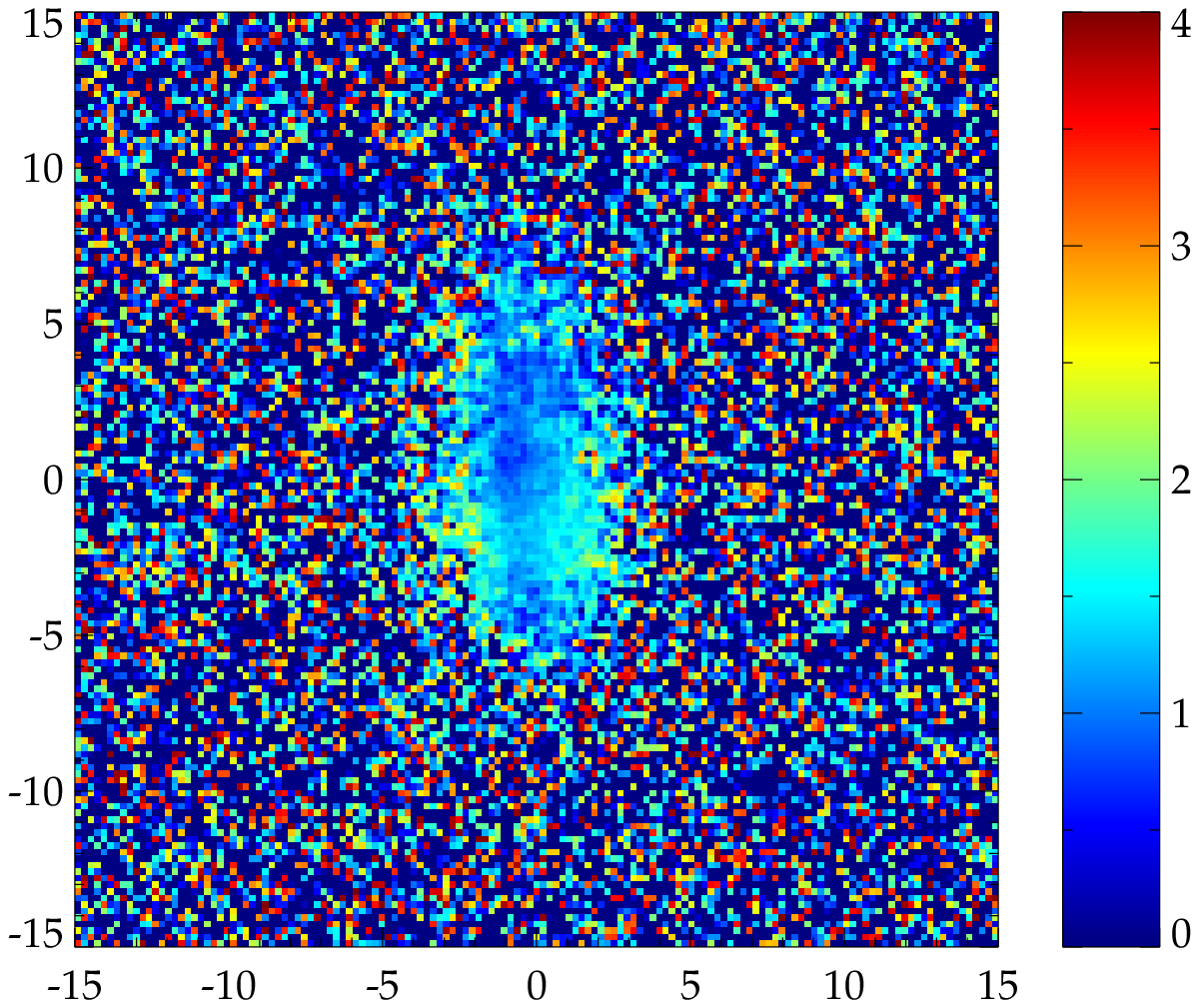}
\includegraphics[height=0.22\textwidth,clip]{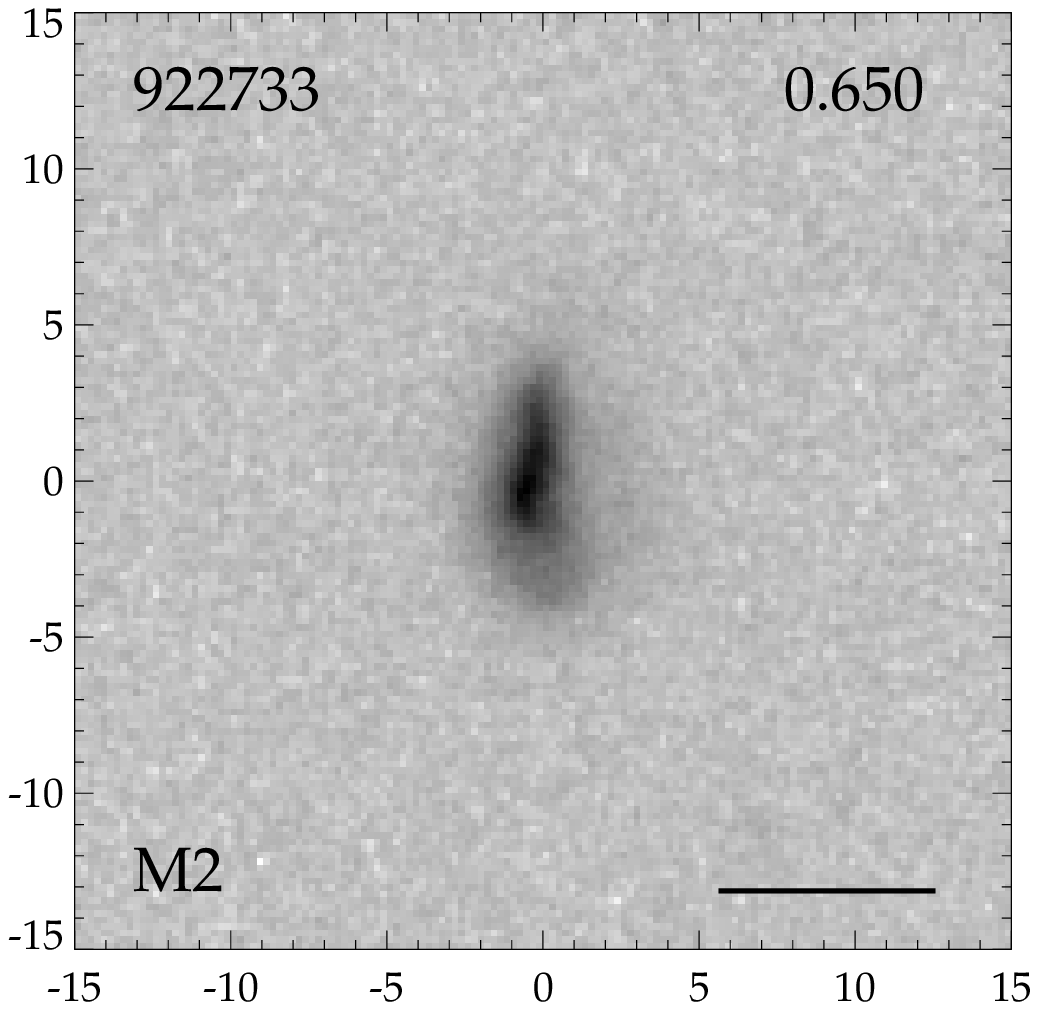} \includegraphics[height=0.22\textwidth,clip]{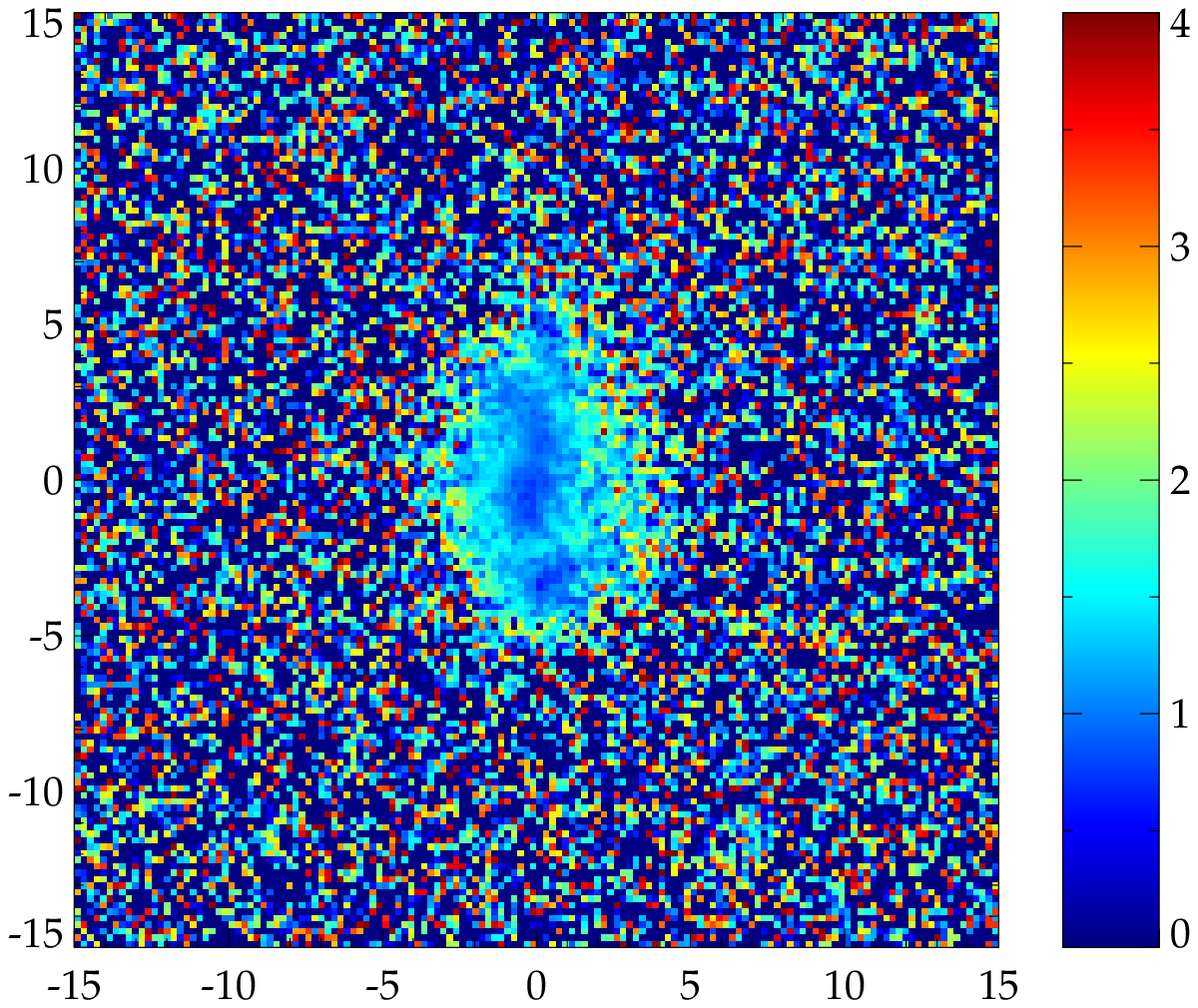}
\includegraphics[height=0.22\textwidth,clip]{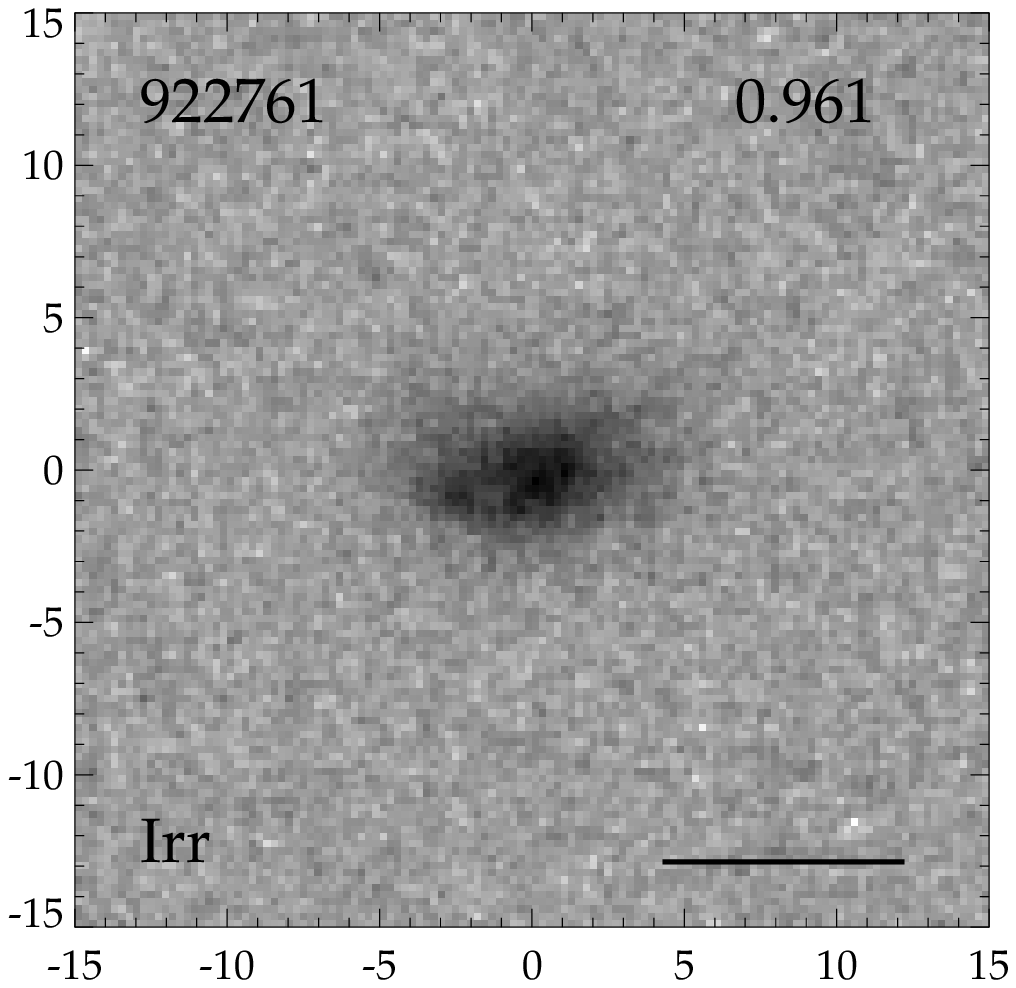} \includegraphics[height=0.22\textwidth,clip]{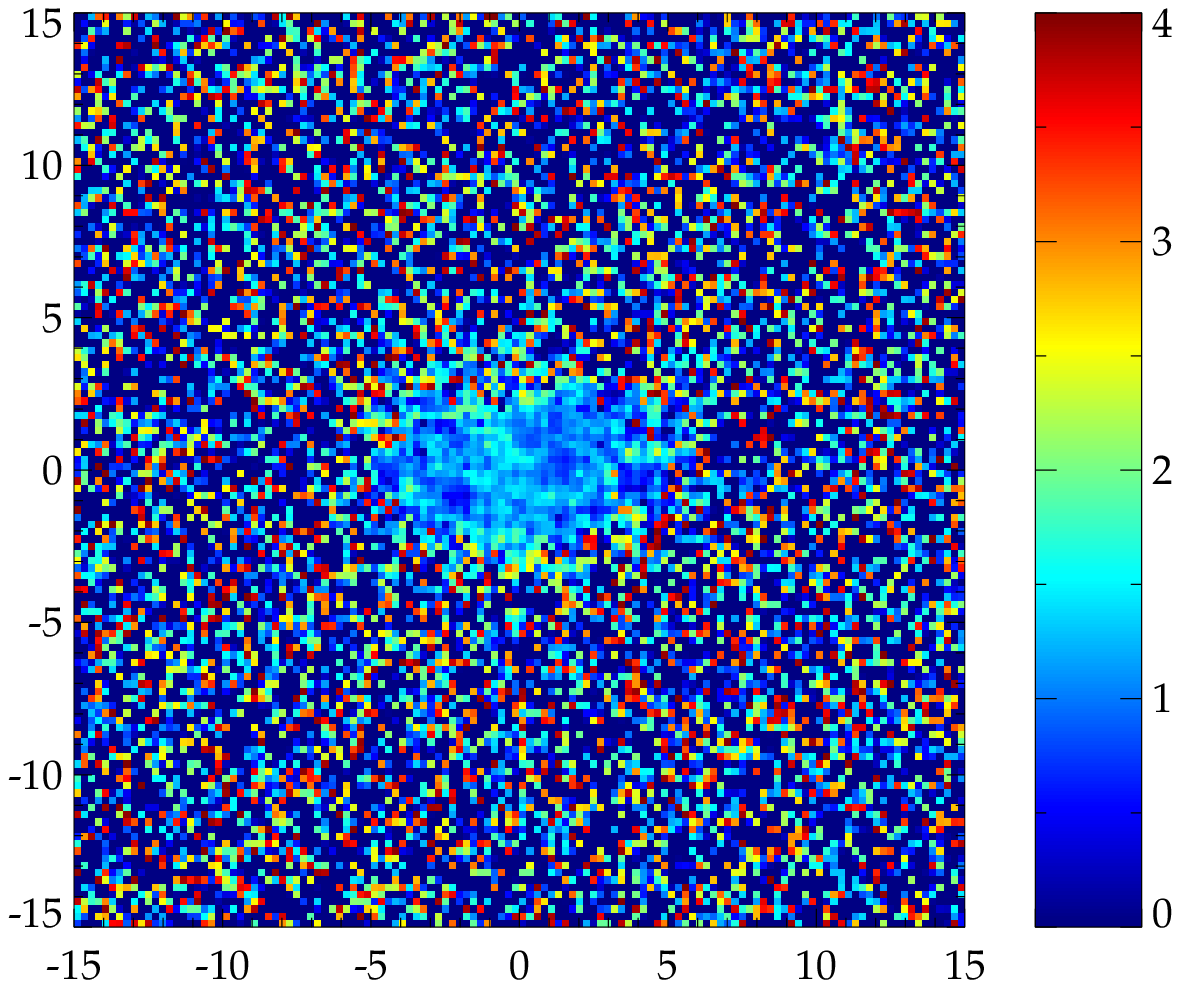}
\includegraphics[height=0.22\textwidth,clip]{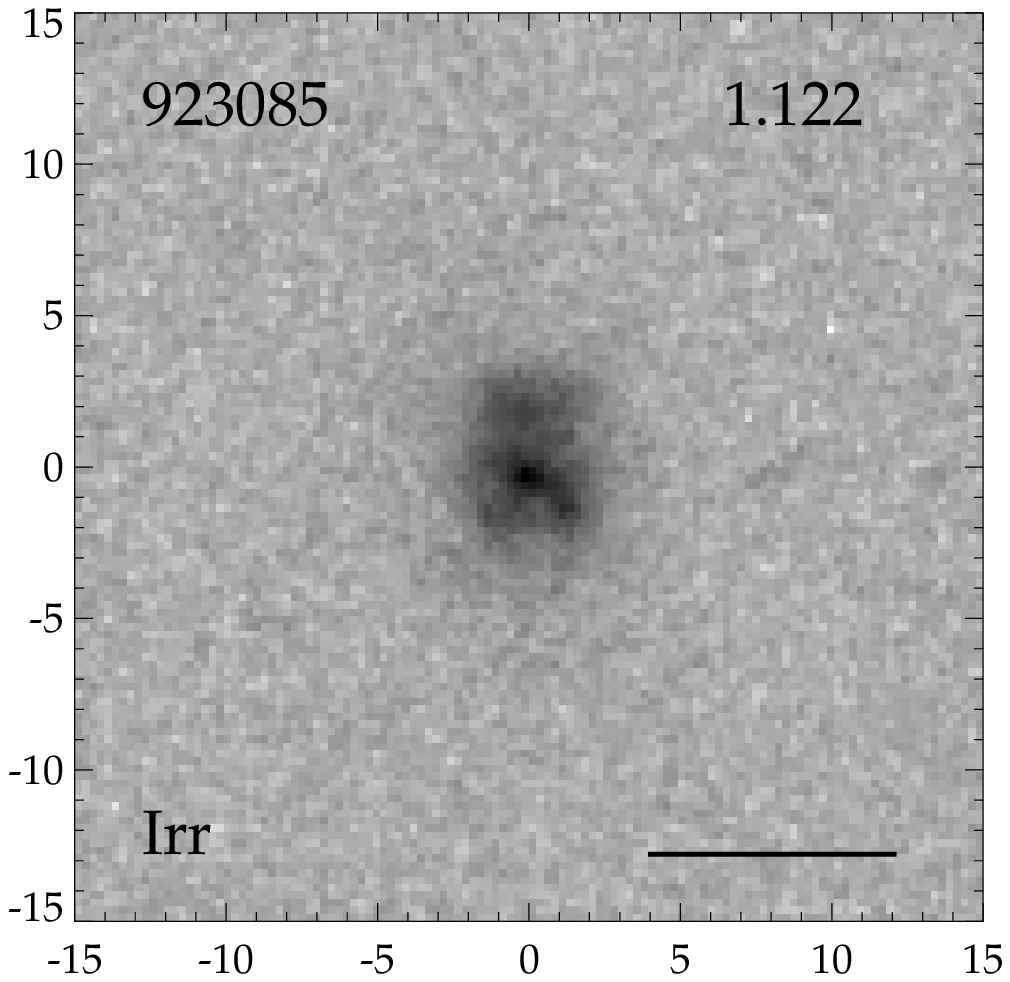} \includegraphics[height=0.22\textwidth,clip]{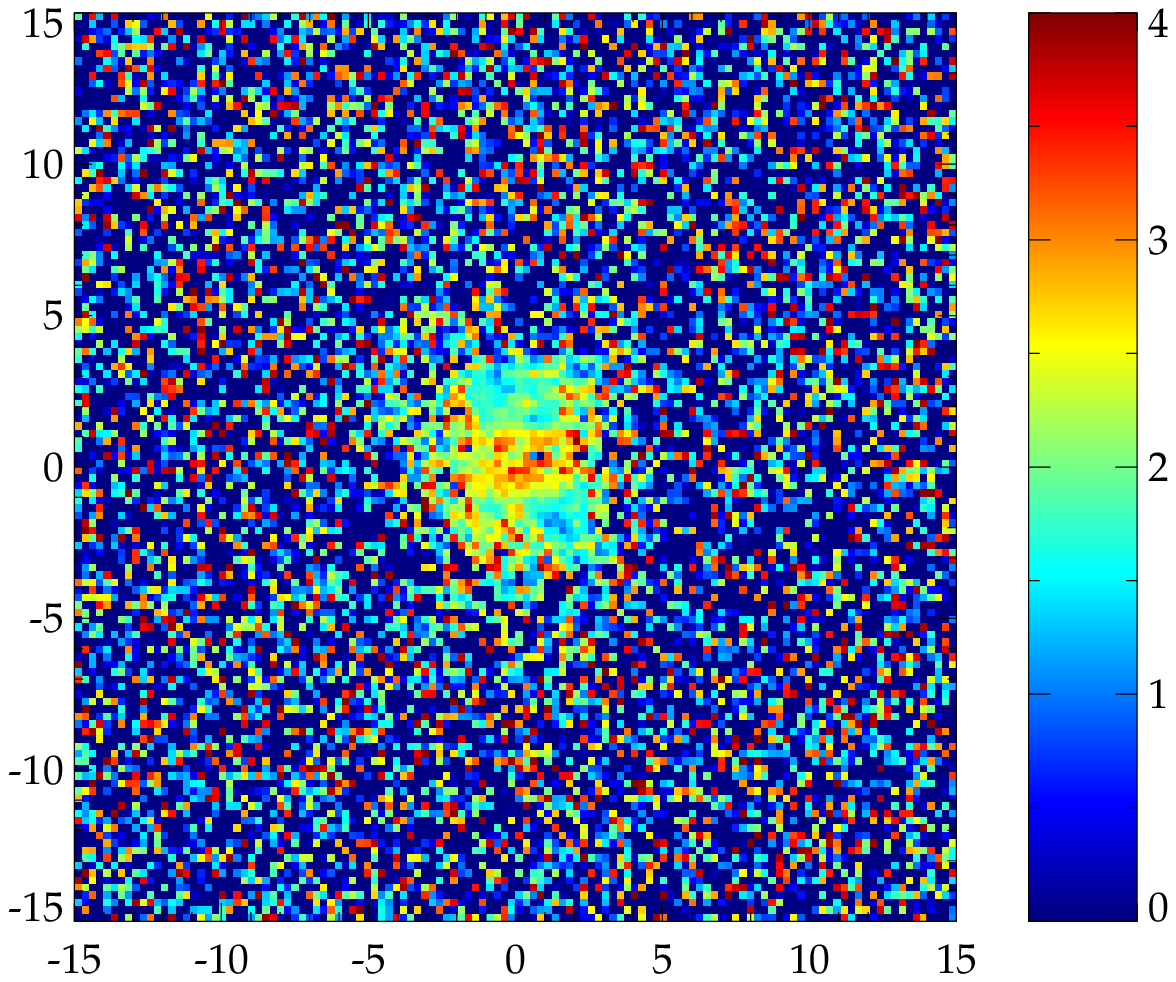}
\includegraphics[height=0.22\textwidth,clip]{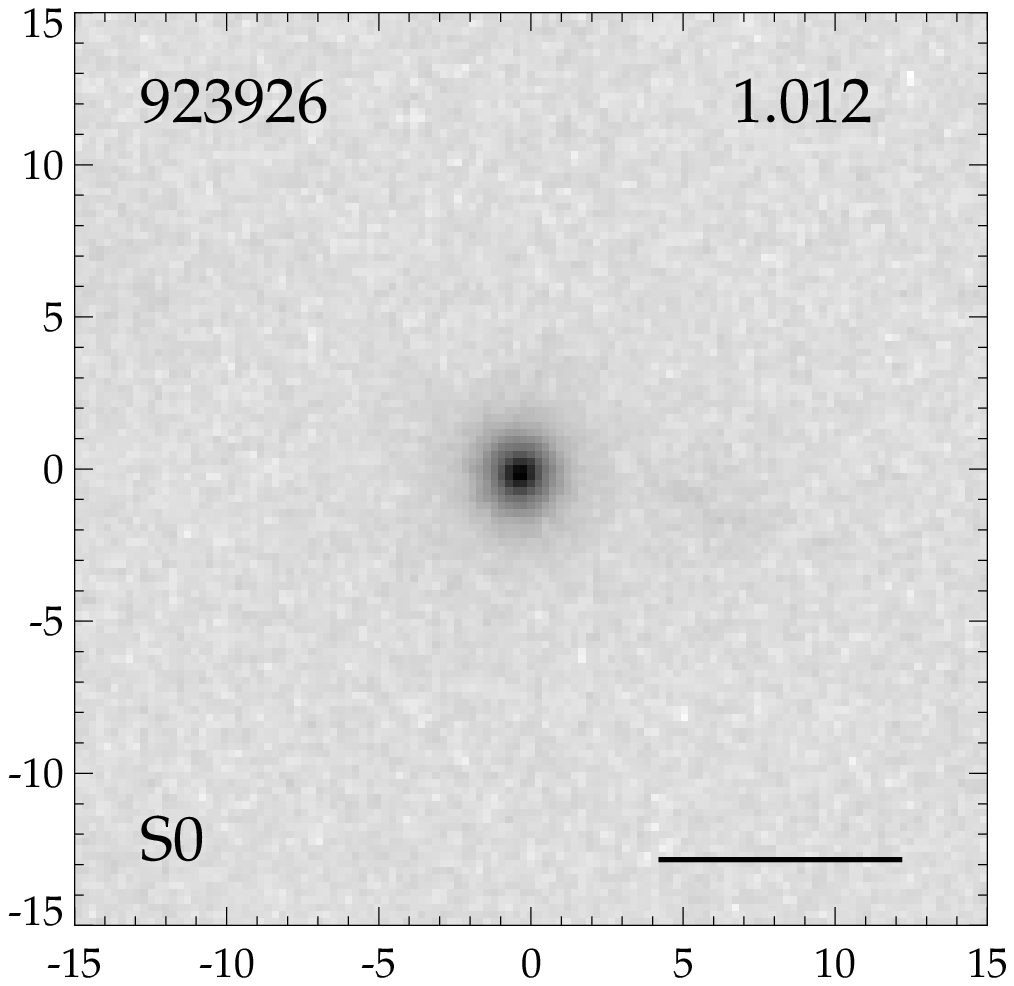} \includegraphics[height=0.22\textwidth,clip]{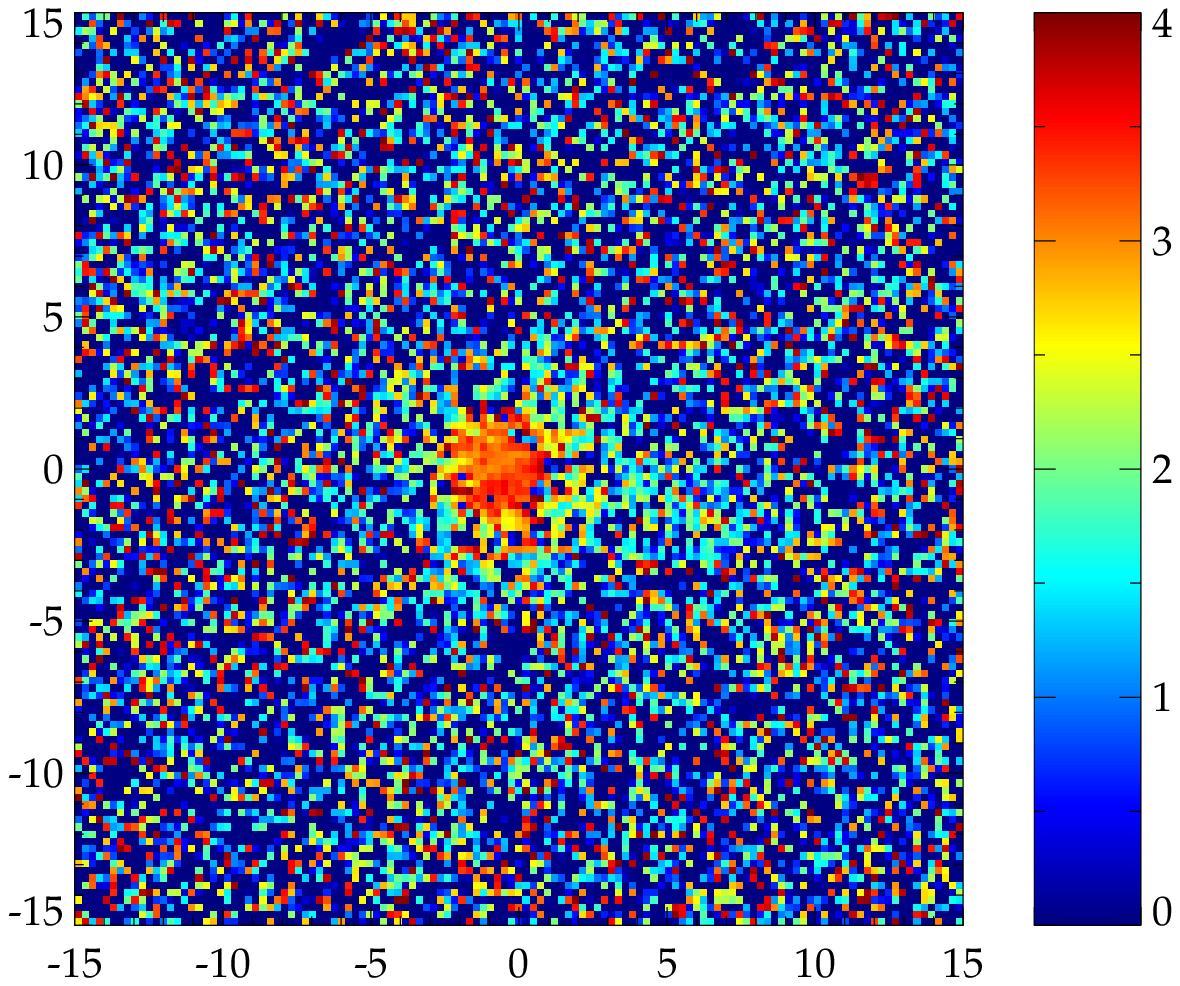}
\caption{Continued.} \end{figure*}

\addtocounter{figure}{-1}
\begin{figure*} \centering
\includegraphics[height=0.22\textwidth,clip]{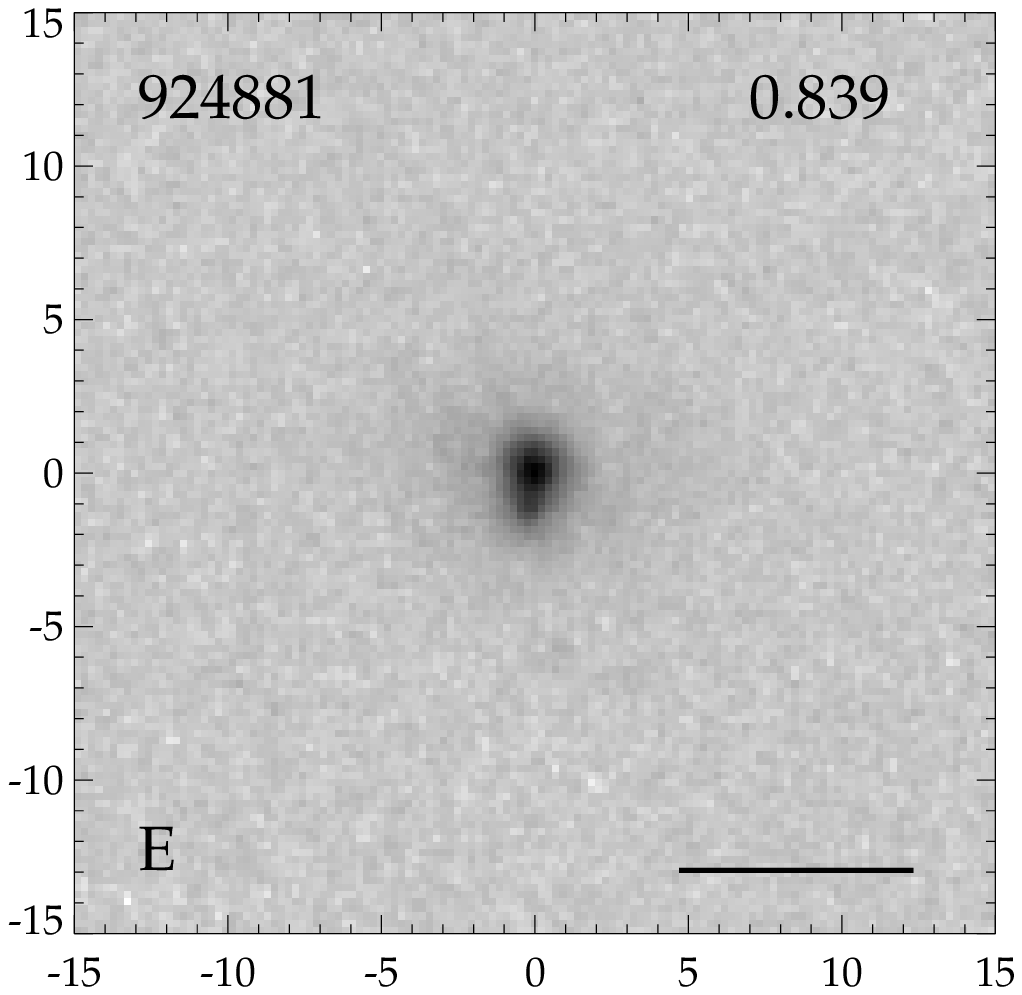} \includegraphics[height=0.22\textwidth,clip]{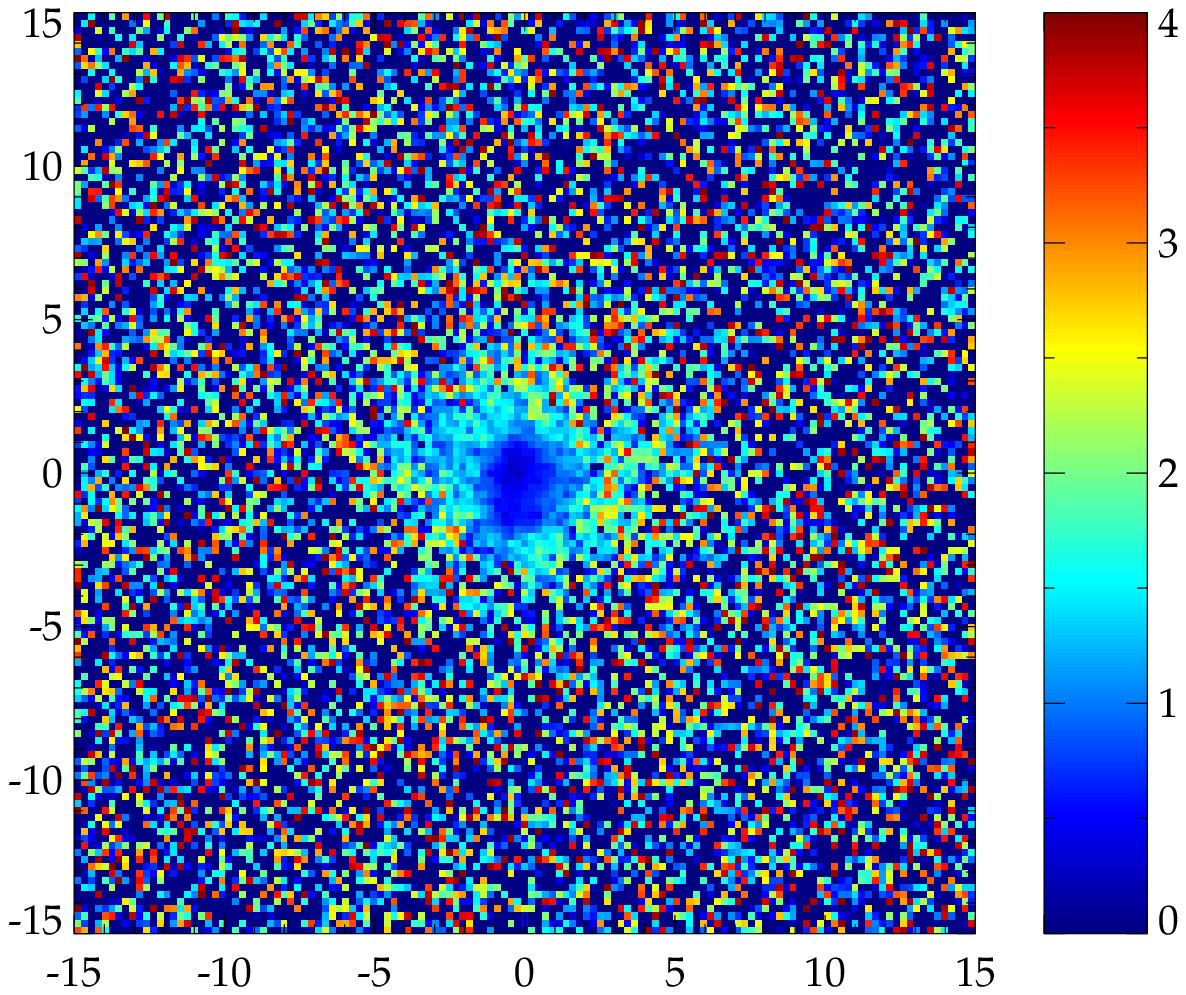}
\includegraphics[height=0.22\textwidth,clip]{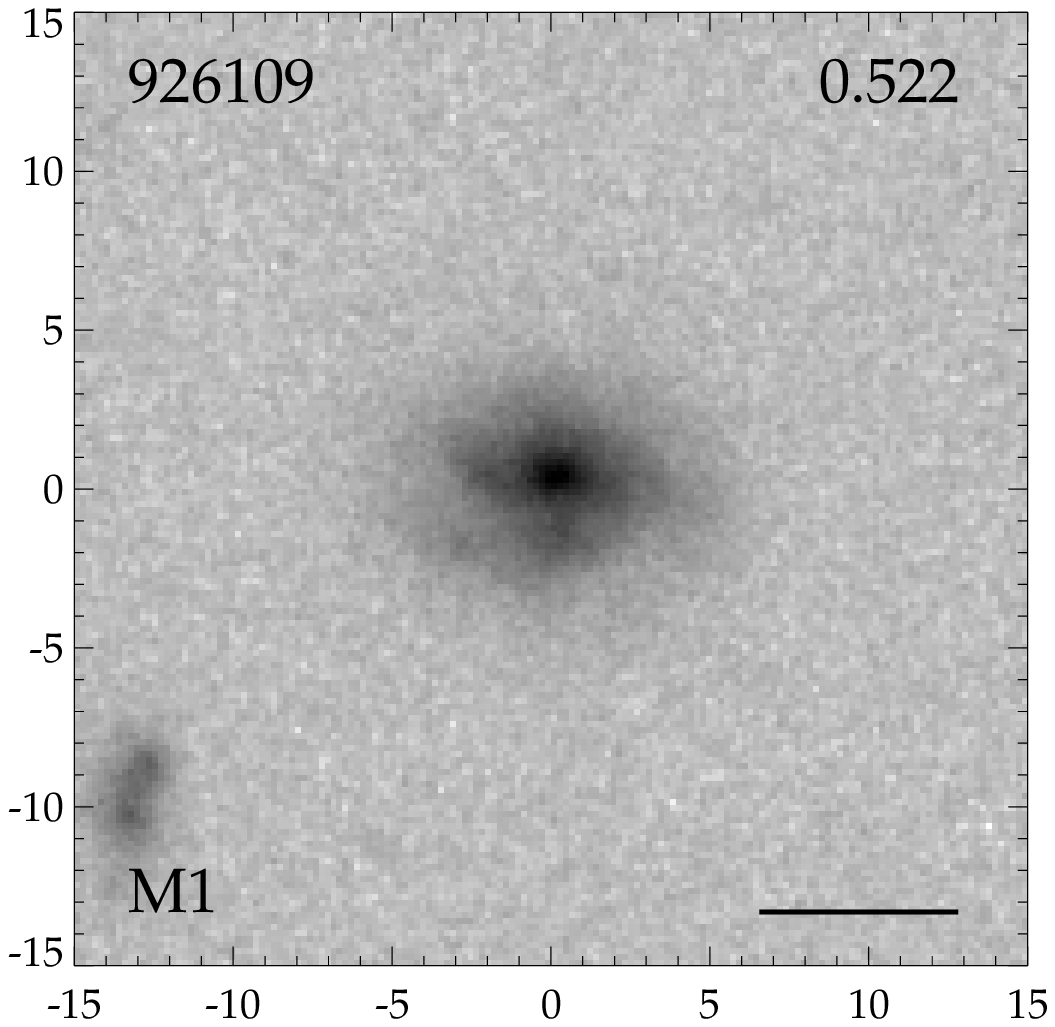} \includegraphics[height=0.22\textwidth,clip]{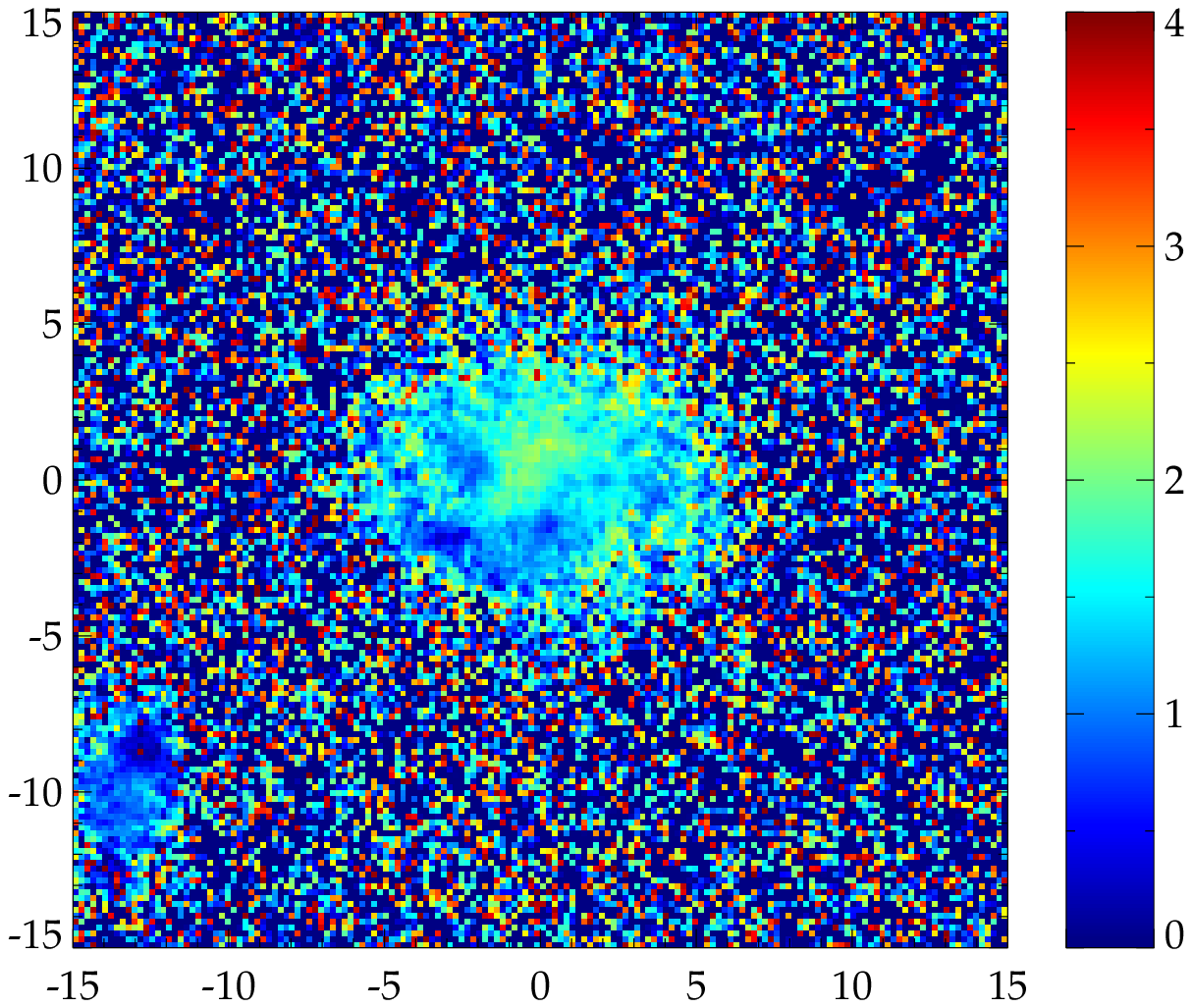}
\includegraphics[height=0.22\textwidth,clip]{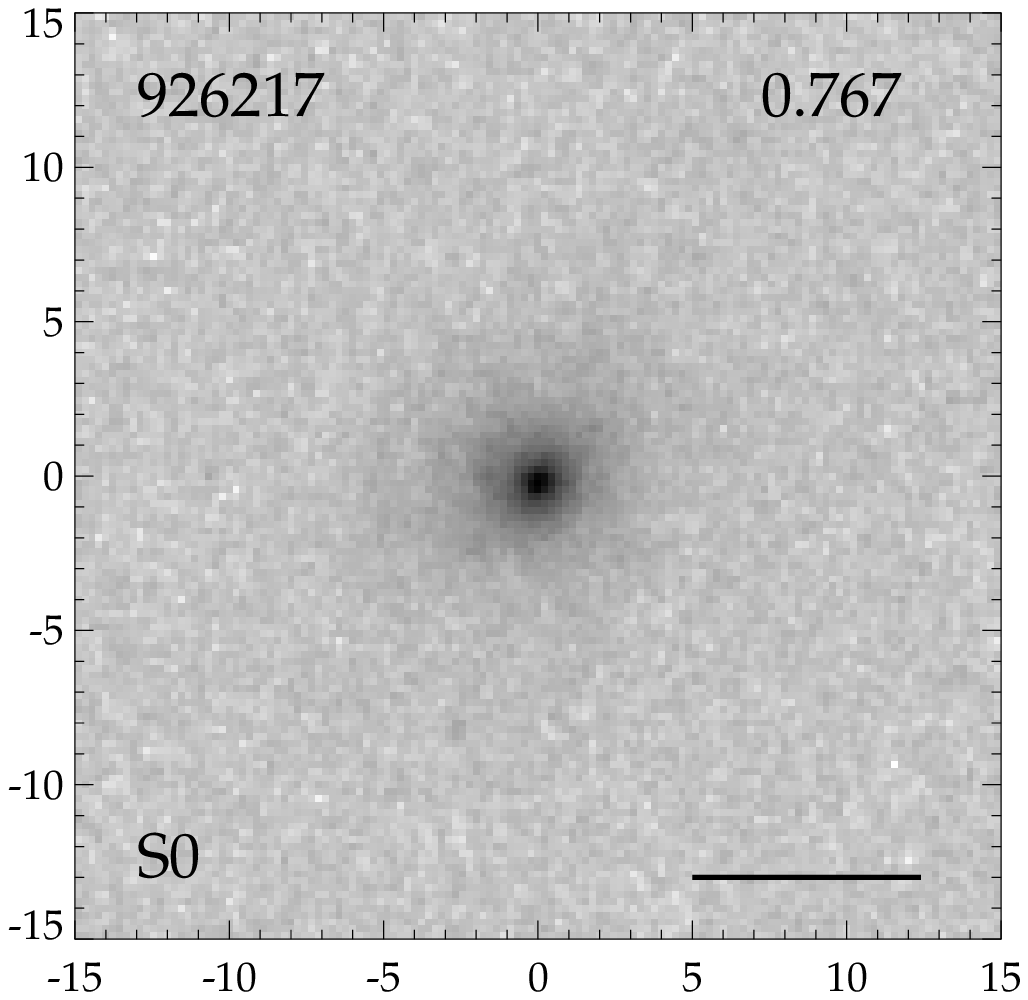} \includegraphics[height=0.22\textwidth,clip]{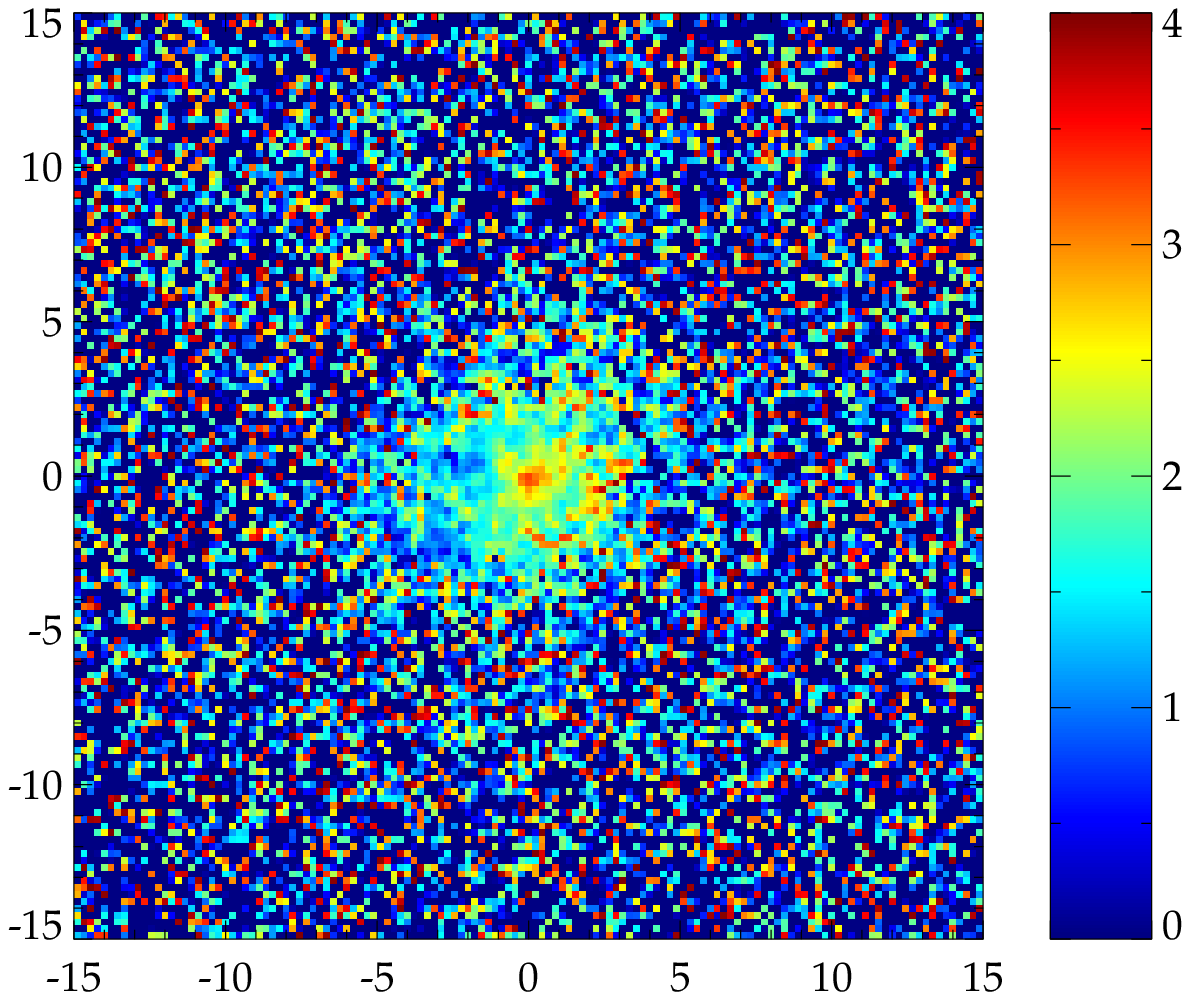} \hspace{0.475\textwidth}
\caption{Continued.
$F850LP$ and the B-z color map images. For each target, the name and redshift are labeled at the top-left and top-right in the $F850LP$ image. Our morphological classification is indicated at the bottom-left of each image. The size of each image is 30$\times$30\kpc~and the bar at the bottom-right of each image indicates 1.0\arcsec. North is up, East is to the left. The intensity scaling is the same (squareroot) in all the grayscale images, while the dynamic range is different for different objects owing to the fact that the max or central intensity can be different for different objects. The color bar shows the B-z color range from 0 to 4 except for object ID 904604, 907361 and 909015 where it ranges from 0 to 6. The blank in object ID 907305 is due to imaging at the edge of the GOODS CDFS region. }
\end{figure*}

\section{Number density evolution}
As has been widely reported, LCGs constitute a very rapidly evolving galaxy population in the intermediate redshift range (Phillips et al. \cite{phillips1997}, Werk et al. \cite{jessica2004}). In order to verify this, we derived a rough estimate of the comoving number density of LCGs in the redshift range $0.5\leq z \leq1.2 $. After correcting for spectroscopic incompleteness using published values of Target Sampling Rate (TSR) by Ilbert et al. (\cite{ilbert2004}), we get a value of $1.86 \times 10^{-3}\unit{Mpc^{-3}}$. Assuming a Poisson distribution yields a 1$\sigma$ error of $\sim$15\% for our sample of 39 galaxies. We must be careful in comparing this number density with other published work because of the different sample selection criterion employed by different groups for choosing LCGs, especially in the absolute magnitude cut off. With this caveat in mind, this number is comparable to the value reported by Phillips et al. (\cite{phillips1997}) as 1.2 $\times 10^{-3}\unit{Mpc^{-3}}$ in the redshift range $0.7\leq z \leq1.0$ with an absolute magnitude cutoff of $M_{B} \leq -20 $. 

In order to look for evolution, we must compare our number density with the known number density of such sources in the local universe. Werk et al. (\cite{jessica2004}) have published the number density of local luminous compact {\it{blue}} galaxies (LCBGs) to be $3.2 \times 10^{-4}\unit{Mpc^{-3}}$ for $\hub$ = $70\hubbleunit$   and an absolute magnitude cutoff of $M_{B} \leq -19.12$. If we restrict their sample to only objects with $M_B \leq -20$ to agree with our sample selection criterion, we get a number density of $1.4 \times 10^{-4}\unit{Mpc^{-3}}$, with the caveat that the Werk et al. sample suffers from small number statistics (only 5 objects) at such an absolute magnitude cutoff.

When we similarly restrict our sample to rest frame $B-V \leq 0.6$ to yield only blue LCGs in the redshift range $0.5 \leq z \leq 1.2$, we are left with 33 of our 39 objects. This yields a comoving number density of $1.57 \times 10^{-3}\unit{Mpc^{-3}}$. A comparison with the number density of LCBGs in the local universe yields a factor $\sim$11 fall in the comoving number density for blue LCGs from intermediate redshifts to the current epoch. The Poisson error on this number will be around $\sim50\%$ which is mainly dominated by the small sample size at redshift zero. With this caveat of large errorbars, we still get a constraint on the number density evolution of blue LCGs which is consistent with earlier published results.

\section{Discussion}

As reported by earlier workers in this field (Guzman et al.~\cite{guzman1997}, Hammer et al. \cite{hammer2001}), by their very nature LCGs are so compact that they yield little morphological information at intermediate redshifts, even using HST/WFPC2 images. Some noteworthy exceptions are the HST followup observations of CNELGs by Koo et al. (\cite{koo1994}), who found evidence for exponential light profiles for these objects. Also, Guzman et al. (\cite{guzman1998}) used HST/WFPC2 images of five CNELGs to reveal the presence of blue high surface brightness knots surrounded by a diffuse ``exponential like'' component. The knots were identified as the location of the current star formation and the diffuse exponential like component was interpreted to be an older underlying population. 

In this work, we have performed quantitative morphological analysis of a sample of 39 LCGs using full 2 dimensional surface brightness profile fitting of the galaxy images. The benefit that we have, which was not available to our predecessors, is the technological advancement offered in the shape of ACS on board the HST. One of the advantages offered by the ACS imaging is the improved pixel scale of 0.05\arcsec/pixel which is a factor of $\sim$2 better than the pixel scale of WFPC2, while at the same time having a larger footprint on the sky, which makes it ideal for large surveys. This better sampling becomes specially important for compact sources such as LCGs whose typical $r_{1/2} \leq$ 0.5\arcsec, in which case the whole extended emission (eg. disk) may be sampled by just a few pixels. This advantage of ACS has been fully leveraged by the HST/ACS GOODS survey which has the unique combination of depth, finer sampling of 0.03\arcsec/pixel and large area coverage which yields a large enough sample of LCGs to do meaningful statistical analysis. There have been other surveys which have some (but not all) of these three qualities, e.g. the Hubble Deep Field (Williams et al.~\cite{williams1996}) was deeper than GOODS but with an area coverage of only $\sim$5\% of GOODS. GEMS (Rix et al.~\cite{rix2004}) has larger area coverage than GOODS and the same pixel scale but is rather shallow (depth of single orbit). The HST/ACS Ultra Deep Field (Beckwith et al.~\cite{beckwith2006}) is much deeper than GOODS, but with area coverage which is much smaller (although see Noeske et al. (\cite{noeske2006}), who derive a reasonable sample size of LCBGs by using photometric redshifts and a less stringent absolute magnitude cutoff $M_B \leq -18.5$). Using the ratio of UDF area to GOODS area and our sample selection criteria for selecting LCGs, we would have got a sample of $\sim$3 LCGs in the HST/ACS UDF.  

In this study, we find that LCGs constitute a significant $\sim$26\% of the $M_{B} \leq -20$ galaxy population in the redshift range $0.5 \leq z \leq 1.2$. We derive SFR values ranging from a few to $\sim$ 65 $M_{\odot}$/year as expected for this class of objects. We establish that our sample LCGs are intermediate mass objects with stellar mass ranging from $9.44 \leq \log_{10}(M/M_{\odot}) \leq 10.96$, with a median mass of $\log_{10}(M/M_{\odot})=10.32$. We also obtain a factor $\sim$11 fall in the comoving number density of blue LCGs from this redshift range to the current epoch, but this number is subject to large uncertainities given the small sample size at zero redshift available from the literature. This implies that LCGs constitute one of the most rapidly evolving galaxy populations at intermediate redshifts. A plausible reason for the rapid decline in the number densities seems to be the extinguishing of the star formation and the associated fading, as pointed out by earlier workers (see Koo et al.~\cite{koo1995}, Guzman et al.~\cite{guzman1998}). The eventual quenching of the starburst might result from supernovae driven galactic winds which might remove any residual gas from the galaxy (Guzman et al.~\cite{guzman1996}). However, the nature of the descendants of these intermediate redshift LCGs will depend in a major way on the timing of this {\it{quenching}} and the possibility of subsequent bursts of star formation. It has been proposed (Koo et al.~\cite{koo1995}, Guzman et al.~\cite{guzman1996}) that CNELGs should fade within a few Gyr.~by as much as 4-7 magnitudes in the rest frame $M_B$ to resemble today's spheroidals. However, one must be careful in comparing the CNELGs with our LCGs. A simple look at the stellar masses derived by us for our sample of LCGs (Table~\ref{table}), make it clear why such a comparision is incompatible. CNELGs are mostly low mass systems, with $M\sim10^9 M_{\odot}$ (Guzman et al.~\cite{guzman1996}), whereas our LCGs are quite massive systems, with some of them having masses comparable to that of the Milky Way already in place by intermediate redshifts. Since they already have masses well in excess of that expected for dwarf galaxies, we can safely rule out the possibility of these LCGs evolving into dwarf spheroidals in the local universe. Given the stellar masses that these objects have at intermediate redshift, and the high star formation rates that we derive for these objects, they are clearly progenitors of intermediate mass objects in the local universe, out of which $\sim$70\% are spirals in the current epoch (Hammer et al.~\cite{hammer2005}).

We find a small but significant $\sim$22\% of the LCGs to be disk dominated systems. The median disk scale length for the disk dominated LCGs is found to be $\sim$2.3\kpc. Interestingly, the colormaps for some of these disk dominated LCGs (eg. 905983, 912744, 915400, 919595) show blue colors on the outer regions of the disk which might be indicative of inside-out disk formation as suggested by Hammer et al. (\cite{hammer2005}). However, equally interesting is the fact that none of these galaxies have a detection in Spitzer. This still allows for a SFR of several $M_{\odot}$/year (see Table~\ref{table}). However, the final fate of these LCGs will depend on when the star formation is quenched.  Given the stellar masses that these objects have already assembled by intermediate redshifts, these disk dominated LCGs are evidently the progenitors of intermediate mass disk galaxies in the local universe, as has been previously suggested by Phillips et al. (\cite{phillips1997}).  

Another relevant input which will have a strong influence on the final form of the descendants of these intermediate redshift LCGs, is the evolution with redshift of the merger rate of galaxies, regarding which there has been wide debate in the community. Parameterizing the fraction of galaxies existing in merger systems at any given redshift as $f(z)=f(0)(1+z)^\alpha$, the reported values of $\alpha$ vary from $0 \leq \alpha \leq 4$ (Le Fevre et al.~\cite{lefevre2000}, Patton et al.~\cite{patton2002}, Conselice et al.~\cite{conselice2003}, Lin et al.~\cite{lin2004}, Bundy et al.~\cite{bundy2004}, Bell et al.~\cite{bell2006}, Lotz et al.~\cite{lotz2006}). Even though the value of the slope $\alpha$ is still debatable, the merger rate estimates are in reasonable agreement at least around redshift 0.5-0.6 (see Bell et al.~\cite{bell2006}).
A large value of $\alpha$ would imply that the merger rate of galaxies was higher at higher redshifts. We have performed a calculation, along the lines of that performed by Le Fevre et al. (\cite{lefevre2000}), using the HST/ACS GOODS dataset and found a rapid evolution in pair fraction of galaxies with redshift out to z$\sim$1.2 (Rawat et al.~\cite{rawat2007}). Even though this debate is far from settled, our work seems to indicate that a large fraction of $L^{*}$ galaxies in the local universe must have undergone a major merger, possibly triggering a new burst of star formation within the last 8\gyr. In the present work we find that a large fraction ($\sim$36\%) of the LCGs are in merging systems, even though only about 1/3rd of these are obvious mergers (M1). These mergers have been identified essentially visually by the presence of multiple maxima of comparable intensity in the rest frame B band images. This was further aided by the use of color maps to distinguish multiple nuclei (generally redder) from HII regions (generally bluer). Our definition of merger is not directly comparable to those reported by other workers such as Lotz et al. (\cite{lotz2006}) who use  a quantitative definition to pick up merging systems. Despite this caveat, our obvious merger fraction (M1) ($\sim12\%$) is quite comparable to the typical merger fraction reported for the general field galaxies by Lotz et al. (\cite{lotz2006}), who find the fraction of galaxies with $M_B\leq -20.5$ that are in merging systems to be $\sim$7\% at intermediate redshifts. If we also include our possible merger systems, this is evidence in support of an unusually high merger rate for LCGs, and points towards a dynamic evolutionary history of LCGs in the intermediate redshift range. It is plausible that LCGs might represent a transient, starbursting (possibly merging) phase in the evolution of a galaxy. In that case, the observed decline in the number density of LCGs from z$\sim$1 to the current epoch can be naturally explained in terms of the decline in the merger rate of galaxies over the same redshift range (Rawat et al.~\cite{rawat2007}). This is in agreement with the scenario suggested by Hammer et al. (\cite{hammer2005}) in which the LCGs are just the merger phase in the hierarchical growth of a galaxy.

The eventual fate of the LCGs undergoing merger is still an open issue. Hammer et al. (\cite{hammer2005}) had suggested that there must be a mechanism at work that rebuilds the disk in a galaxy after a major merger event, perhaps fueled by infalling gas left over after the merger. They had argued that if the fate of every LCG is to end up as an early type galaxy, it will overproduce the number of early types in the local universe (see Table 2 in Hammer et al.~\cite{hammer2005}). In this scenario, we can envisage that some of the LCGs that are seen in merger phase will go on to form a disk as they evolve to z=0. Recent work on numerical simulations of galaxy mergers (Robertson et al. \cite{robertson2005}) indicates that gas rich mergers at high redshift can lead to the formation of rotationally supported disks in merger remnants if feedback mechanisms limit the conversion of gas into stars. This gas rich merger driven disk galaxy formation at high redshifts degenerates into merger driven elliptical galaxy formation at lower redshifts, as star formation depletes the available gas and makes gas rich mergers less likely.

Another support for this scenario comes from the new stringent constraints being put on the evolution of early type galaxies (ETGs) since $z\sim1$ by Cimatti et al. (\cite{cimatti2006}). They have proposed that at every redshift, there is a critical luminosity (or mass) threshold above which virtually all ETGs seem to be already in place. This is also known as the downsizing scenario. Now, given the fact that our sample LCGs are intermediate mass objects with a mass range $9.44 \leq Log_{10}(M/M_{\odot}) \leq 10.96$ (see Table~\ref{table}), this does not seem to favor their evolution into early types at the current epoch, simply because most of the ETGs in this mass range were already in place latest by intermediate redshifts (see Fig. 2 in Cimatti et al. \cite{cimatti2006}).       

Interestingly, Lotz et al. (\cite{lotz2006}) have claimed a large factor $\sim$2.5 increase in the number density of early types E/S0/Sa from redshift z$\sim$1.1 to z$\sim$0.3 for a volume limited sample of $M_B \leq -20.5$. However, a simple look at their Table 2 shows that most of their {\it{evolution}} comes from the highest redshift bin, with the lower redshift data consistent with little or no evolution out to a redshift of $\sim$1. 
Their result is in fact similar to the result published by Smith et al. (\cite{smith2005}), who find the fraction of early type galaxies to remain constant with $f_{E+S0}=0.4\pm0.1$ till $z \sim 1$ in the low density regions. 
This is also consistent with Cimatti et al.(\cite{cimatti2006}) result that most of the massive (luminous) ETGs are already in place by intermediate redshifts. Now, given the fact that LCGs constitute only $\sim$1/4 of the field galaxy population with $M_B\leq-20$, it is entirely possible that a large fraction of LCGs might still evolve into ETGs by the current epoch, without upsetting the constant $f_{E+S0}$ fraction constraint discussed above within their quoted errorbars. So, even though we cannot {\it{rule out}} an evolutionary scenario of intermediate redshift LCGs evolving into early types in the current epoch (given the uncertainties of the early type fraction evolution with redshift), the current results on ETG fraction evolution do not seem to encourage a large scale evolution into ETGs in this luminosity range, over the redshift range under consideration. This result will become stronger, as the errorbars on the $f_{E+S0}$ evolution with redshift improve with larger surveys becoming available.  

If we combine all the evidence listed above, we envisage that it is plausible that some of the LCGs that are undergoing mergers at intermediate redshifts might go on to form a disk from gas left over from the merger event, by the time they evolve to the current epoch.

Puech et al. (\cite{puech2006}) have recently reported on the morphological mix of a sample of 17 intermediate redshift LCGs using the technique of mapping the velocity field of a galaxy with the Integral Field Unit (IFU) spectrograph FLAMES/GIRAFFE on the VLT. They report a figure of $\sim$18\% galaxies being consistent with rotating disks compared to the $\sim$22\% in this work. Furthermore, they report that $\sim$82\% of the LCGs in their sample have perturbed or complex kinematics, with roughly half of them being probable mergers, a number which again agrees with our result designating $\sim$36\% of the LCGs as merging systems. The techniques applied by Puech et al. are entirely different from ours: \\a) They probe ionised gas, whereas we probe stellar mass, \\b) They use ground based spectroscopy while we use space based imaging. \\The fact that we get results that are comparable gives us increased confidence in our results.

Noeske et al. (\cite{noeske2006}) have recently reported on the morphology of LCGs from the Hubble Ultra Deep Field in the redshift range 0.2 to 1.3. They find typical scale lengths of the extended components to be $\leq$ 2\kpc, and conclude that these objects will evolve into small disk galaxies and low mass spheroidal or irregular galaxies in the local universe. However, their sample selection criteria are not fully compatible with those of Hammer et al. (\cite{hammer2001}), which we have used in this work, and which might favor progenitors of larger local galaxies. We find that the median disk scale length for the disk dominated LCGs in our work is $\sim$2.3\kpc~(see Table~\ref{table}). These LCGs are intermediate mass objects and given the limits on their SFR of a few $M_{\odot}$/year, are likely to be the progenitors of intermediate mass disk galaxies in the local universe.

\section{Conclusions}
An effort was made to obtain quantitative morphological parameters for a representative sample of 39 intermediate redshift ($0.5 \leq z \leq 1.2$) luminous compact galaxies (LCGs) and to classify them based on their rest frame B band morphology and B-z color maps using multiband imaging data from the HST/ACS GOODS survey. The main result that we obtain for their morphological mix is as follows: Mergers: $\sim$36\%, Disk dominated: $\sim$22\%, S0: $\sim$20\%, Early types: $\sim$7\%, Irr/tadpole: $\sim$15\%.

We find that LCGs constitute $\sim$26\% of the $M_{B} \leq -20$ galaxy population in the redshift range $0.5 \leq z \leq 1.2$. We also obtain a factor $\sim$11 fall in the comoving number density of blue LCGs from the redshift range $0.5 \leq z \leq 1.2$ to the current epoch, using published values for the comoving number density of blue LCGs in the local universe (Werk et al. \cite{jessica2004}). We establish that our sample LCGs are intermediate mass objects with stellar mass ranging from $9.44 \leq \log_{10}(M/M_{\odot}) \leq 10.96$, with a median mass of $\log_{10}(M/M_{\odot})=10.32$. We also derive SFR values ranging from a few to $\sim$ 65 $M_{\odot}$/year as expected for this class of objects.

We find that the median disk scale length for the disk dominated LCGs in our work is $\sim$2.3\kpc. Some of the disk dominated LCGs show blue colors on the outer regions of the disks which might be indicative of inside-out disk formation as suggested by Hammer et al. (\cite{hammer2005}). 
These disk dominated LCGs are intermediate mass objects, and given the star formation rates that we derive for these objects, they are clearly progenitors of intermediate mass disk galaxies in the local universe.

We also establish that we cannot {\it{rule out}} an evolutionary scenario comprising of intermediate redshift LCGs evolving into early types in the current epoch, given the typical errobars on the early type fraction evolution available from literature. Despite this, we note that the current results on ETG fraction evolution, do not seem to encourage a large scale evolution into ETGs in this luminosity range, over the redshift range under consideration.
We envisage that some of the LCGs that are classified as merging systems, might go on to rebuild their disks and evolve into disk galaxies in the local universe.

\begin{acknowledgements}

A.R. would like to thank CSIR for PhD. funding, and the French embassy in India for providing financial assistance in the form of {\it Sandwich Thesis Scholarship 2004}. We would also like to acknowledge the Centre Franco-Indien pour la Promotion de la Recherche Avancee (CEFIPRA) for financial assistance under project number 2804-1. We would like to thank the anonymous referee for his/her insightful comments and suggestions which have gone a long way in enriching this work. Special thanks to Harry Ferguson for kindly providing the GOODS completeness curves, to Mathieu Puech for kind help with his IDL codes and to X. Z. Zheng and Swara Ravindranath for helpful discussions. 

\end{acknowledgements}

\begin{figure}[h]
\centering
\includegraphics[width=0.5\textwidth]{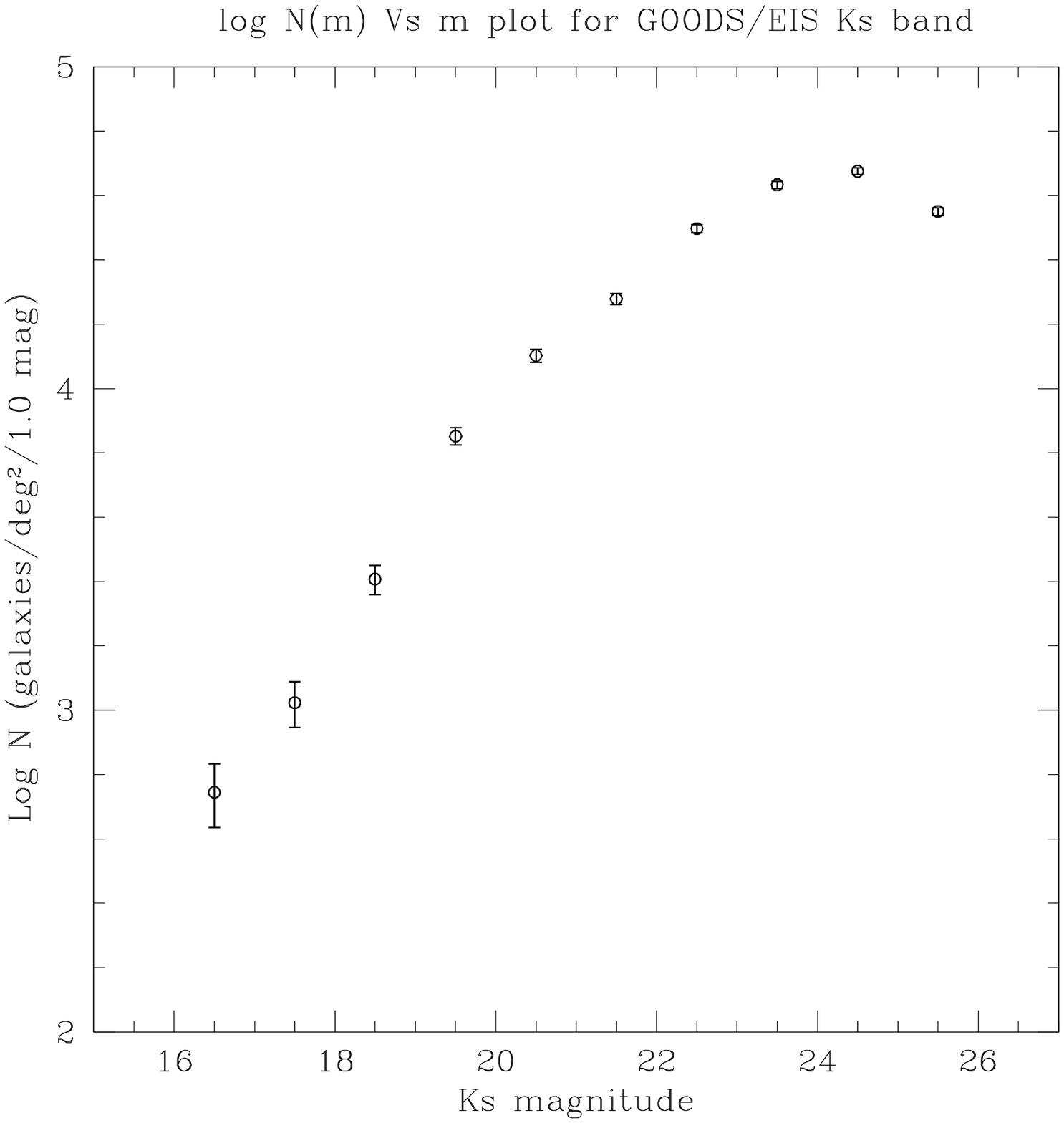}
\caption[]
{$K_s$ band number counts for the VLT GOODS/EIS}
\label{numbercounts}
\end{figure}

\begin{center}
{\bf\large Appendix A: Determination of absolute magnitudes} 
\end{center}
We have used the technique of fitting template galaxy spectrum to a set of photometric datapoints for each galaxy to obtain the spectral energy distribution (SED) of the object, which can then be used for deriving the K-correction. Obviously, as in the case of any fitting procedure, the greater the number of photometric datapoints, the better we will be able to constrain the galaxy SED. Hence, in addition to using the four band HST/ACS photometric catalog for the GOODS/CDFS sources, we obtain two more photometric datapoints using the J \& $K_s$ imaging data from the ESO GOODS/EIS release. SExtractor based photometry for the J \& $K_s$ fields was performed by us using the photometric zeropoints provided as part of the GOODS/EIS release: v1.0. The photometric catalogs for the J and $K_s$ band in this field have $\sim$8000 sources each. The differential number counts for sources in the $K_s$ band is given in Fig. \ref{numbercounts}. 

In the absence of detailed completeness curves for the $K_s$ band source catalog, the differential number counts can be used as a useful indicator for the completeness of the $K_s$ band catalog. As we can see from Fig. \ref{numbercounts} the number counts are steadily rising till $K_s$ $\sim$24 magnitude, and it starts to taper off for fainter magnitudes. Hence we can assume that the $K_s$ band catalog suffers from incompleteness for sources fainter than $K_s$ $\sim$24 magnitude. Since in our analysis work, we are only interested in galaxies having spectroscopic redshifts from the VVDS survey, all of which have $I_{AB} \leq 24$, we will not miss any object in the $K_s$ band, unless it has an extremely unusually blue color of $(I-K_s) \leq$ 0.

Having compiled the list of all the objects in the CDFS having photometry in 6 bands plus spectroscopic redshifts from the VVDS as mentioned above, we derived the absolute magnitudes for those objects as follows.

The transformation equation relating {\it observed frame} photometry with {\it rest frame} absolute magnitudes is...
\begin{equation}
~~~~~~~~~~~~~~~~~~~~~~~ m_{B} = M_{B} + DM(z) + k_{B}(z) + A_{B} 
\end{equation}
where, \\
$m_{B}$         : apparent magnitude in B band\\
$M_{B}$         : absolute magnitude in B band\\
$DM(z)$         : distance modulus\\
$k_{B}(z)$      : the K-correction in B band\\
$A_{B}$         : Galactic extinction correction in B band

For the purpose of fitting template galaxy spectra, we used a code developed by Hammer et al. (\cite{hammer2001}). To derive absolute magnitude at a given {\it{rest-frame}} wavelength $\lambda_{rest}$, we consider two {\it{observed}} frame wavelengths $\lambda_{obs1}$ \& $\lambda_{obs2}$, such that $\lambda_{obs1}/(1+z)$ \& $\lambda_{obs2}/(1+z)$ encloses the rest-frame $\lambda_{rest}$, where z is the redshift of the object.
To this set of galaxy observed frame magnitudes, we fit a galaxy spectral template derived from stellar population synthesis models (Bruzual and Charlot~\cite{bruzual1993};\cite{bruzual2003}). For our fitting purpose, we used an exponentially decreasing SFR model with e-folding timescale of $\tau=0.5\gyr$. We then vary the age of this model to optimize the fit to observed frame magnitudes. The template has been optimized to minimize the residuals between the actual galaxy magnitudes and the galaxy magnitudes reconstructed from the galaxy SED fit. We then use an interpolation to calculate the flux at the given rest-frame $\lambda_{rest}$.
This code then uses a specified cosmology to calculate the distance modulus and output the absolute magnitudes directly. We used a flat $\Lambda$CDM cosmology with $\Omega_{m}$=0.3 and $\Omega_{\Lambda}$=0.7 and $H_{0}=70\hubbleunit$.

We used the Galactic extinction maps published by Schlegel et al. (\cite{schlegel1998}) to estimate the Galactic extinction in each of the four HST/ACS photometric bands for the CDFS. Since the CDFS is a high Galactic latitude field, we found that the extinction values were quite small; being of the order of photometric accuracy of our catalog i.e. $\sim$1-2\%. Hence we decided not to use these extinction values for our final estimation of the absolute magnitudes.    \\

\begin{center}
{\bf\large Appendix B: {\it{Red Halo}} in the HST/ACS F850LP PSF} 
\end{center}
It has been well documented and reported that at redder wavelengths, all HST/ACS CCD detectors suffer from an extended halo due to scattered light (Sirianni et al.~\cite{sirianni2005}, Gilliland and Riess~\cite{gilliland2002}). This effect is particularly severe in the F850LP filter and depends upon the color of the PSF star being observed. This happens due to the fact that the CCD becomes more transparent to photons of longer wavelength. This means that for a redder star, the scattering of light into the halo is more severe compared to a bluer star. This variation of the PSF with the color of the PSF star can be quite large and can potentially affect the structural parameters that we derive for our sample galaxies. 

\begin{figure}[h]
\centering
\includegraphics[width=0.5\textwidth]{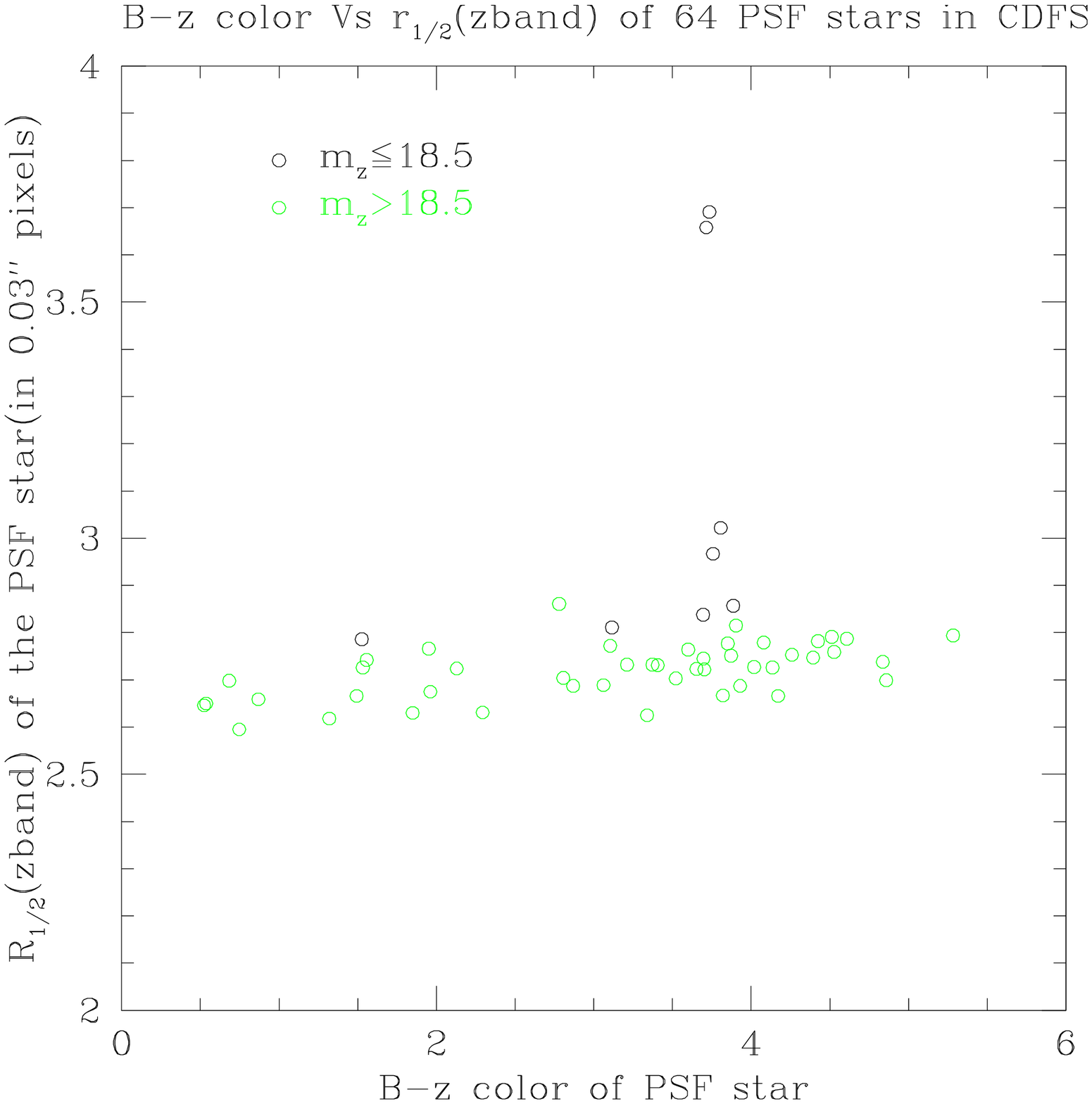}
\caption[]
{Variation of half light radius $R_{1/2}$ of the PSF stars (in the F850LP filter) used by us against the B-z color of the PSF stars. Different symbols have been used for faint ($m_z \geq 18.5$) and bright ($m_z \leq 18.5$) stars using an arbitrary definition of what constitutes a {\it{bright}} star. Notice that at a given color, the brighter stars tend to have larger $R_{1/2}$ (even though there is a considerable amount of scatter).}
\label{halo.color}
\end{figure}

In order to check for this possibility, we plotted the F850LP filter half light radius $R_{1/2}$ of the PSF stars used by us against the B-z color of the PSF star. This is shown in Fig.~\ref{halo.color}. Different symbols have been used for faint ($m_z \geq 18.5$) and bright ($m_z \leq 18.5$) stars using an arbitrary definition of what constitutes a {\it{bright}} star.

Notice in Fig.~\ref{halo.color} that at a given color, the brighter stars tend to have larger $R_{1/2}$. Also notice the small gradient towards increasing $R_{1/2}$ for redder stars. However, it is seen that even though there is a general trend for redder stars to have slightly larger $R_{1/2}$, this effect is $\leq$ 5\%. The only catastrophically large $R_{1/2}$ is seen in relatively bright stars. From Fig.~\ref{halo.color} we see that only four stars in our whole sample show this anomolous behaviour. We also find that all of the ``peaks'' that we had seen in Fig.~\ref{psf} belong to these same red and bright PSF stars. Once these four rogue stars were identified, there were removed from our list of PSF stars and were not used in any of our analysis.

\end{document}